# Temperature-dependent structure of 1-propanol/water mixtures: X-ray diffraction experiments and computer simulations at low and high alcohol contents

Ildikó Pethes[1,a], László Pusztai[a,b], Koji Ohara[c], László Temleitner[a]

[a]Wigner Research Centre for Physics, Konkoly Thege út 29-33., H-1121 Budapest, Hungary

[b]International Research Organization for Advanced Science and Technology (IROAST), Kumamoto University, 2-39-1 Kurokami, Chuo-ku, Kumamoto 860-8555, Japan

[c]Diffraction and Scattering Division, Japan Synchrotron Radiation Research Institute (JASRI/Spring-8), 1-1-1 Kouto, Sayo-cho, Sayo-gun, Hyogo 679-5198, Japan

**Abstract**

Aqueous mixtures of 1-propanol have been investigated by high-energy synchrotron X-ray diffraction upon cooling. X-ray weighted total scattering structure factors of 6 mixtures, from 8 mol% to 89 mol% alcohol content, as well as that of pure 1-propanol are reported from room temperature down to the freezing points of the liquids. Molecular dynamics simulations have been performed, in order to interpret measured data. The all atom OPLS-AA potential model was used for 1-propanol, combined with both the SPC/E and the TIP4P/2005 water models: both combinations provide a semi-quantitative description of the measured total structure factors at low and high alcohol contents, while the agreement is qualitative for the mixture with 71 mol% of 1-propanol. From the simulated particle configurations, partial radial distribution functions were calculated. Furthermore, detailed description of the hydrogen bonded network is provided, in terms of hydrogen bond numbers, analysis of proton donor-acceptor ratios, size distributions of hydrogen bonded clusters and ring size statistics. Strong temperature dependence of the percolation threshold, as well as of the participation of the number of doubly hydrogen bonded molecules in cyclic entities, has been found for the mixture with 89 mol% of 1-propanol. Above an alcohol content of 20 mol%, 5-fold rings are the most frequent cyclic entities, with a strong temperature dependence in terms of the number of rings.

---

[1] Corresponding author: e-mail: pethes.ildiko@wigner.hu

## 1. Introduction

The structure of aqueous alcohol mixtures is still an extensively studied topic nowadays. In addition to the great importance of these liquids in the industry, their interesting properties make them popular among researchers. Short chain alcohols (less than 4 carbon atoms in the alkyl chain, namely methanol, ethanol, 1-propanol and 2-propanol) are miscible with water over the entire concentration range at room temperature. Although their aqueous solutions do not show macroscopic phase separation, local anisotropy, micro-heterogeneities (aggregates of water or alcohol molecules) may be present (see e.g. [1, 2]). Characteristic changes of thermodynamic and dielectric properties with alcohol concentration (e.g. [3-5, and refs therein]) suggest a structural origin, e.g. the variation of hydrogen bonded (HB) aggregates formed in the mixture.

The longest *n*-alcohol that is miscible with water over the whole concentration range is *n*-propanol (1-propanol, *n*-propyl alcohol, propan-1-ol). The concentration dependence of several physicochemical properties of 1-propanol – water mixtures indicates different microstructures at low, medium and high alcohol concentrations. Data have been reported for excess activation free energy, enthalpy, entropy [3], excess heat of mixing, excess density, excess viscosity [5, and references therein], speed of sound, isentropic compressibility [6], viscosity, self-diffusion coefficient, spin-spin relaxation rate [4], spin-lattice relaxation rate [7] and terahertz spectroscopy measurements [8]. The structure of 1-propanol – water mixtures at room temperature was investigated by small angle neutron- [9] and X-ray scattering [10], wide-angle X-ray scattering [11], X-ray Raman scattering [12], near-infrared absorption spectra [13] and X-ray diffraction [14]. Besides experiments, several molecular dynamics studies have been published in the last two decades, e. g. [5, 14-19].

According to the above studies, at low alcohol concentrations the tetrahedral structure of water is dominant, (isolated) alcohol molecules are surrounded by water molecules (large cages are constructed by hydrophobic hydration). As alcohol concentration increases the hydration shells overlap, and alcohol molecules start to aggregate. In the medium concentration range (about 0.1 – 0.7 alcohol mole fraction) 1-propanol chains and water clusters are suggested to coexist within the extended alcohol – water network. At high alcohol concentrations, 1-propanol chains dominate, their structure being similar to that of the pure alcohol, and water molecules are dispersed in the mixture.

The evolution of the structure of alcohol – water mixtures with decreasing temperature has been examined in methanol – water, ethanol – water and 2-propanol – water mixtures [20-28]. An increase of the number of small hydrogen-bonded (H-bonded) rings upon cooling was reported in water-rich methanol – water [22], ethanol – water [23] and 2-propanol – water [25] mixtures. Enhanced micro-segregation between methanol and water clusters was suggested on the basis of neutron diffraction data on methanol – water [21] and ethanol – water [24] mixtures. Enhancing mixing was found at medium concentrations in methanol – water mixtures [26]. Percolating of the H-bonded network was reported up to higher alcohol concentrations as the temperature decreases in 2-propanol – water [27] and ethanol – water [28] mixtures. However, on the temperature dependence of the structure of 1-propanol – water mixtures, according to our best of knowledge, only a small angle neutron scattering experiment can be found in the literature [9]. In that study a mixture with 0.114 mole fraction of 1-propanol was measured between 278 – 298 K, and an increased level of aggregation with decreasing temperature was found.

By the present study, we wish to complete the series of temperature-dependent X-ray diffraction experiments of our group on short chain alcohol-water mixtures, by investigating 1-propanol – water mixtures, from room temperature down to the freezing points of the liquids. For an interpretation of the measured data, molecular dynamics simulations are also performed. Following a thorough comparison between measured and simulated total scattering structure factors (TSSF), the structure of the mixtures will be described based on the simulation results. Partial radial distribution functions, H-bond distributions and properties of the H-bonded network are examined and reported.

## 2. Experimental details

*n*-propyl alcohol (1-propanol) was purchased from Sigma-Aldrich Co. Apart from pure 1-propanol, 6 different 1-propanol – water mixtures have been prepared, with 8, 20, 30, 40, 71 and 89 mol% propanol contents. Nominal and exact compositions are collected in Table 1.

X-ray diffraction experiments have been performed at the BL04B2 [29] high-energy X-ray diffraction beamline of the Japan Synchrotron Radiation Research Institute (SPring-8, Hyogo, Japan). Photon energy was 61.117 keV, which corresponds to a wavelength 0.2029 Å. The experimental setup, as well as the data evaluation process, were identical to what have recently been reported for 2-propanol – water mixtures [27]. Each measurement was performed after an equilibration time (15 to 30 min) at each temperature, as shown in Table 1 and Fig. 1.

### 3. Molecular dynamics simulations

Classical molecular dynamics (MD) simulations were performed by the GROMACS software package (version 2018.2) [31]. The all atom OPLS-AA [32] model was used for 1-propanol, and the SPC/E [33] and TIP4P/2005 [34] models were taken for water molecules. Cubic simulation boxes contained 4000 molecules. Initial configurations were obtained by placing the molecules into the simulation box randomly. Simulation box sizes were chosen according to the room temperature density from Ref. [35]. At first, energy minimization was performed, using the steepest-descent method, after that the equations of motion were integrated via the leapfrog algorithm, applying a 0.5 fs time step. Following a 2.2 ns 'heat-treatment' step at 320 K, to avoid the aggregation of molecules, the system was cooled down to the investigated temperature ($T_{inv}$) with a 20 K/ns cooling rate, using the 'simulated annealing' option of the GROMACS software. 10 ns long NpT simulations were carried out at all investigated temperature for equilibration and to determine the density. Finally, 20 ns long NVT (production) runs were performed, during which trajectories were saved (in every 2 ps). 1001 configurations (20 ps apart) were used for calculating partial radial distribution functions (PRDF, $g_{ij}(r)$), by the 'gmx_rdf' program of the GROMACS software, from which functions the total scattering structure factor were composed. More details about the MD simulations can be found in the Supplementary Material (SM). Simulated temperature/concentration points are marked in Fig. 1. Simulated densities are collected in Tables S4 and S5, and are shown in Fig. S2.

The self-diffusion coefficients of water and 1-propanol molecules were calculated from the trajectories by calculating the mean-square displacements (MSD) of the centres of mass of the molecules and using the Einstein-relationship (Eq. 1):

$$6 D_A t = \lim_{t \to \infty} \langle \| r_i(t) - r_i(0) \|^2 \rangle_{i \in A} \quad . \tag{1}$$

Here *A* refers to either 1-propanol or water molecules, the averaging is performed over all molecules of the given species. 20 ns long trajectories were used by restarting the MSD-calculation every 200 ps. The MSD-*t* curves, without the first and last 10%, were fitted for the determination of $D_A$. The 'gmx msd' program of the GROMACS package was applied for the calculations. The statistical uncertainties of the values of the diffusion constants are usually around 5%.

Every 100[th] (altogether 101) configurations were used for scrutinizing H-bonding properties. H-bonds were defined using a geometric definition: two molecules were identified as H-bonded if the intermolecular distance between an oxygen and a neighboring hydrogen atom is less than 2.5 Å, and the O...O-% in-house software, based on the HBTOPOLOGY code [36, 37].

## 4. Results and discussion

The temperature/concentration points considered are shown in Fig. 1. Some of the measurement points fall below the solid-liquid coexistence curve of Ref. [30]; however, in the X-ray diffraction (XRD) patterns, no evidence for the presence of crystalline phase was found. These concentration/temperature points may be assigned to a supercooled liquid (or perhaps an amorphous solid) phase. In Fig. 1, temperatures at which Bragg-peaks appeared are also marked. For the 1-propanol – water mixtures with 89 mol% alcohol this temperature value is above the solid-liquid curve of Manakov et al. [30]: this finding challenges somewhat the accuracy of the literature data just mentioned [30].

*4.1. Validation of the interatomic potentials used against experimental densities and self-diffusion coefficients*

Before going into a detailed discussion of structural properties, we compare simulation data to two frequently used quantities: density and self-diffusion coefficient. Unfortunately, no comprehensive temperature dependent experimental studies have been performed on 1-propanol – water mixtures. We therefore compare simulation results for the temperatures where there are measured data, and then make an attempt to rationalize simulated densities and diffusion coefficients (and in particular, trends observable in their behavior).

The concentration dependence of the simulated densities and self-diffusion coefficients are shown in Figs. 2 and 3. Available literature data are also shown. Simulated density values are in excellent agreement with experimental ones at 298 K, and the 273 K simulated values are within 2% of the ones measured at 278 K. Concerning densities, there is no significant difference between two water models considered.

The self-diffusion coefficient of 1-propanol molecules at 298 K follows the experimentally observed tendency: it has a minimum around 20 – 30 mol% 1-propanol concentration. Individual values are also reproduced pretty well, particularly in the water-rich region. The self-diffusion coefficient of water molecules ($D_W$) in the experiments decreases continuously with increasing 1-propanol content. On the other hand, simulations predict a weak upward turn of $D_W$ in the alcohol-rich region. Individual values are reproduced satisfactorily in the water-rich region. What really is noteworthy is that in terms of the diffusion coefficients, the TIP4P/2005 water potential clearly outperforms the SPC/E one.

The temperature dependencies of the simulated self-diffusion coefficients are also shown in Figs S3 and S4. Except for pure 1-propanol at temperatures $T \leq 180$ K the self-diffusion coefficients are more than 4% of the room temperature values, indicating that molecules are mobile enough to consider the system as liquid. For pure 1-propanol at temperatures $T \leq 180$ K the self-diffusion coefficients are very low (less than 0.1% of the room temperature value): the molecules barely move, and the system is most probably in an amorphous solid phase. (As in this study we intend to focus on the liquid state, thus we didn't analyze the H-bond distributions at $T \leq 180$ K.)

In summary, concerning densities the SPC/E water force field seems to perform very slightly better than TIP4P/2005 one, while the latter does much better (by many tens of percents) in terms of the self-diffusion coefficients. For this reason, we believe that the TIP4P/2005 water potential provides more reliable results at this stage.

*4.2. Total scattering structure factors*

Experimental XRD total scattering structure factors are shown at three selected temperatures for $x_P = 0.2$ in Fig. 4. The total set of the measured $F(Q)$ functions are presented in the SM, Figs. S5-S18.

Four regions may be distinguished for investigating temperature and concentration dependence:

a) below 1 Å$^{-1}$

b) the first peak region, around 1.5 Å$^{-1}$

c) the second peak region, around 2.8 Å$^{-1}$

d) a broad third peak region, around 5 – 5.5 Å$^{-1}$.

At higher $Q$ values additional maxima can be observed, mostly at low alcohol concentrations, which peaks flatten and smear as alcohol concentration increases. Their position and height is difficult to determine except for the 8 mol% 1-propanol content sample. Moreover, experimental uncertainties increase with $Q$. For these reasons, these peaks will not be examined in more detail.

a) At the lowest alcohol concentration, $F(Q)$ flat below 1 Å$^{-1}$. As 1-propanol concentration increases, first some extra intensity appears (below 0.5 Å$^{-1}$ at $x_P = 0.2$) that evolves into a pre-peak for the alcohol-rich samples. Considering that measured TSSF-s behave reasonably both below $x_P = 0.2$ and above $x_P = 0.4$, we believe that the 'small angle scattering'-like intensities are genuine in the concentration region in between. The position of the pre-peak shifts toward higher $Q$ values as alcohol concentration increases and/or temperature decreases. The height of this small pre-peak increases with increasing alcohol concentration and decreases with decreasing temperature.

b) The amplitude of the first (or main) peak increases with increasing alcohol concentration and decreasing temperature. The position of the peak shifts toward smaller $Q$ values at higher alcohol contents and moves toward higher $Q$ values at lower temperatures.

c) The position of the 2$^{nd}$ peak changes similarly to that of the 1$^{st}$ maximum. The amplitude of the 2$^{nd}$ peak decreases as the alcohol content increases, and increases with decreasing temperature. In pure 1-propanol this maximum is nearly missing, its amplitude is small and does not change with temperature.

d) The broad 3$^{rd}$ maximum hardly changes with temperature, its position shifts toward higher $Q$ values and its amplitude increases with increasing alcohol concentration. (These properties are summarized in Table 2.)

All the TSSF-s obtained from MD simulations using both the TIP4P/2005 and SPC/E water models are presented in Figs. S5-S20. Here (in Fig. 4) only the structure factors obtained with the TIP4P/2005 water model are shown. There is a semi-quantitative agreement with most individual experimental result, although there are differences in the values and ratios of the peaks heights and also, in terms of the exact positions of the extrema. Larger discrepancies are found only in the case of the sample with 71 mol% 1-propanol content. Trends with decreasing temperature are reproduced exceptionally well (even for the $x_P = 0.71$ mixture). There are only small differences between the total structure factors obtained with TIP4P/2005 and SPC/E water models: $R$-factors, representing a quantitative measure of the difference, for the former are slightly better, particularly for the higher water content mixtures. For the definition of the $R$-factor see the SM. The calculated $R$-factor values are collected in Table S6.

We can then conclude that although agreement between the experimental and simulated total scattering structure factors is not perfect in every individual case, observable tendencies are the same. We therefore will exploit simulated particle configurations for revealing the effect of changing temperature and concentration on the structure.

*4.3. Partial radial distribution functions*

Temperature dependence of the partial radial distribution functions (PRDFs) related to H-bonding is shown for $x_P = 0.2$ in Figure 5, as obtained by using the TIP4P/2005 water model. Remaining PRDFs for this concentration are displayed in Figs. S21 and S22. Similar behavior can be obtained using the SPC/E model (Fig.S35). The temperature dependence of the H-bond related PRDFs for the other concentrations is presented in Figs. S23-S28 and S34-S39, while concentration dependence is shown in Figs. S29 and S30.

First maxima of the oxygen-oxygen and oxygen-hydrogen PRDFs (around 2.8 Å and 1.85 Å, respectively) correspond to H-bonding. These peaks became sharper as temperature decreases: their heights increase while widths decrease. The first minima following the peak becomes deeper upon cooling. The position of the peak does not change significantly.

With increasing alcohol concentration, the heights of the first maxima increase, while the position of the peak remains unchanged.

PRDF-s of a homogenous liquid are supposed to oscillate around unity. In the figures presented above, several PRDF-s oscillate around a slope (the long range behavior is enlarged in Figs. S31-S33). The same phenomenon was observed by Perera in Ref. [18] by MD simulations, using the SPC/E water and TraPPe 1-propanol model. The author proposed an explanation involving 'domain-ordering' where water and alcohol domains are segregated. Our simple interpretation would be that molecules of 1-propanol have a quite bulky alkyl-chain, so that the 'excluded volume' effect is considerable. The space occupied by the alkyl-chain cannot be reached by neighboring molecules: this is the reason why some PRDF-s exhibit 'slopes' while reaching their asymptotic limits. This kind of short range behavior may be just a part of the more complex domain ordering described in Ref. [18]: in the present work, we were not able to investigate this matter further, since our (otherwise, quite large) computer models are too small for studying long range (and large scale) phenomena.

*4.4. H-bond numbers*

H-bonded molecules are identified using a geometric definition in which the intermolecular O...H bond length is less than 2.5 Å, and the O...O-H angle is smaller than 30 degrees. The results of the H-bond analysis obtained by simulations using the TIP4P/2005 water model will be discussed in the following; the results obtained by the SPC/E model, which are qualitatively the same, are presented in the Supplementary Material (Figs. S63-S95).

The temperature dependence of the average number of H-bonds per molecule ($N_{Hb}$) is shown in Fig. S40, while the distribution of the average H-bond number can be found in Fig. S41. For the calculation of $N_{Hb}$ all molecules (1-propanol and water) were considered. The average number of H-bonds increases monotonously as the temperature decreases: it is 5-14% higher at the lowest investigated temperature than at room temperature. The increase is more significant at higher 1-propanol concentrations, see Fig. 6, where the numbers are normalized by the 298 K values. The number of H-bonds is higher at higher water concentrations: this is natural, as water molecules have more possibilities of forming them. The most frequent H-bond number is 4 for the water-rich ($x_P \leq 0.4$) mixtures, while the population of molecules with 2 H-bonds increases with increasing alcohol content. In the alcohol-rich mixtures ($x_P \geq 0.71$) the most frequent H-bond number is 2. The ratio of the molecules with 3 H-bonded pairs are around 30% in all water-rich mixtures, and decreases with alcohol-concentration in the alcohol-rich mixtures.

The relative frequency of molecules participating in 0, 1 and ≥ 2 H-bonds as a function of concentration and temperature is shown in Fig. S42. The 'solitary' molecules, that have no H-bond, may also be called as monomers. The ratio of monomers (Fig. S42a,b) is below 5% at all temperatures and concentrations, and it decreases with decreasing 1-propanol concentration and temperature. Molecules with a single H-bond (Fig. S42c,d) are in the end of chains, or two of these molecules can form a dimer. The ratio of these molecules is between 2 and 27%: it increases with increasing alcohol concentration and temperature. Most molecules have at least 2 H-bonds (Fig. S42e,f): the ratio of these molecules is more than 70% and it decreases with increasing alcohol concentration and increasing temperature.

The number of H-bonds between different kinds of molecules is presented in Fig. S43, whereas the distributions of H-bond numbers for the different pairs are shown at some selected concentrations and temperatures in Fig. S44. The number of water molecules around water ($N_{WW}$, Fig. S43a) is around 3.5 in the $x_P = 0.08$ sample and only 0.5 in the $x_P = 0.89$ mixture. $N_{WW}$ increases with temperature in the water-rich mixtures, but it is nearly independent of temperature in the $x_P = 0.89$ mixture. The number of 1-propanol molecules around water molecules ($N_{WP}$, Fig. S43b) is around 0.2 in the $x_P = 0.08$ sample and tends to 3 in the $x_P = 0.89$ mixture. The increase of $N_{WP}$ with temperature is more pronounced at higher alcohol contents. The total number of H-bonded molecules (water and 1-propanol together) around water ($N_W=N_{WP}+N_{WW}$, Fig. S43c) increases with decreasing temperature, and it is higher by about 5-12% at the lowest investigated temperature than at $T = 298$ K (see also Fig. S45a).

The vast majority (≥ 97%) of water molecules participate in ≥ 2 H-bonds, see Fig. S46. The number of water monomers is below 0.15%, while the ratio of water molecules with 1 H-bond is between 0.1 and 2.5%.

The environment of 1-propanol molecules consists of more than 2 water ($N_{PW}$, Fig. S43d) and only 0.1 1-propanol molecules ($N_{PP}$, Fig S43e) on average in the $x_P = 0.08$ mixture and about 1.5 1-propanol and only 0.3 water molecules on average in the $x_P = 0.89$ mixture. Both the number of 1-propanol and the number of water molecules increase in these environments as temperature decreases. (In frozen pure 1-propanol, at T ≤ 180 K, $N_{PP} = 2$.) The total H-bond number of 1-propanol molecules ($N_P=N_{PW}+N_{PP}$, Fig S43f) increases from 2.2 to 2.4 in the $x_P = 0.08$ mixture and from 1.7 to 2 in pure 1-propanol as the temperature decreases from room temperature to the lowest investigated temperature. The increase in terms of $N_P$ is around 6-14%, higher for higher alcohol-content mixtures (see also Fig. S45b).

0.1 – 4.5% of 1-propanol molecules are monomers, 8 – 27% of them have 1 H-bond, see Fig. S47. Both ratios increase with increasing alcohol concentration and decrease with decreasing temperature. 1-propanol molecules with at least 2 H-bonds are 69 – 91% of the 1-propanol molecules: the higher values are characteristic to lower alcohol concentrations and lower temperatures.

The concentration dependence of the number of H-bonded pairs/molecule (at some selected temperatures) are presented in Fig. 7. The number of 1-propanol molecules increases while the number of water molecules decreases as the alcohol concentration increases both around water and 1-propanol. However, the curves are not linear: the number of 1-propanol molecules ($N_{WP}$ and $N_{PP}$) is less than it should be if it was proportional to concentration. The number of water molecules around water molecules ($N_{WW}$) is higher. The $N_{PW}$ curves (water molecules around 1-propanol molecules) are S-shaped: the deviation from linear is slightly positive at high, while being slightly negative at low alcohol concentrations.

Both water and 1-propanol molecules can participate in H-bonds as proton donors and as acceptors. Proton donor molecules are defined as the molecules that are connected by their (hydroxilic) hydrogen atoms, while acceptor molecules participate in H-bonds by their oxygen atoms. Molecules are in $N_D$ D – $N_A$ A state (or shortly $N_D$ – $N_A$ donor-acceptor state) if they participate in $N_D$ H-bond as donors and in $N_A$ H-bond as acceptors. The relative frequency of donor-acceptor states of 1-propanol and water molecules at 298 K are shown in Fig. 8. For 1-propanol molecules the most frequent donor-acceptor state is 1D-1A (i.e., when a molecule participates in 1 H-bond as donor and 1 H-bond as acceptor) over the entire concentration range, but at low alcohol concentrations the 1D-2A state is nearly as frequent. For water the most frequent donor-acceptor state is 2D-2A at high water concentrations, and 2D-1A state at low water concentrations.

The concentration and temperature dependence of the most frequent donor-acceptor states are shown in Figs. S48 and S49 for 1-propanol and for water molecules, respectively. Concerning 1-propanol molecules, the frequency of 1D-2A and 0D-2A state decreases as the number of available donor molecules decreases (together with the decreasing number of water molecules as water concentration decreases). As 1-propanol concentration increases the occurrence of 1D-1A and 1D-0A states increases. The 0D-1A state has a maximum around $x_P = 0.3$. The numbers of alcohol molecules in the 0D-1A, 1D-0A and 0D-2A states decrease with decreasing temperature, while the frequencies of 1D-1A and 1D-2A states increase at the same time.

2D-2A and 1D-2A water states are most frequent at low alcohol concentrations, while the occurrence of 2D-1A and 1D-1A states increase with increasing alcohol content. The frequency of the 2D-2A state increases as the temperature decreases, whilst the 2D-1A, 1D-2A and 1D-1A states are less frequent at lower temperatures.

Concentration and temperature dependence of the number of H-bonds per 1-propanol and water molecules as donors and as acceptors are shown in Fig. S50, while the ratio of donor and acceptor H-bonds are presented in Fig. 9. In the pure liquids the number of donor and acceptor H-bonds are necessarily equal: in case of water they are almost 2, while for 1-propanol they are near 1. In water-rich mixtures the number of acceptor H-bonds of 1-propanol molecules is nearly 2 times higher than the number of donor H-bonds. At high 1-propanol concentrations the donor/acceptor ratio of 1-propanol molecules is higher than 0.9, the number of acceptor H-bonds being only slightly higher than the number of donor H-bonds. For water molecules the ratio of donor and acceptor H-bonds is close to 1 in the $x_P = 0.08$ mixture and nearly 1.5 in the $x_P = 0.89$ mixture. The ratio of donor/acceptor H-bonds (both for water and 1-propanol molecules) depends mainly on the alcohol concentration and changes only slightly with temperature: at lower temperatures the ratio is closer to 1.

*4.5. Monomers, clusters, cluster size distributions, percolation*

The vast majority of molecules have at least one H-bond. Molecules connected via H-bonds form H-bonded clusters: two molecules belong to the same cluster if they are connected by a chain of H-bonded molecules. The size of a cluster ($n_c$) is defined as the number of molecules belonging to it. The system is percolating if the number of molecules in the largest cluster is in the order of the number of molecules in the simulation box.

A special kind of 'clusters' are the ones with one member only: these are solitary molecules, without a single H-bond connected to them. The number of such 'clusters', i.e., of monomers, is also studied here as a function of temperature and composition.

The concentration and temperature dependence of the H-bonded network can be investigated for the total system, i.e., considering all possible kinds of H-bonds (water – water, 1-propanol

– 1-propanol and 1-propanol – water also), and also, for the water and 1-propanol subsystems separately, in which case only the H-bonds between like molecules are taken into account.

The concentration and temperature dependence of the number of monomers is shown in Fig. S51 for the total system, as well as for the two subsystems. Concerning 1-propanol and water molecules together, the number of monomers is low, it increases with increasing alcohol concentration and decreases with decreasing temperature. The vast majority of these monomers are 1-propanol molecules (see also Figs. S42, S46 and S47).

The cluster size distribution of H-bonded clusters at 298 K in is shown in Fig. 10 (all H-bonds considered here, irrespective of the kind type of molecule). The size of the largest cluster decreases as 1-propanol concentration increases. The largest size is close to 4000, the number of molecules in the simulation box, for $x_P \leq 0.71$, between 100-2000 for $x_P = 0.89$, and around 100 for pure 1-propanol. The temperature dependence of the cluster size distribution is presented in Figs. 11 and S52. The size of the largest clusters increases with decreasing temperature for all concentrations. The system is percolating for $x_P \leq 0.71$ at all temperatures, and also for $x_P = 0.89$ at $T \leq 273$ K, cf. Fig. 11. In pure 1-propanol small clusters dominate: the largest clusters contain less than 1000 molecules (around 500-600), even at $T = 230$ K.

In the water subsystem (only the H-bonds between water molecules are counted) the number of water monomers increases with increasing 1-propanol concentration and decreases with decreasing temperature (see Fig. S51c,d). In water-rich mixtures the ratio of monomers is low, below 1-2%. In 1-propanol – rich mixtures the number of water monomers increases rapidly with concentration: 20% of water molecules are monomers in the $x_P = 0.71$ mixture and 60% of them in the $x_P = 0.89$ mixture. (Considering that in the total system the number of water monomers is very low, the high number of the water monomers in the water subsystem means that water molecules are surrounded by 1-propanol molecules only.) Apart from monomers, small water clusters dominate in the 1-propanol – rich mixtures, see Fig. S53. In the water-rich mixtures (see Fig. S54) the size of the largest water cluster is about equal to the number of water molecules in the mixture: the water subsystem percolates.

In the 1-propanol subsystem (only H-bonds between 1-propanol molecules are considered) the number of monomers (see Fig. S51e,f) decreases with increasing 1-propanol concentration, from 90% (in the $x_P = 0.08$ mixture) to below 5% in pure 1-propanol at $T = 298$ K. In the water-rich mixtures the ratio of monomers depends only slightly on temperature, while in the 1-propanol-rich mixtures it decreases visibly with decreasing temperature: it is below 1% in pure 1-propanol at $T = 230$ K. In water-rich mixtures 1-propanol molecules can only be found in small clusters (besides the monomers): the largest clusters contain only 12-16 molecules even at low temperatures in the $x_P = 0.4$ mixture (see Fig. S55). The size of the largest 1-propanol cluster increases slightly with 1-propanol concentration, but it is still below 100 molecules (3% of the number of 1-propanol molecules) even in the $x_P = 0.89$ mixture (Fig. S56). The presence of medium-size 1-propanol clusters can be observed only in pure 1-propanol, where the size of the largest cluster increases with decreasing temperature and reaches 15% of the number of molecules at $T = 230$ K.

*4.6. Cycles, primitive rings, ring size distributions*

H-bonded chains whose last and first molecules are H-bonded together are called cyclic entities (cycles). Molecules participating in cycles need to have at least two H-bonds. Note that any molecule can participate in more than one cycle.

The number of molecules that participate in cycles is compared to the number of molecules having at least 2 H-bonds (ratio of molecules in cycles and molecules in chains) in Fig. 12. In water-rich mixtures more than 98% of molecules with ≥ 2 H-bonds participate in cycles. The ratio of molecules in cycles decreases with increasing alcohol concentration, so that in pure 1-propanol only 10% of molecules in chains participate in cycles. The ratio of molecules in cycles increases with decreasing temperature, which change is most significant for the $x_P = 0.89$ mixture: at 298 K only 40% of chain molecules participate in cycles, whilst at 230 K this ratio is nearly 80%.

The size of a cyclic entity can be defined as the shortest path through the H-bonded molecules that are in the cycle. A cyclic entity is called as a primitive ring (or shortly as ring) hereafter if it cannot be decomposed into smaller rings. The ring search algorithm developed by Chihaia [37] is used in this work. Rings with sizes up to 8 will be considered from this point on.

The average number of rings per molecules is shown in Fig. S57. The temperature dependence of the number of rings appears more pronounced in Fig. S58, where the numbers of rings are normalized by the $T = 298$ K values. In the $x_P = 0.08$ mixture the number of rings is higher than the number of molecules, which means that quite a few molecules must participate in more than one rings. As 1-propanol concentration increases the number of rings per molecule decreases:

it is around 0.5 for $x_P = 0.4$, 0.1 – 0.15 for $x_P = 0.71$, and below 0.01 in pure 1-propanol. The number of rings increases as temperature decreases in $x_P \leq 0.71$ with about the same rate. In the $x_P = 0.89$ mixture at first it increases with decreasing temperature, but around $T = 263$ K it reaches a maximum value, and slowly decreases as the temperature continues to decrease. In pure 1-propanol the number of rings (that is very low already at room temperature) decreases as the temperature decreases: propanol molecules with two H-bonds seem to tend to form linear/branched chains at low T.

In the calculations presented above, all possible rings were taken into account. Rings can be classified according to the molecules which form them: rings that contain only water molecules (water-rings), only 1-propanol molecules (1-propanol rings) and water and 1-propanol molecules together (mixed rings). The concentration dependence of these 3 types of rings at 298 K is shown in Fig. 13, whilst the temperature dependence of the ratios of these ring types is presented for 3 concentrations in Fig. S59. The ratio of the different ring types is nearly independent of temperature. Purely water rings are the most frequent at $x_P = 0.08$ (more than 80% of the rings), and their ratio decreases as 1-propanol content increases. At $x_P \geq 0.4$, mixed rings (rings containing both water and 1-propanol molecules) are the most abundant. Purely 1-propanol rings are rare: their ratio is only 12% even in the $x_P = 0.89$ mixture, where the number of rings/molecule is around 0.03, the number of rings in the 4000 molecule simulation box is around 130-150, and the number of purely 1-propanol rings is around 12-16.

The temperature dependence of water rings ($N_R^W$) and mixed rings ($N_R^{mix}$) are shown in Figs. S60 and S61. The number of rings increases with decreasing temperature for purely water and mixed rings, although the increase in terms of the number of rings upon cooling is more pronounced for the mixed rings.

The distribution of ring sizes is shown in Fig. 14. In the $x_P = 0.08$ mixture six-membered rings are the most frequent, similarly to pure water and methanol-water mixtures with low methanol concentrations [22,40]. As 1-propanol concentration increases, five-membered rings become the most abundant (note in pure 1-propanol, 4 and 5-membered rings are roughly equally frequent, but the number of rings in pure 1-propanol is very low). This change may be connected to the relatively large size of the n-alkyl chain. As temperature decreases, the number of rings increases for all ring sizes, except for 4-fold rings in $x_P = 0.89$ and 1.0, where the number of 4-membered rings decreases.

A more detailed analyses of the ring types can be performed by investigating the rings containing $R_P$ 1-propanol and $R_W$ water molecules ($R_P - R_W$ in short). The temperature and concentration dependence of the occurrences of such ring types is shown in Fig. 15. The most popular rings in the $x_P = 0.08$ mixture are the 0-5 and 0-6 types (no 1-propanol and 5 and 6 water molecules, respectively). As 1-propanol concentration increases, the relative frequency of rings containing more 1-propanol molecules (at first the 1-4 and 1-5 types, then in alcohol-rich mixtures the 1-3, 2-2, 2-3, 3-2 types) increases.

Participation of 1-propanol and water molecules in primitive rings can be compared to the molar fraction of the different molecules in the mixtures. 1-propanol and water molecules in rings with given sizes are identified, then the ratio of 1-propanol and water molecules in the rings is divided by the ratio of 1-propanol and water molecules in the mixture. The temperature and concentration dependence of the participation of the 1-propanol/water molecules in primitive rings is shown in Fig. S62. At all concentration and temperature values the presence of water molecules in these rings is more pronounced than it can be expected from the composition. In most cases the ratio is around 1/3, which means that 3 times more water molecules participate in these rings than in the entire mixture on average. This finding should be taken cautiously, since water molecules can form more H-bonds than alcohol molecules, which distorts the conclusion somewhat. The relative frequency of 1-propanol molecules in 3 to 6 membered rings decreases as alcohol concentration increases: primitive rings contain relatively more and more water than 1-propanol. The concentration dependence is strongest for the 3-membered rings and is weakening for larger ring sizes.

## 5. Summary and Conclusions

Aqueous mixtures of 1-propanol have been studied by high-energy synchrotron X-ray diffraction upon cooling. X-ray weighted total scattering structure factors of 6 mixtures, from 8 mol% to 89 mol% alcohol content, as well as that of pure 1-propanol are reported from room temperature down to the freezing points of the liquids.

Molecular dynamics simulations have been performed, in order to interpret measured data. The all atom OPLS-AA potential model was used for 1-propanol, combined with both the SPC/E and the TIP4P/2005 water models: both combinations provide a semi-quantitative description of the measured total structure factors at low and high alcohol contents, while the agreement is qualitative for the mixture with 71 mol% of 1-propanol. Based on comparisons with available experimental data on density and self-diffusion coefficients, results for the TIP4P/2005 water model are presented in the main text.

A detailed description of the hydrogen bonded network has been provided, in terms of hydrogen bond numbers, analysis of proton donor-acceptor ratios, size distributions of hydrogen bonded clusters and ring size statistics. The following findings are thought to be noteworthy:

(i) Most molecules have at least 2 H-bonds: the ratio of these molecules is more than 70% and it decreases with increasing alcohol concentration and increasing temperature.

(ii) The ratio of donor/acceptor H-bonds (both for water and 1-propanol molecules) depends mainly on the alcohol concentration and changes only slightly with temperature: at lower temperatures the ratio is closer to 1.

(iii) The size of the largest clusters increases with decreasing temperature for all concentrations. The system is percolating for $x_P \leq 0.71$ at all examined temperatures (room temperature and below), and also for $x_P = 0.89$ at $T \leq 273$ K.

(iv) A strong temperature dependence of the percolation threshold, as well as of the participation of the number of doubly hydrogen bonded molecules in cyclic entities, has been found for the mixture with 89 mol% of 1-propanol.

(v) At and above an alcohol content of 20 mol%, 5-fold rings are the most frequent cyclic entities, with a strong temperature dependence in terms of the number of rings.

**Acknowledgments**

IP, LP and LT acknowledge financial support from the National Research, Development and Innovation Office (NKFIH), under grant no. KH 130425 and mobility grant No. TÉT_16-1-2016-0202. Synchrotron radiation experiments were performed on the BL04B2 of SPring-8, with the approval of the Japan Synchrotron Radiation Research Institute (JASRI) (Proposal Nos. 2018B1210 and 2019A1517). LT is grateful to the János Bolyai Research Scholarship of17 the Hungarian Academy of Sciences. For the careful preparation of the mixtures Ms. A. Szuja (Centre for Energy Research, Hungary) is gratefully acknowledged. We thank Dr. Imre Bakó for sharing the computer code used for hydrogen bond analysis.

**References**

[1] S. Dixit, J. Crain, W.C.K. Poon, J.L. Finney, A.K. Soper, Molecular segregation observed in a concentrated alcohol–water solution, Nature 416 (2002) 829–832. doi:10.1038/416829a.

[2] S. Choi, S. Parameswaran, J.-H. Choi, Understanding alcohol aggregates and the water hydrogen bond network towards miscibility in alcohol solutions: graph theoretical analysis, Phys. Chem. Chem. Phys. 22 (2020) 17181–17195. doi:10.1039/D0CP01991G.

[3] T. Sato, A. Chiba, R. Nozaki, Dielectric relaxation mechanism and dynamical structures of the alcohol/water mixtures, J. Mol. Liq. 101 (2002) 99–111. doi:10.1016/S0167-7322(02)00085-5.

[4] M.Z. Jora, M.V.C. Cardoso, E. Sabadini, Correlation between viscosity, diffusion coefficient and spin-spin relaxation rate in $^1$H NMR of water-alcohols solutions, J. Mol. Liq. 238 (2017) 341–346. doi:10.1016/j.molliq.2017.05.006.

[5] E.J.W. Wensink, A.C. Hoffmann, P.J. van Maaren, D. van der Spoel, Dynamic properties of water/alcohol mixtures studied by computer simulation, J. Chem. Phys. 119 (2003) 7308–7317. doi:10.1063/1.1607918.

[6] W. Marczak, M. Spurek, Compressibility and Volume Effects of Mixing of 1-Propanol with Heavy Water, J. Solution Chem. 33 (2004) 99–116. doi:10.1023/B:JOSL.0000030278.21608.a1.

[7] K. Yoshida, A. Kitajo, T. Yamaguchi, $^{17}$O NMR relaxation study of dynamics of water molecules in aqueous mixtures of methanol, ethanol, and 1-propanol over a temperature range of 283–403 K, J. Mol. Liq. 125 (2006) 158–163. doi:10.1016/j.molliq.2005.11.009.

[8] R. Li, C. D'Agostino, J. McGregor, M.D. Mantle, J.A. Zeitler, L.F. Gladden, Mesoscopic Structuring and Dynamics of Alcohol/Water Solutions Probed by Terahertz Time-Domain Spectroscopy and Pulsed Field Gradient Nuclear Magnetic Resonance, J. Phys. Chem. B 118 (2014) 10156–10166. doi:10.1021/jp502799x.

[9] G. D'Arrigo, J. Teixeira, Small-angle neutron scattering study of $D_2O$–alcohol solutions, J. Chem. Soc., Faraday Trans. 86 (1990) 1503–1509. doi:10.1039/FT9908601503.

[10] H. Hayashi, K. Nishikawa, T. Iijima, Small-angle x-ray scattering study of fluctuations in 1-propanol-water and 2-propanol-water systems, J. Phys. Chem. 94 (1990) 8334–8338. doi:10.1021/j100384a062.

[11] T. Takamuku, H. Maruyama, K. Watanabe, T. Yamaguchi, Structure of 1-Propanol–Water Mixtures Investigated by Large-Angle X-ray Scattering Technique, J. Solution Chem. 33 (2004) 641–660. doi:10.1023/B:JOSL.0000043631.21673.8b.

[12] I. Juurinen, T. Pylkkänen, C.J. Sahle, L. Simonelli, K. Hämäläinen, S. Huotari, et al., Effect of the Hydrophobic Alcohol Chain Length on the Hydrogen-Bond Network of Water, J. Phys. Chem. B 118 (2014) 8750–8755. doi:10.1021/jp5045332.

[13] P. Tomza, M.A. Czarnecki, Microheterogeneity in binary mixtures of propyl alcohols with water: NIR spectroscopic, two-dimensional correlation and multivariate curve resolution study, J. Mol. Liq. 209 (2015) 115–120. doi:10.1016/j.molliq.2015.05.033.

[14] J.G. Méndez-Bermúdez, H. Dominguez, L. Temleitner, L. Pusztai, On the Structure Factors of Aqueous Mixtures of 1-Propanol and 2-Propanol: X-Ray Diffraction Experiments and Molecular Dynamics Simulations, Phys. Status Solidi B 255 (2018) 1800215. doi:10.1002/pssb.201800215.

[15] A.B. Roney, B. Space, E.W. Castner, R.L. Napoleon, P.B. Moore, A Molecular Dynamics Study of Aggregation Phenomena in Aqueous n-Propanol, J. Phys. Chem. B 108 (2004) 7389–7401. doi:10.1021/jp037922j.

[16] J.G. Méndez-Bermúdez, H. Dominguez, L. Pusztai, S. Guba, B. Horváth, I. Szalai, Composition and temperature dependence of the dielectric constant of 1-propanol/water mixtures: Experiment and molecular dynamics simulations, J. Mol. Liq. 219 (2016) 354–358. doi:10.1016/j.molliq.2016.02.053.

[17] M. Ogawa, Y. Ishii, N. Ohtori, Dynamic Behavior of Mesoscopic Concentration Fluctuations in an Aqueous Solution of 1-Propanol by MD Simulation, Chem. Lett. 45 (2016) 98–100. doi:10.1246/cl.150952.

[18] A. Perera, Molecular emulsions: from charge order to domain order, Phys. Chem. Chem. Phys. 19 (2017) 28275–28285. doi:10.1039/C7CP05727J.

[19] E.G. Pérez, D. González-Salgado, E. Lomba, Molecular dynamics simulations of aqueous solutions of short chain alcohols. Excess properties and the temperature of maximum density, Fluid Phase Equilib. 528 (2021) 112840. doi:10.1016/j.fluid.2020.112840.

[20] T. Takamuku, K. Saisho, S. Nozawa, T. Yamaguchi, X-ray diffraction studies on methanol–water, ethanol–water, and 2-propanol–water mixtures at low temperatures, J. Mol. Liq. 119 (2005) 133–146. doi:10.1016/j.molliq.2004.10.020.

[21] L. Dougan, R. Hargreaves, S.P. Bates, J.L. Finney, V. Réat, A.K. Soper, et al., Segregation in aqueous methanol enhanced by cooling and compression, J. Chem. Phys. 122 (2005) 174514. doi:10.1063/1.1888405.

[22] I. Bakó, L. Pusztai, L. Temleitner, Decreasing temperature enhances the formation of sixfold hydrogen bonded rings in water-rich water-methanol mixtures, Sci. Rep. 7 (2017) 1073. doi:10.1038/s41598-017-01095-7.

[23] S. Pothoczki, L. Pusztai, I. Bakó, Variations of the Hydrogen Bonding and Hydrogen-Bonded Network in Ethanol–Water Mixtures on Cooling, J. Phys. Chem. B 122 (2018) 6790–6800. doi:10.1021/acs.jpcb.8b02493.

[24] S. Lenton, N.H. Rhys, J.J. Towey, A.K. Soper, L. Dougan, Temperature-Dependent Segregation in Alcohol–Water Binary Mixtures Is Driven by Water Clustering, J. Phys. Chem. B 122 (2018) 7884–7894. doi:10.1021/acs.jpcb.8b03543.

[25] S. Pothoczki, L. Pusztai, I. Bakó, Molecular Dynamics Simulation Studies of the Temperature-Dependent Structure and Dynamics of Isopropanol–Water Liquid Mixtures at Low Alcohol Content, J. Phys. Chem. B 123 (2019) 7599–7610. doi:10.1021/acs.jpcb.9b05631.

[26] I. Pethes, L. Pusztai, K. Ohara, S. Kohara, J. Darpentigny, L. Temleitner, Temperature-dependent structure of methanol-water mixtures on cooling: X-ray and neutron diffraction and molecular dynamics simulations, J. Mol. Liq. 314 (2020) 113664. doi:10.1016/j.molliq.2020.113664.

[27] S. Pothoczki, I. Pethes, L. Pusztai, L. Temleitner, D. Csókás, S. Kohara, et al., Hydrogen bonding and percolation in propan-2-ol – Water liquid mixtures: X-ray diffraction experiments and computer simulations, J. Mol. Liq. 329 (2021) 115592. doi:10.1016/j.molliq.2021.115592.

[28] S. Pothoczki, I. Pethes, L. Pusztai, L. Temleitner, K. Ohara, I. Bakó, Properties of Hydrogen-Bonded Networks in Ethanol–Water Liquid Mixtures as a Function of Temperature: Diffraction Experiments and Computer Simulations, J. Phys. Chem. B 125 (2021) 6272–6279. doi:10.1021/acs.jpcb.1c03122.

[29] S. Kohara, K. Suzuya, Y. Kashihara, N. Matsumoto, N. Umesaki, I. Sakai, A horizontal two-axis diffractometer for high-energy X-ray diffraction using synchrotron radiation on bending magnet beamline BL04B2 at SPring-8, Nucl. Instruments Methods Phys. Res. Sect. A Accel. Spectrometers, Detect. Assoc. Equip. 467–468 (2001) 1030–1033. doi:10.1016/S0168-9002(01)00630-1.

[30] A.Y. Manakov, L.S. Aladko, A.G. Ogienko, A.I. Ancharov, Hydrate formation in the system n-propanol–water, J. Therm. Anal. Calorim. 111 (2013) 885–890. doi:10.1007/s10973-012-2246-1.

[31] M.J. Abraham, T. Murtola, R. Schulz, S. Páll, J.C. Smith, B. Hess, et al., GROMACS: High performance molecular simulations through multi-level parallelism from laptops to supercomputers, SoftwareX 1–2 (2015) 19–25. doi:10.1016/j.softx.2015.06.001.

[32] W.L. Jorgensen, D.S. Maxwell, J. Tirado-Rives, Development and Testing of the OPLS All-Atom Force Field on Conformational Energetics and Properties of Organic Liquids, J. Am. Chem. Soc. 118 (1996) 11225–11236. doi:10.1021/ja9621760.

[33] H.J.C. Berendsen, J.R. Grigera, T.P. Straatsma, The missing term in effective pair potentials, J. Phys. Chem. 91 (1987) 6269–6271. doi:10.1021/j100308a038.

[34] J.L.F. Abascal, C. Vega, A general purpose model for the condensed phases of water: TIP4P/2005, J. Chem. Phys. 123 (2005) 234505. doi:10.1063/1.2121687.

[35] F.-M. Pang, C.-E. Seng, T.-T. Teng, M.H. Ibrahim, Densities and viscosities of aqueous solutions of 1-propanol and 2-propanol at temperatures from 293.15 K to 333.15 K, J. Mol. Liq. 136 (2007) 71–78. doi:10.1016/j.molliq.2007.01.003.

[36] I. Bakó, T. Megyes, S. Bálint, T. Grósz, V. Chihaia, Water–methanol mixtures: topology of hydrogen bonded network, Phys. Chem. Chem. Phys. 10 (2008) 5004. doi:10.1039/b808326f.

[37] V. Chihaia, S. Adams, W.F. Kuhs, Molecular dynamics simulations of properties of a (001) methane clathrate hydrate surface, Chem. Phys. 317 (2005) 208–225. doi:10.1016/j.chemphys.2005.05.024.

[38] N. Tsvetov, A. Sadaeva, M. Toikka, I. Zvereva, A. Toikka, Excess molar heat capacity for the binary system n-propyl alcohol + water in the temperature range 278.15–358.15 K: new data and application for excess enthalpy calculation, J. Therm. Anal. Calorim. 142 (2020) 1977–1987. doi:10.1007/s10973-020-09605-y.

[39] E. Hawlicka, R. Grabowski, Self-diffusion in water-alcohol systems. 3. 1-Propanol-water solutions of sodium iodide, J. Phys. Chem. 96 (1992) 1554–1557. doi:10.1021/j100183a013.

[40] I. Bakó, Á. Bencsura, K. Hermannson, S. Bálint, T. Grósz, V. Chihaia, et al., Hydrogen bond network topology in liquid water and methanol: a graph theory approach, Phys. Chem. Chem. Phys. 15 (2013) 15163. doi:10.1039/c3cp52271g.

**Tables**

**Table 1** Investigated 1-propanol – water mixtures: nominal and exact compositions, temperatures examined in X-ray diffraction measurement and in molecular dynamics simulations. In the MD simulations every temperature/concentration point visited by XRD measurements was examined and some additional temperatures also, only these latter are shown in this Table.

| Nominal 1-propanol content, $x_P$ | Measured 1-propanol content of samples investigated by XRD [mol%] | Temperatures examined by XRD [K] | Additional temperatures examined (only) in MD simulations [K] |
|---|---|---|---|
| 0.08 | 8.005 | 298, 268 | 273 |
| 0.2 | 20.006 | 298, 273, 268, 263, 253 | |
| 0.3 | 29.999 | 298, 273, 268, 263, 253 | |
| 0.4 | 40.068 | 298, 273, 268, 263, 253, 243 | |
| 0.71 | 71.093 | 298, 263, 243 | 273, 253 |
| 0.89 | 89.263 | 298, 263, 230 | 273, 253, 243 |
| 1.0 | 100 | 298, 263, 230, 180, 150 | 273, 253, 243 |

**Table 2** Temperature and concentration dependence of the pre-peak and the first 3 peaks of the total structure factor in the experiments and in the simulations. (→(right): shifts right to higher Q values; ←(left) : shifts left to lower Q values; ↑(up) increases; ↓(down) decreases; 0: does not change or just slightly, or uncertain). The order of the symbols represents the behavior in experiment / simulation with TIP4P/2005 water / simulation with SPC/E water.

|  | Increasing alcohol concentration | Decreasing temperature |
| --- | --- | --- |
| Pre-peak position | → / → / → | → / → / 0 |
| Pre-peak amplitude | ↑ / ↑ / ↑ | ↓ / ↓ / ↓ |
| 1st peak position | ← / ← / ← | → / → / → |
| 1st peak amplitude | ↑ / ↑ / ↑ | ↑ / ↑ / ↑ |
| 2nd peak position | ← / ← / ← or 0 | → / → / → |
| 2nd peak amplitude | ↓ / ↓ / ↓ | ↑ / ↑ / ↑ |
| 3rd peak position | → / → / → | 0 / 0 / 0 |
| 3rd peak amplitude | ↑ / ↑ / ↑ | 0 / ↑ / ↑ |

**Figures**

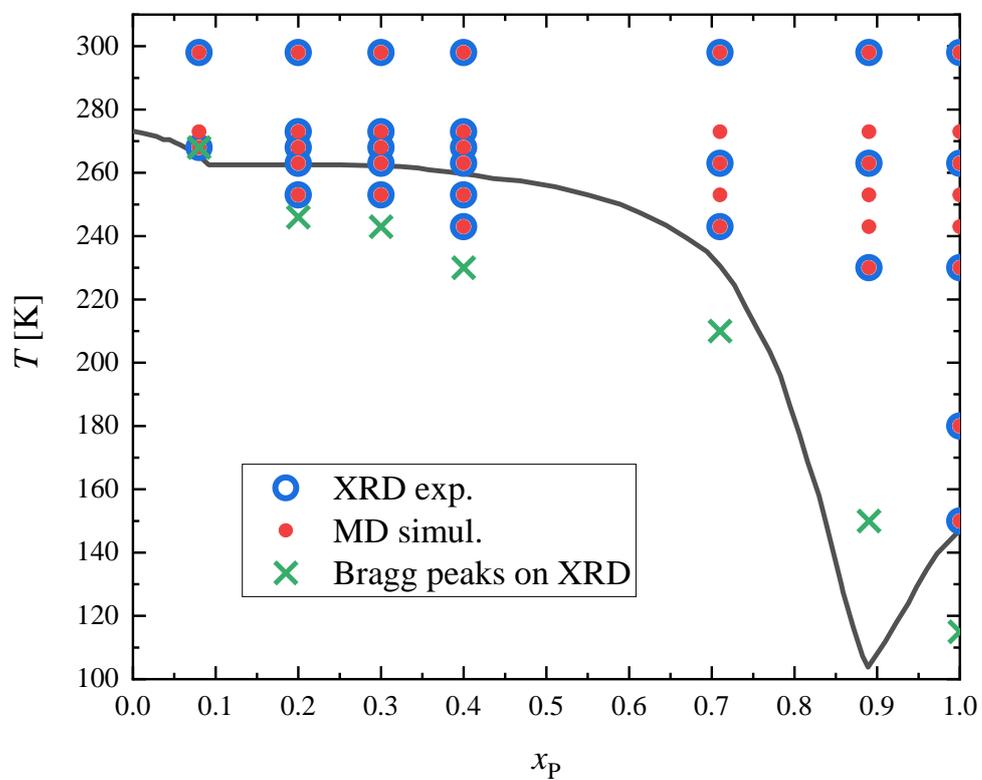

**Figure 1** Phase diagram for 1-propanol – water mixtures at $p = 1$ bar [30]. The thick line is the solid-liquid coexistence curve. Temperatures where X-ray diffraction experiments (open blue circles) and molecular dynamics simulations (solid red circles) were performed at different 1-propanol concentrations are also shown. Temperatures at which Bragg-peaks appeared during the X-ray diffraction measurements are marked with green 'x'.

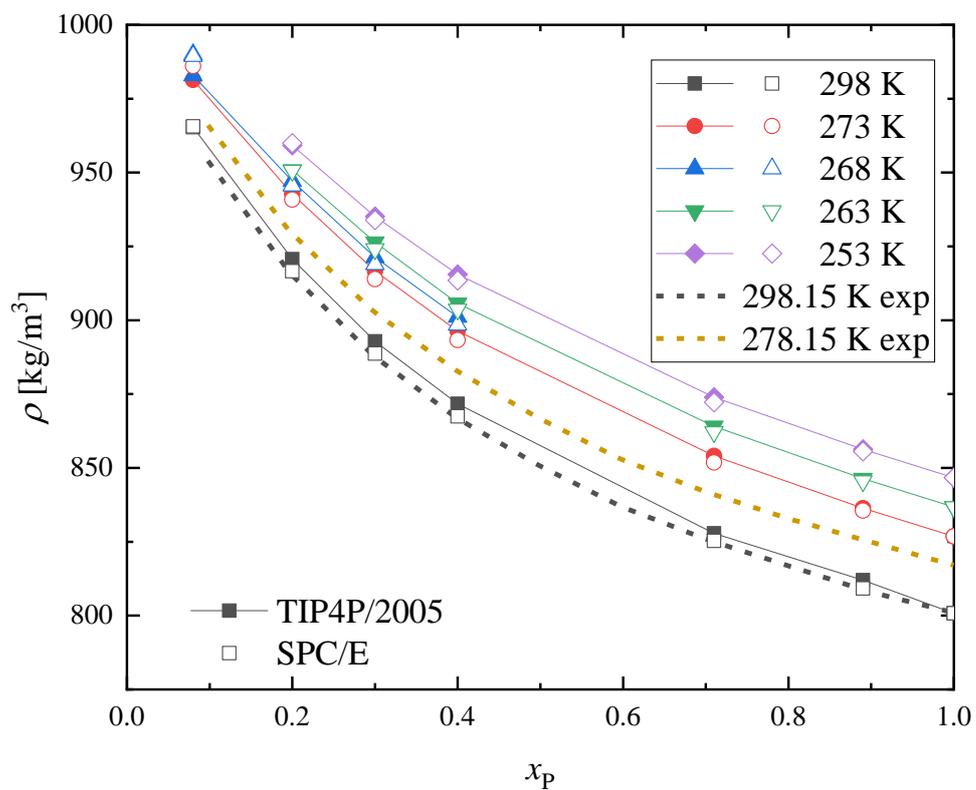

**Figure 2** Concentration dependence of density of 1-propanol – water mixtures obtained by MD simulations using the TIP4P/2005 (solid symbols) and SPC/E (open symbols) water models. Experimental data from Ref [38] are also shown for comparison (dashed lines).

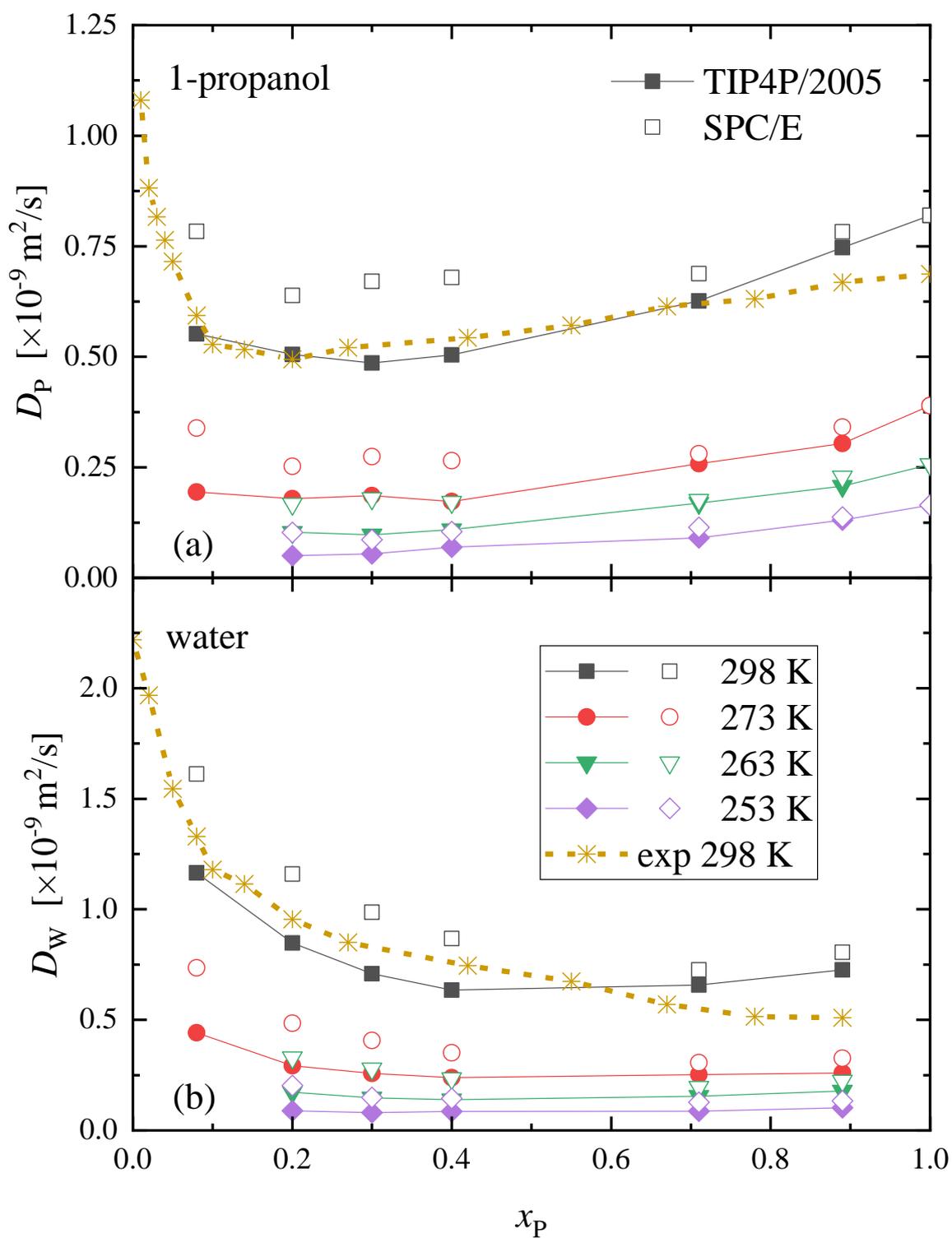

**Figure 3** Concentration dependence of the self-diffusion coefficients of (a) 1-propanol and (b) water molecules, as obtained by MD simulations using the TIP4P/2005 (solid symbols) and SPC/E (open symbols) water models. Experimental data are from Ref. [39].

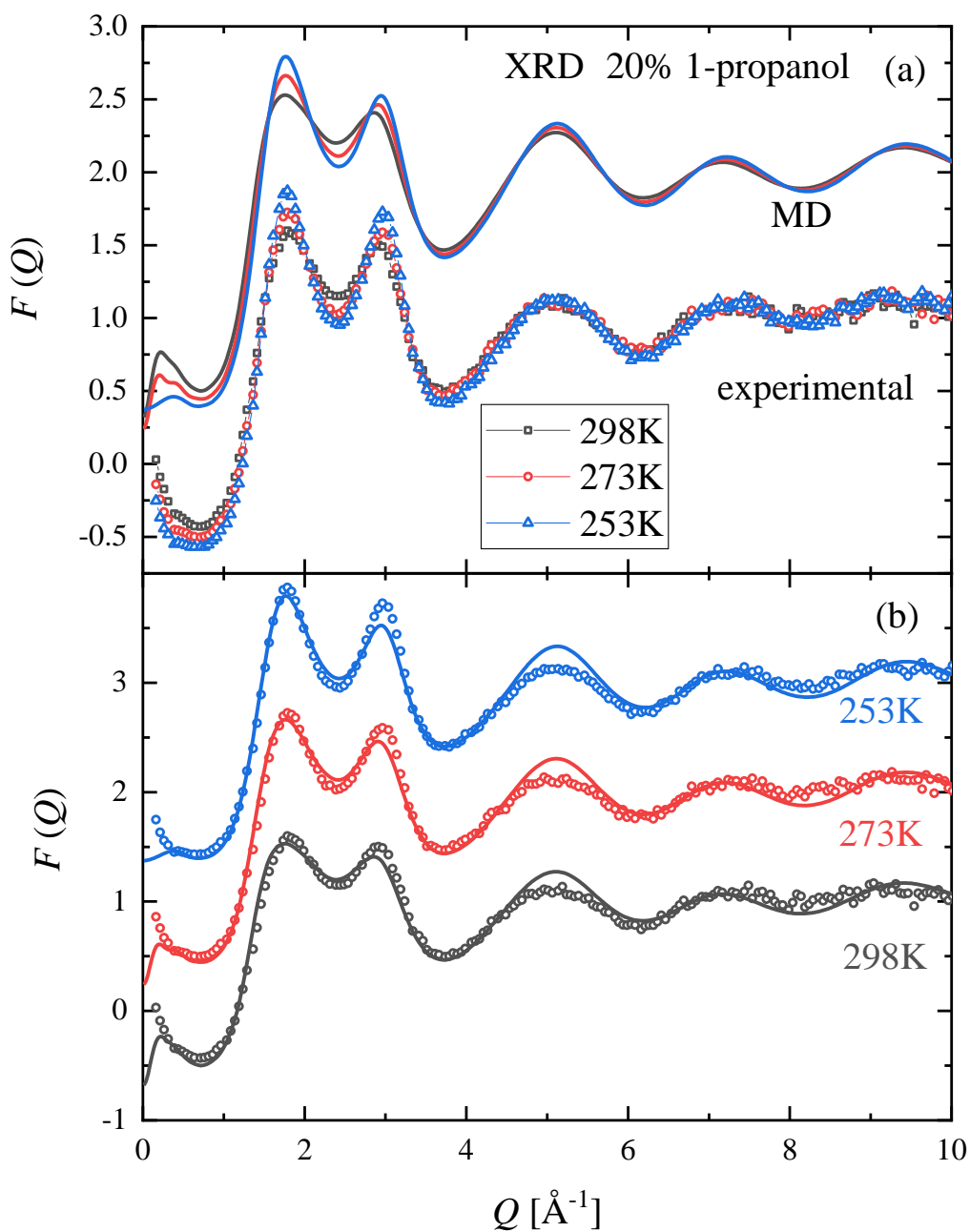

**Figure 4** Temperature dependence of measured (symbols) and simulated (lines) XRD structure factors for the mixture with 20 mol% 1-propanol content. Comparisons of (a) trends observed upon cooling, (b) measured and simulated curves at three selected temperatures: 298 K (black lines and symbols), 273 K (red) and 253 K (blue). The simulated curves were obtained by using the TIP4P/2005 water model. (Curves are shifted for clarity.)

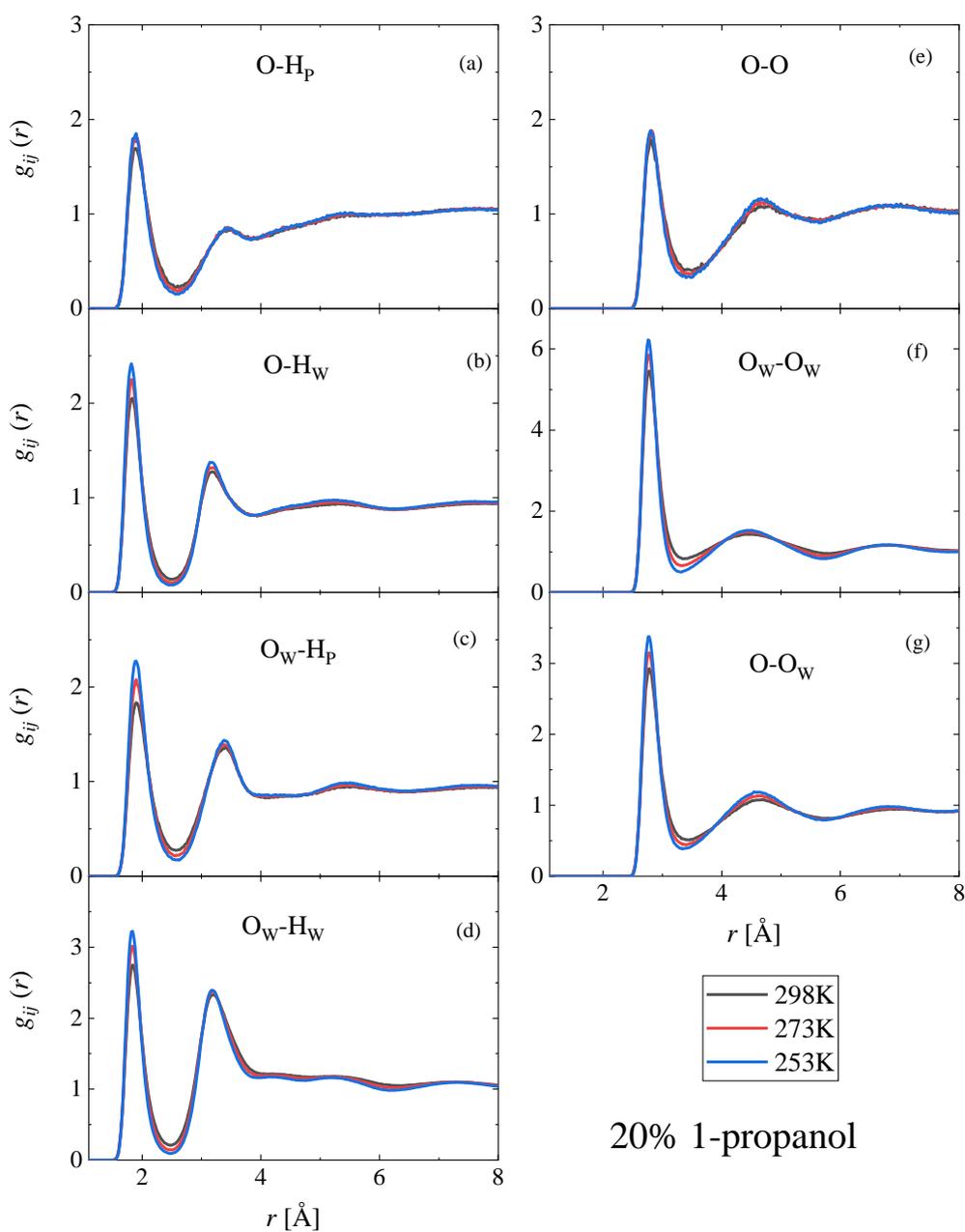

**Figure 5** Temperature dependence of simulated partial radial distribution functions of the mixture with 20 mol% 1-propanol content. H-bonding related partials are shown: (a) 1-propanol O (denoted as O) – hydroxyl H of 1-propanol (denoted as $H_P$), (b) 1-propanol O – water H (denoted as $H_W$), (c) water O (denoted as $O_W$) – hydroxyl H of 1-propanol, (d) water O – water H, (e) 1-propanol O – 1-propanol O, (f) water O – water O, (g) 1-propanol O – water O. The curves were obtained by using the TIP4P/2005 water model.

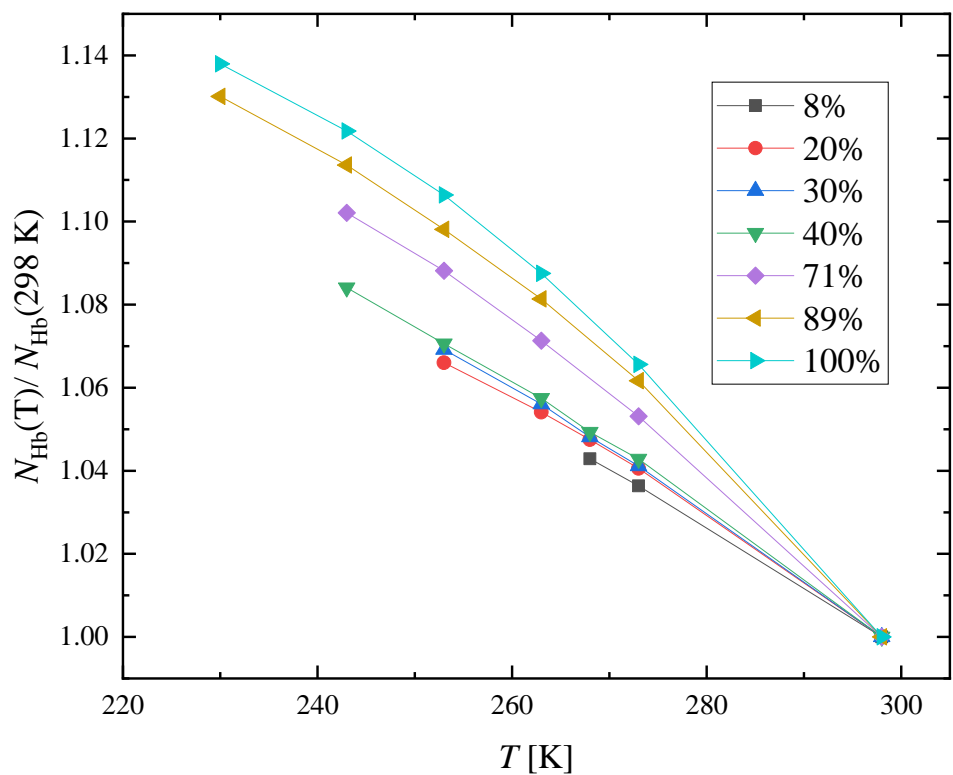

**Figure 6** Temperature dependence of the average number of hydrogen bonds per molecule, normalized to the 298 K value at different 1-propanol concentrations.

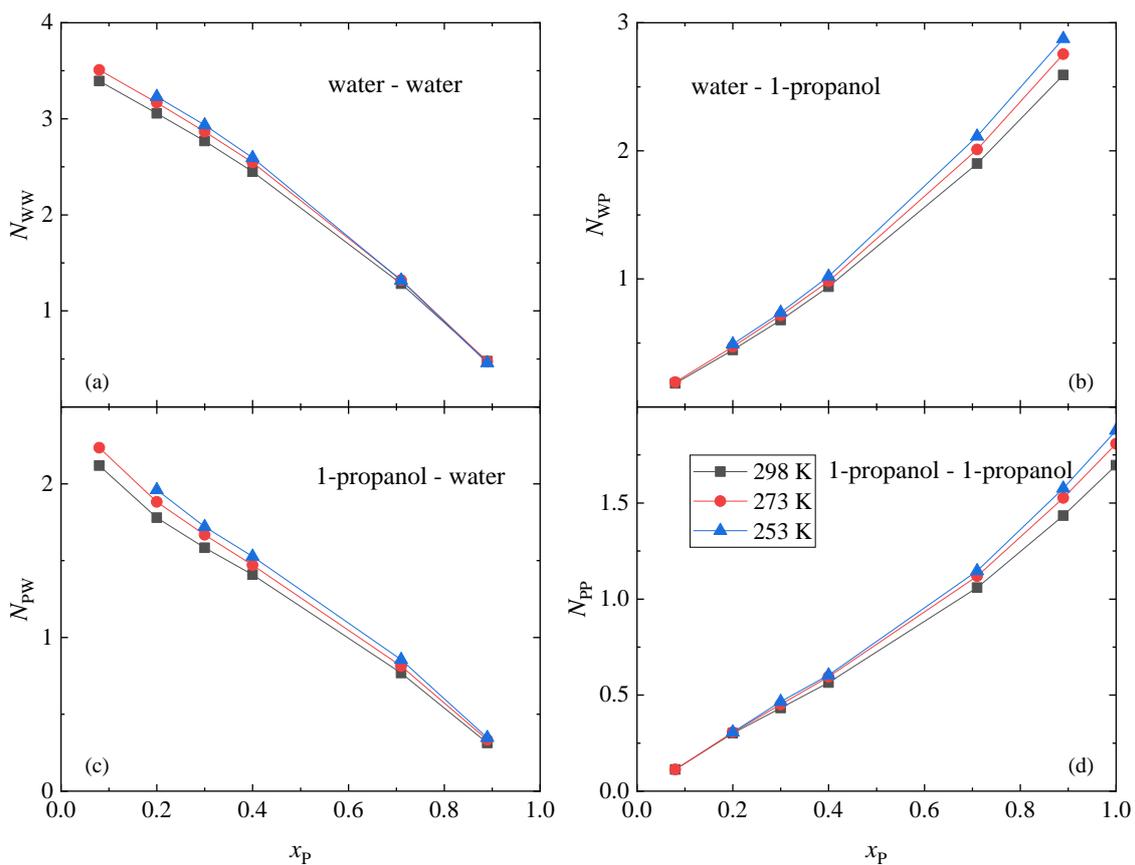

**Figure 7** Concentration dependence of the number of hydrogen bonds at different temperatures, as obtained by MD simulations using the TIP4P/2005 water model: (a) average number of H-bonded water molecules around water, (b) average number of H-bonded 1-propanol molecules around water, (c) average number of H-bonded water molecules around 1-propanol, (d) average number of H-bonded 1-propanol molecules around 1-propanol.

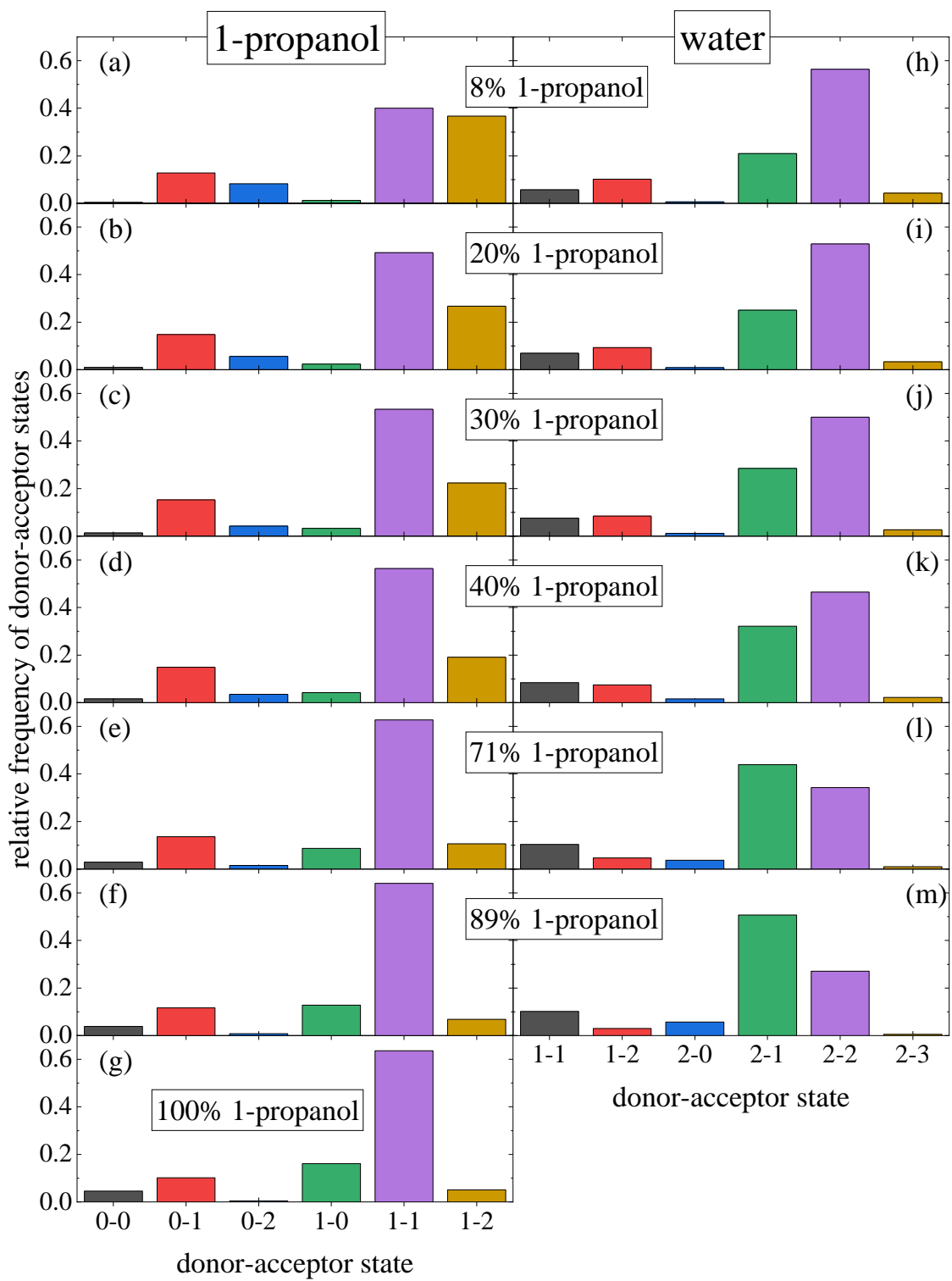

**Figure 8** Relative frequencies of the different donor-acceptor states of (a-g) 1-propanol and (h-m) water molecules in the 1-propanol – water mixtures at 298 K, as obtained by MD simulations using the TIP4P/2005 water model.

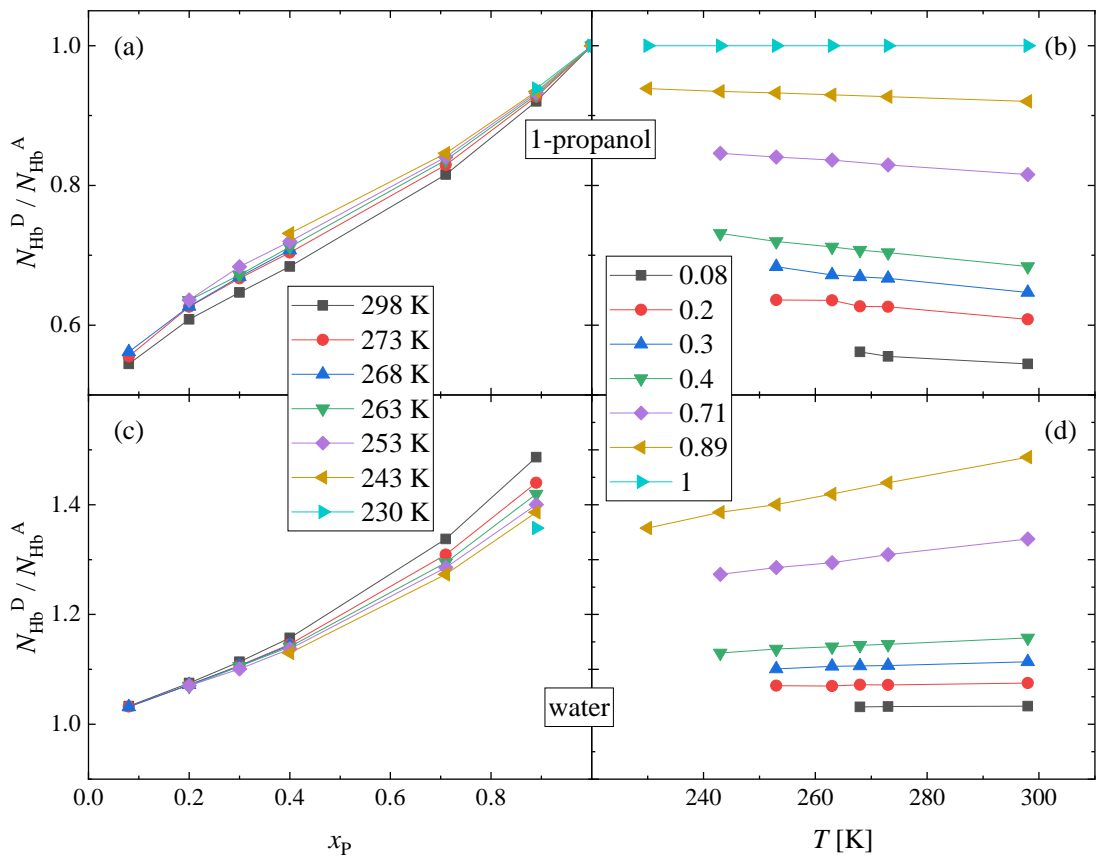

**Figure 9** Concentration (left panels) and temperature (right panels) dependence of the donor/acceptor ratio of (a,b) 1-propanol and (c,d) water molecules.

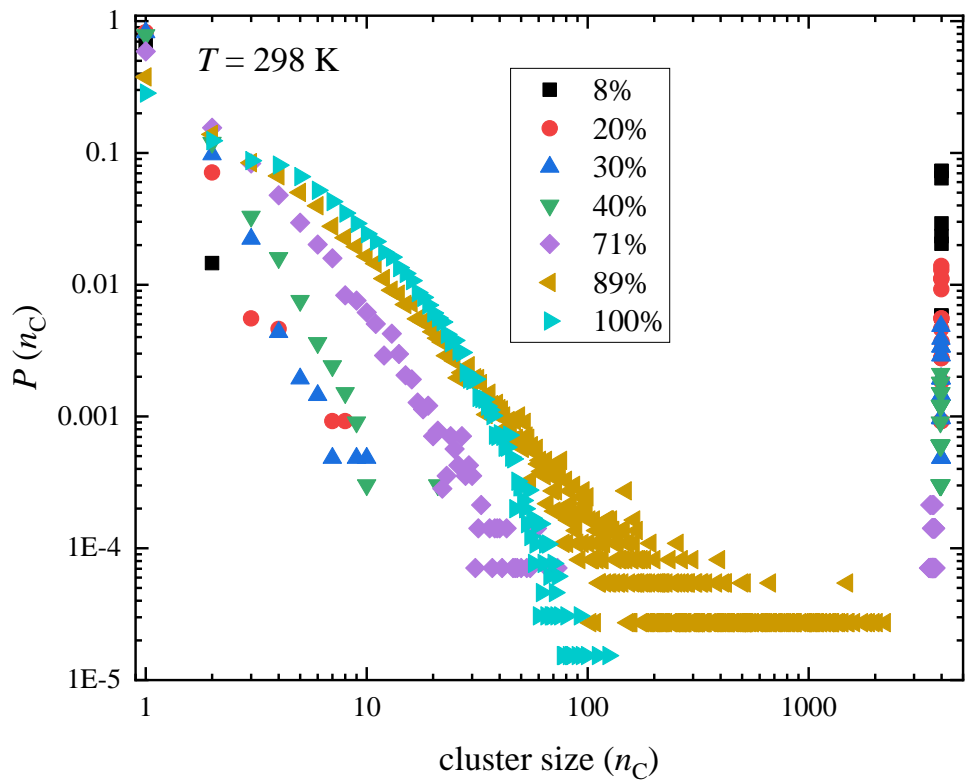

**Figure 10** Cluster size distributions in 1-propanol – water mixtures at $T$ = 298 K, considering H-bonds between any types of molecules.

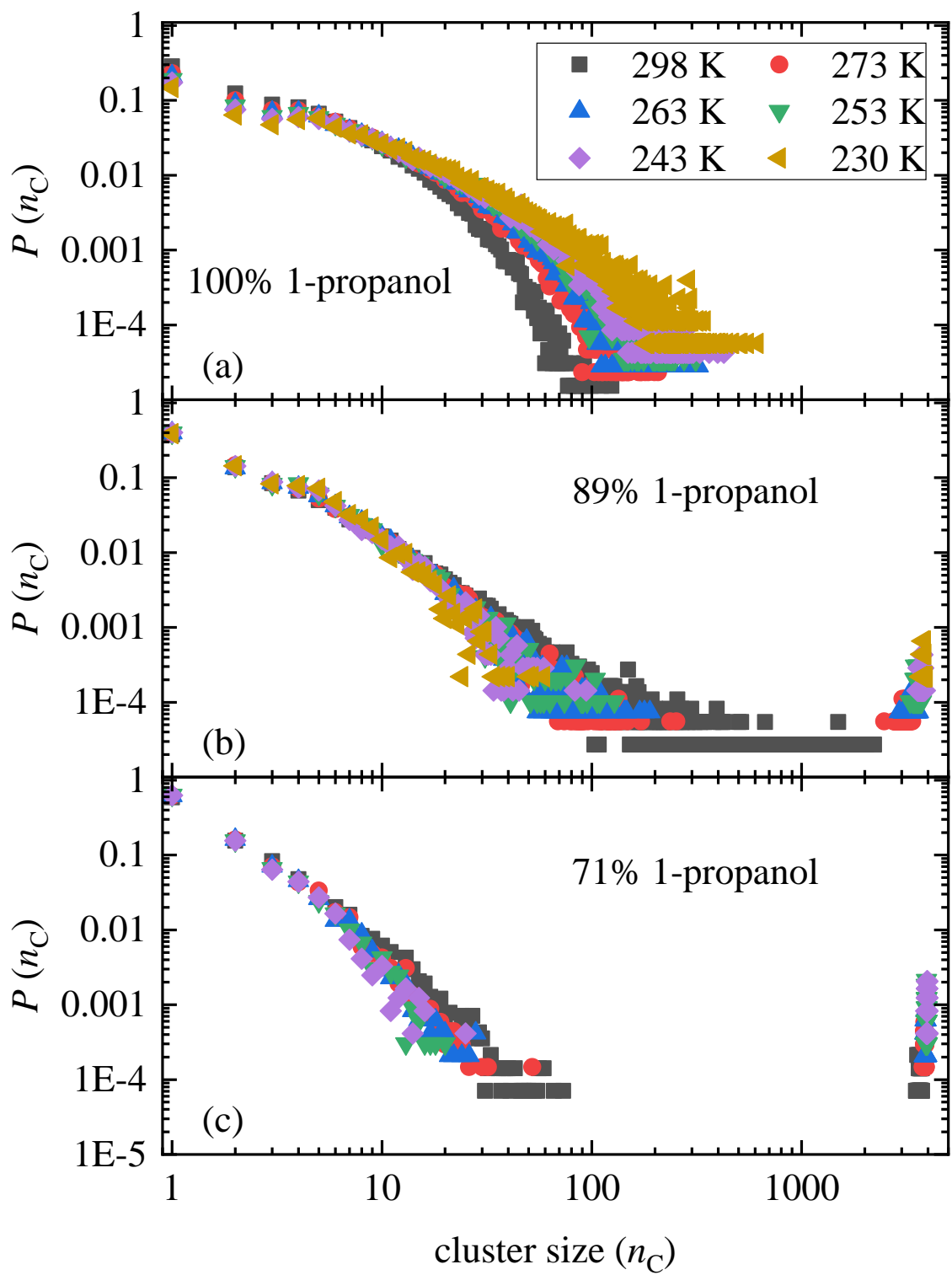

**Figure 11** Cluster size distributions in alcohol-rich 1-propanol – water mixtures at different temperatures, considering H-bonds between any types of molecules.

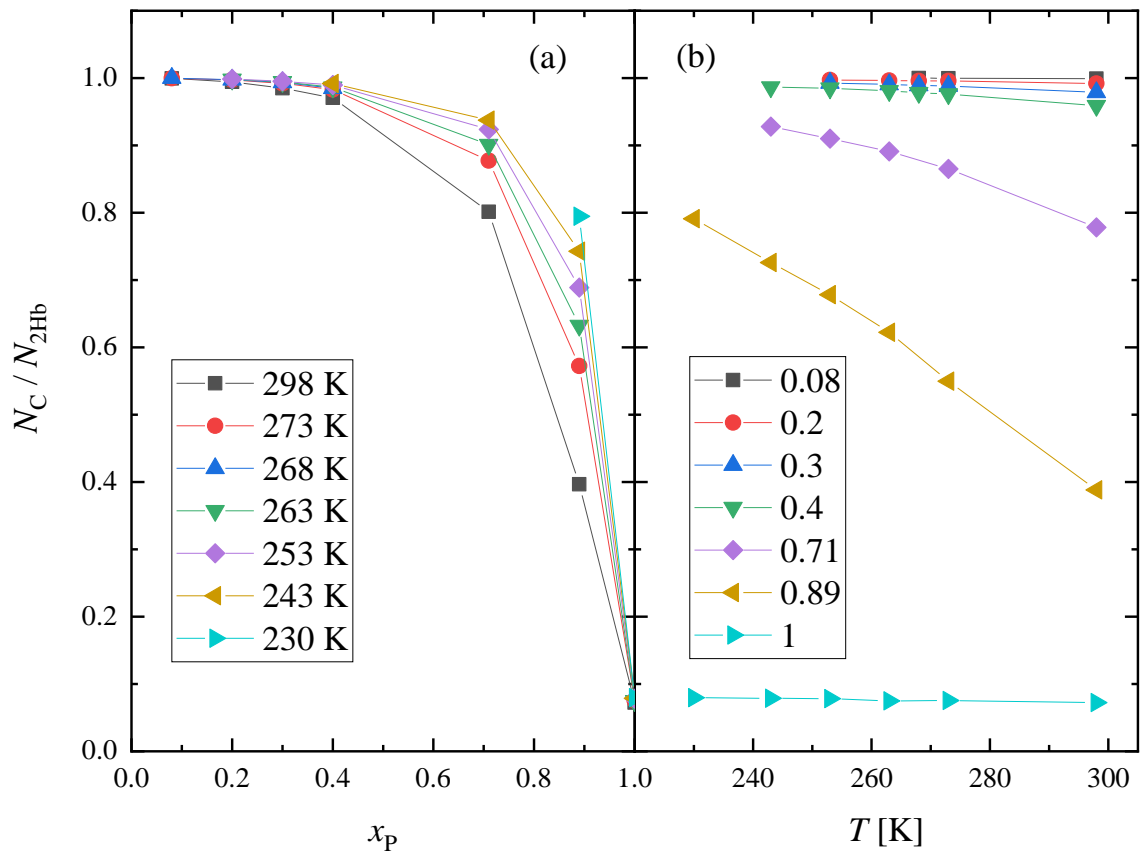

**Figure 12** (a) Concentration and (b) temperature dependence of the ratio of molecules participating in cycles ($N_C$) and molecules with at least 2 H-bonds ($N_{2Hb}$). (All molecules and all bond types are taken into account.)

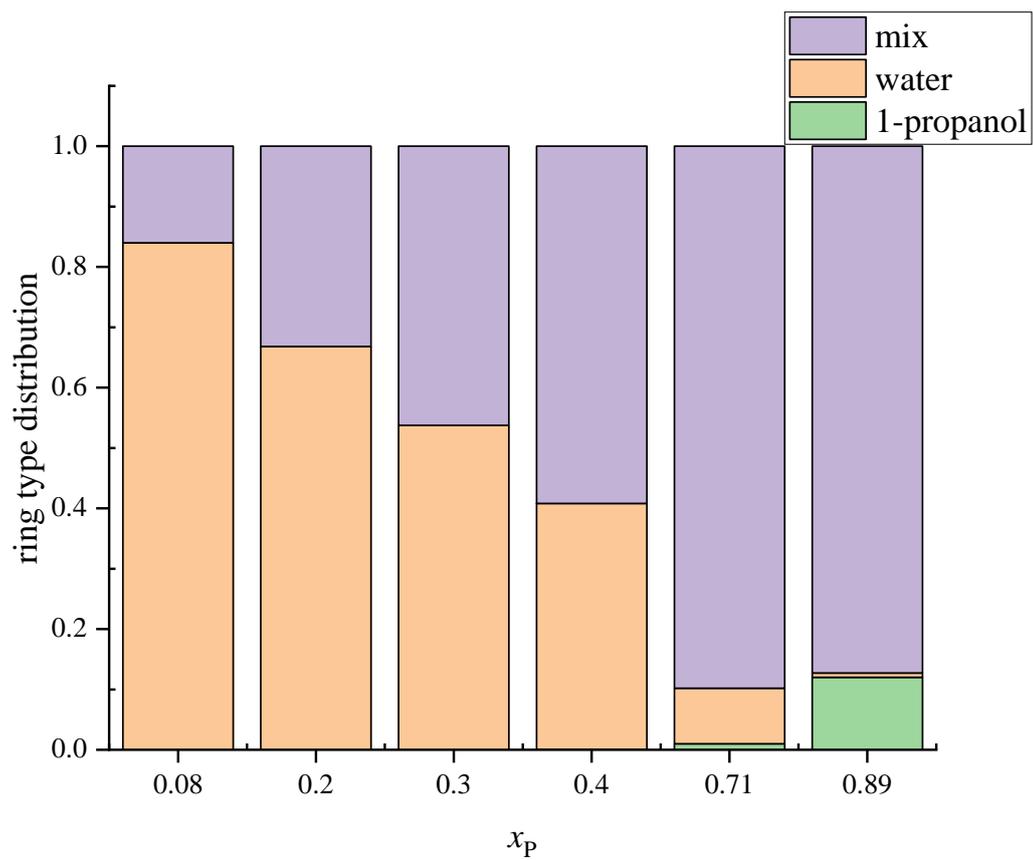

**Figure 13** Concentration dependence of the distribution of different ring types (rings containing only water molecules, only 1-propanol molecules, and both water and 1-propanol molecules) at 298 K.

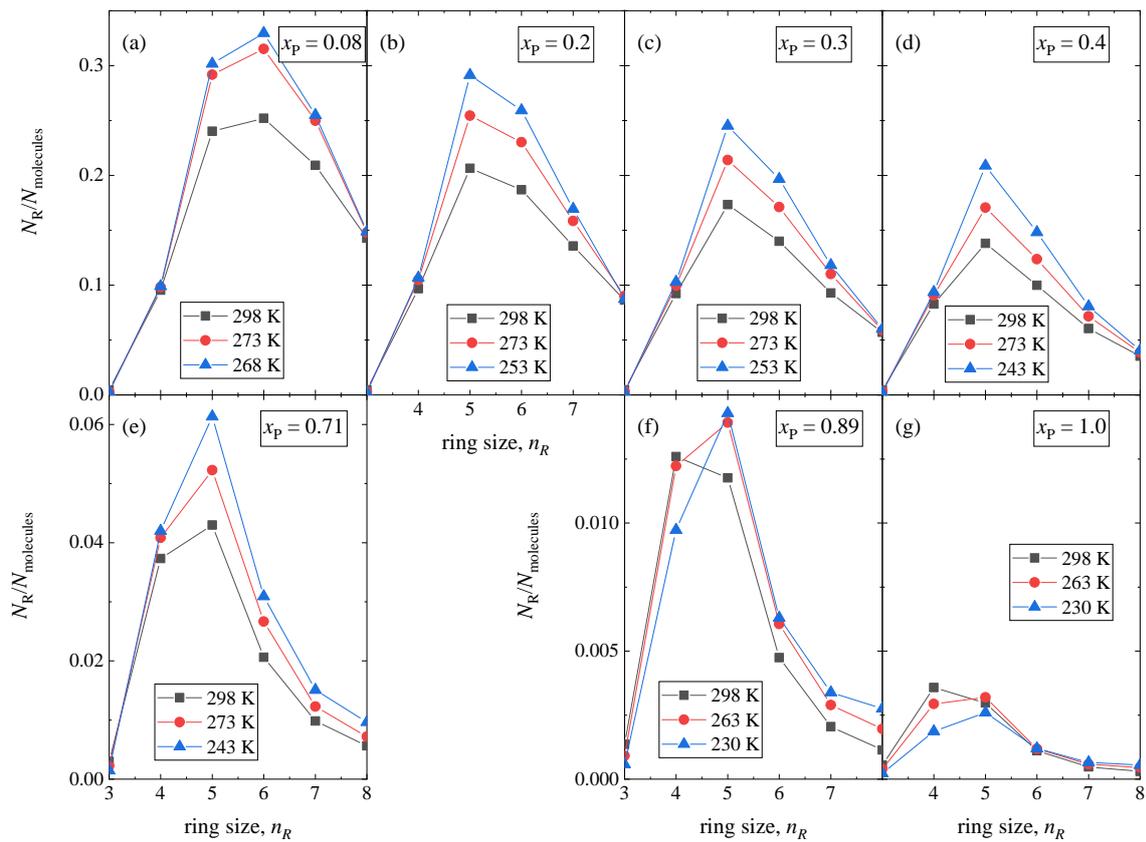

**Figure 14** Ring size distributions normalized by the number of molecules, as a function of temperature at various 1-propanol concentrations.

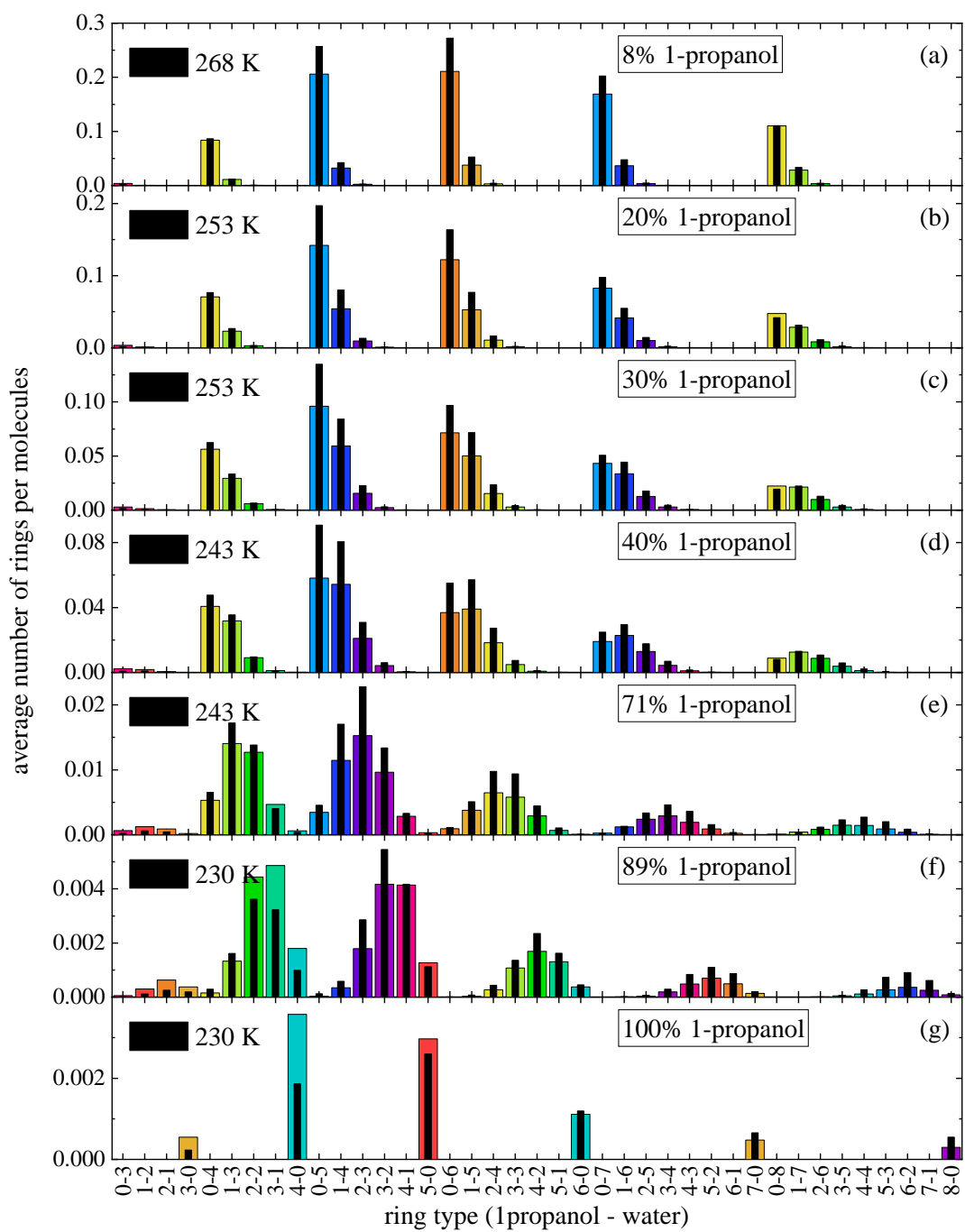

**Figure 15** Temperature and concentration dependence of the ring type distributions in 1-propanol – water mixtures. The numbers of rings are normalized by the number of molecules in the simulation box. Colored bars show the values at 298 K, whilst the black bars are the values at the lowest temperature investigated for the given composition: (a) 268 K, (b,c) 253 K, (d,e) 243 K, (f, g) 230 K.

# Supplementary Material

# Temperature-dependent structure of 1-propanol/water mixtures: X-ray diffraction experiments and computer simulations at low and high alcohol contents

Ildikó Pethes[1,a], László Pusztai[a,b], Koji Ohara[c], László Temleitner[a]

[a]Wigner Research Centre for Physics, Konkoly Thege út 29-33., H-1121 Budapest, Hungary
[b]International Research Organization for Advanced Science and Technology (IROAST), Kumamoto University, 2-39-1 Kurokami, Chuo-ku, Kumamoto 860-8555, Japan
[c]Diffraction and Scattering Division, Japan Synchrotron Radiation Research Institute (JASRI/Spring-8), 1-1-1 Kouto, Sayo-cho, Sayo-gun, Hyogo 679-5198, Japan

---

1 Corresponding author: e-mail: pethes.ildiko@wigner.hu

**Molecular dynamics simulations**

Classical molecular dynamics (MD) simulations were performed by the GROMACS software package (version 2018.2) [1]. Simulations were conducted in cubic simulation boxes with periodic boundary conditions. The number of 1-propanol and water molecules together in the simulation boxes was 4000. The initial box sizes were determined from the room temperature densities [2].

The all atom OPLS-AA [3] force field was applied for 1-propanol, whilst for water two models, the SPC/E [4] and the TIP4P/2005 [5] were tested. A schematic representation of the all-atom model of 1-propanol molecule is shown in Fig. S1. Non-bonded interactions were described by the 12-6 Lennard-Jones interaction and the Coulomb potential (see Eq. 1):

$$V_{ij}^{NB}(r_{ij}) = \frac{1}{4\pi\varepsilon_0}\frac{q_i q_j}{r_{ij}} + 4\varepsilon_{ij}\left[\left(\frac{\sigma_{ij}}{r_{ij}}\right)^{12} - \left(\frac{\sigma_{ij}}{r_{ij}}\right)^{6}\right], \tag{1}$$

where $r_{ij}$ is the distance between particles $i$ and $j$, $q_i$ and $q_j$ are the partial charges on these particles, $\varepsilon_0$ is the vacuum permittivity, and $\varepsilon_{ij}$ and $\sigma_{ij}$ represent the energy and distance parameters of the LJ potential. The LJ parameters ($\varepsilon_{ii}$ and $\sigma_{ii}$) and the partial charges applied ($q_i$) to the different atoms are collected in Table S1. The $\varepsilon_{ij}$ and $\sigma_{ij}$ parameters between unlike atoms are calculated as the geometric average of the homoatomic parameters (geometric combination rule, in accordance with the OPLS/AA force field).

Intramolecular non-bonded interactions between first and second neighbors were neglected, whereas between third neighbors (atoms separated by 3 bonds, e.g. $C_2 - H_P$ bond) they were reduced by a factor of 2. The intramolecular (or bonded) forces considered here are the bond-stretching (2-body), angle bending (3-body) and the dihedral angle torsion (4-body) interactions. Bond lengths in 1-propanol molecules were fixed by using the LINCS [6] algorithm, while bond angles and torsional angles were flexible. The rigid water geometry was handled by the SETTLE algorithm [7]. Bond lengths, equilibrium angles and force constants are given in Table S2.

The smoothed particle-mesh Ewald (SPME) method [8, 9] was used for treating Coulomb interactions, using a 20 Å cutoff in real space. Non-bonded LJ interactions were cut-off at 20 Å, with added long-range corrections to energy and pressure [10].

Initial configurations were obtained by placing the molecules into the simulation box randomly. At first an energy minimization was performed using the steepest-descent method. In the following simulation steps the equations of motion were integrated via the leapfrog algorithm, the time step applied was 0.5 fs. Energy minimization was followed by a 2.2 ns heat-treatment at $T = 320$ K to avoid the aggregation of the molecules. From this high temperature the system was "cooled down" with a 20 K/ns cooling rate using the 'simulated annealing' option of the GROMACS software. At each investigated temperature a short equilibration was performed first, followed by a long NpT (constant pressure and temperature) run, from which densities are calculated. That was followed by a second equilibration at constant volume and temperature (NVT ensemble). Finally, a 20 ns long NVT production run was performed. Trajectories were saved at every 2 ps. The simulation stages are collected in Table S3, together with the parameters of the applied barostat and thermostats.

Every 10$^{th}$ collected configuration was used to calculate partial radial distribution functions (PRDF, $g_{ij}(r)$), using the 'gmx_rdf' programme of the GROMACS software. The model structure factor can be obtained from the PRDFs, according to the Faber-Ziman formalism [16], by the following equations:

$$S_{ij}(Q) - 1 = \frac{4\pi\rho_0}{Q} \int_0^\infty r\big(g_{ij}(r) - 1\big)\sin(Qr)\mathrm{d}r, \tag{2}$$

where $Q$ is the amplitude of the scattering vector, and $\rho_0$ is the average number density.
The XRD ($F(Q)$) total scattering structure factors can be composed from the partial structure factors $S_{ij}(Q)$ as:

$$F(Q) = \sum_{i \leq j} w_{ij}(Q) S_{ij}(Q), \tag{3}$$

where $w_{ij}$ denotes the X-ray scattering weights, that are given by equation (4):

$$w_{ij}(Q) = (2 - \delta_{ij}) \frac{c_i c_j f_i(Q) f_j(Q)}{\sum_{ij} c_i c_j f_i(Q) f_j(Q)}. \tag{4}$$

Here $\delta_{ij}$ is the Kronecker delta, $c_i$ denotes atomic concentrations, $f_i(Q)$ is the atomic form factor. The total structure factors obtained from simulations were compared with the measured curves by calculating the goodness-of-fit ($R$-factor) values:

$$R = \frac{\sqrt{\sum_i \big(F_{mod}(Q_i) - F_{exp}(Q_i)\big)^2}}{\sqrt{\sum_i \big(F_{exp}(Q_i)\big)^2}} \tag{5}$$

where $Q_i$ denote the experimental points, 'mod' indicates the simulated and 'exp' the experimental curves.

**References**

[1] M.J. Abraham, T. Murtola, R. Schulz, S. Páll, J.C. Smith, B. Hess, E. Lindahl, GROMACS: High performance molecular simulations through multi-level parallelism from laptops to supercomputers, SoftwareX 1–2 (2015) 19–25. doi:10.1016/j.softx.2015.06.001.

[2] F.-M. Pang, C.-E. Seng, T.-T. Teng, M.H. Ibrahim, Densities and viscosities of aqueous solutions of 1-propanol and 2-propanol at temperatures from 293.15 K to 333.15 K, J. Mol. Liq. 136 (2007) 71–78. doi:10.1016/j.molliq.2007.01.003.

[3] W.L. Jorgensen, D.S. Maxwell, J. Tirado-Rives, Development and Testing of the OPLS All-Atom Force Field on Conformational Energetics and Properties of Organic Liquids, J. Am. Chem. Soc. 118 (1996) 11225–11236. doi:10.1021/ja9621760.

[4] H.J.C. Berendsen, J.R. Grigera, T.P. Straatsma, The missing term in effective pair potentials, J. Phys. Chem. 91 (1987) 6269–6271. doi:10.1021/j100308a038.

[5] J.L.F. Abascal, C. Vega, A general purpose model for the condensed phases of water: TIP4P/2005, J. Chem. Phys. 123 (2005) 234505. doi:10.1063/1.2121687.

[6] B. Hess, H. Bekker, H.J.C. Berendsen, J.G.E.M. Fraaije, LINCS: A linear constraint solver for molecular simulations, J. Comput. Chem. 18 (1997) 1463–1472. doi:10.1002/(SICI)1096-987X(199709)18:12<1463::AID-JCC4>3.0.CO;2-H.

[7] S. Miyamoto, P.A. Kollman, SETTLE: An analytical version of the SHAKE and RATTLE algorithm for rigid water models, J. Comput. Chem. 13 (1992) 952–962. doi:10.1002/jcc.540130805.

[8] T. Darden, D. York, L. Pedersen, Particle mesh Ewald: An N·log( N ) method for Ewald sums in large systems, J. Chem. Phys. 98 (1993) 10089–10092. doi:10.1063/1.464397.

[9] U. Essmann, L. Perera, M.L. Berkowitz, T. Darden, H. Lee, L.G. Pedersen, A smooth particle mesh Ewald method, J. Chem. Phys. 103 (1995) 8577–8593. doi:10.1063/1.470117.

[10] M.P. Allen, D.J. Tildesley, Computer Simulation of Liquids, Oxford University Press, Oxford, 1987.

[11] H.J.C. Berendsen, J.P.M. Postma, W.F. van Gunsteren, A. DiNola, J.R. Haak, Molecular dynamics with coupling to an external bath, J. Chem. Phys. 81 (1984) 3684–3690. doi:10.1063/1.448118.

[12] S. Nosé, A molecular dynamics method for simulations in the canonical ensemble, Mol. Phys. 52 (1984) 255–268. doi:10.1080/00268978400101201.

[13] W.G. Hoover, Canonical dynamics: Equilibrium phase-space distributions, Phys. Rev. A 31 (1985) 1695–1697. doi:10.1103/PhysRevA.31.1695.

[14] M. Parrinello, A. Rahman, Polymorphic transitions in single crystals: A new molecular dynamics method, J. Appl. Phys. 52 (1981) 7182–7190. doi:10.1063/1.328693.

[15] S. Nosé, M.L. Klein, Constant pressure molecular dynamics for molecular systems, Mol. Phys. 50 (1983) 1055–1076. doi:10.1080/00268978300102851.

[16] T.E. Faber, J.M. Ziman, A theory of the electrical properties of liquid metals, Philos. Mag. 11 (1965) 153–173. doi:10.1080/14786436508211931.

# Tables

**Table S1** Non-bonded force field parameters. Notation of the atoms in 1-propanol is shown in Fig. S1. Atom type names are used in the bond, angle and dihedral types. In the TIP4P/2005 water model there is a fourth (virtual) site (M). It is situated along the bisector of the $H_W$-$O_W$-$H_W$ angle and coplanar with the oxygen and hydrogens. The negative charge is placed on site M.

| atom | type name | $q$ [e] | $\sigma_{ii}$ [nm] | $\varepsilon_{ii}$ [kJ mol$^{-1}$] |
|---|---|---|---|---|
| \multicolumn{5}{c}{OPLS/AA 1-propanol [3]} | | | | |
| $C_1$ | C | 0.145 | 0.35 | 0.276144 |
| $C_2$ | C | -0.12 | 0.35 | 0.276144 |
| $C_3$ | C | -0.18 | 0.35 | 0.276144 |
| O | O | -0.683 | 0.312 | 0.71128 |
| $H_1, H_2, H_3$ | H | 0.06 | 0.25 | 0.12552 |
| $H_P$ | $H_P$ | 0.418 | 0 | 0 |
| \multicolumn{5}{c}{SPC/E water [4]} | | | | |
| $O_W$ | | -0.8476 | 0.3166 | 0.6502 |
| $H_W$ | | 0.4238 | 0 | 0 |
| \multicolumn{5}{c}{TIP4P/2005 water [5]} | | | | |
| $O_W$ | | 0 | 0.3159 | 0.7749 |
| $H_W$ | | 0.5564 | 0 | 0 |
| M | | -1.1128 | 0 | 0 |

**Table S2** Equilibrium bond lengths, angle bending parameters, and dihedral angle torsion force constants. Bond angle vibrations are represented by harmonic potentials, $\theta^0_{ijk}$ is the equilibrium angle and $k^a_{ijk}$ is the force constant. Dihedral torsion angles in the OPLS/AA force field are given as the first three terms of a Fourier series: $V(\varphi_{ijkl})=1/2(F_1(1+\cos\varphi_{ijkl})+F_2(1-\cos2\varphi_{ijkl})+F_3(1+\cos3\varphi_{ijkl}))$, where $\varphi_{ijkl}$ is the angle between the *ijk* and *jkl* planes. $\varphi_{ijkl} = 0$ corresponds to the 'cis' conformation (*i* and *l* are on the same side).

| Bond type | Bond length [nm] | | |
|---|---|---|---|
| C-H | 0.109 | | |
| C-C | 0.1529 | | |
| C-O | 0.141 | | |
| O-$H_P$ | 0.0945 | | |
| $O_W$-$H_W$ SPC/E | 0.1 | | |
| $O_W$-$H_W$ TIP4P/2005 | 0.09572 | | |
| $O_W$-M TIP4P/2005 | 0.01546 | | |
| Angle type | $\theta^0_{ijk}$ [degree] | $k^a_{ijk}$ [kJ mol$^{-1}$ rad$^{-2}$] | |
| H-C-H | 107.8 | 276.144 | |
| C-O-$H_P$ | 108.5 | 460.24 | |
| H-C-C | 110.7 | 313.800 | |
| C-C-O | 109.5 | 418.4 | |
| C-C-C | 112.7 | 488.273 | |
| H-C-O | 109.5 | 292.88 | |
| $H_W$-$O_W$-$H_W$ SPC/E | 109.47 | -- (rigid) | |
| $H_W$-$O_W$-$H_W$ TIP4P/2005 | 104.52 | -- (rigid) | |
| Dihedral type | $F_1$ [kJ mol$^{-1}$] | $F_2$ [kJ mol$^{-1}$] | $F_3$ [kJ mol$^{-1}$] |
| H-C-O-$H_P$ | 0 | 0 | 1.8828 |
| H-C-C-O | 0 | 0 | 1.9581 |
| C-C-O-$H_P$ | -1.4895 | -0.728016 | 2.058528 |
| C-C-C-O | 7.1588 | -2.092 | 2.773992 |
| H-C-C-H | 0 | 0 | 1.2552 |
| C-C-C-H | 0 | 0 | 1.2552 |

**Table S3** Simulation stages after the energy minimization. In every step the leap-frog algorithm was used with a time step of 0.5 fs. ($T_{inv}$ denotes the investigated temperature.)

|  | Temperature [K] | Ensemble | Run time [ns] | Thermostat / time constant [ps] | Barostat / time constant [ps] |
|---|---|---|---|---|---|
| High temperature equilibration | 320 K | NVT | 0.2 | Berendsen[a] / 0.1 | - |
| Heat treatment | 320 K | NpT | 2 | Nose-Hoover[b] / 0.5 | Parrinello-Rahman[c] / 2.0 |
| Cooling down | 320 K → $T_{inv}$ | NpT | 50 ps / K | Berendsen / 0.1 | Berendsen[a] / 0.3 |
| Equilibration | $T_{inv}$ | NpT | 0.2 | Berendsen / 0.1 | Berendsen / 0.3 |
| Equilibration and density determination | $T_{inv}$ | NpT | 10 | Nose-Hoover / 0.5 | Parrinello-Rahman / 2.0 |
| Equilibration | $T_{inv}$ | NVT | 2 | Nose-Hoover / 0.5 | - |
| Production | $T_{inv}$ | NVT | 20 | Nose-Hoover / 0.5 | - |

[a] Ref [11]
[b] Refs. [12,13]
[c] Refs. [14,15]

**Table S4** Densities (in kg/m³) of 1-propanol – water mixtures obtained by MD simulations using the TIP4P/2005 water model.

| | Temperature [K] | | | | | | | | | |
|---|---|---|---|---|---|---|---|---|---|---|
| $x_P$ | 298 | 273 | 268 | 263 | 253 | 243 | 230 | 180 | 150 | 115 |
| 0.08 | 965 | 981 | 983 | | | | | | | |
| 0.2 | 921 | 943 | 947 | 951 | 959 | | | | | |
| 0.3 | 893 | 917 | 921 | 926 | 935 | | | | | |
| 0.4 | 872 | 896 | 901 | 906 | 915 | 925 | | | | |
| 0.71 | 828 | 854 | | 864 | 874 | 884 | | | | |
| 0.89 | 812 | 836 | | 846 | 856 | 866 | 879 | | | |
| 1 | 801 | 827 | | 837 | 847 | 856 | 869 | 917 | 937 | 954 |

**Table S5** Densities (in kg/m$^3$) of the 1-propanol – water mixtures obtained by MD simulations using the SPC/E water model.

| $x_P$ | Temperature [K] | | | | | | | | | |
|---|---|---|---|---|---|---|---|---|---|---|
| | 298 | 273 | 268 | 263 | 253 | 243 | 230 | 180 | 150 | 115 |
| 0.08 | 966 | 986 | 990 | | | | | | | |
| 0.2 | 917 | 941 | 946 | 951 | 960 | | | | | |
| 0.3 | 889 | 914 | 919 | 924 | 934 | | | | | |
| 0.4 | 867 | 893 | 898 | 904 | 913 | 923 | | | | |
| 0.71 | 825 | 852 | | 862 | 872 | 882 | | | | |
| 0.89 | 809 | 836 | | 846 | 856 | 865 | 878 | | | |
| 1 | 801 | 827 | | 837 | 847 | 856 | 869 | 917 | 937 | 954 |

**Table S6** Goodness of the fit values ($R$-factors, in %) calculated from the comparison of the experimental and simulated XRD structure factors of 1-propanol – water mixtures. Bold (upper) numbers are from simulations using the TIP4P/2005 water model, italic (lower) numbers are obtained using the SPC/E water model.

| $x_P$ | Temperature [K] | | | | | | | | | |
|---|---|---|---|---|---|---|---|---|---|---|
| | 298 | 273 | 268 | 263 | 253 | 243 | 230 | 180 | 150 | 115 |
| 0.08 | **19.7** | | **15.6** | | | | | | | |
| | *21.7* | | *19.7* | | | | | | | |
| 0.2 | **18.8** | **18.6** | **16.1** | **16.6** | **16.7** | | | | | |
| | *21.2* | *21.7* | *20.1* | *20.4* | *20.3* | | | | | |
| 0.3 | **20.6** | **20.2** | **18.7** | **18.4** | **17.8** | | | | | |
| | *23.5* | *23.0* | *21.0* | *21.2* | *20.5* | | | | | |
| 0.4 | **19.4** | **19.8** | **18.9** | **17.0** | **17.0** | **18.3** | | | | |
| | *22.7* | *21.8* | *20.4* | *19.1* | *18.8* | *20.0* | | | | |
| 0.71 | **33.9** | | | **31.2** | | **29.8** | | | | |
| | *35.0* | | | *31.7* | | *30.7* | | | | |
| 0.89 | **17.7** | | **17.4** | | | | **17.7** | | | |
| | *18.4* | | *17.8* | | | | *17.8* | | | |
| 1 | 18.3 | | 16.7 | | | | | 17.4 | 20.0 | 17.7 | 21.0 |

**Figures**

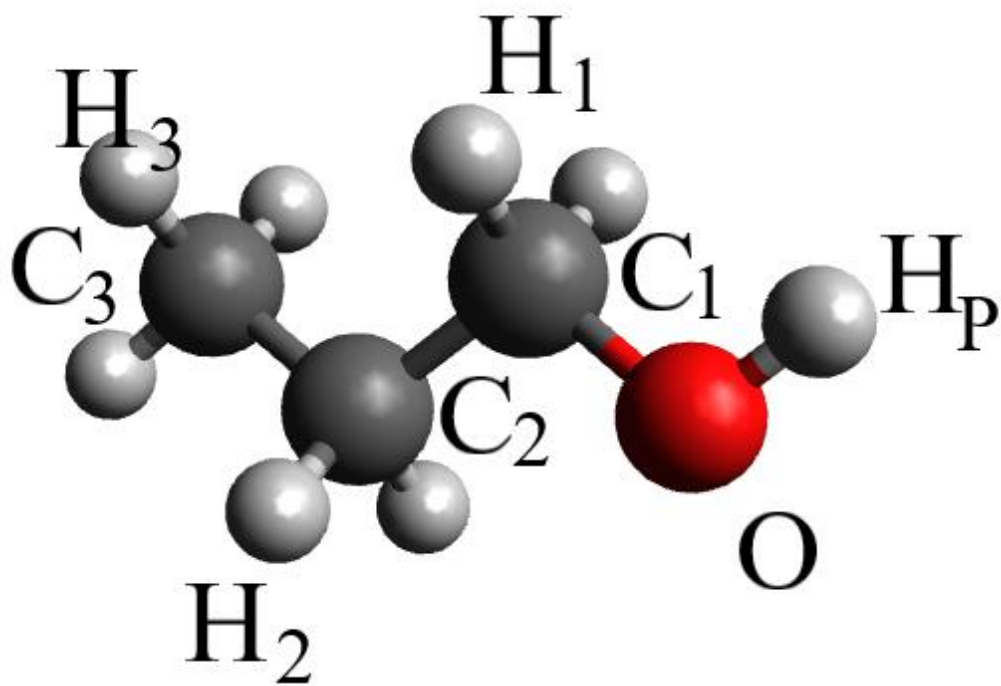

**Figure S1** Schematic representation of the all-atom model of the 1-propanol molecule.

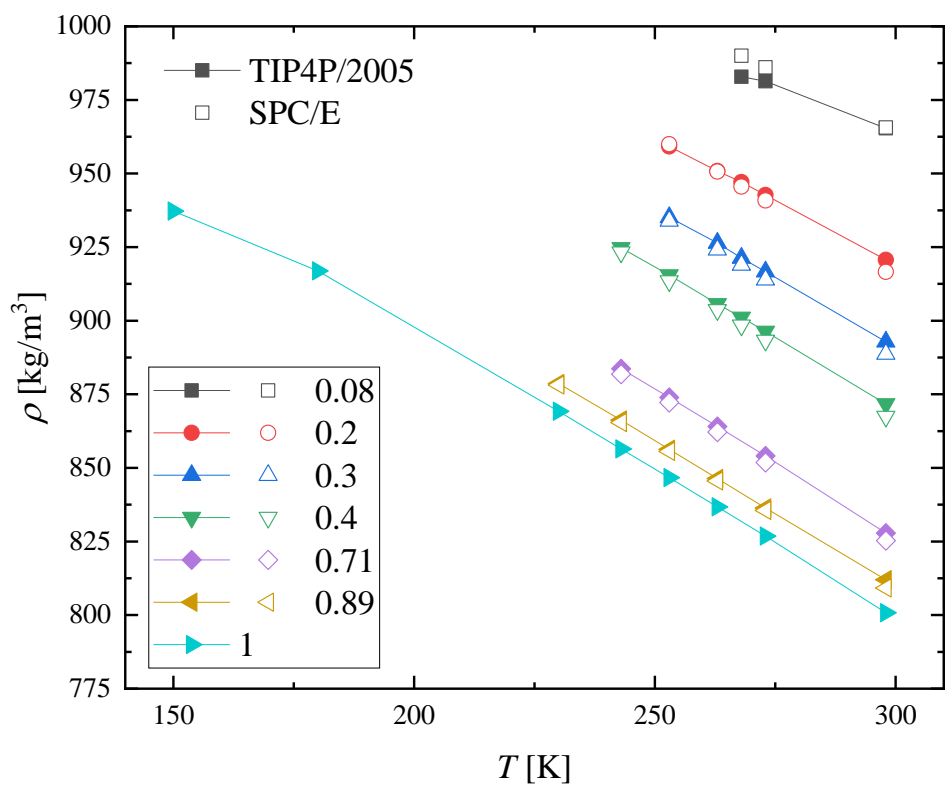

**Figure S2** Temperature dependence of the densities of 1-propanol – water mixtures obtained by MD simulations using the TIP4P/2005 (solid symbols) and the SPC/E (open symbols) water models. (The lines are just guides to the eye.)

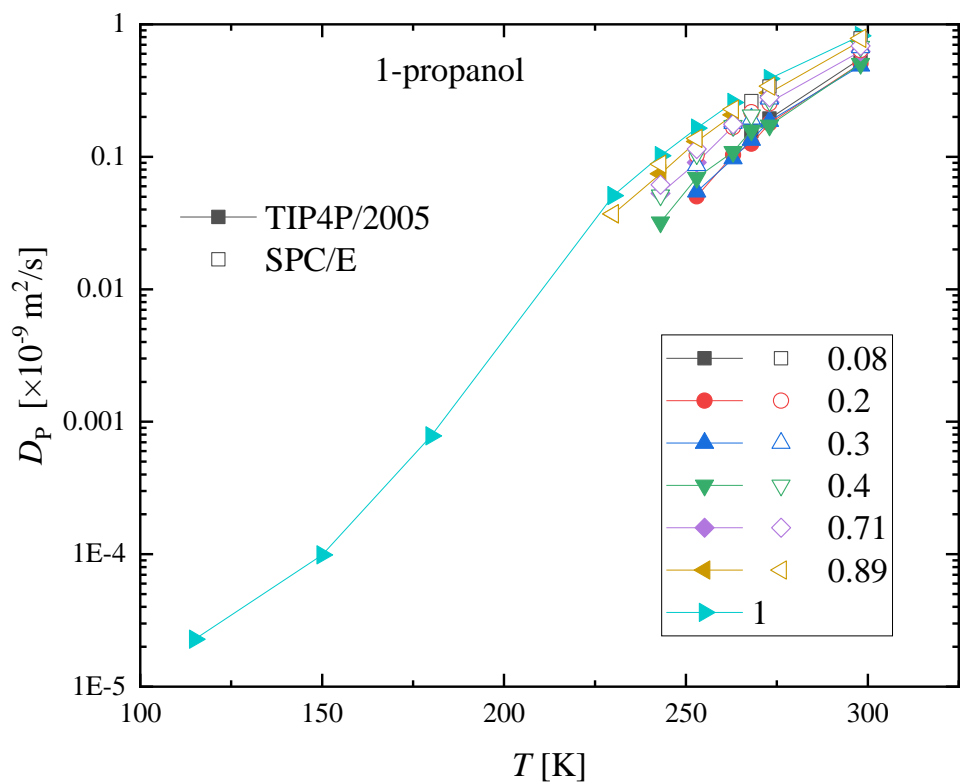

**Figure S3** Temperature dependence of the self-diffusion coefficient of 1-propanol molecules in 1-propanol – water mixtures, as obtained by MD simulations using the TIP4P/2005 (solid symbols) and the SPC/E (open symbols) water models. (The lines are just guides to the eye.)

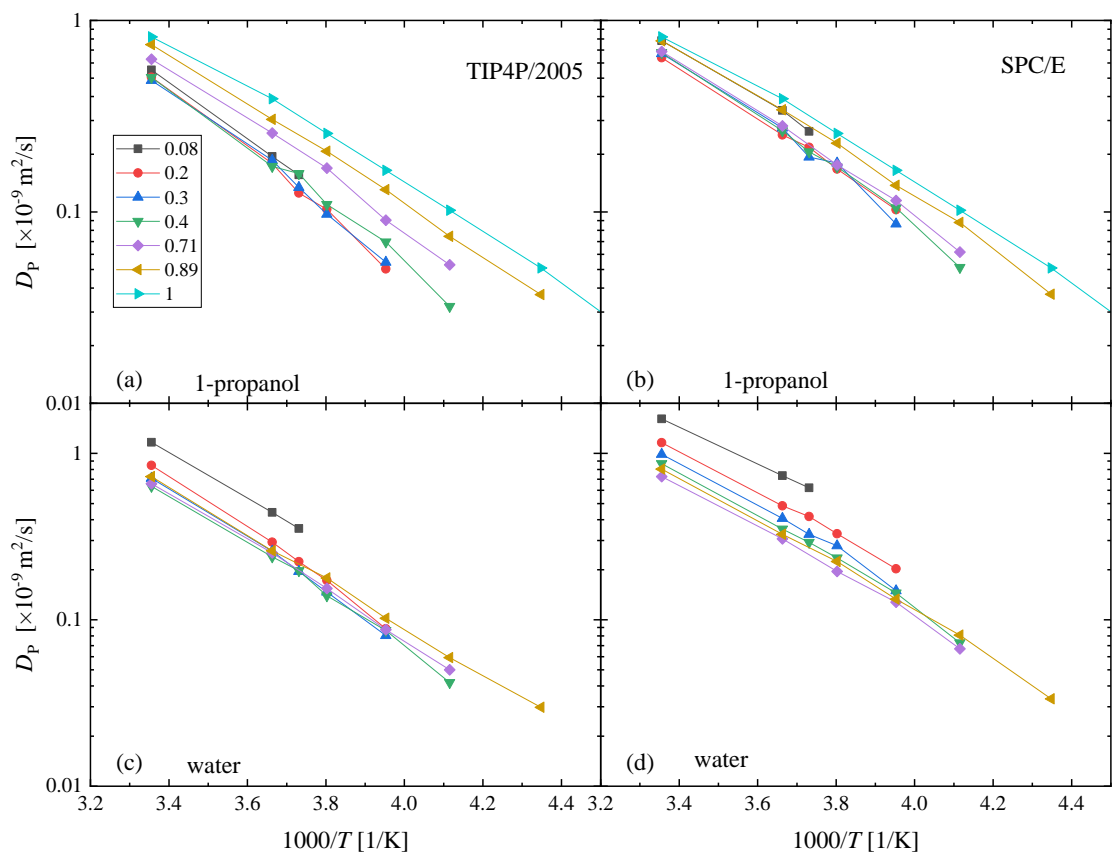

**Figure S4** Arrhenius-type plots of the self-diffusion coefficients of (a,b) 1-propanol and (c,d) water molecules in 1-propanol – water mixtures, as obtained by MD simulations using the (a,c) TIP4P/2005 and the (b,d) SPC/E water models.

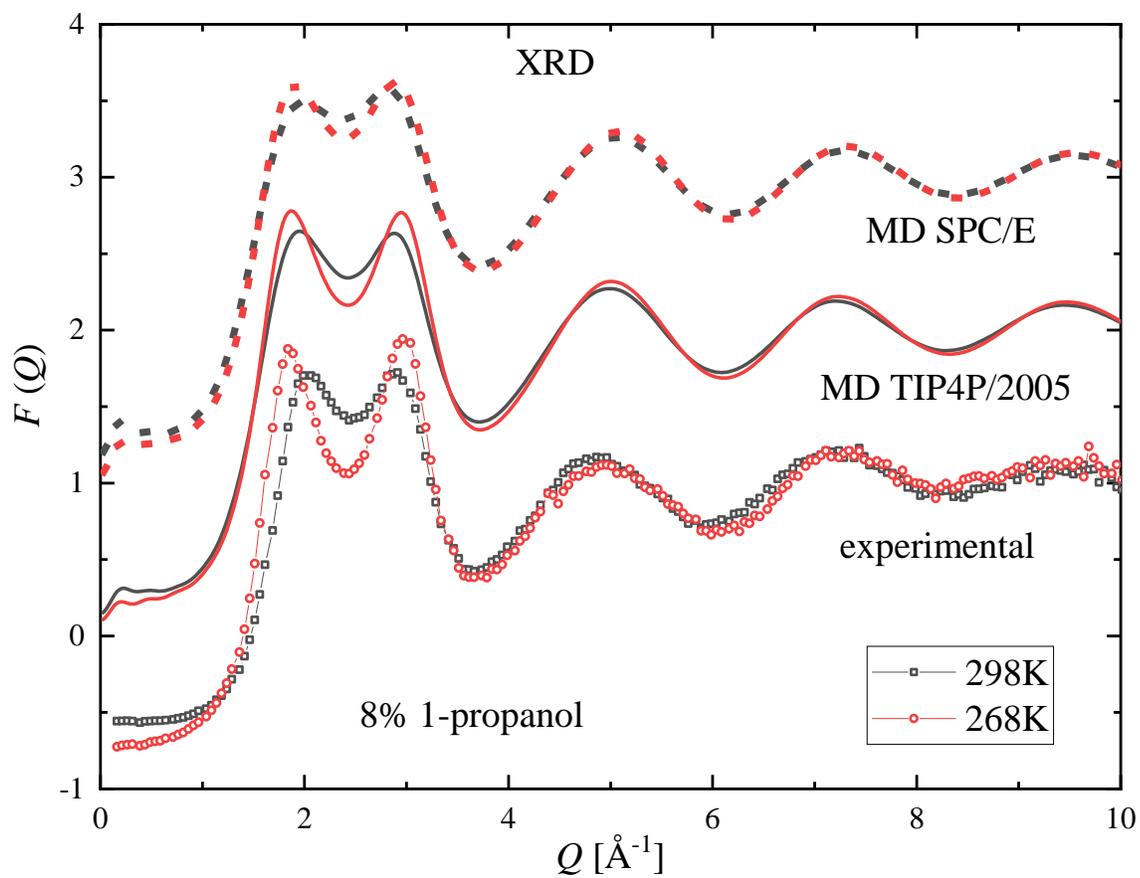

**Figure S5** Temperature dependence of measured (symbols) and simulated (lines) XRD structure factors of the 1-propanol – water mixture with 8 mol % 1-propanol content: comparison of trends observed upon cooling. Black lines and symbols: 298 K, red: 268 K. Simulated curves were obtained by using the TIP4P/2005 (solid lines) and SPC/E (dashed lines) water models. (Curves are shifted for clarity.)

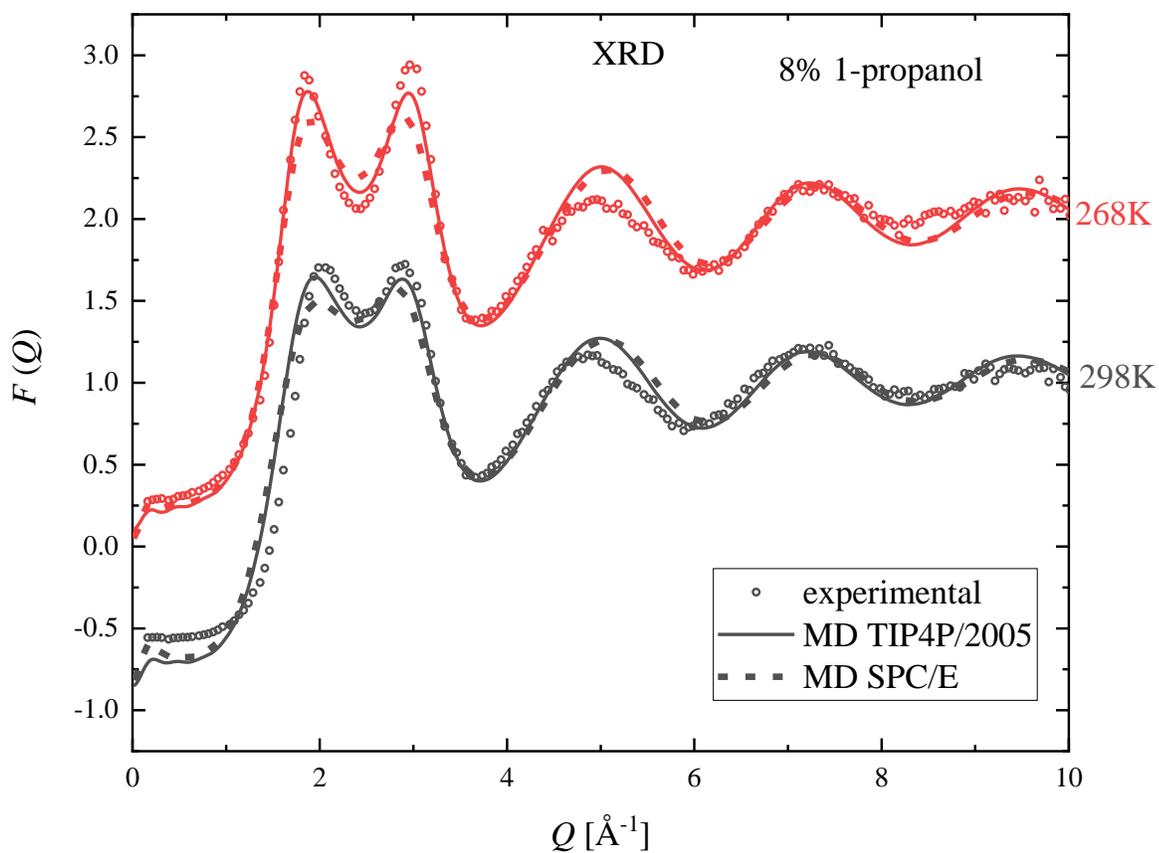

**Figure S6** Comparison of XRD structure factors obtained from experiments (symbols) and simulations using TIP4P/2005 (solid lines) and SPC/E (dashed lines) water models for the 1-propanol – water mixture with 8 mol % 1-propanol content. (Curves are shifted for clarity.)

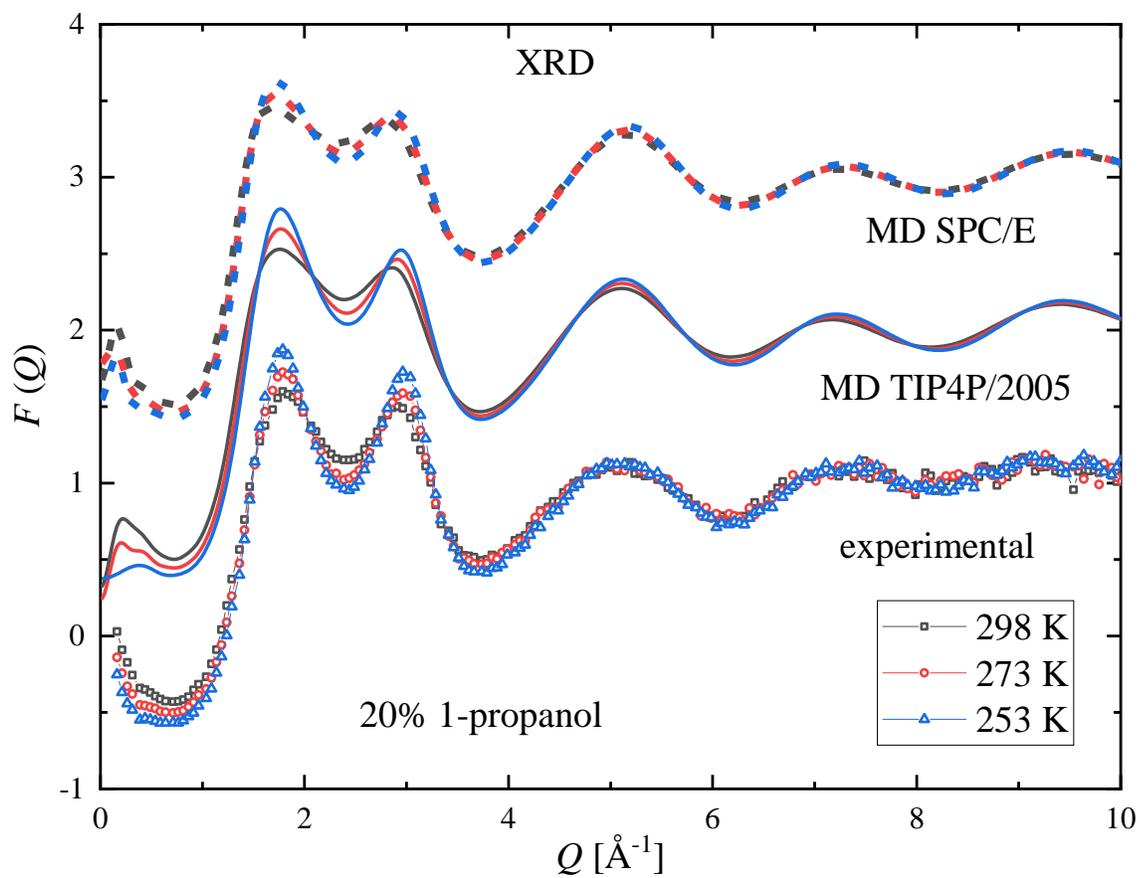

**Figure S7** Temperature dependence of measured (symbols) and simulated (lines) XRD structure factors for the 1-propanol – water mixture with 20 mol % 1-propanol content: comparison of trends observed upon cooling. Black lines and symbols: 298 K, red: 273 K, blue: 253 K. Simulated curves were obtained by using the TIP4P/2005 (solid lines) and SPC/E (dashed lines) water models. (Curves are shifted for clarity.)

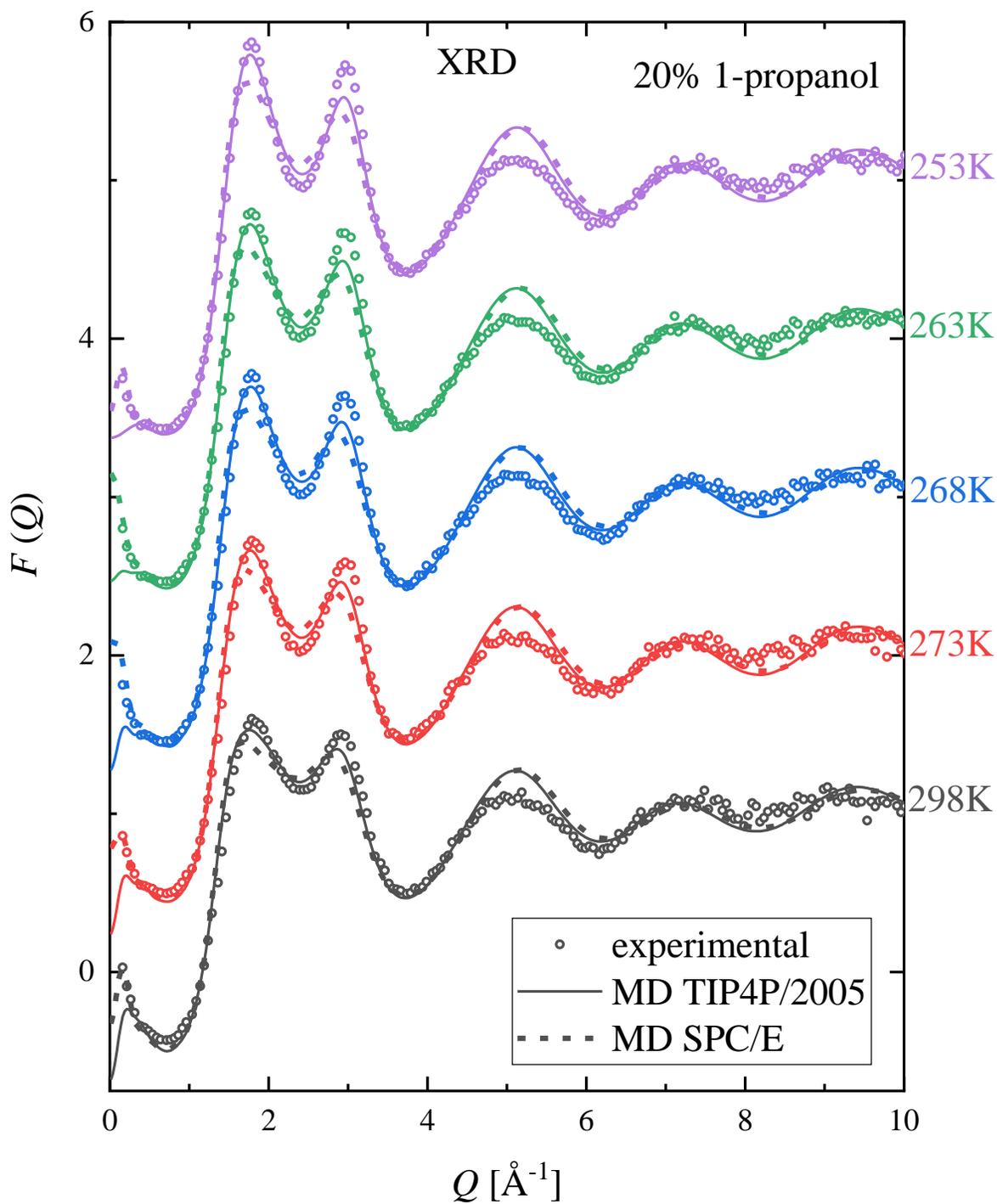

**Figure S8** Comparison of XRD structure factors obtained from experiments (symbols) and simulations using TIP4P/2005 (solid lines) and SPC/E (dashed lines) water models for the 1-propanol – water mixture with 20 mol % 1-propanol content. (Curves are shifted for clarity.)

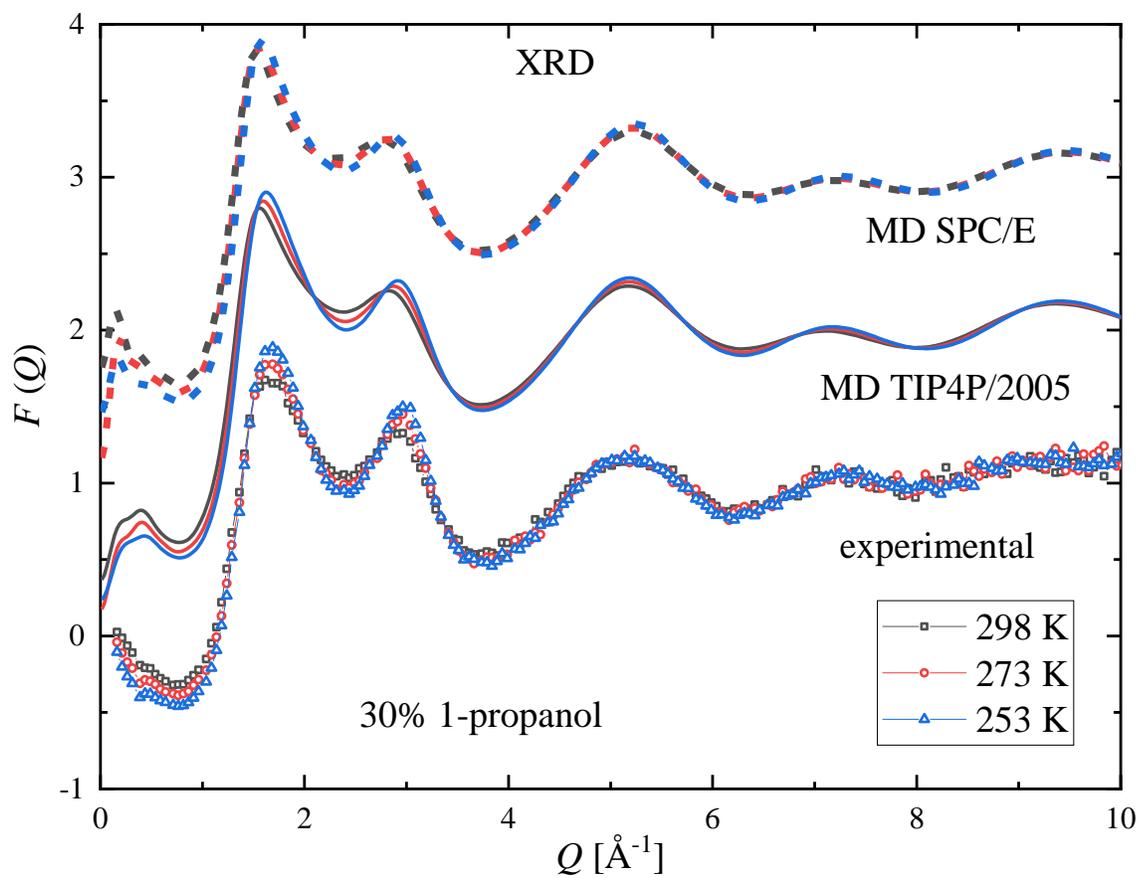

**Figure S9** Temperature dependence of measured (symbols) and simulated (lines) XRD structure factors for the 1-propanol – water mixture with 30 mol % 1-propanol content: comparison of trends observed upon cooling. Black lines and symbols: 298 K, red: 273 K, blue: 253 K. Simulated curves were obtained by using the TIP4P/2005 (solid lines) and SPC/E (dashed lines) water models. (Curves are shifted for clarity.)

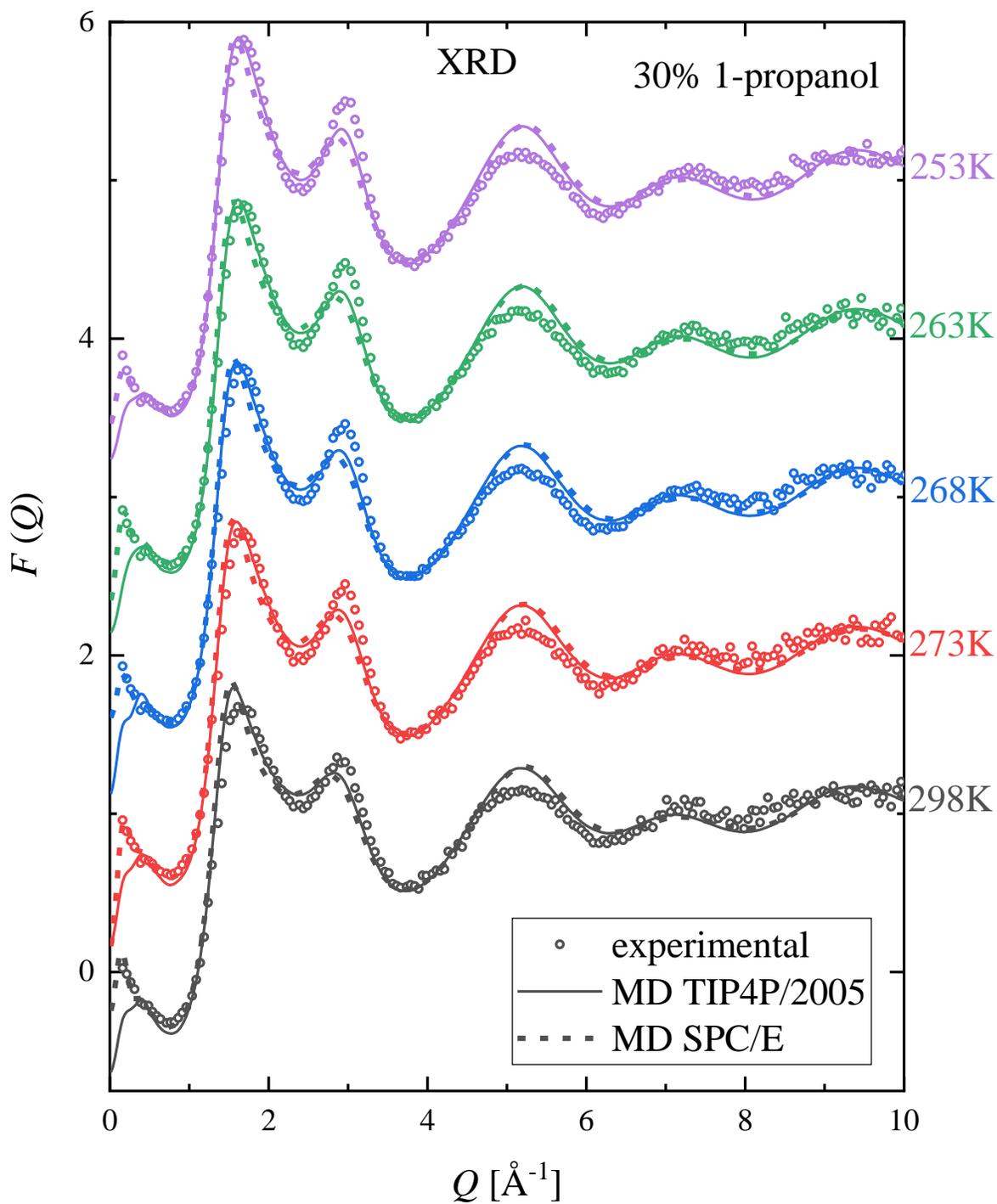

**Figure S10** Comparison of XRD structure factors obtained from experiments (symbols) and simulations using TIP4P/2005 (solid lines) and SPC/E (dashed lines) water models for the 1-propanol – water mixture with 30 mol % 1-propanol content. (Curves are shifted for clarity.)

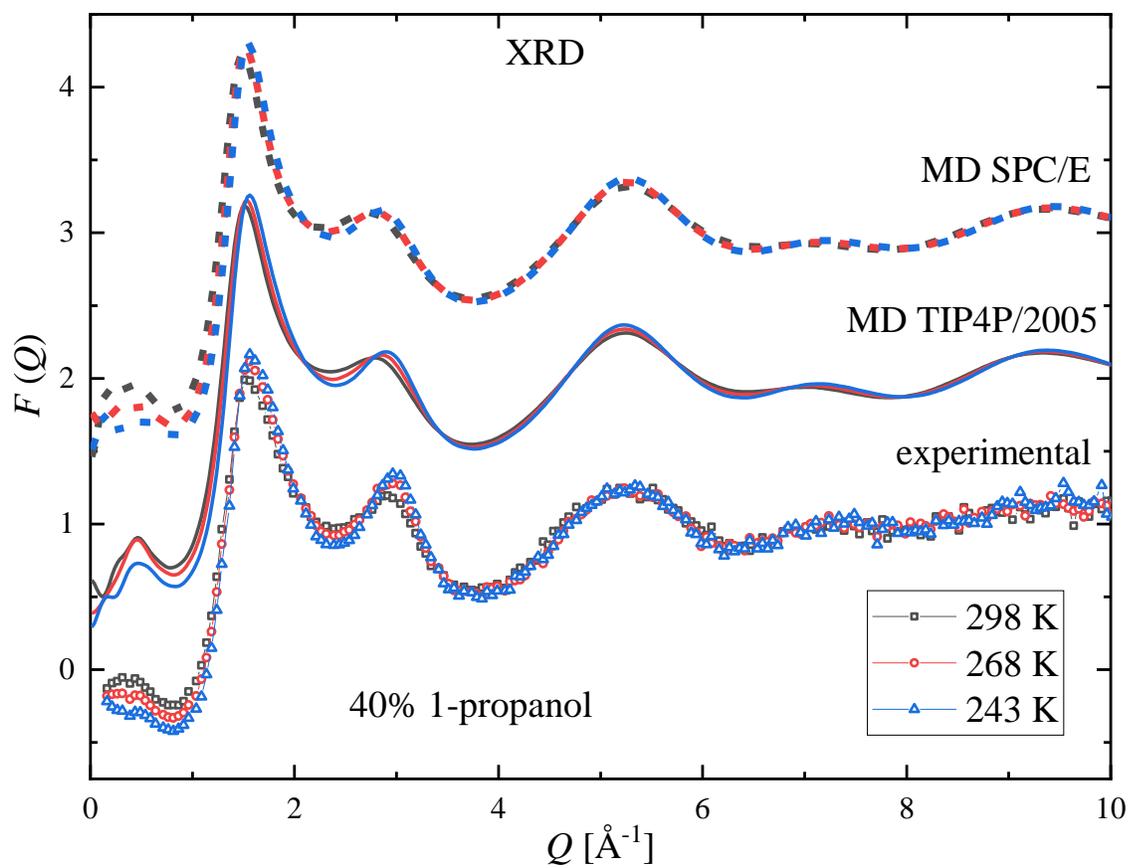

**Figure S11** Temperature dependence of measured (symbols) and simulated (lines) XRD structure factors for the 1-propanol – water mixture with 40 mol % 1-propanol content: comparison of trends observed upon cooling. Black lines and symbols: 298 K, red: 268 K, blue: 243 K. Simulated curves were obtained by using the TIP4P/2005 (solid lines) and SPC/E (dashed lines) water models. (Curves are shifted for clarity.)

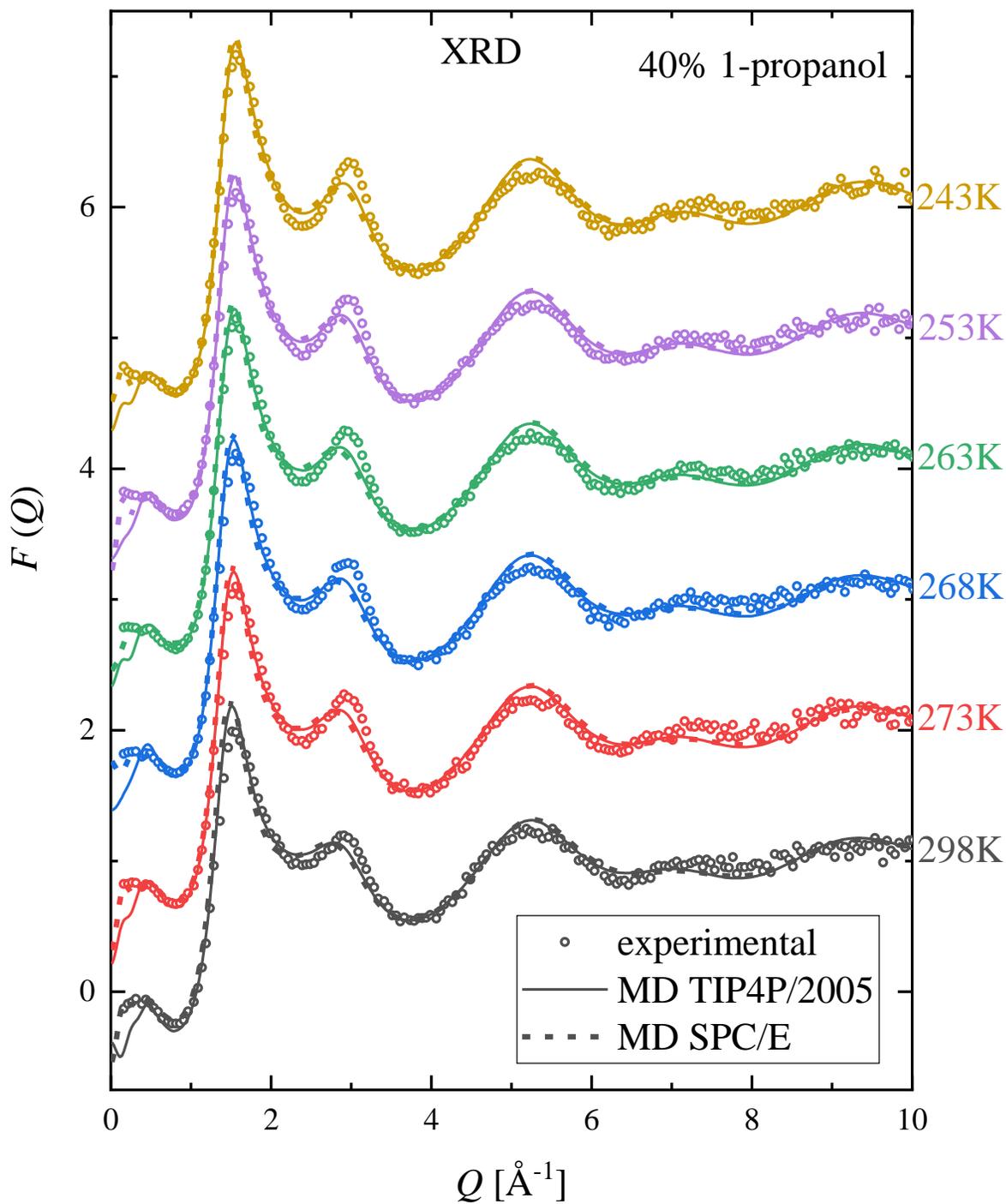

**Figure S12** Comparison of XRD structure factors obtained from experiments (symbols) and simulations using TIP4P/2005 (solid lines) and SPC/E (dashed lines) water models for the 1-propanol – water mixture with 40 mol % 1-propanol content. (Curves are shifted for clarity.)

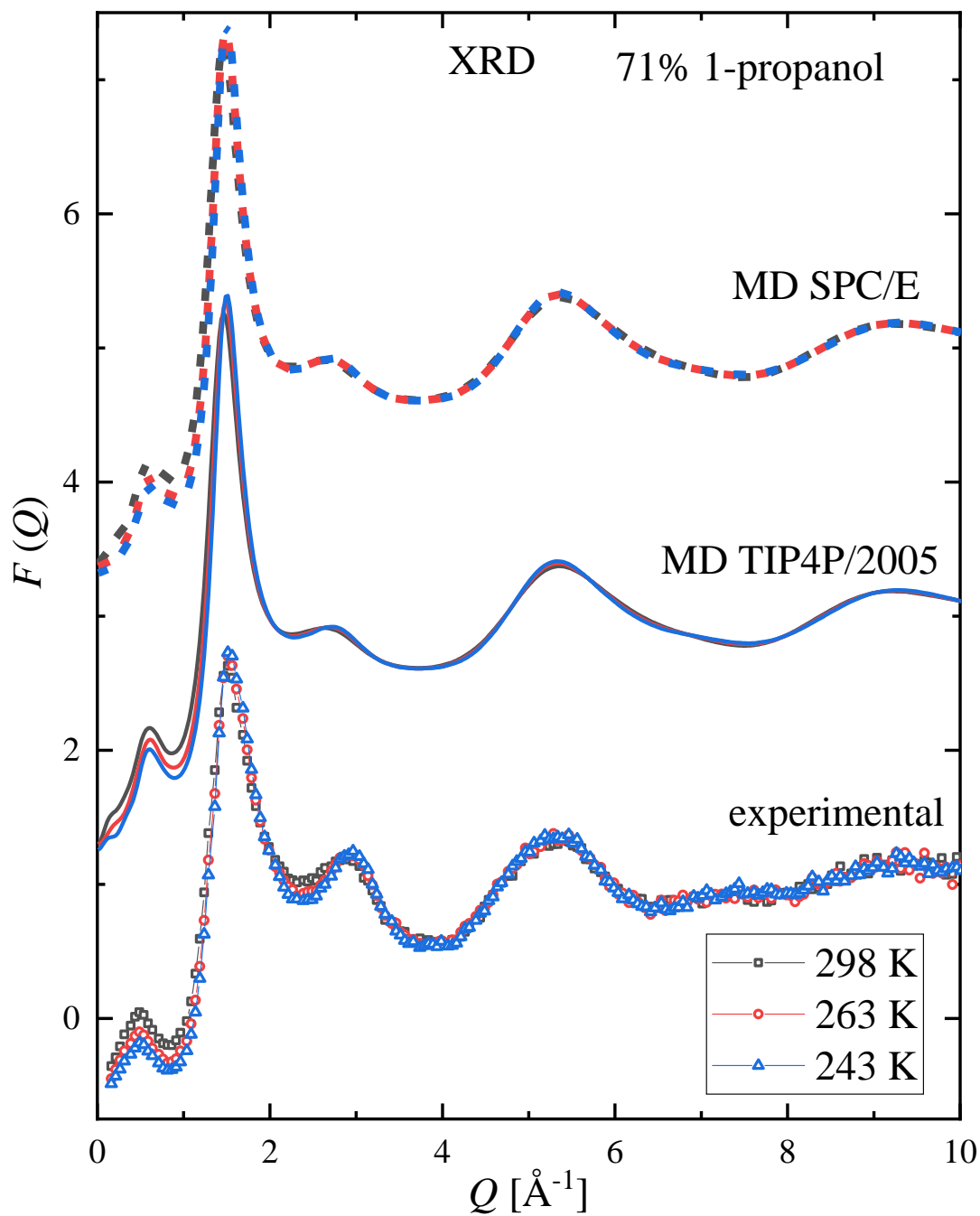

**Figure S13** Temperature dependence of measured (symbols) and simulated (lines) XRD structure factors for the 1-propanol – water mixture with 71 mol % 1-propanol content: comparison of trends observed upon cooling. Black lines and symbols: 298 K, red: 263 K, blue: 243 K. Simulated curves were obtained by using the TIP4P/2005 (solid lines) and SPC/E (dashed lines) water models. (Curves are shifted for clarity.)

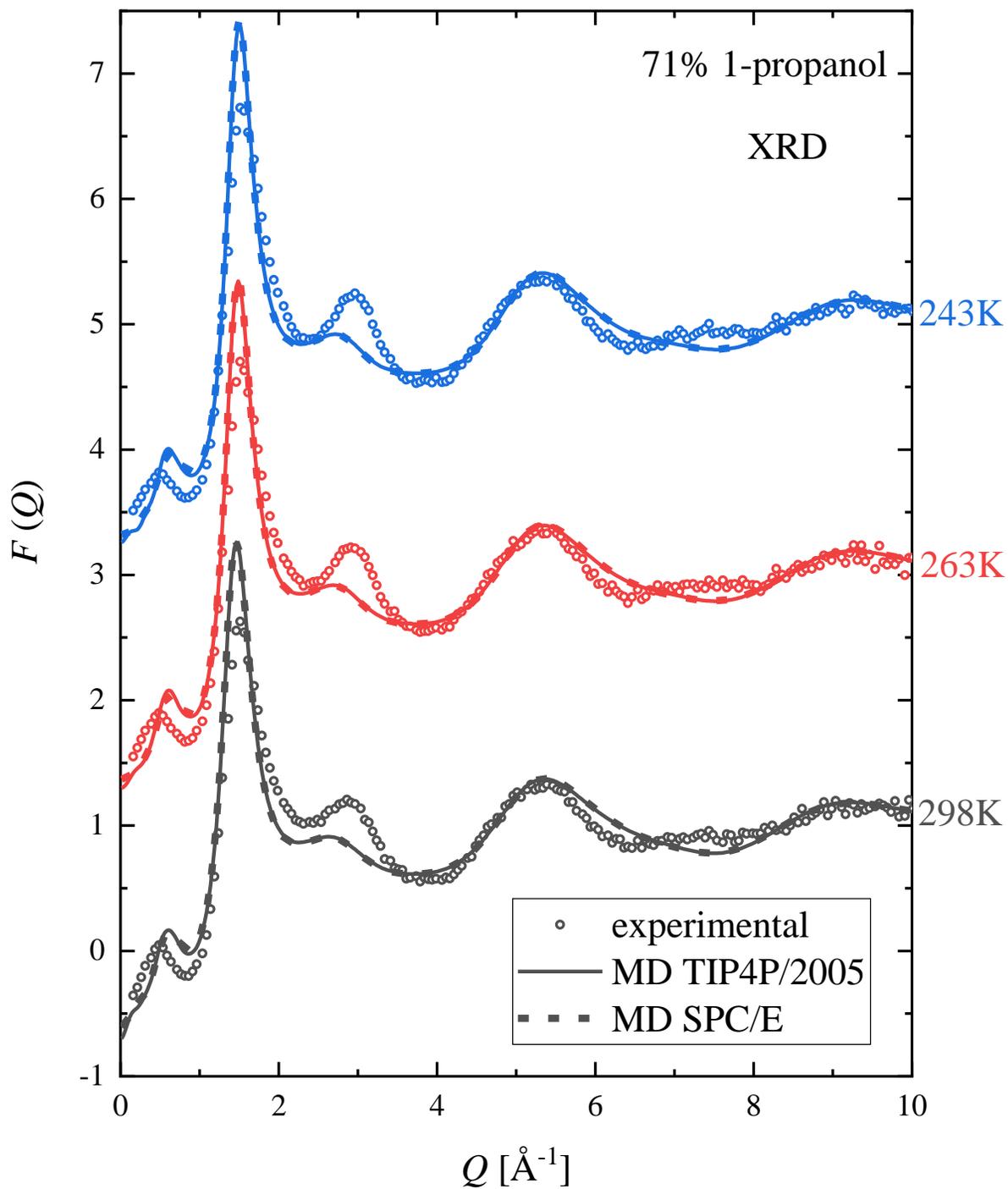

**Figure S14** Comparison of XRD structure factors obtained from experiments (symbols) and simulations using TIP4P/2005 (solid lines) and SPC/E (dashed lines) water models for the 1-propanol – water mixture with 71 mol % 1-propanol content. (Curves are shifted for clarity.)

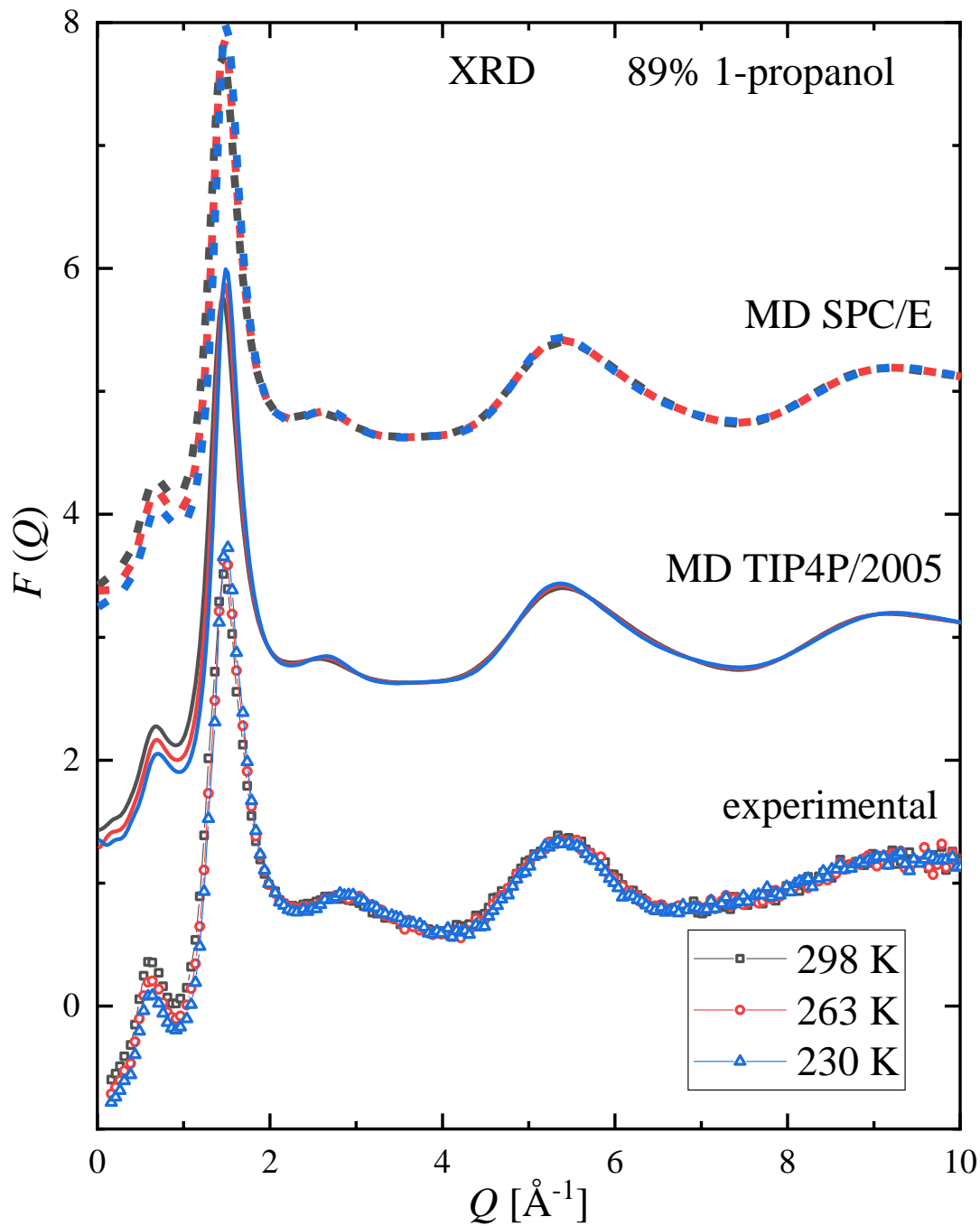

**Figure S15** Temperature dependence of measured (symbols) and simulated (lines) XRD structure factors for the 1-propanol – water mixture with 89 mol % 1-propanol content: comparison of trends observed upon cooling. Black lines and symbols: 298 K, red: 263 K, blue: 230 K. Simulated curves were obtained by using the TIP4P/2005 (solid lines) and SPC/E (dashed lines) water models. (Curves are shifted for clarity.)

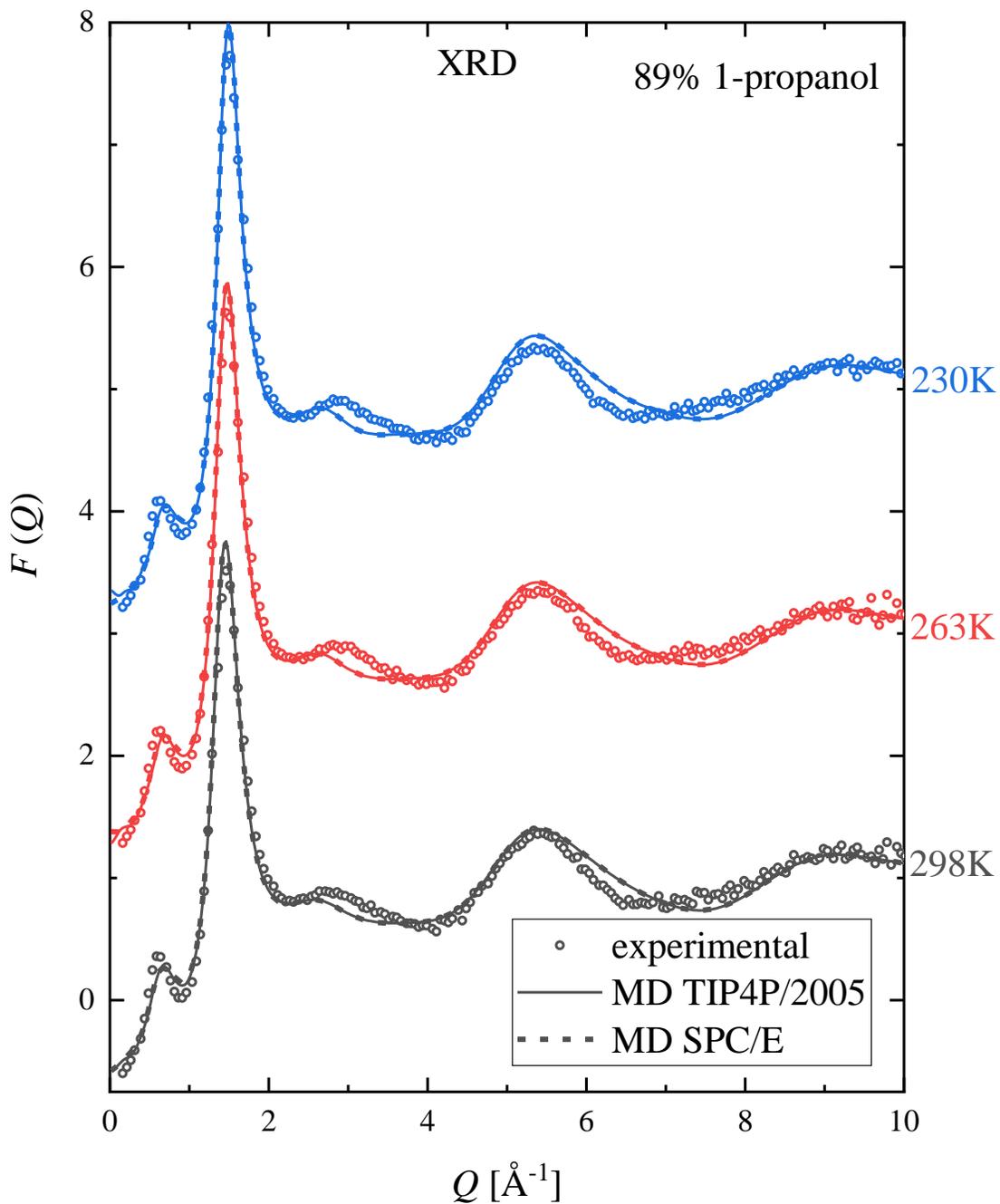

**Figure S16** Comparison of XRD structure factors obtained from experiments (symbols) and simulations using TIP4P/2005 (solid lines) and SPC/E (dashed lines) water models for the 1-propanol – water mixture with 89 mol % 1-propanol content. (Curves are shifted for clarity.)

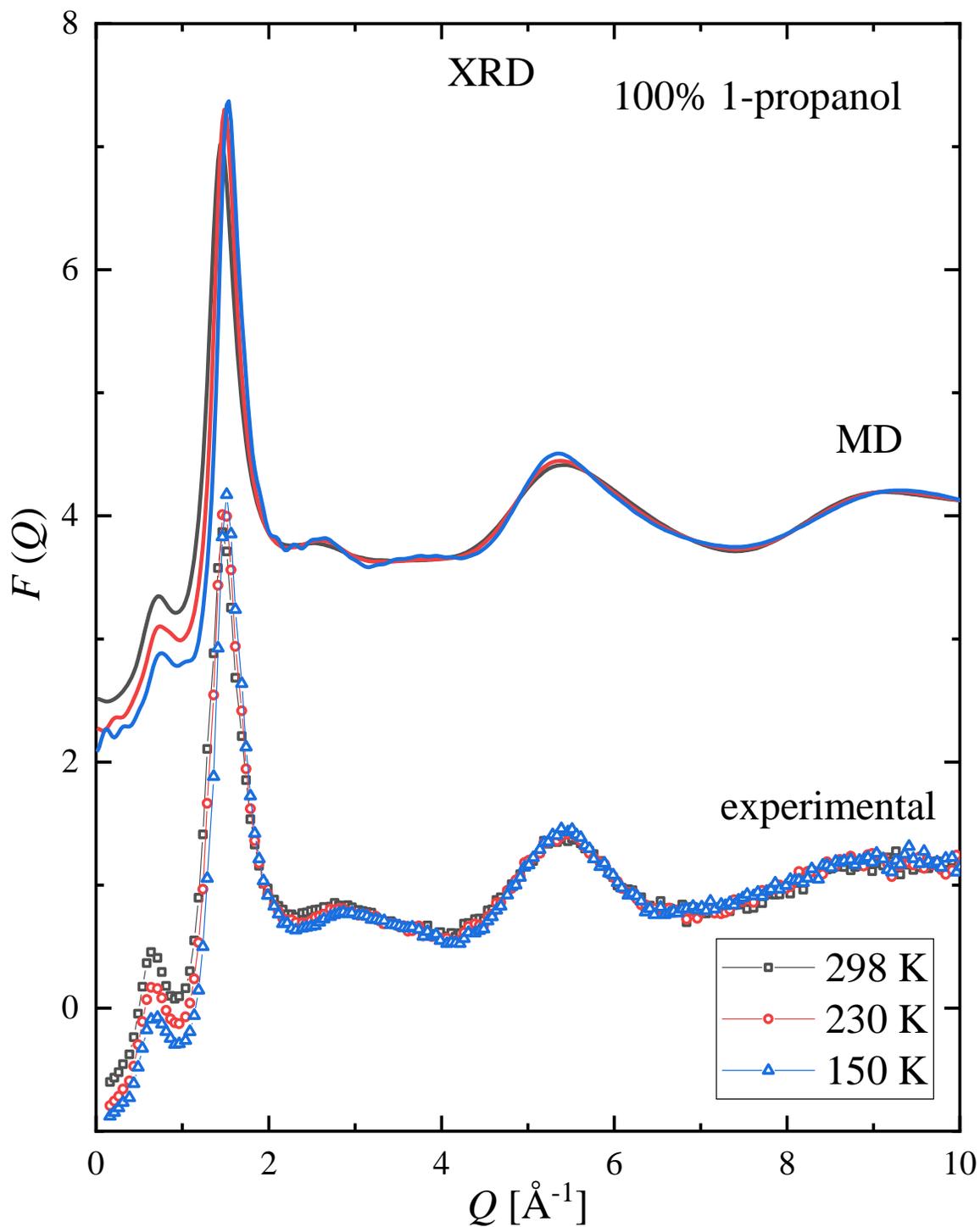

**Figure S17** Temperature dependence of measured (symbols) and simulated (lines) XRD structure factors for pure 1-propanol: comparison of trends observed upon cooling. Black lines and symbols: 298 K, red: 230 K, blue: 150 K. (Curves are shifted for clarity.)

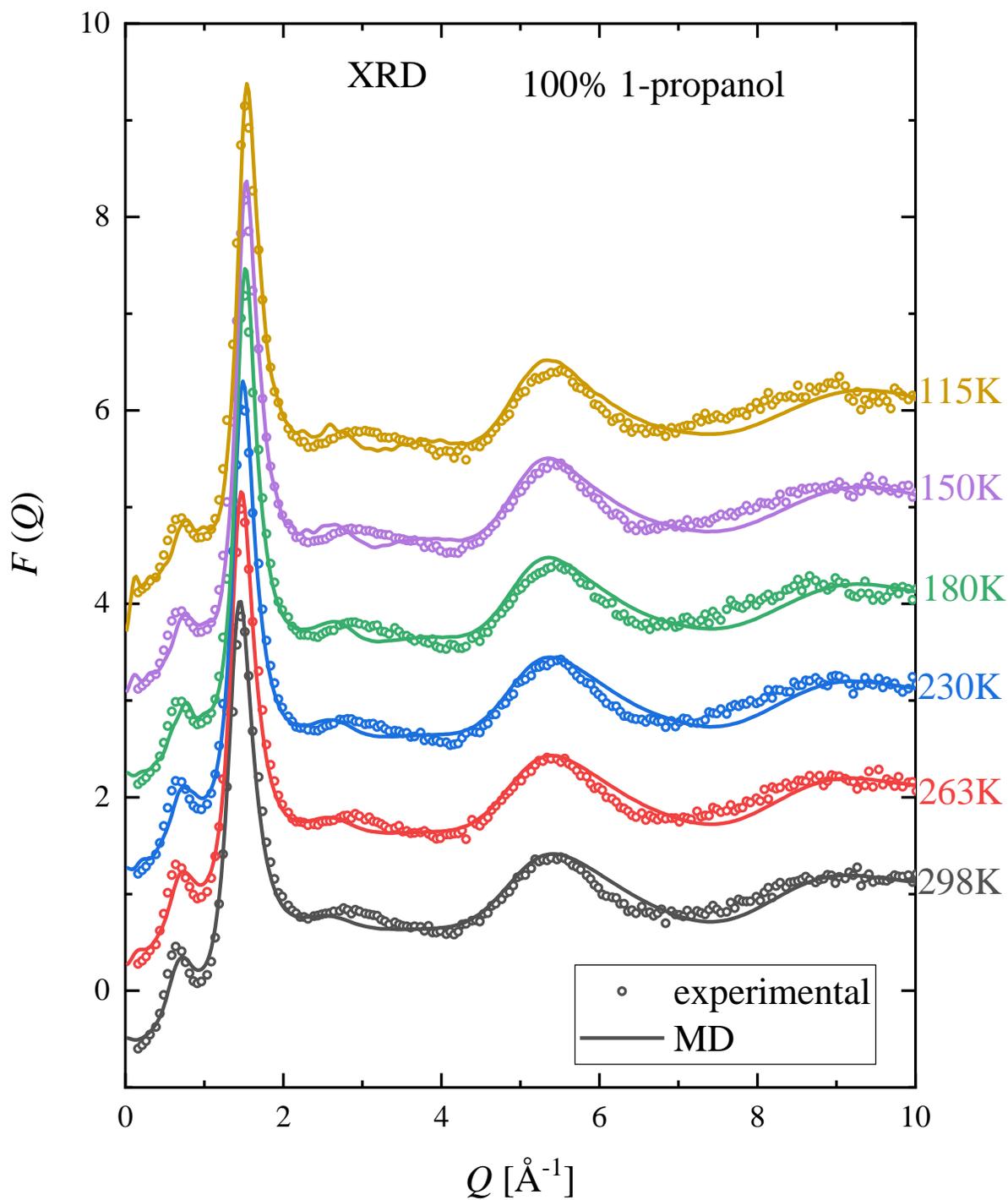

**Figure S18** Comparison of measured (symbols) and simulated (lines) XRD structure factors for pure 1-propanol over a wide range of temperatures. (Curves are shifted for clarity.)

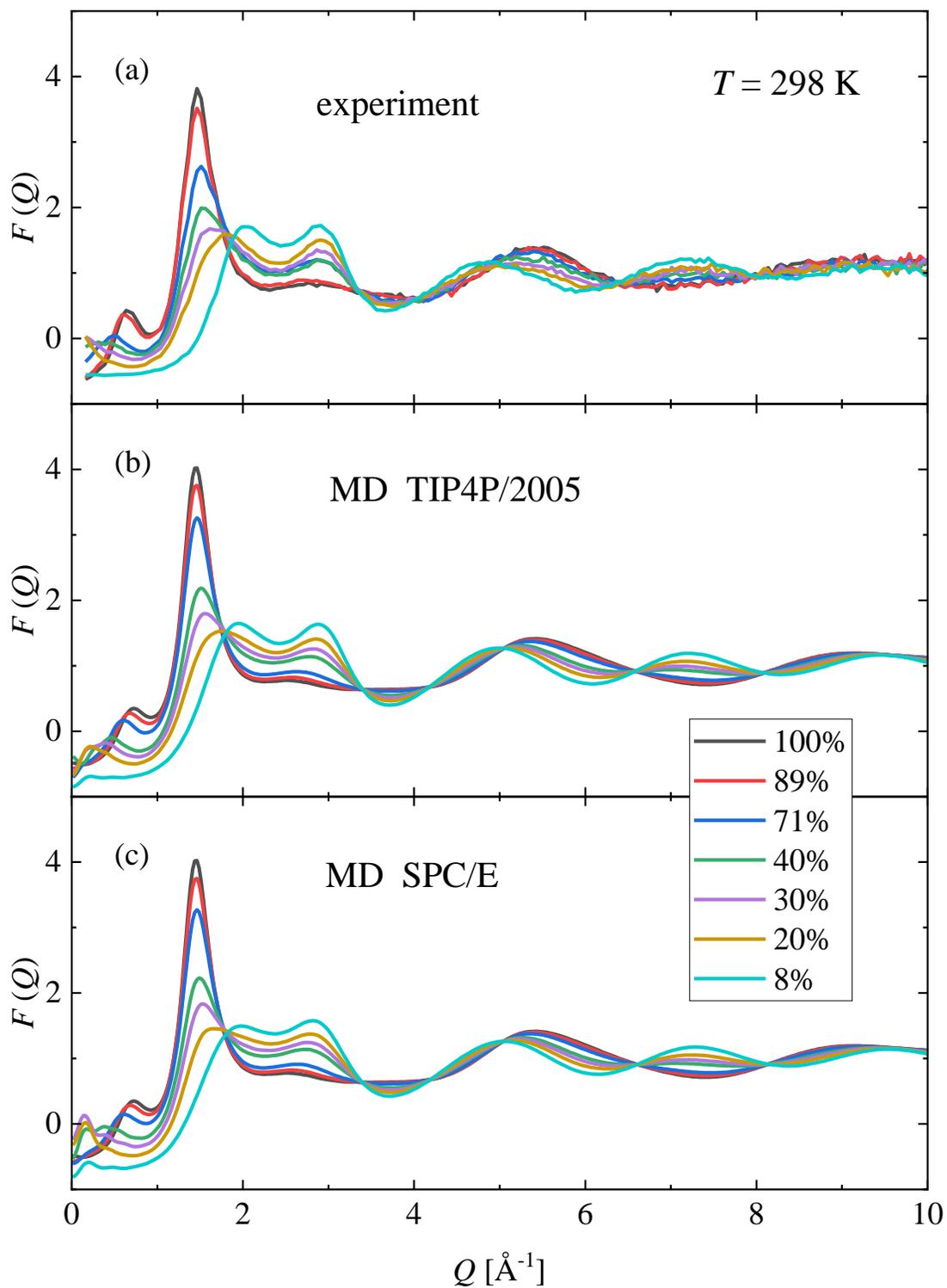

**Figure S19** Concentration dependence of XRD structure factors at $T = 298$ K obtained by (a) experiments and (b,c) simulations using the (b) TIP4P/2005 and (c) SPC/E water models.

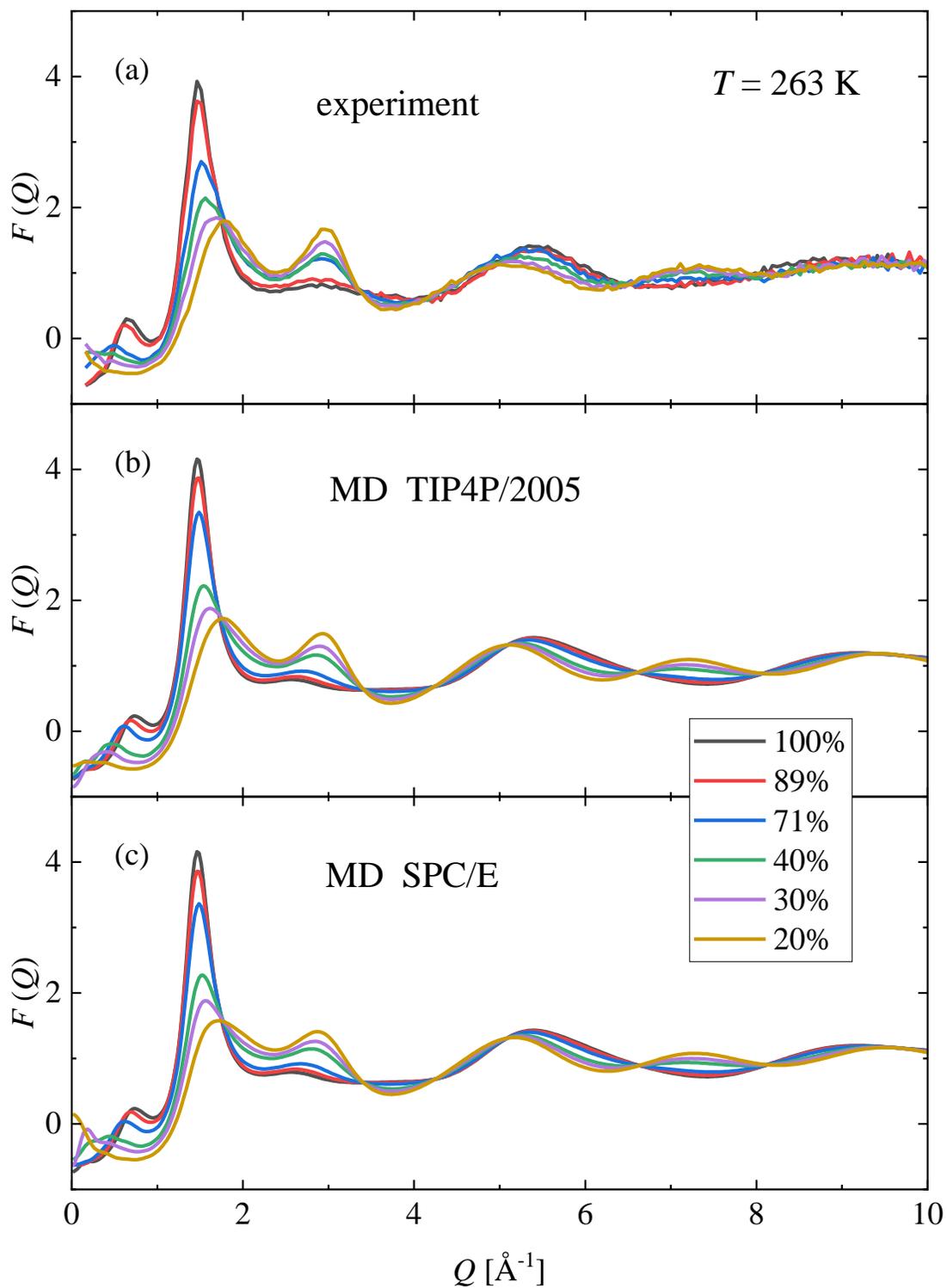

**Figure S20** Concentration dependence of XRD structure factors at $T = 263$ K obtained by (a) experiments and (b, c) simulations using the (b) TIP4P/2005 and (c) SPC/E water models.

**Partial radial distribution functions obtained from MD simulations by using TIP4P/2005 water**

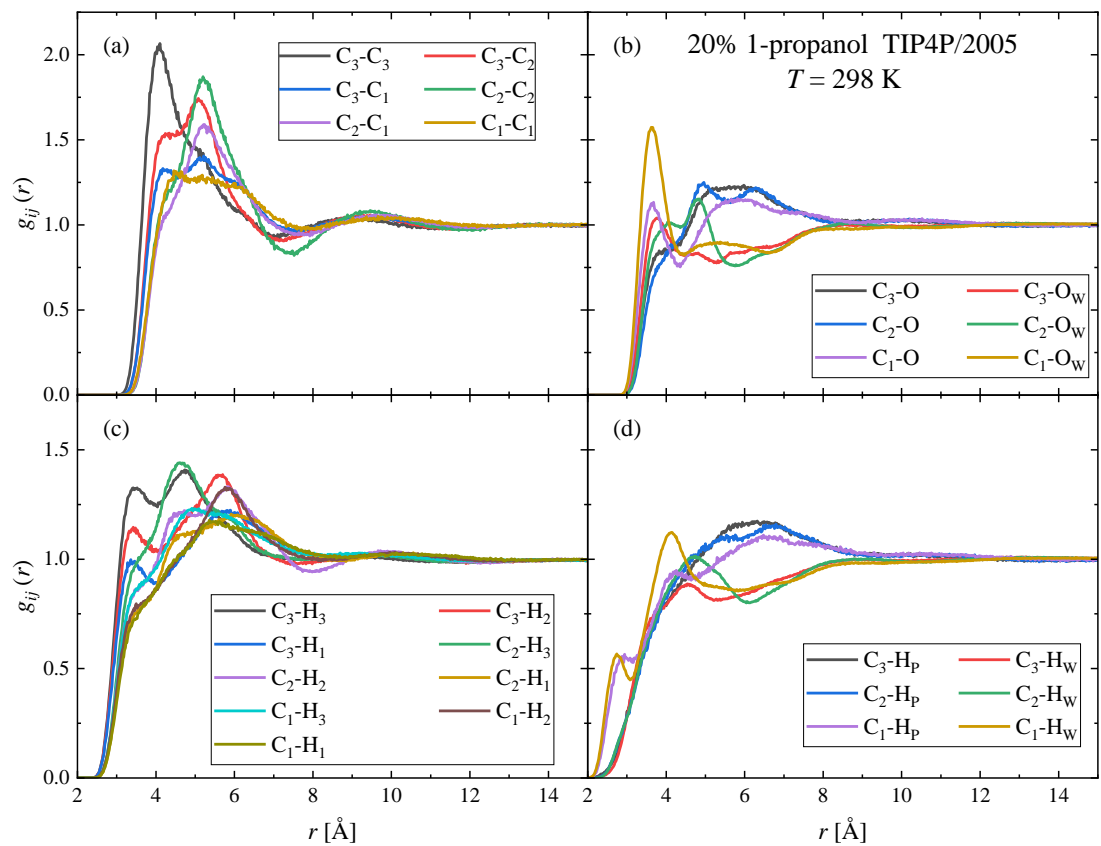

**Figure S21** Partial radial distribution functions of 20 mol% 1-propanol – water mixtures at 298 K, as obtained by MD simulation using the TIP4P/2005 water models. (C-related partials.)

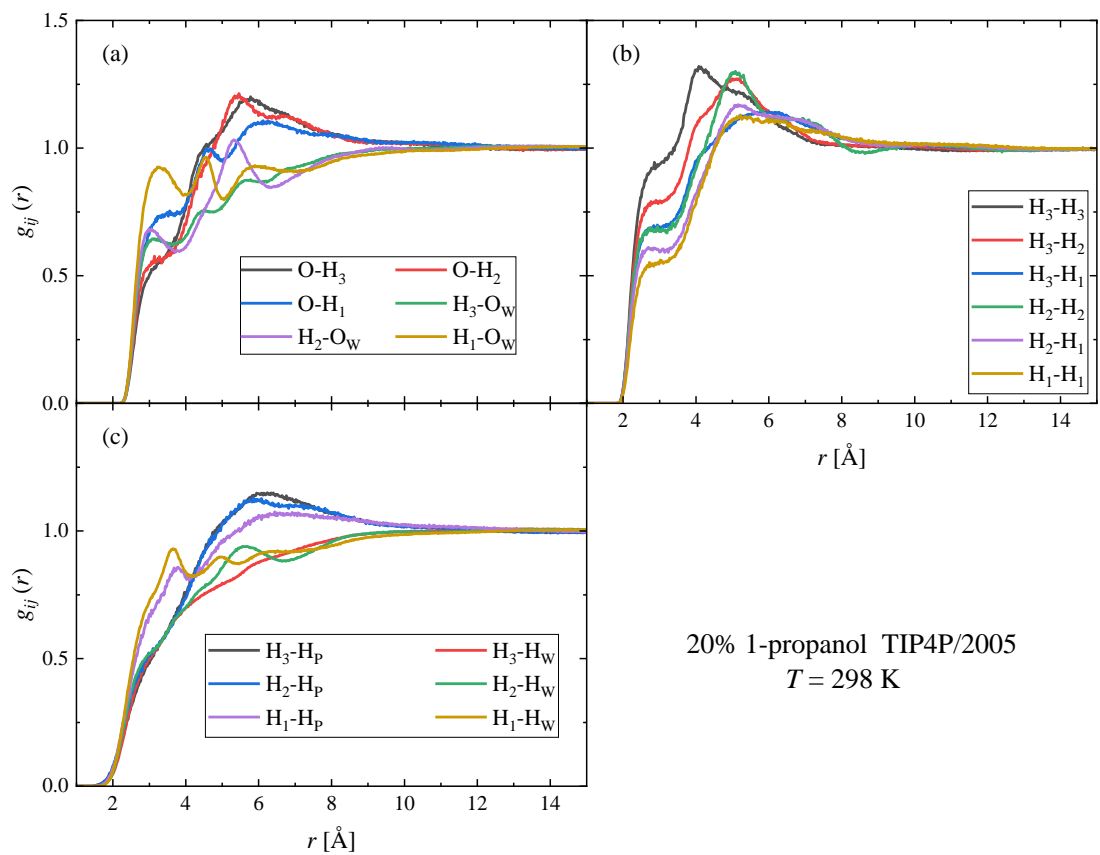

**Figure S22** Partial radial distribution functions of 20 mol% 1-propanol – water mixtures at 298 K, as obtained by MD simulation using the TIP4P/2005 water models (remaining partials).

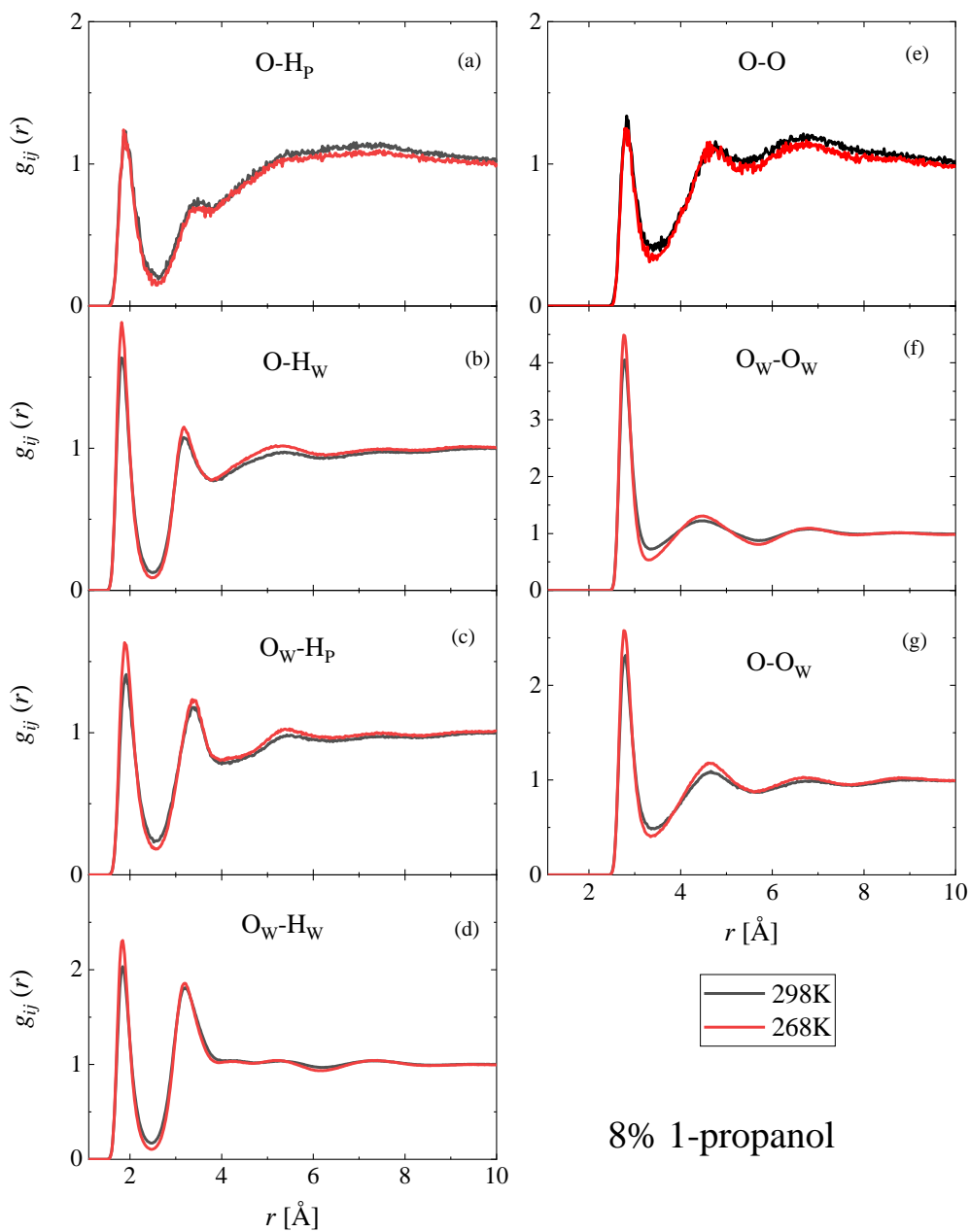

**Figure S23** Temperature dependence of simulated partial radial distribution functions of the 1-propanol – water mixture with 8 mol% 1-propanol content. The H-bonding related partials are shown: (a) 1-propanol O (denoted as O) – hydroxyl H of 1-propanol (denoted as $H_P$), (b) 1-propanol O – water H (denoted as $H_W$), (c) water O (denoted as $O_W$) – hydroxyl H of 1-propanol, (d) water O – water H, (e) 1-propanol O – 1-propanol O, (f) water O – water O, (g) 1-propanol O – water O. The curves were obtained by using the TIP4P/2005 water model.

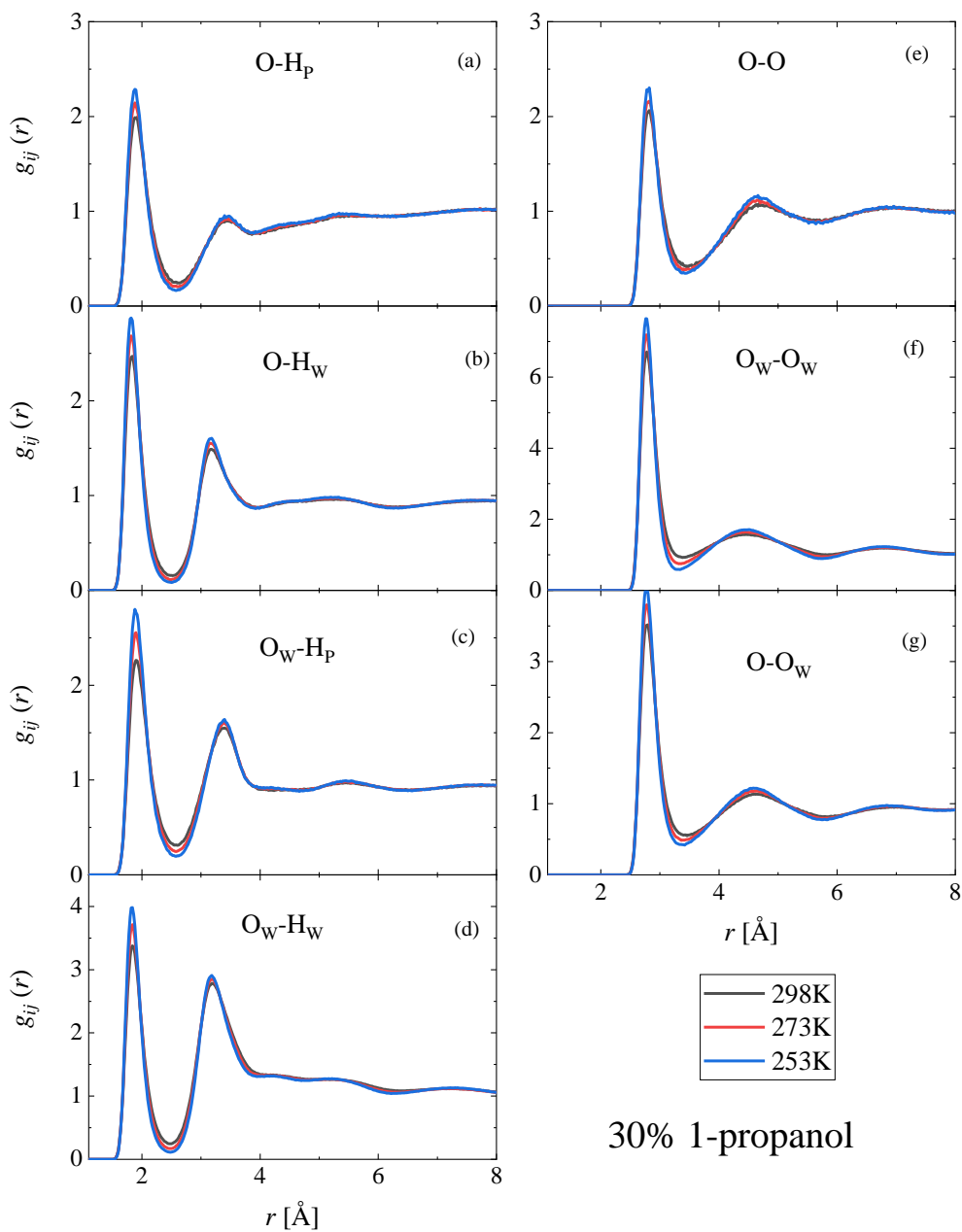

**Figure S24** Temperature dependence of simulated partial radial distribution functions of the 1-propanol – water mixture with 30 mol% 1-propanol content. The H-bonding related partials are shown: (a) 1-propanol O (denoted as O) – hydroxyl H of 1-propanol (denoted as $H_P$), (b) 1-propanol O – water H (denoted as $H_W$), (c) water O (denoted as $O_W$) – hydroxyl H of 1-propanol, (d) water O – water H, (e) 1-propanol O – 1-propanol O, (f) water O – water O, (g) 1-propanol O – water O. The curves were obtained by using the TIP4P/2005 water model.

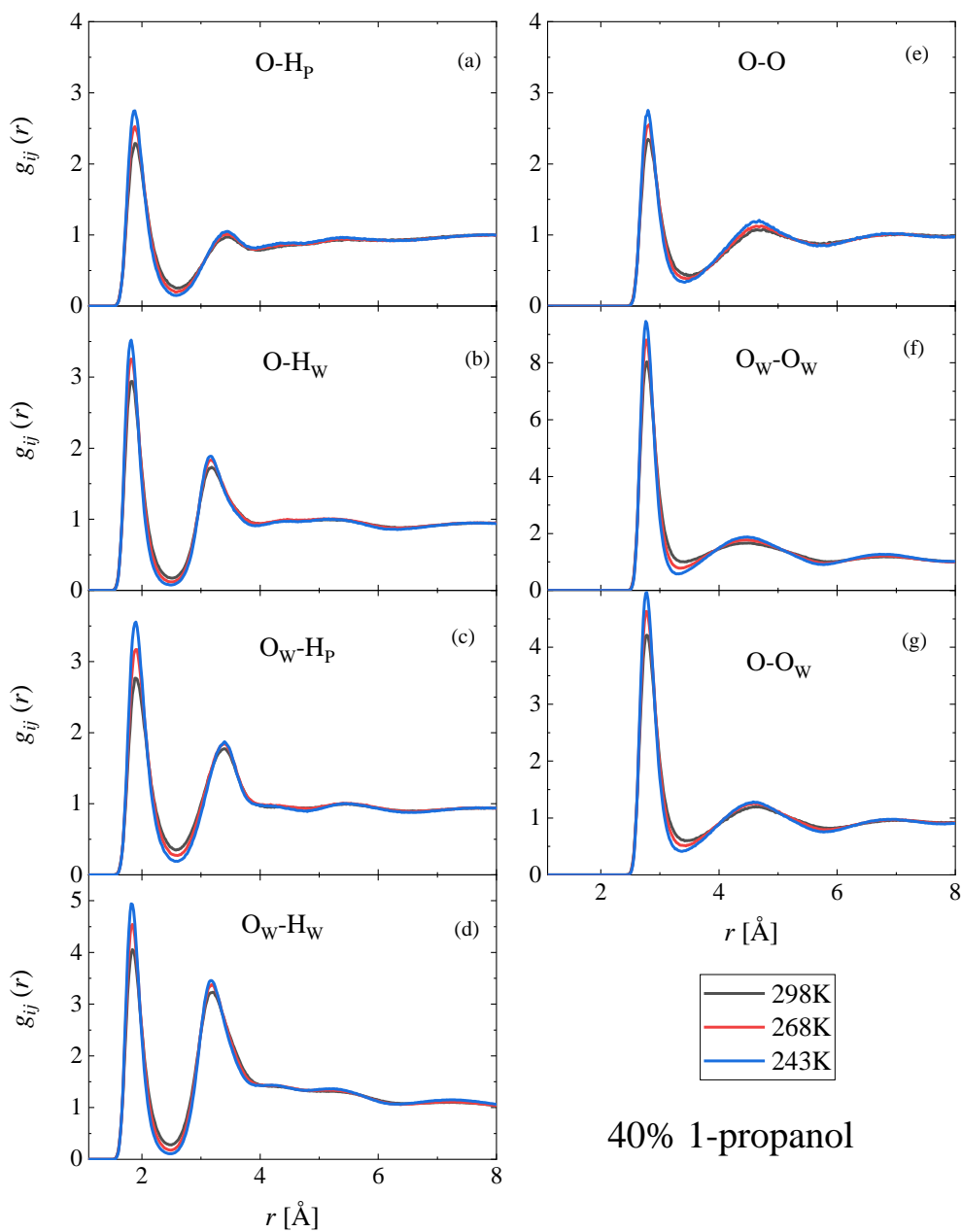

**Figure S25** Temperature dependence of simulated partial radial distribution functions of the 1-propanol – water mixture with 40 mol% 1-propanol content. The H-bonding related partials are shown: (a) 1-propanol O (denoted as O) – hydroxyl H of 1-propanol (denoted as $H_P$), (b) 1-propanol O – water H (denoted as $H_W$), (c) water O (denoted as $O_W$) – hydroxyl H of 1-propanol, (d) water O – water H, (e) 1-propanol O – 1-propanol O, (f) water O – water O, (g) 1-propanol O – water O. The curves were obtained by using the TIP4P/2005 water model.

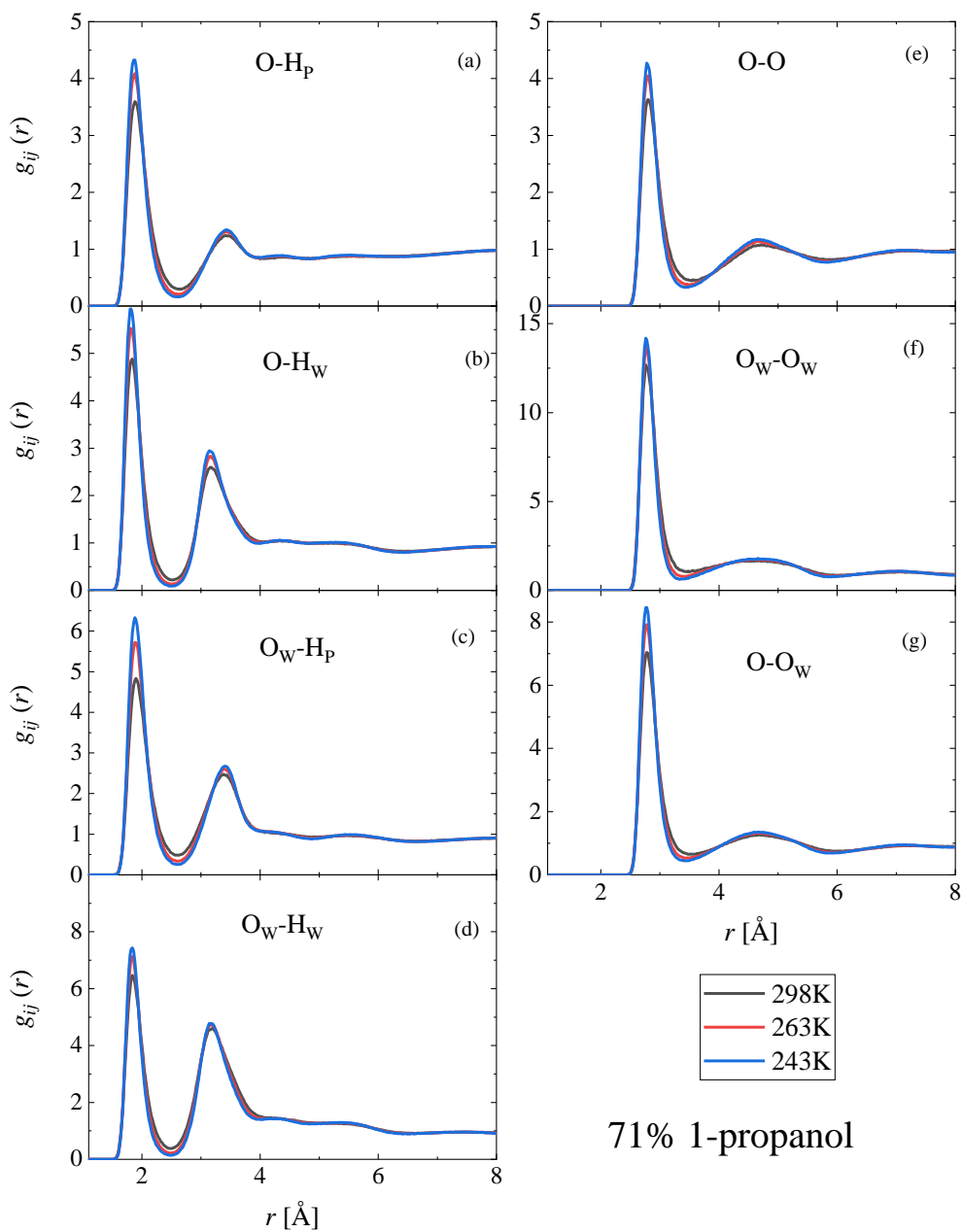

**Figure S26** Temperature dependence of simulated partial radial distribution functions of the 1-propanol – water mixture with 71 mol% 1-propanol content. The H-bonding related partials are shown: (a) 1-propanol O (denoted as O) – hydroxyl H of 1-propanol (denoted as $H_P$), (b) 1-propanol O – water H (denoted as $H_W$), (c) water O (denoted as $O_W$) – hydroxyl H of 1-propanol, (d) water O – water H, (e) 1-propanol O – 1-propanol O, (f) water O – water O, (g) 1-propanol O – water O. The curves were obtained by using the TIP4P/2005 water model.

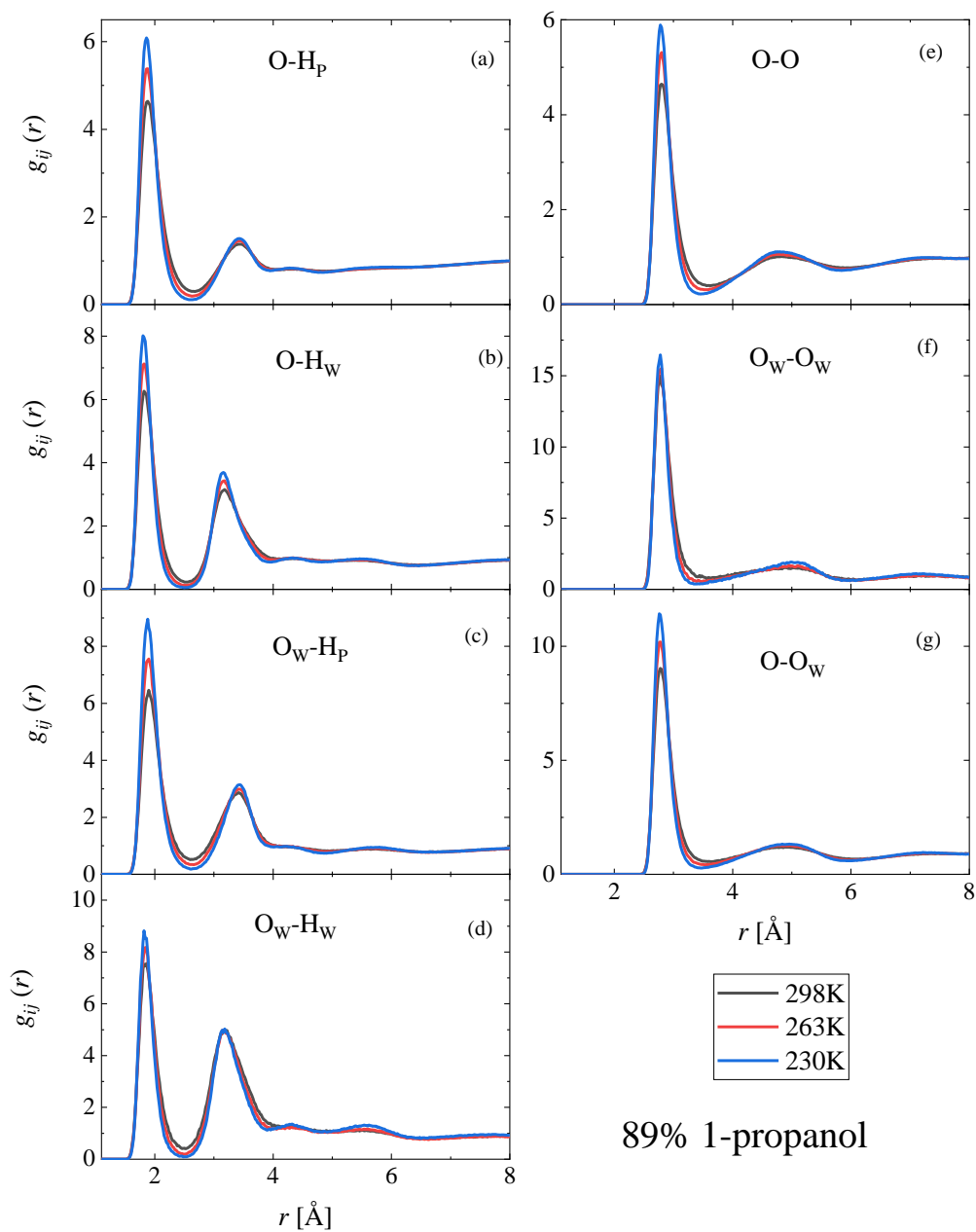

**Figure S27** Temperature dependence of simulated partial radial distribution functions of the 1-propanol – water mixture with 89 mol% 1-propanol content. The H-bonding related partials are shown: (a) 1-propanol O (denoted as O) – hydroxyl H of 1-propanol (denoted as $H_P$), (b) 1-propanol O – water H (denoted as $H_W$), (c) water O (denoted as $O_W$) – hydroxyl H of 1-propanol, (d) water O – water H, (e) 1-propanol O – 1-propanol O, (f) water O – water O, (g) 1-propanol O – water O. The curves were obtained by using the TIP4P/2005 water model.

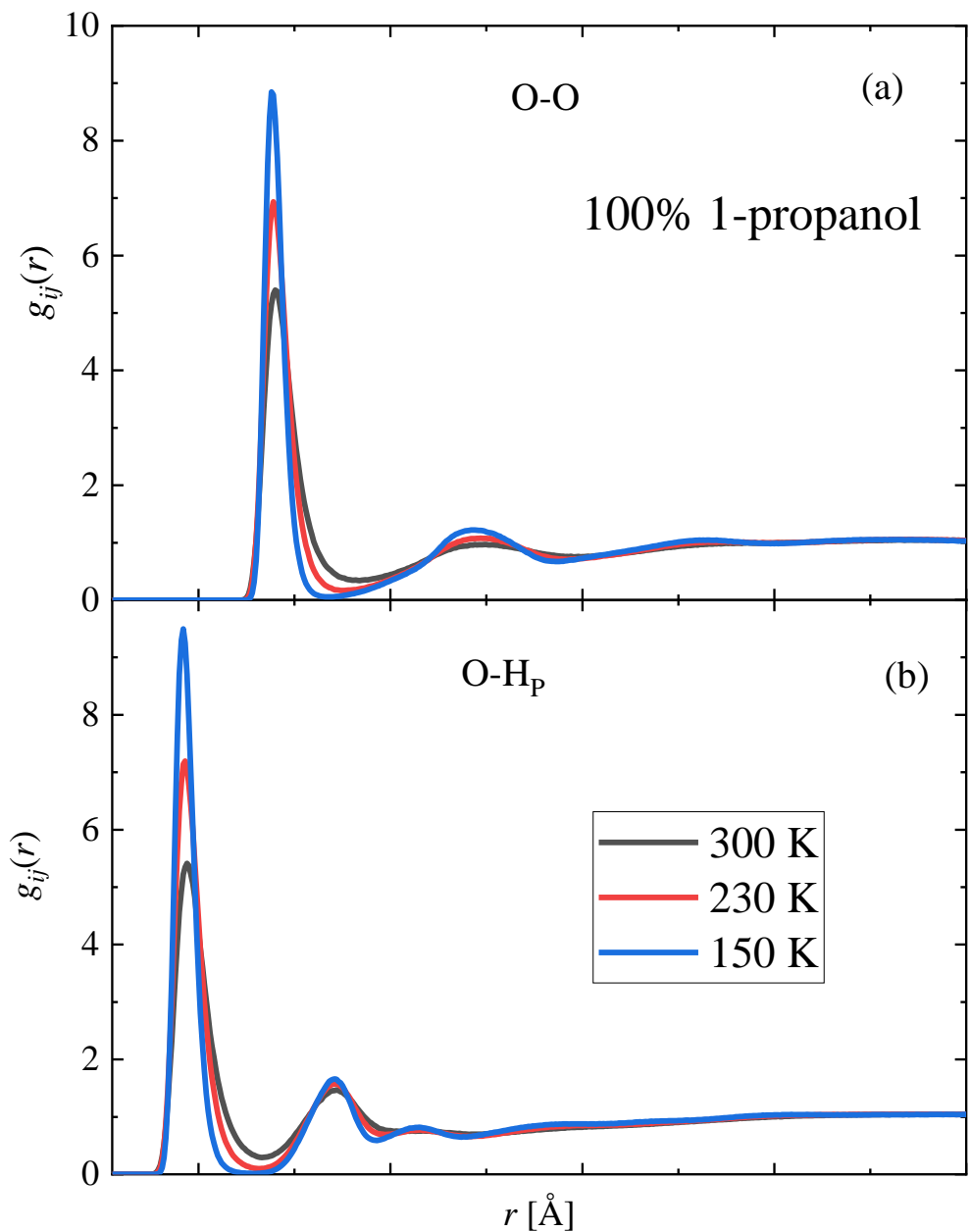

**Figure S28** Temperature dependence of simulated partial radial distribution functions of pure 1-propanol. The H-bonding related partials are shown: (a) O – O, (b) O – hydroxyl H of 1-propanol (denoted as $H_P$).

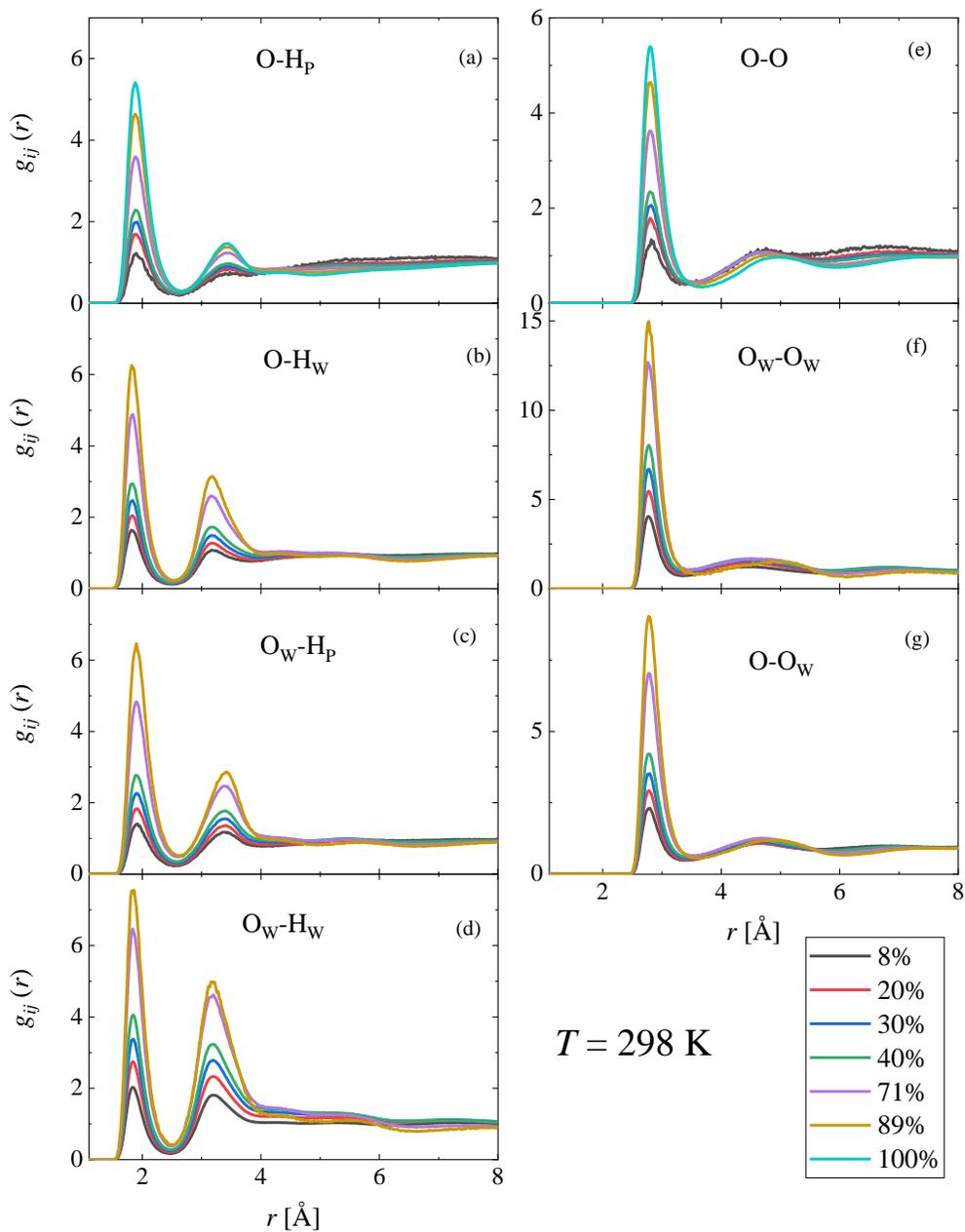

**Figure S29** Concentration dependence of simulated partial radial distribution functions of 1-propanol – water mixtures at 298 K. The H-bonding related partials are shown: (a) 1-propanol O (denoted as O) – hydroxyl H of 1-propanol (denoted as $H_P$), (b) 1-propanol O – water H (denoted as $H_W$), (c) water O (denoted as $O_W$) – hydroxyl H of 1-propanol, (d) water O – water H, (e) 1-propanol O – 1-propanol O, (f) water O – water O, (g) 1-propanol O – water O. The curves were obtained by using the TIP4P/2005 water model.

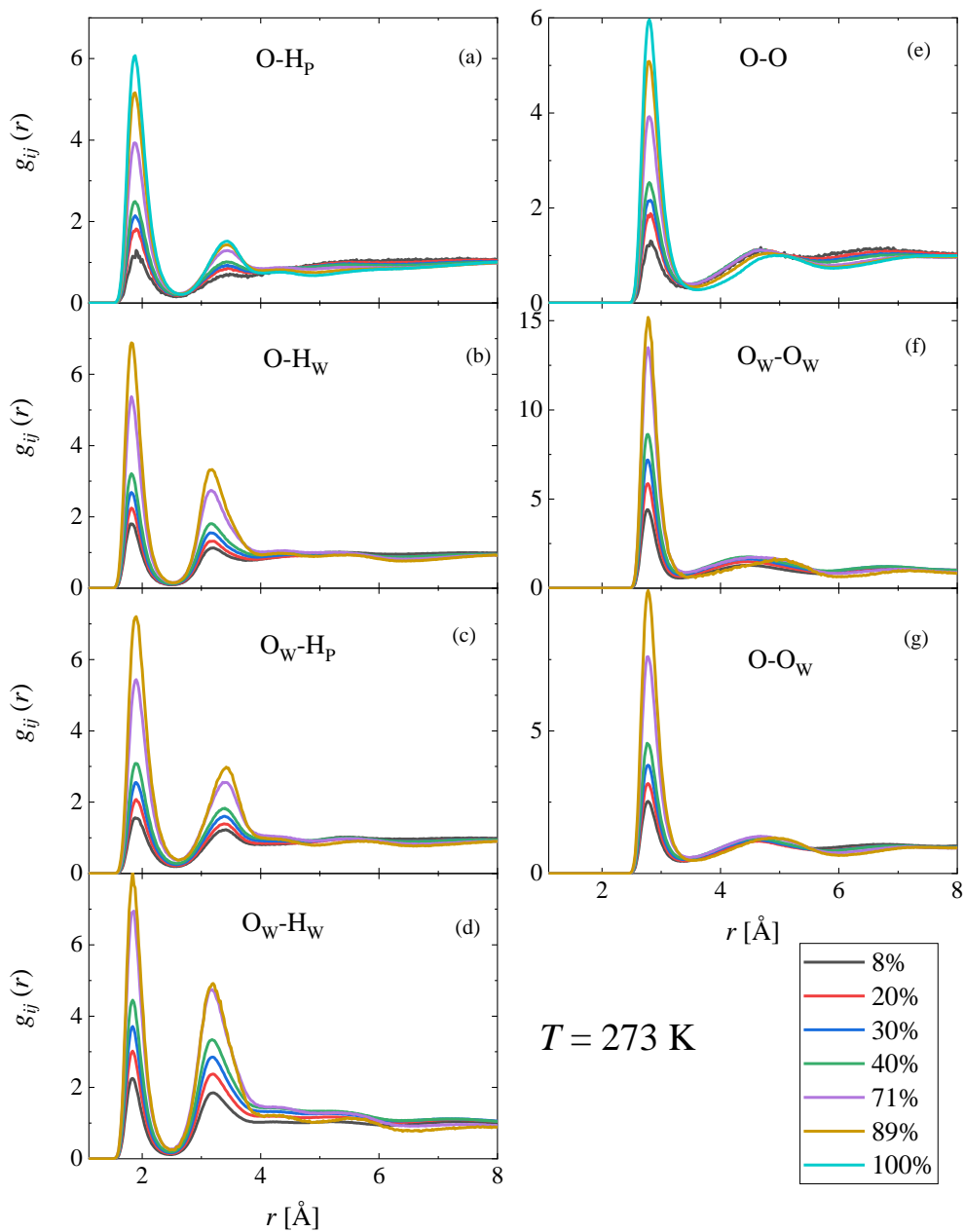

**Figure S30** Concentration dependence of simulated partial radial distribution functions of 1-propanol – water mixtures at 273 K. The H-bonding related partials are shown: (a) 1-propanol O (denoted as O) – hydroxyl H of 1-propanol (denoted as $H_P$), (b) 1-propanol O – water H (denoted as $H_W$), (c) water O (denoted as $O_W$) – hydroxyl H of 1-propanol, (d) water O – water H, (e) 1-propanol O – 1-propanol O, (f) water O – water O, (g) 1-propanol O – water O. The curves were obtained by using the TIP4P/2005 water model.

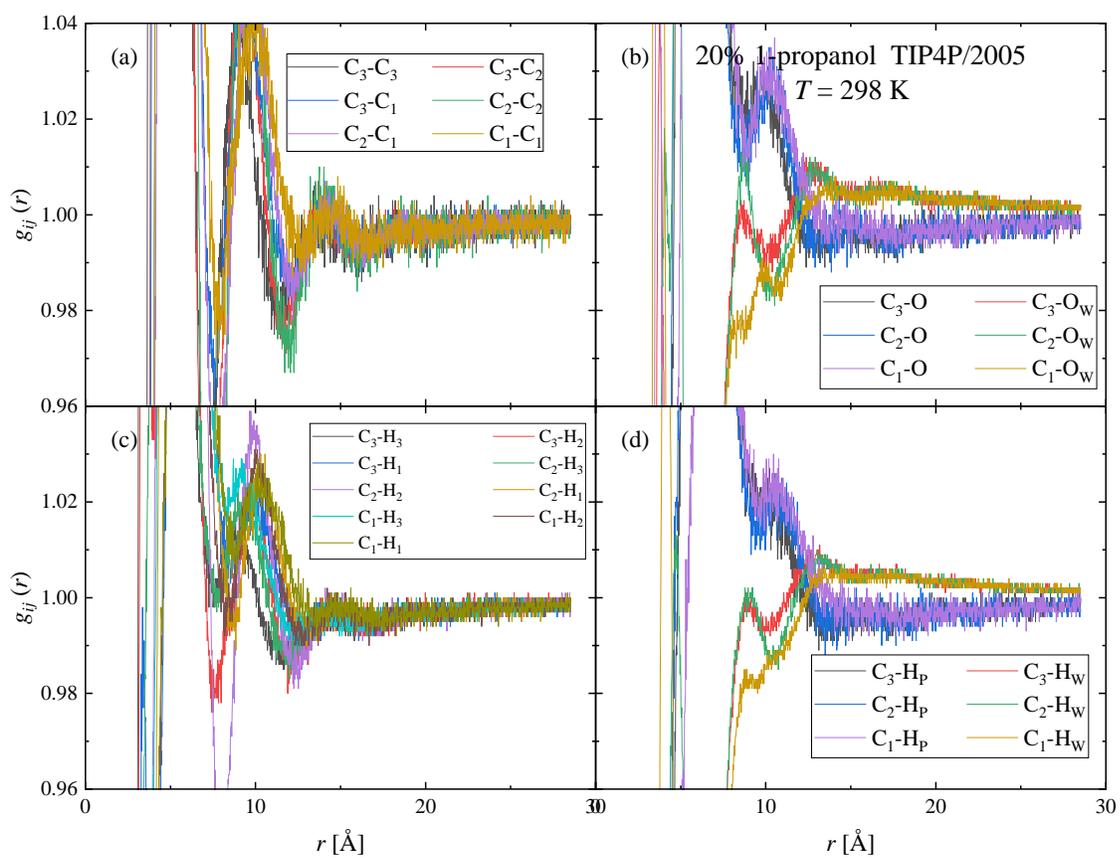

**Figure S31** Long range behavior of the partial radial distribution functions of 20 mol% 1-propanol – water mixtures at 298 K, as obtained by MD simulation using the TIP4P/2005 water model. (C-related partials.)

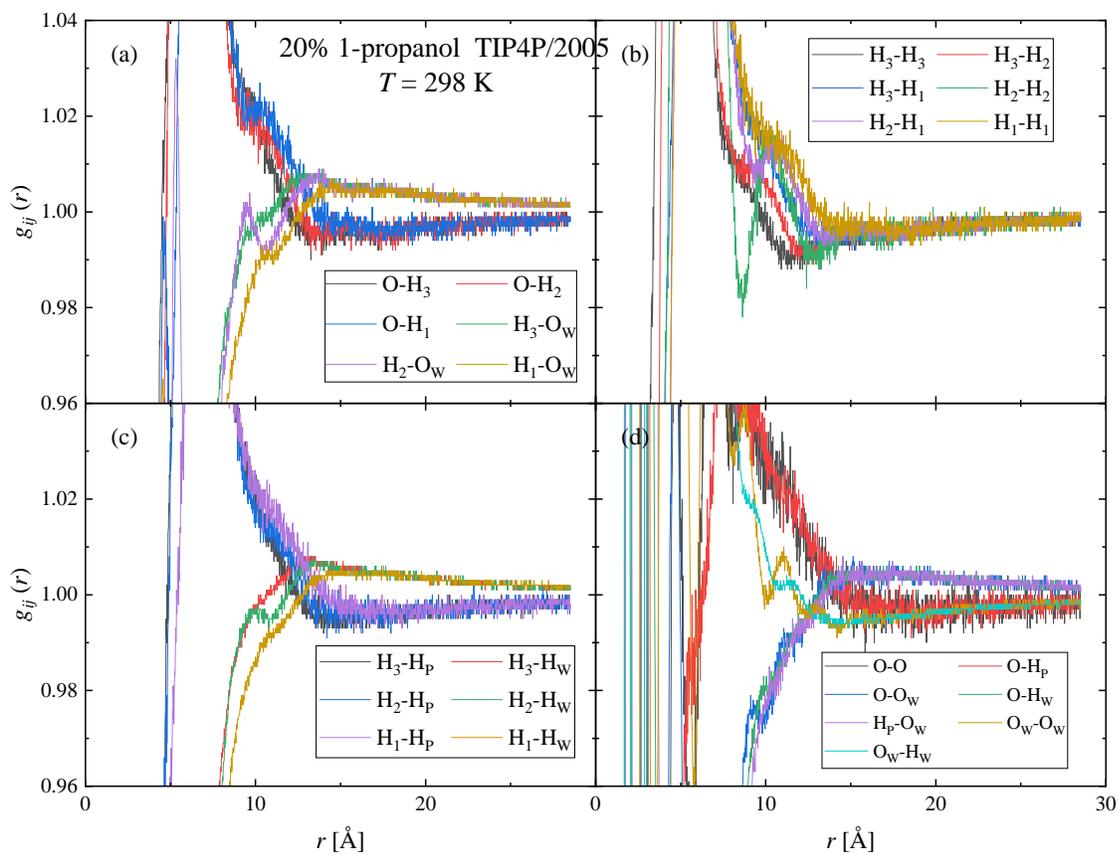

**Figure S32** Long range behavior of the partial radial distribution functions of 20 mol% 1-propanol – water mixtures at 298 K, as obtained by MD simulation using the TIP4P/2005 water model.

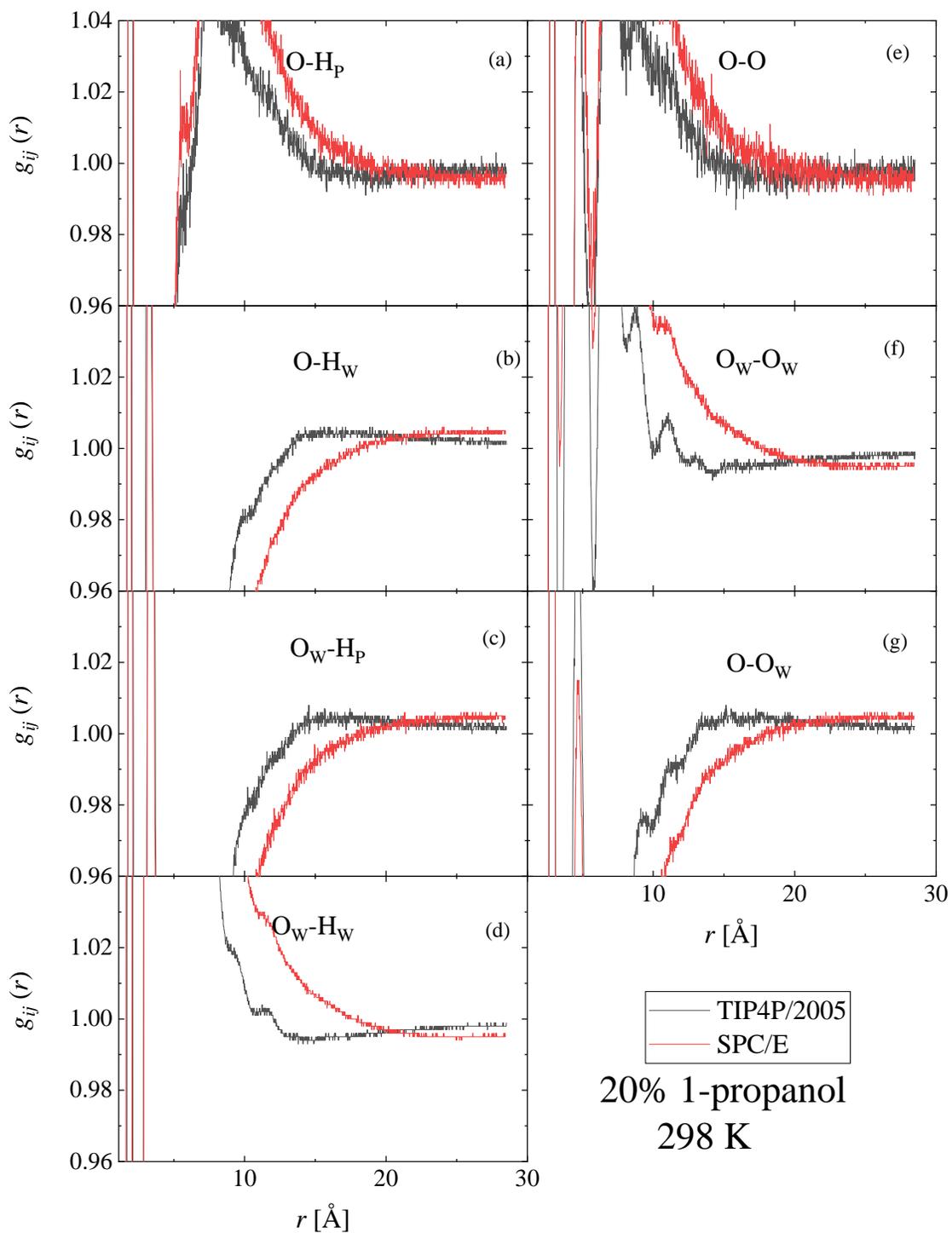

**Figure S33** Comparison of the long range behavior of H-bond related partial radial distribution functions obtained by using the TIP4P/2005 and SPC/E water models.

**Partial radial distribution functions obtained from MD simulations by using SPC/E water**

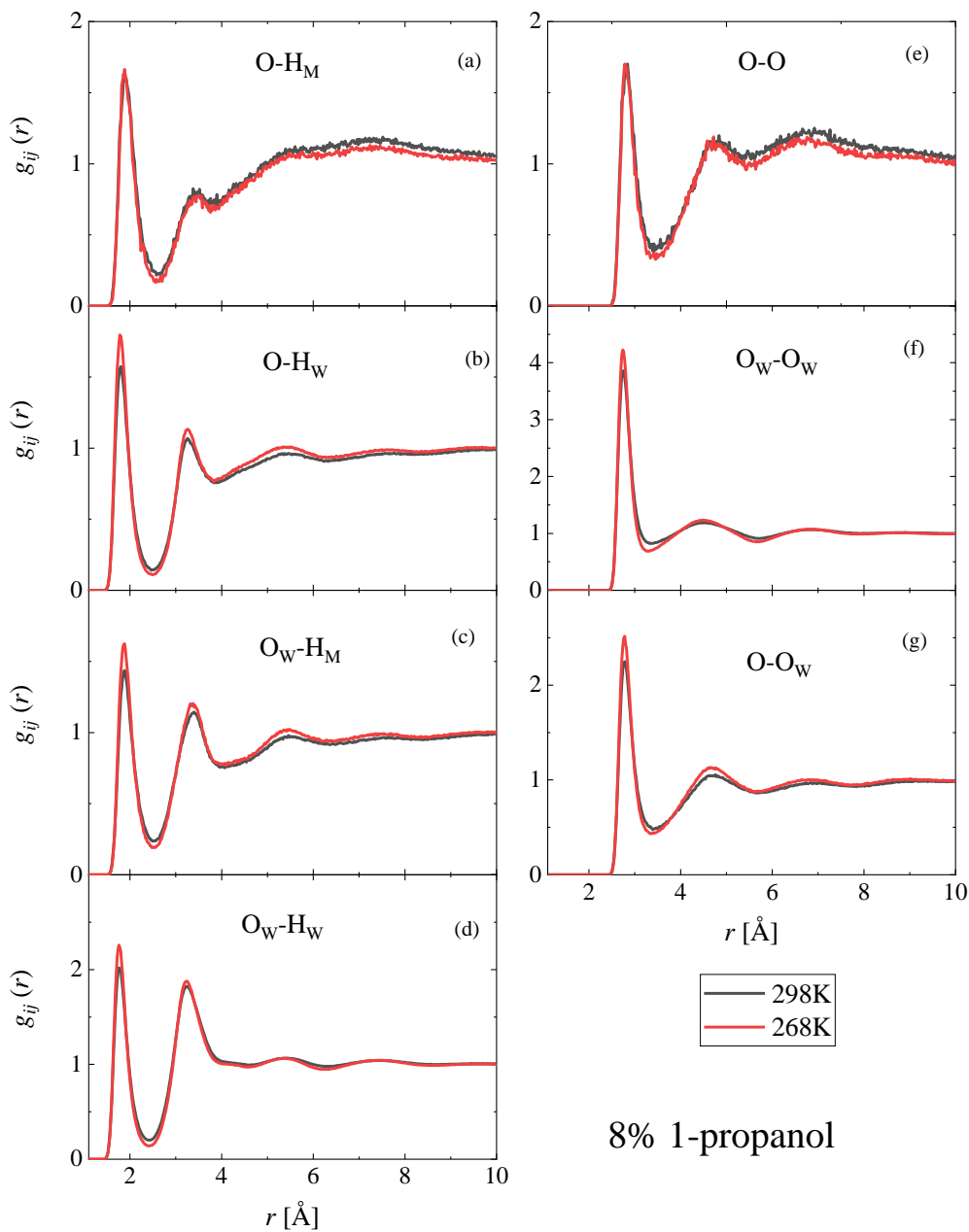

**Figure S34** Temperature dependence of simulated partial radial distribution functions of the 1-propanol – water mixture with 8 mol% 1-propanol content. The H-bonding related partials are shown: (a) 1-propanol O (denoted as O) – hydroxyl H of 1-propanol (denoted as $H_P$), (b) 1-propanol O – water H (denoted as $H_W$), (c) water O (denoted as $O_W$) – hydroxyl H of 1-propanol, (d) water O – water H, (e) 1-propanol O – 1-propanol O, (f) water O – water O, (g) 1-propanol O – water O. The curves were obtained by using the SPC/E water model.

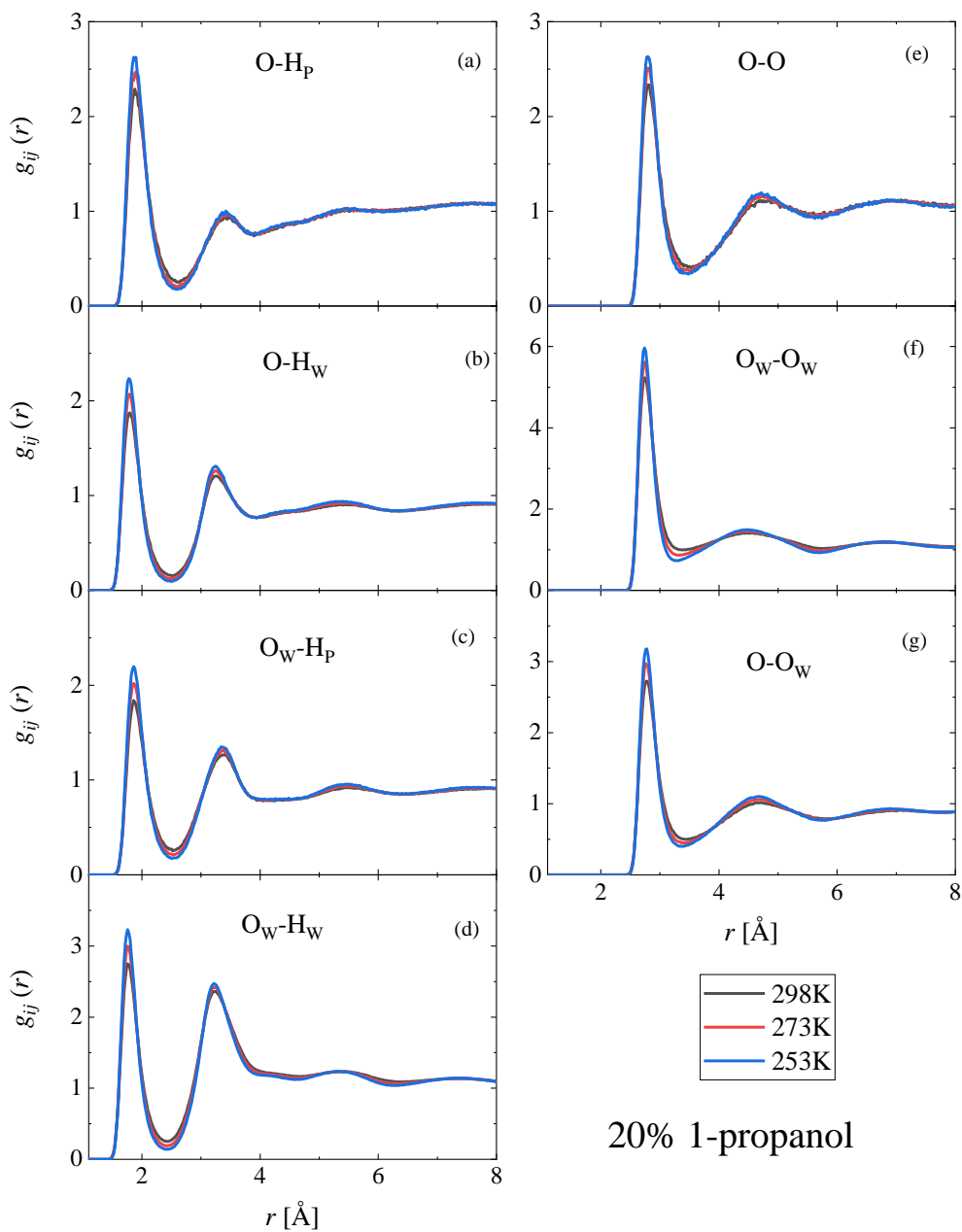

**Figure S35** Temperature dependence of simulated partial radial distribution functions of the 1-propanol – water mixture with 20 mol% 1-propanol content. The H-bonding related partials are shown: (a) 1-propanol O (denoted as O) – hydroxyl H of 1-propanol (denoted as $H_P$), (b) 1-propanol O – water H (denoted as $H_W$), (c) water O (denoted as $O_W$) – hydroxyl H of 1-propanol, (d) water O – water H, (e) 1-propanol O – 1-propanol O, (f) water O – water O, (g) 1-propanol O – water O. The curves were obtained by using the SPC/E water model.

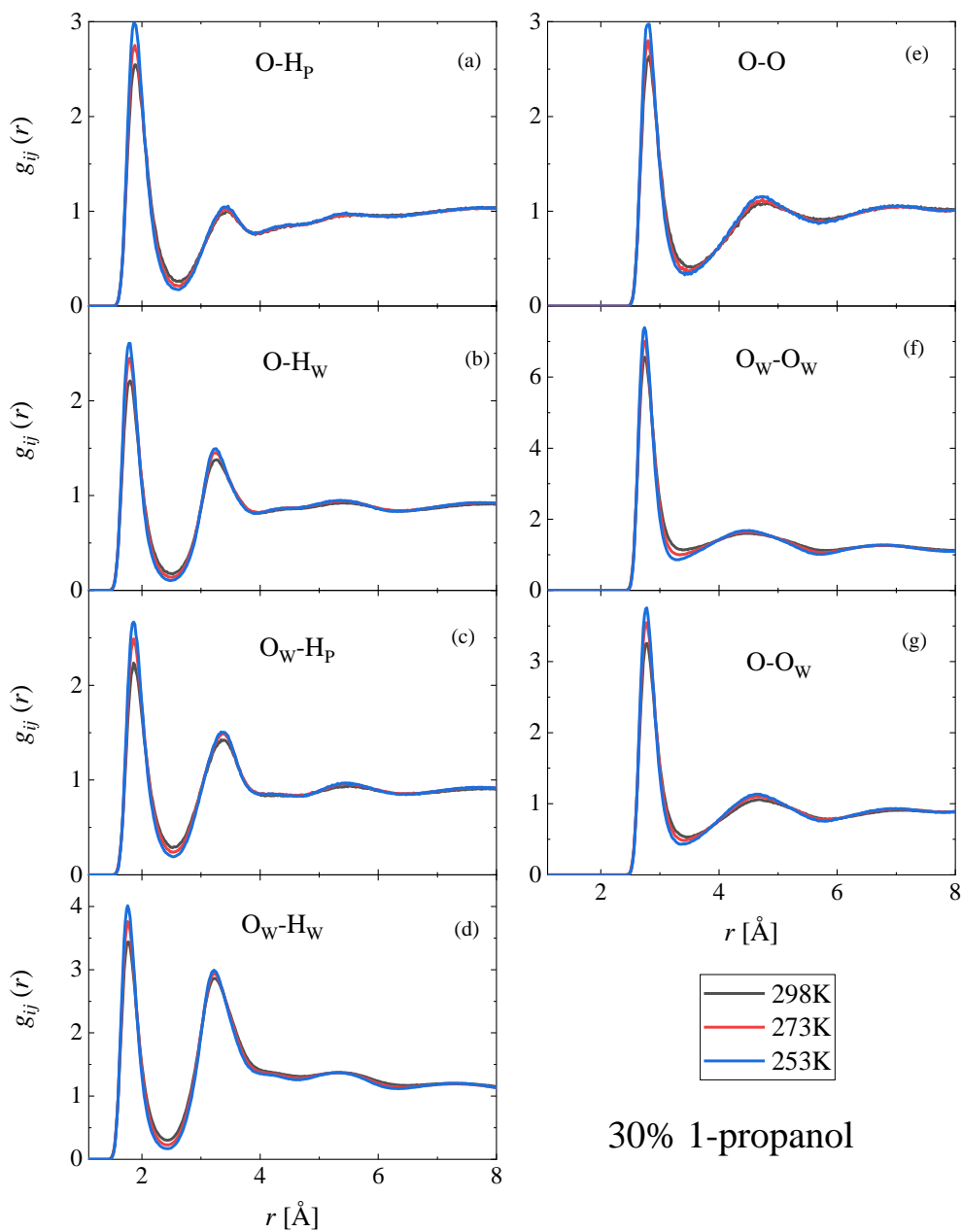

**Figure S36** Temperature dependence of simulated partial radial distribution functions of the 1-propanol – water mixture with 30 mol% 1-propanol content. The H-bonding related partials are shown: (a) 1-propanol O (denoted as O) – hydroxyl H of 1-propanol (denoted as $H_P$), (b) 1-propanol O – water H (denoted as $H_W$), (c) water O (denoted as $O_W$) – hydroxyl H of 1-propanol, (d) water O – water H, (e) 1-propanol O – 1-propanol O, (f) water O – water O, (g) 1-propanol O – water O. The curves were obtained by using the SPC/E water model.

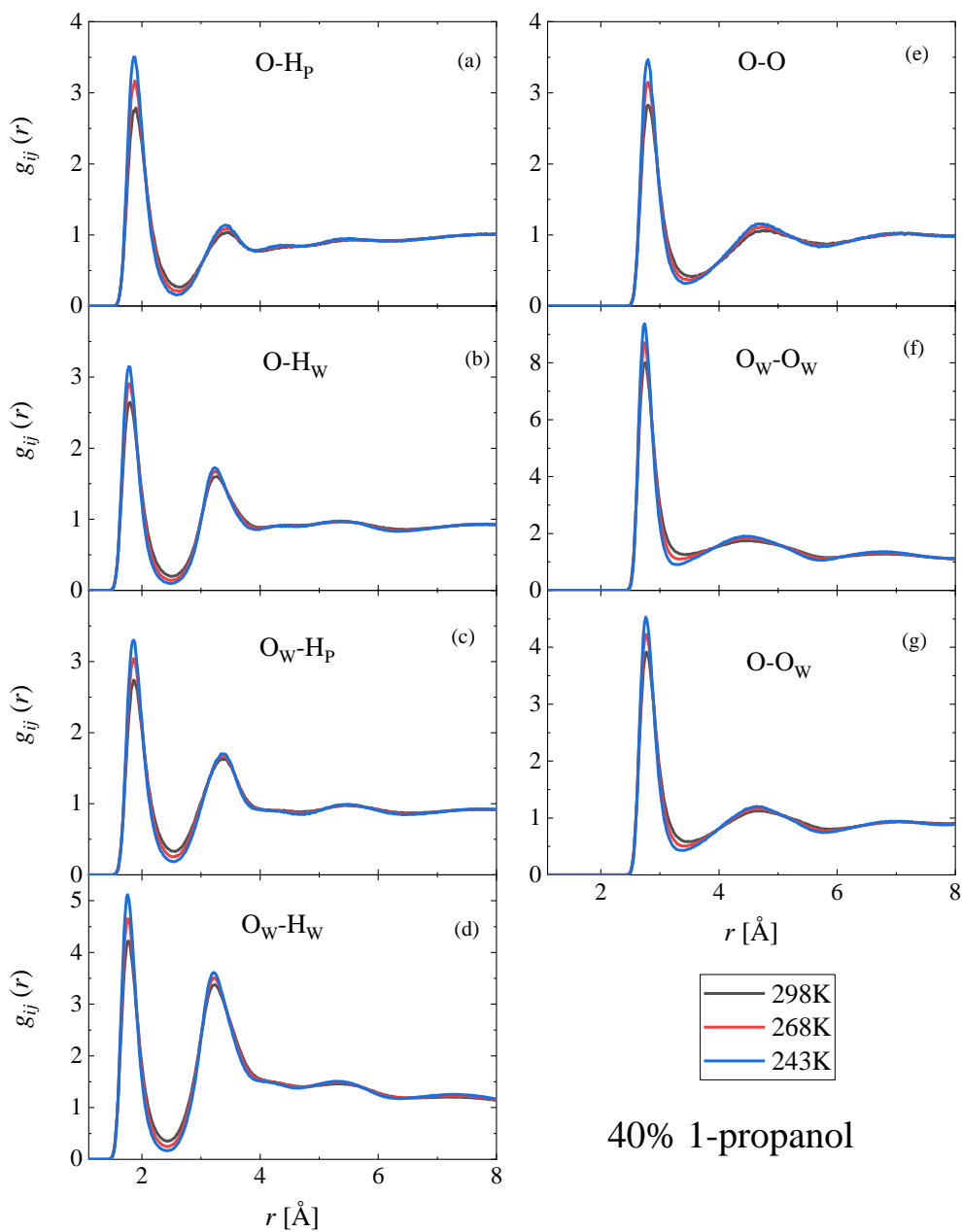

**Figure S37** Temperature dependence of simulated partial radial distribution functions of the 1-propanol – water mixture with 40 mol% 1-propanol content. The H-bonding related partials are shown: (a) 1-propanol O (denoted as O) – hydroxyl H of 1-propanol (denoted as $H_P$), (b) 1-propanol O – water H (denoted as $H_W$), (c) water O (denoted as $O_W$) – hydroxyl H of 1-propanol, (d) water O – water H, (e) 1-propanol O – 1-propanol O, (f) water O – water O, (g) 1-propanol O – water O. The curves were obtained by using the SPC/E water model.

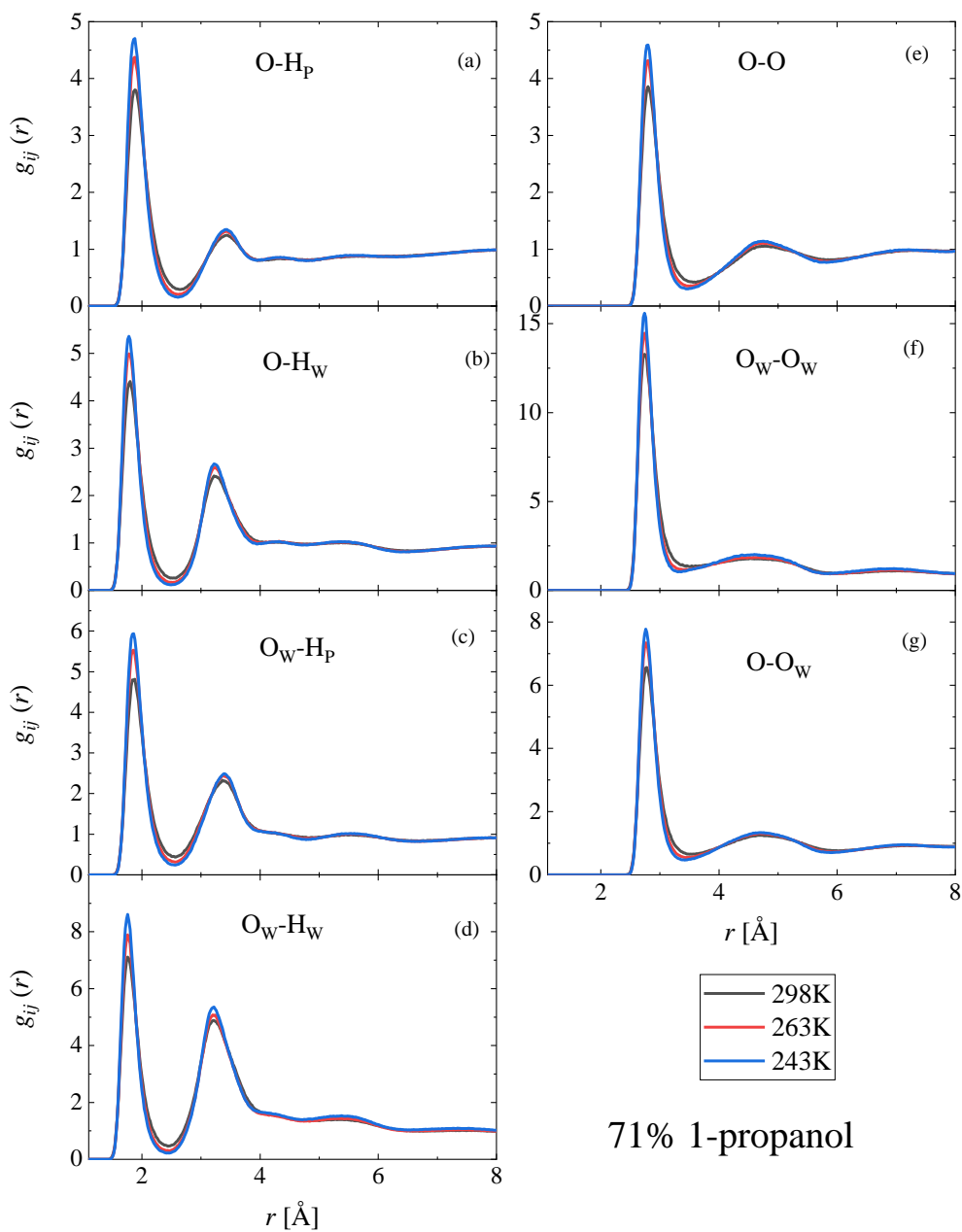

**Figure S38** Temperature dependence of simulated partial radial distribution functions of the 1-propanol – water mixture with 71 mol% 1-propanol content. The H-bonding related partials are shown: (a) 1-propanol O (denoted as O) – hydroxyl H of 1-propanol (denoted as $H_P$), (b) 1-propanol O – water H (denoted as $H_W$), (c) water O (denoted as $O_W$) – hydroxyl H of 1-propanol, (d) water O – water H, (e) 1-propanol O – 1-propanol O, (f) water O – water O, (g) 1-propanol O – water O. The curves were obtained by using the SPC/E water model.

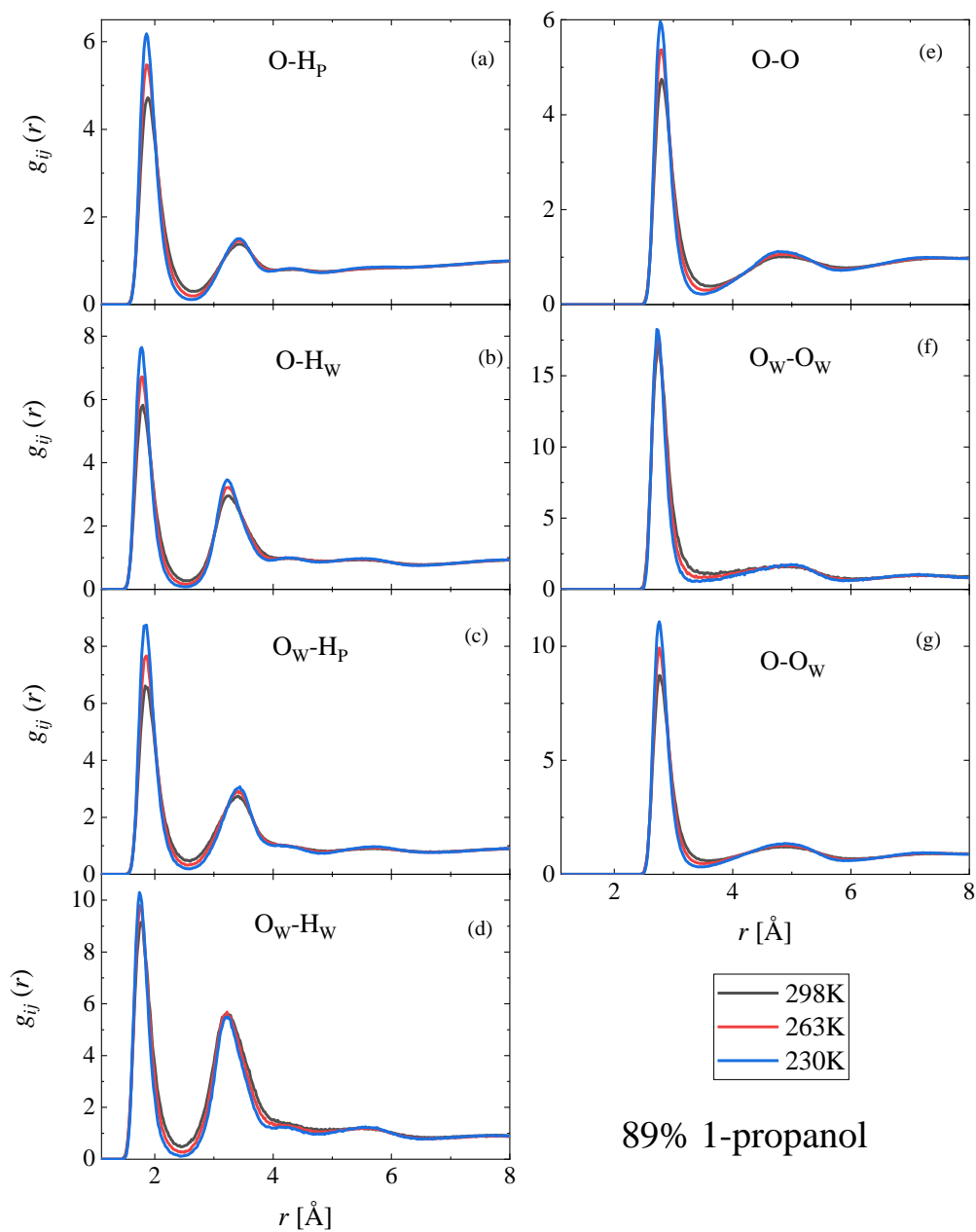

**Figure S39** Temperature dependence of simulated partial radial distribution functions of the 1-propanol – water mixture with 89 mol% 1-propanol content. The H-bonding related partials are shown: (a) 1-propanol O (denoted as O) – hydroxyl H of 1-propanol (denoted as $H_P$), (b) 1-propanol O – water H (denoted as $H_W$), (c) water O (denoted as $O_W$) – hydroxyl H of 1-propanol, (d) water O – water H, (e) 1-propanol O – 1-propanol O, (f) water O – water O, (g) 1-propanol O – water O. The curves were obtained by using the SPC/E water model.

**H-bond related figures obtained by simulations using the TIP4P/2005 water model**

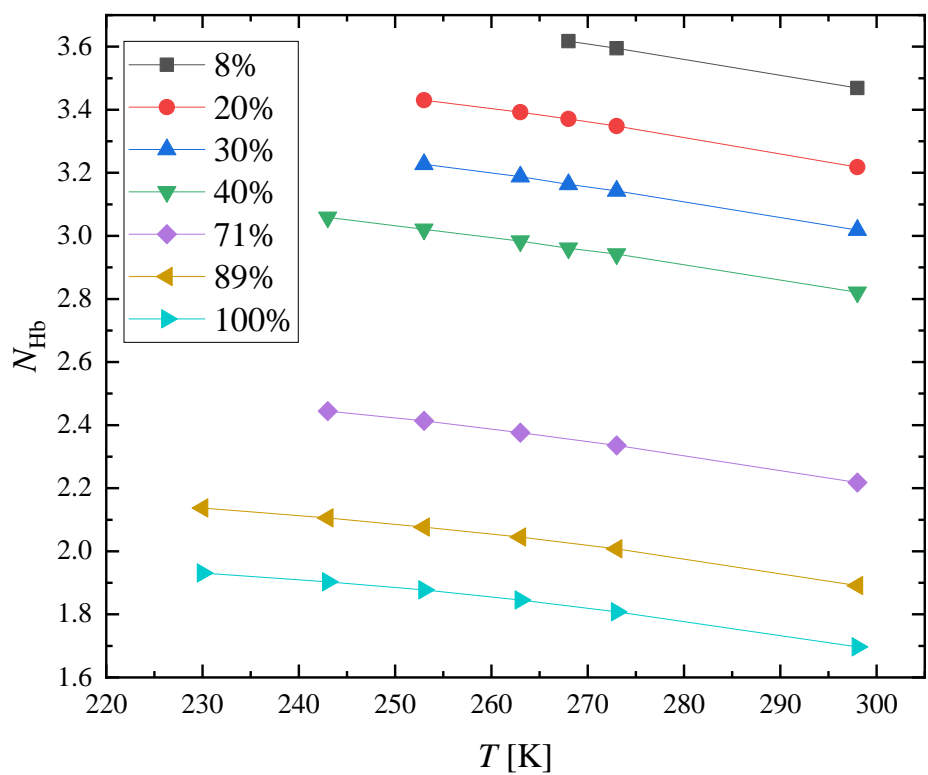

**Figure S40** Average number of hydrogen bonds per molecule in 1-propanol – water mixtures as a function of temperature at different concentrations, as obtained by MD simulations using the TIP4P/2005 water model.

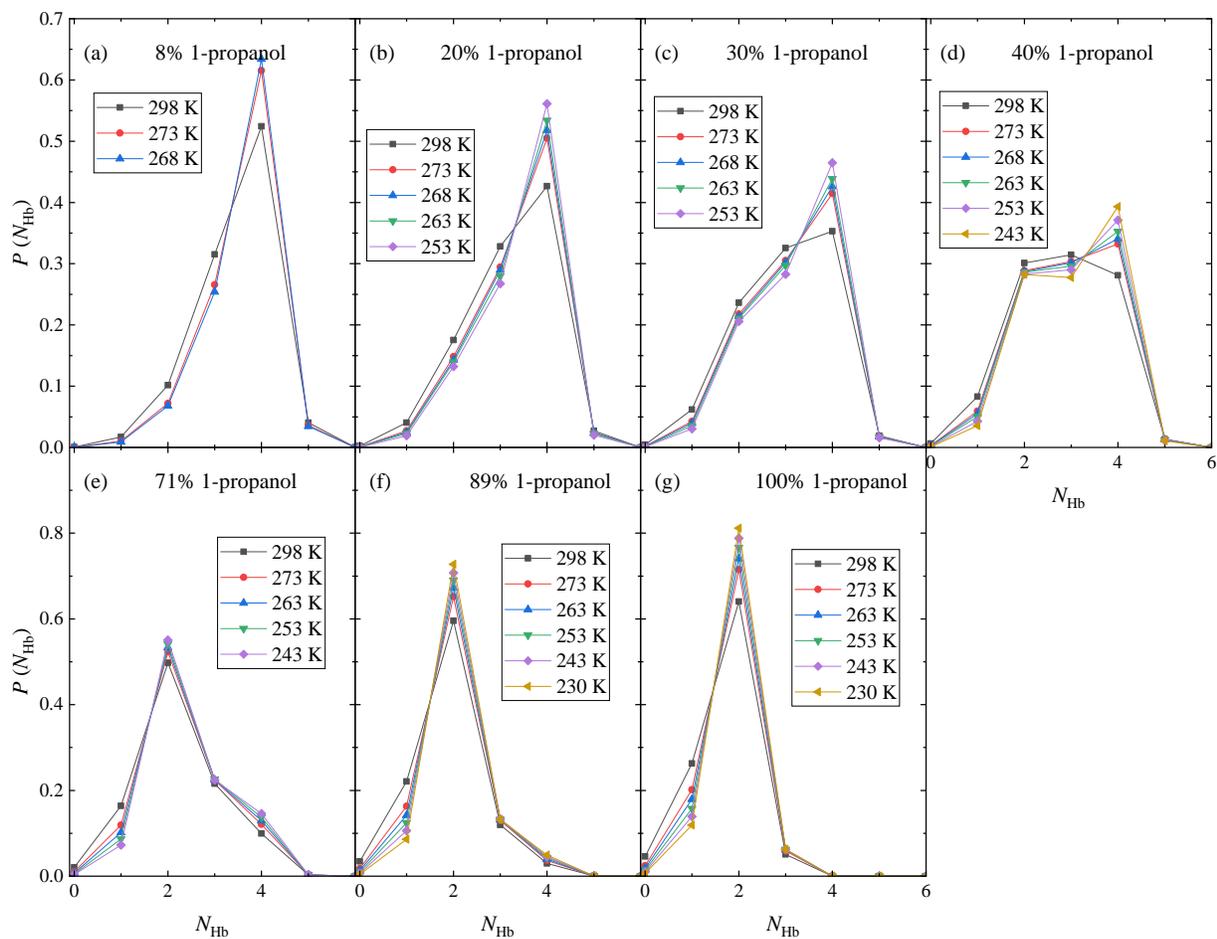

**Figure S41** Distribution of the number of H-bonds at different temperatures and 1-propanol concentrations in 1-propanol - water mixtures, as obtained by MD simulations using the TIP4P/2005 water model. (a) 8 %, (b) 20%, (c) 30%, (d) 40%, (e) 71%, (f) 89% and (g) 100% 1-propanol content.

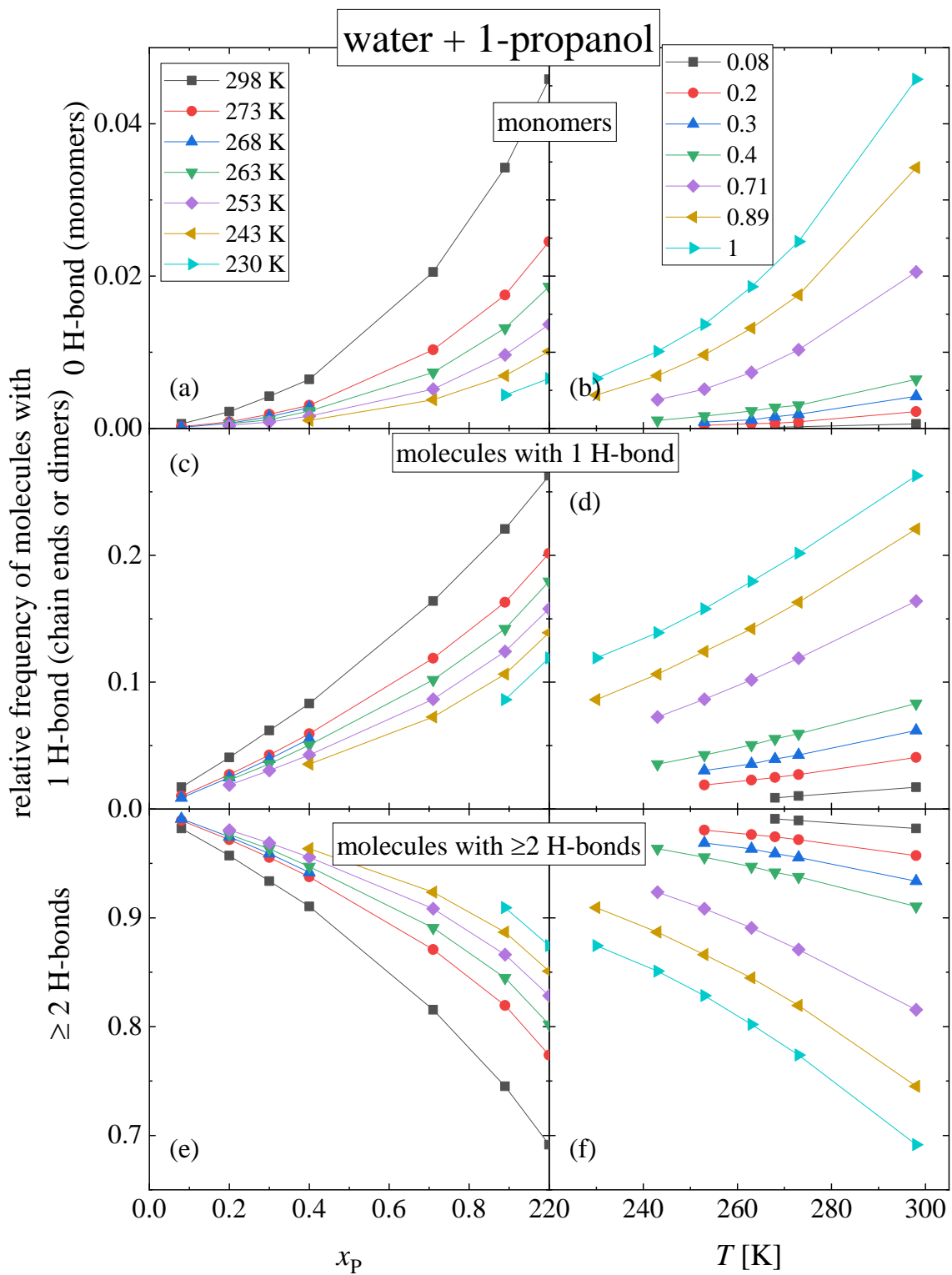

**Figure S42** (a,c,e) Concentration and (b,d,f) temperature dependences of the relative frequencies of molecules with (a,b) 0, (c,d) 1 or (e,f) ≥ 2 H-bonds. All type of molecules and all type of bonds (bonds between like and unlike molecules as well) are taken into account.

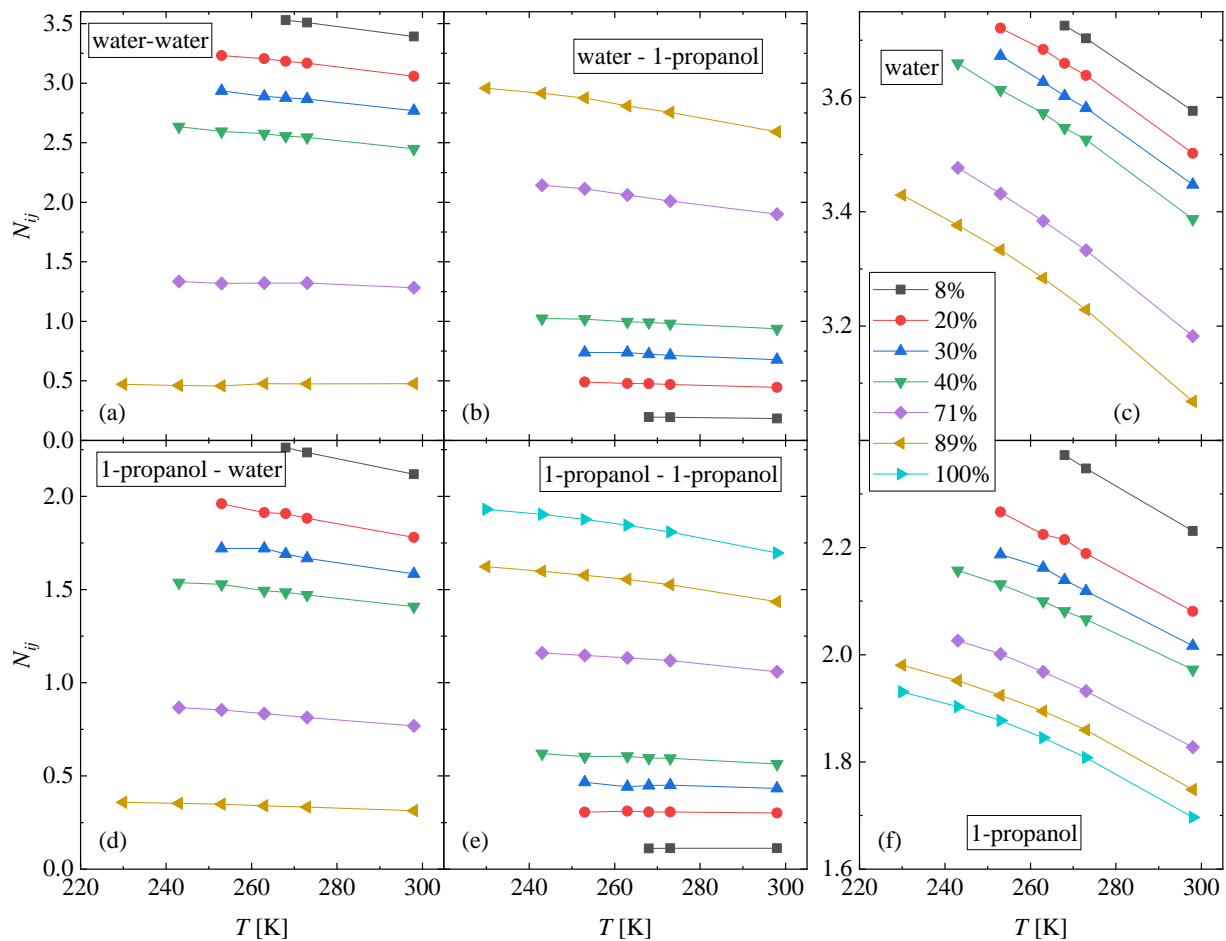

**Figure S43** Temperature dependence of the number of hydrogen bonds at different concentrations, as obtained by MD simulations using the TIP4P/2005 water model: (a) average number of H-bonded water molecules around water, (b) average number of H-bonded 1-propanol molecules around water, (c) average number of H-bonded (water and 1-propanol) molecules around water, (d) average number of H-bonded water molecules around 1-propanol, (e) average number of H-bonded 1-propanol molecules around 1-propanol, (f) average number of H-bonded (water and 1-propanol) molecules around 1-propanol.

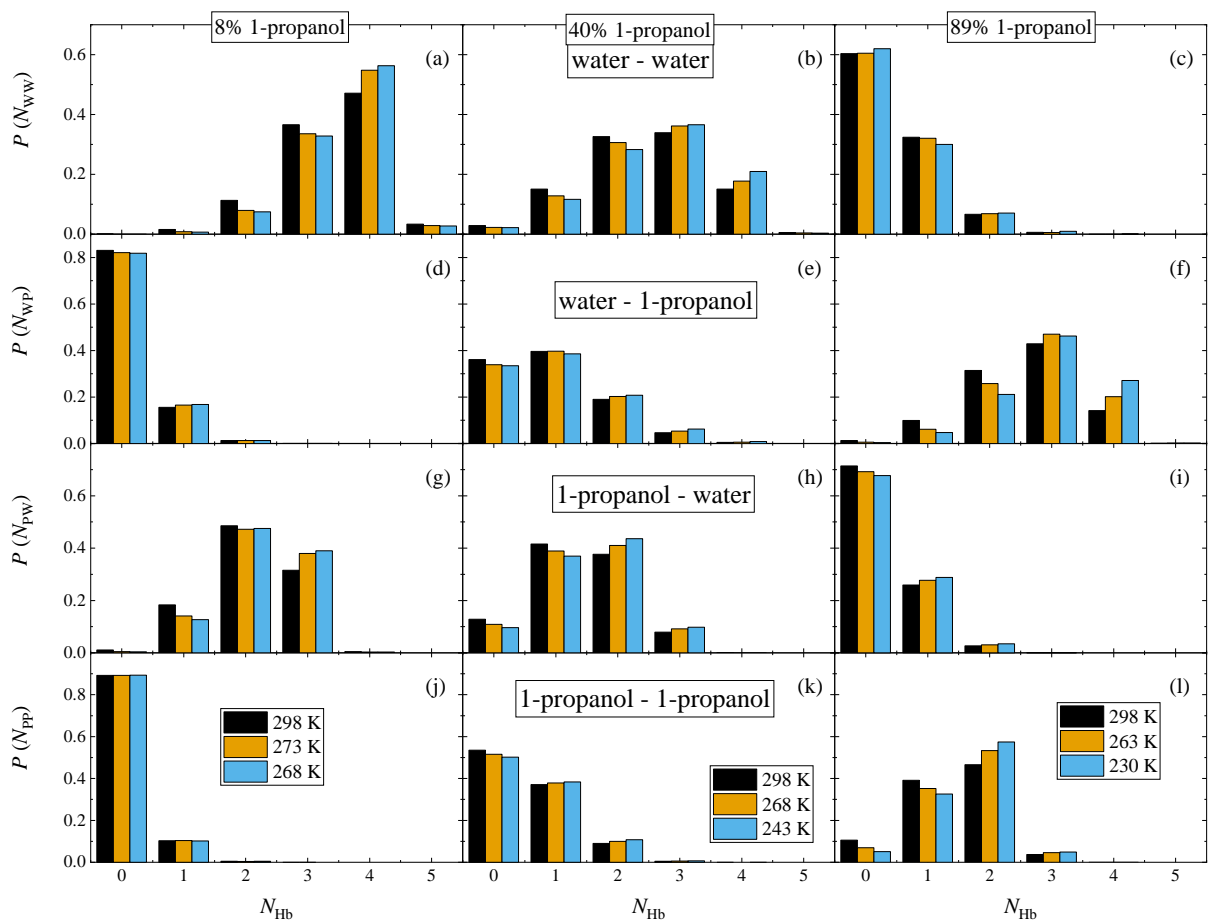

**Figure S44** Distribution of the number of hydrogen bonds at selected temperatures and 1-propanol concentrations (at (a,d,g,i) 8 mol%, (b,e,h,k) 40 mol% and (c,f,l) 89 mol% 1-propanol content), as obtained by MD simulations using the TIP4P/2005 water model: (a,b,c) H-bonded water molecules around water, (d,e,f) H-bonded 1-propanol molecules around water, (g,h,i) H-bonded water molecules around 1-propanol, (j,k,l) H-bonded 1-propanol molecules around 1-propanol.

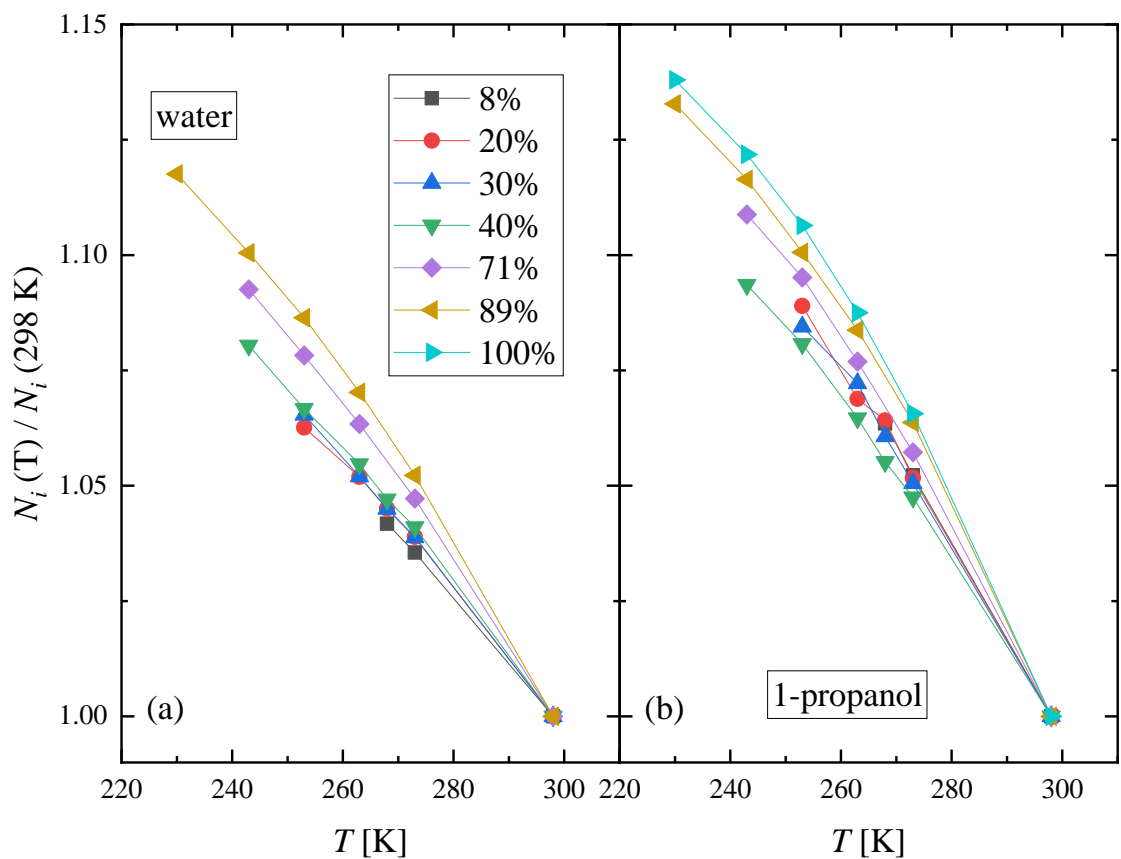

**Figure S45** Temperature dependence of the average number of H-bonded molecules (water and 1-propanol) around (a) water and (b) 1-propanol normalized to the 298 K value at different 1-propanol concentrations.

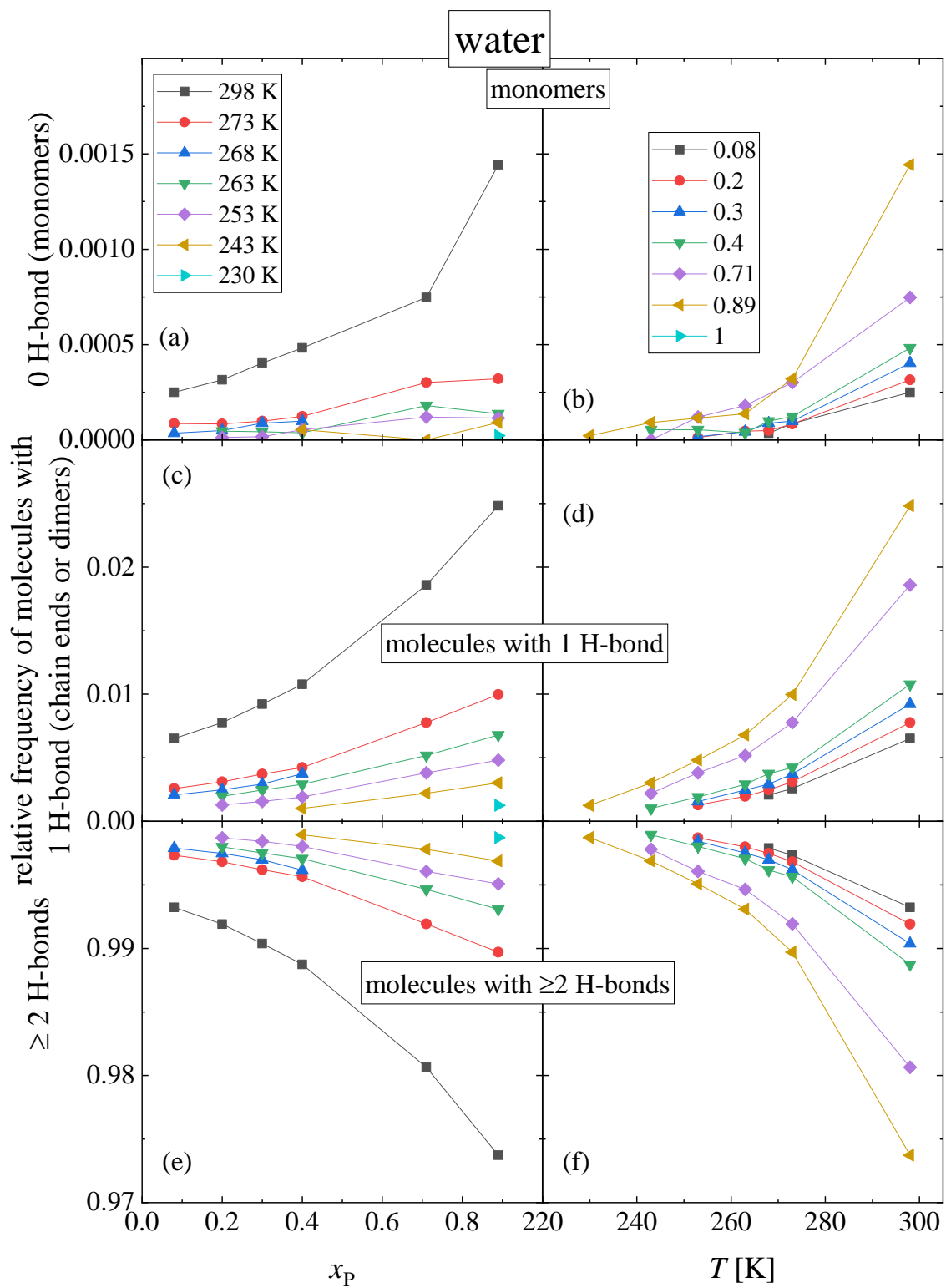

**Figure S46** (a,c,e) Concentration and (b,d,f) temperature dependence of the relative frequencies of water molecules with (a,b) 0, (c,d) 1 or (e,f) ≥ 2 H-bonds. All types of bonds (water - water and water – 1-propanol bonds, as well) are taken into account.

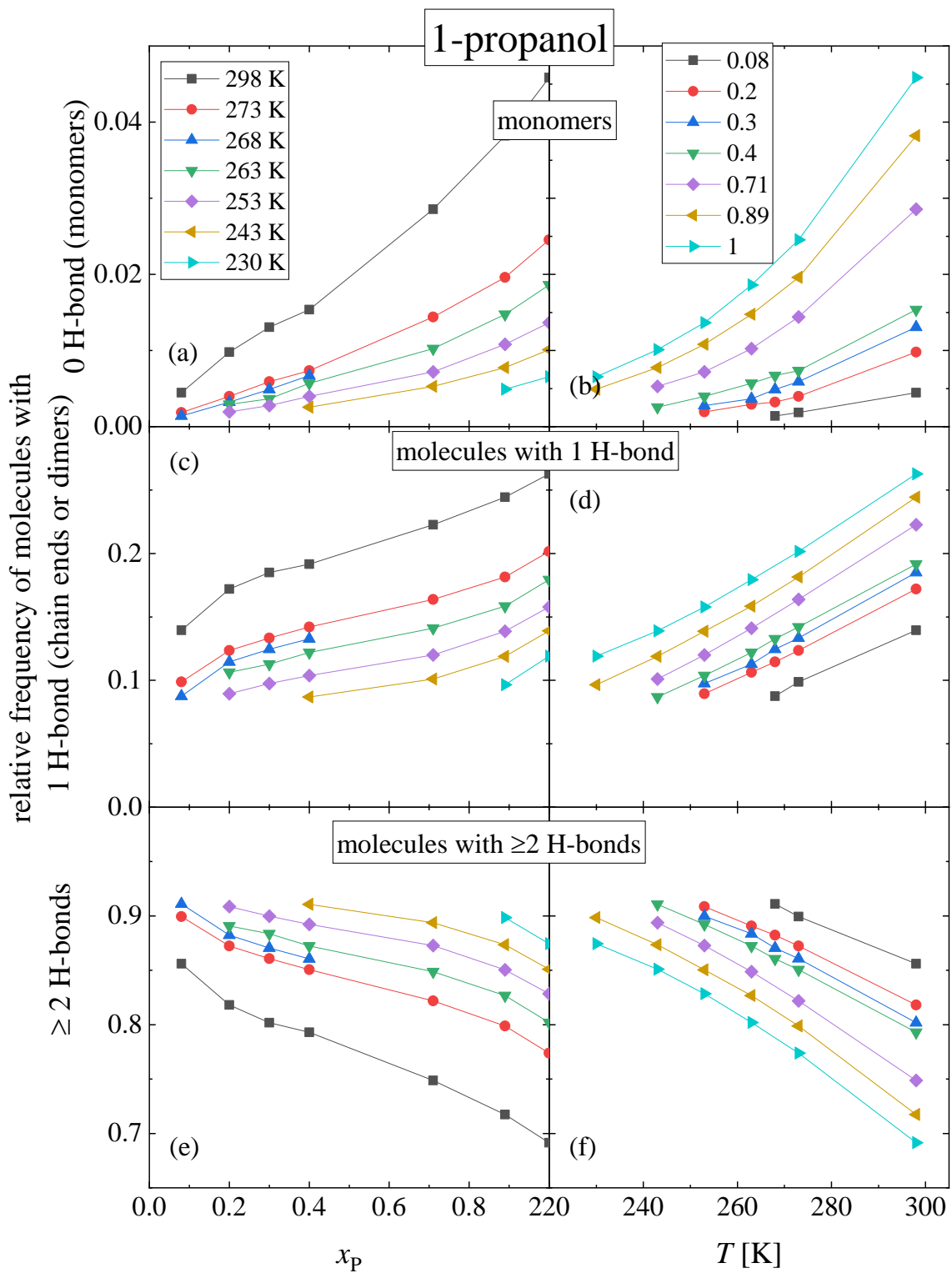

**Figure S47** (a,c,e) Concentration and (b,d,f) temperature dependence of the relative frequencies of 1-propanol molecules with (a,b) 0, (c,d) 1 or (e,f) ≥ 2 H-bonds. All types of bonds (1-propanol – 1-propanol and 1-propanol – water bonds as well) are taken into account.

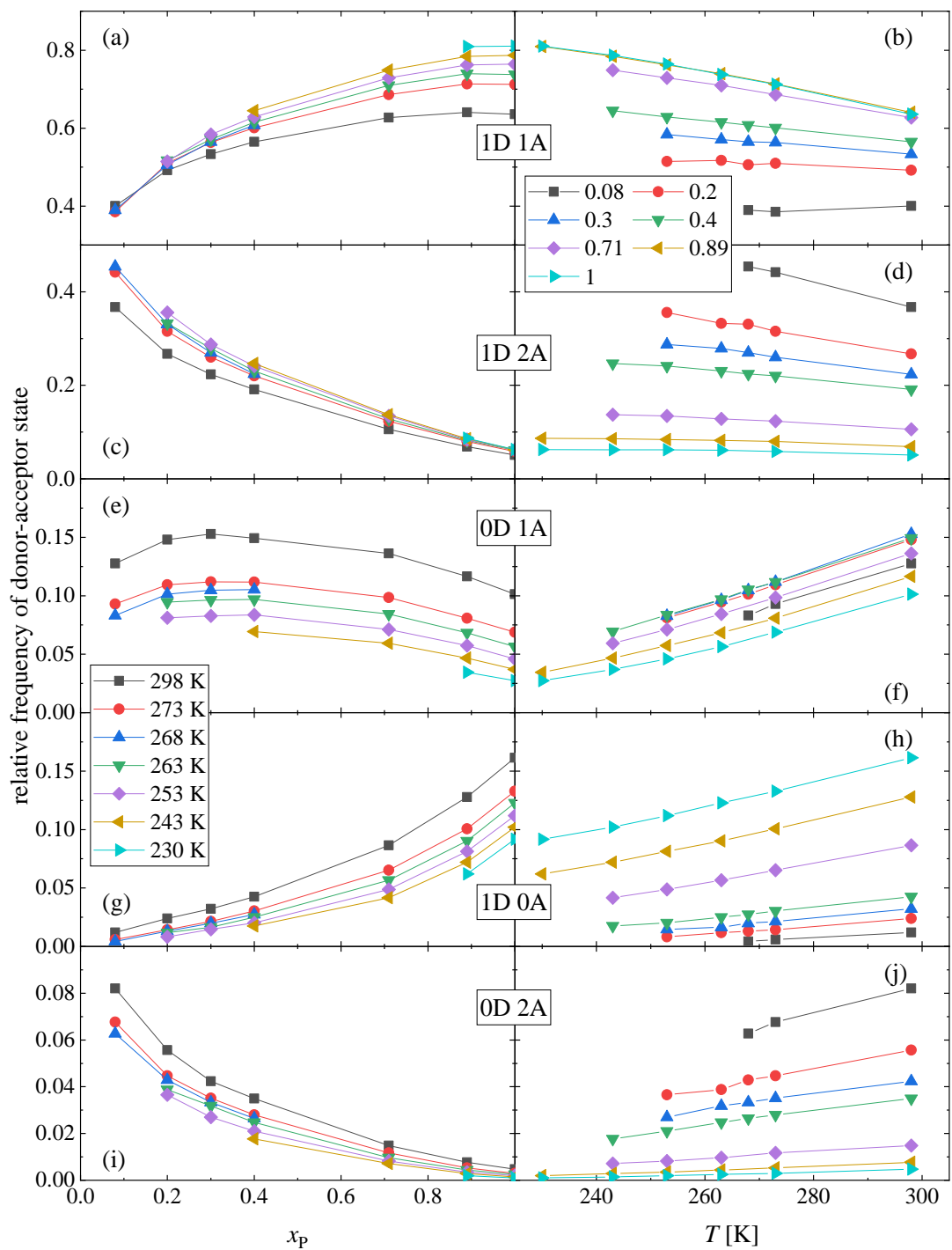

**Figure S48** (left side) Concentration dependence and (right side) temperature dependence of the most frequent donor-acceptor states of 1-propanol molecules: (a,b) 1 donor 1 acceptor state, (c,d) 1 donor 2 acceptor state, (e,f) 0 donor 1 acceptor state, (g,h) 1 donor 0 acceptor state, and (i,j) 0 donor 2 acceptor state.

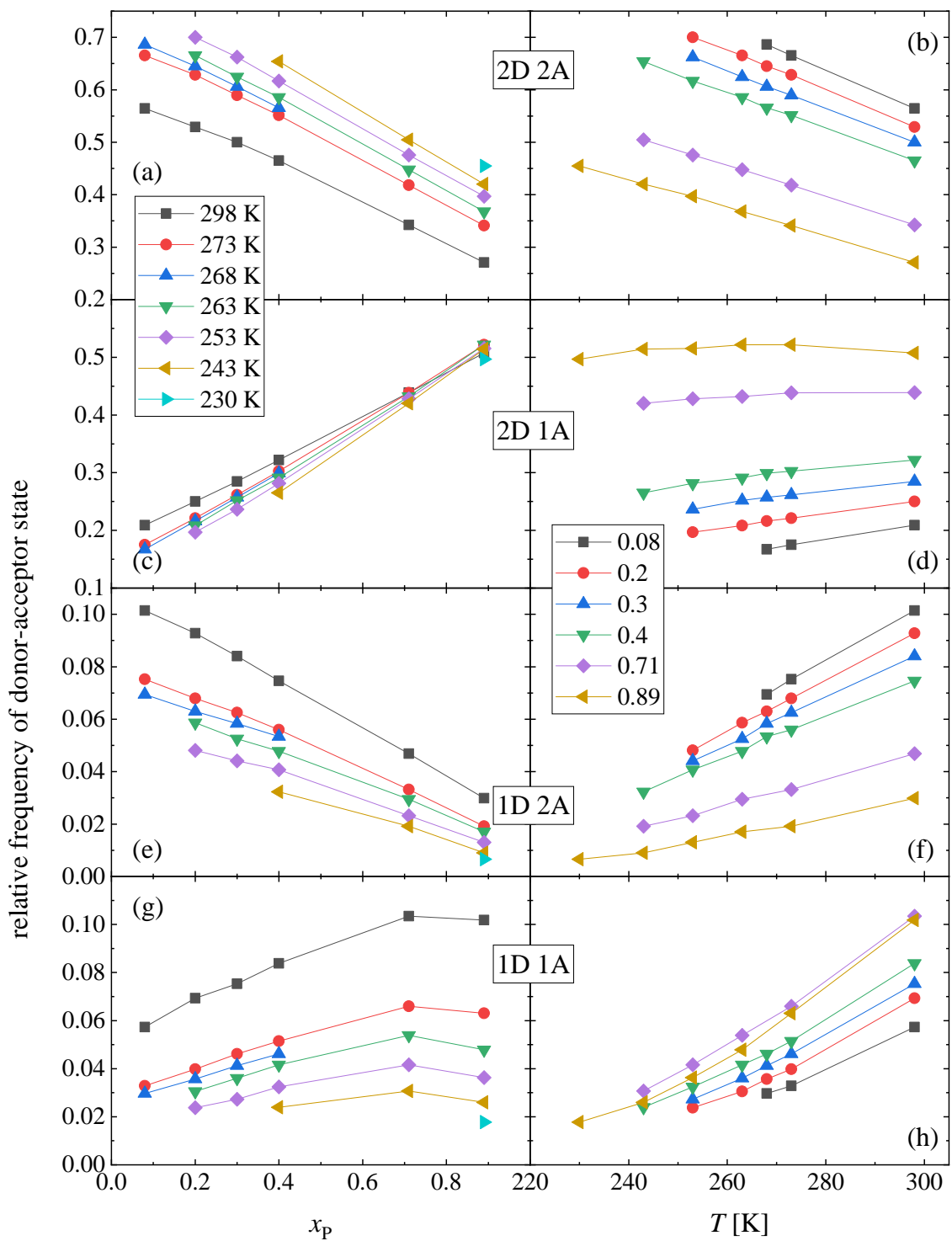

**Figure S49** (left side) Concentration dependence and (right side) temperature dependence of the most frequent donor-acceptor states of water molecules: (a,b) 2 donor 2 acceptor state, (c,d) 2 donor 1 acceptor state, (e,f) 1 donor 2 acceptor state, and (g,h) 1 donor 1 acceptor state.

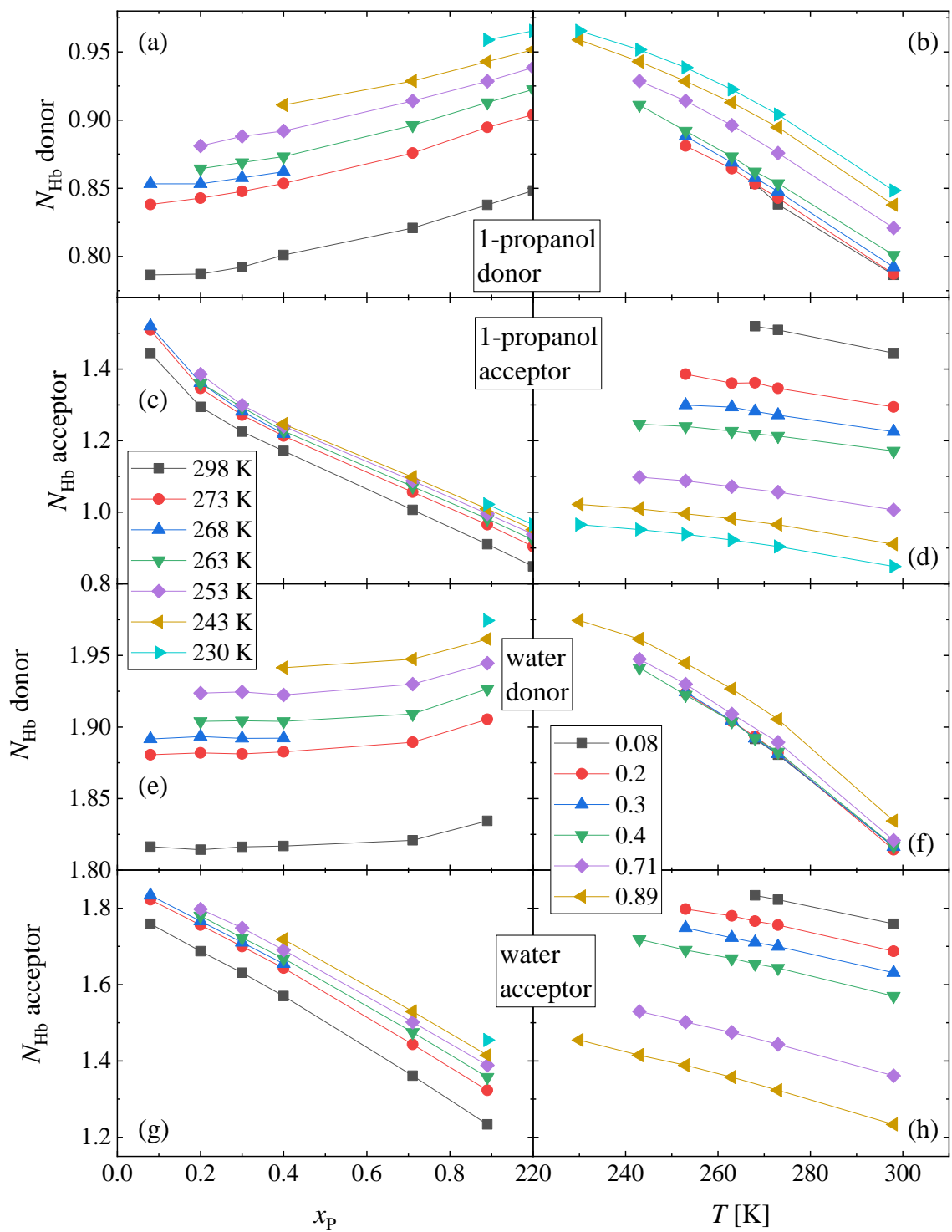

**Figure S50** (left side) Concentration and (right side) temperature dependence of the average number of H-bonds of (a-d) 1-propanol and (e-h) water molecules as (a,b,e,f,) donors and as (c,d,g,h) acceptors.

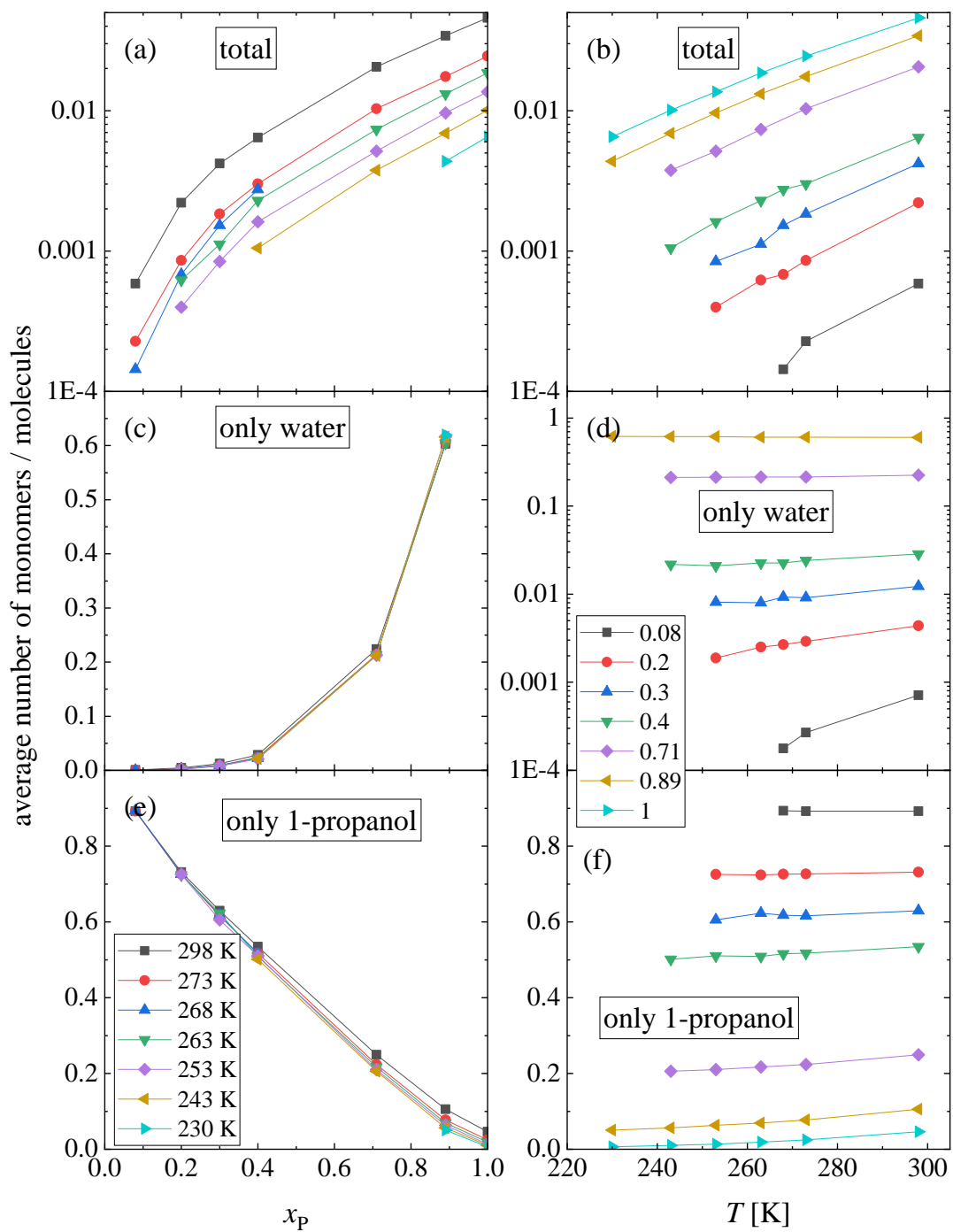

**Figure S51** (left side, a,c,e) Concentration and (right side, b,d,f) temperature dependence of the average number of monomer molecules concerning (a,b) all molecules, (c,d) water subsystem, (e,f) 1-propanol subsystem. The number of monomers is normalized by the number of the corresponding molecules in the simulation box. (Note that the $y$-scales of parts (c, e, f) are linear, while the others are logarithmic!)

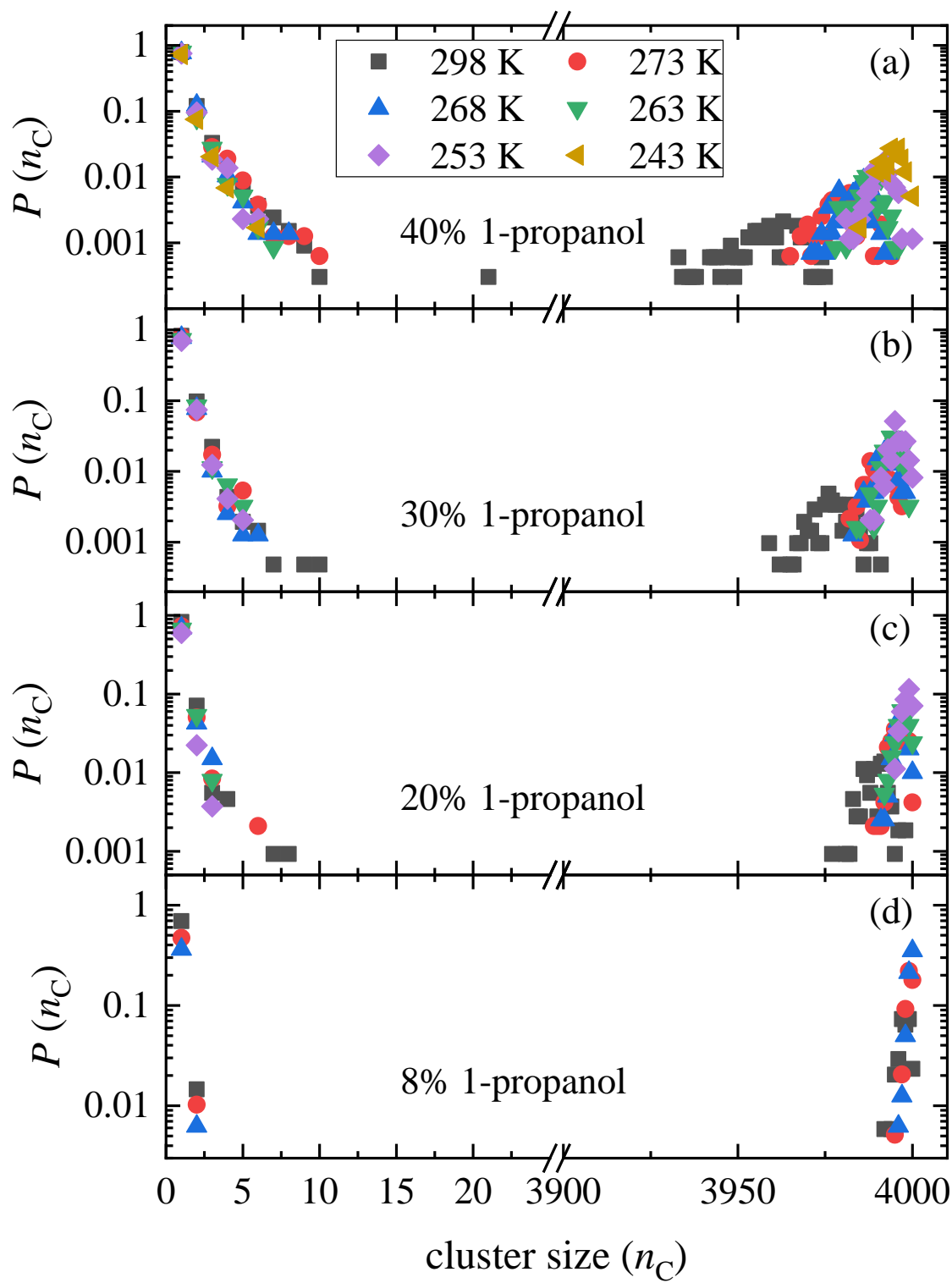

**Figure S52** Cluster size distributions in water-rich 1-propanol – water mixtures at different temperatures concerning H-bonds between any types of molecules.

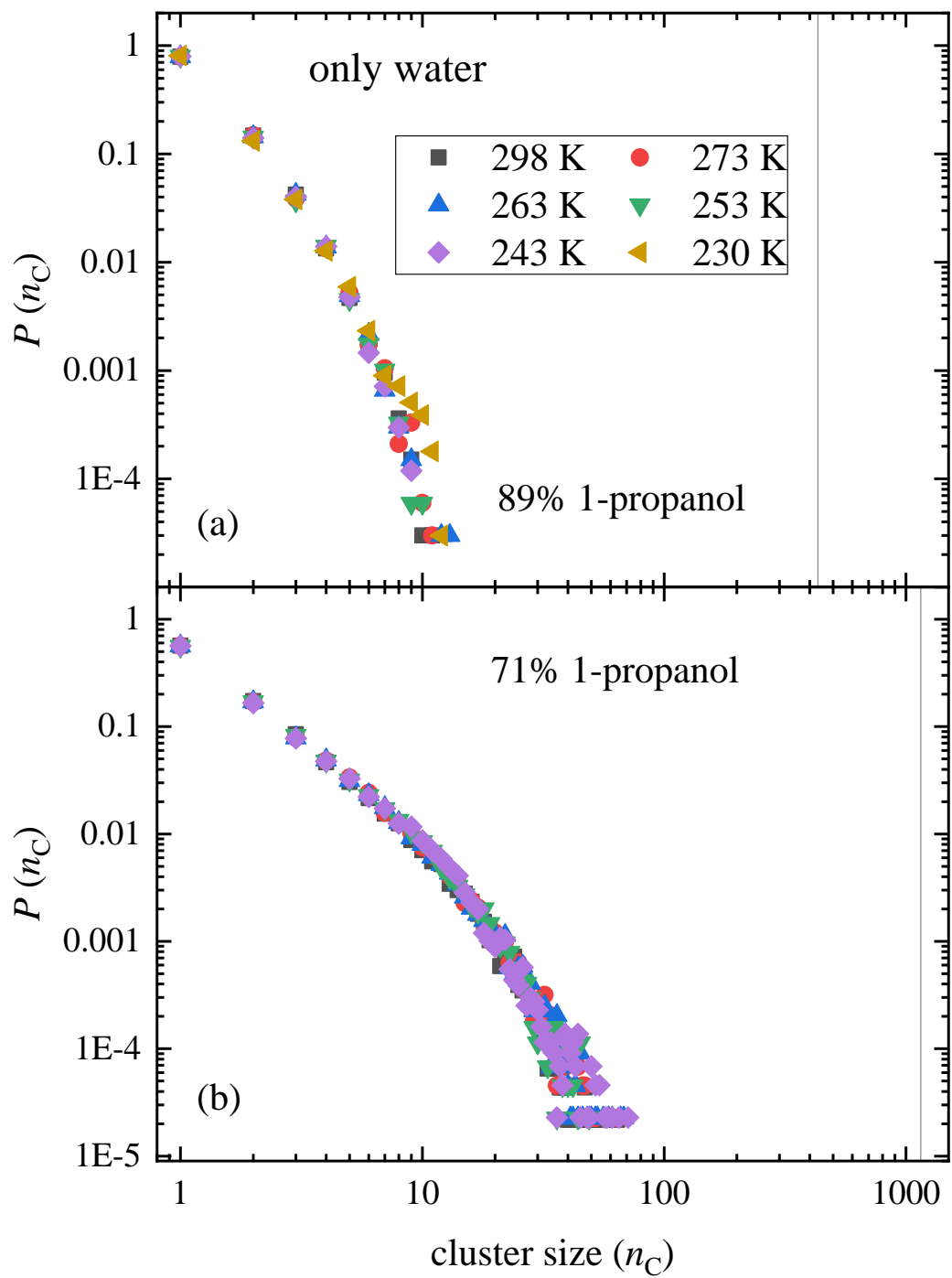

**Figure S53** Cluster size distributions within the water subsystem in 1-propanol – rich 1-propanol-water mixtures at different temperatures (H-bonds between water molecules are considered only.) The vertical lines show the numbers of water molecules in the simulation boxes.

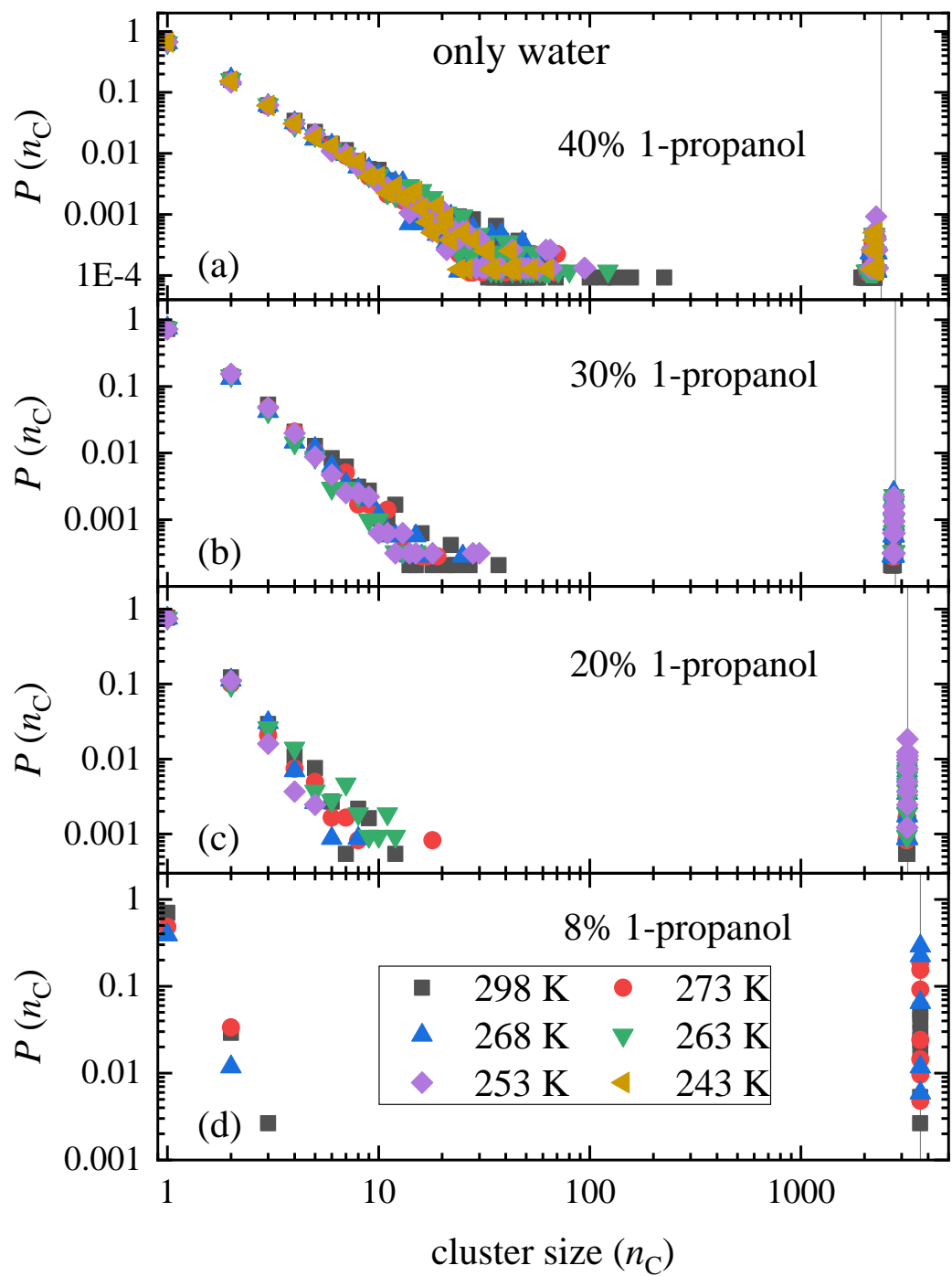

**Figure S54** Cluster size distributions within the water subsystem in water-rich 1-propanol – water mixtures at different temperatures (H-bonds between water molecules are considered only.) The vertical lines show the numbers of water molecules in the simulation boxes.

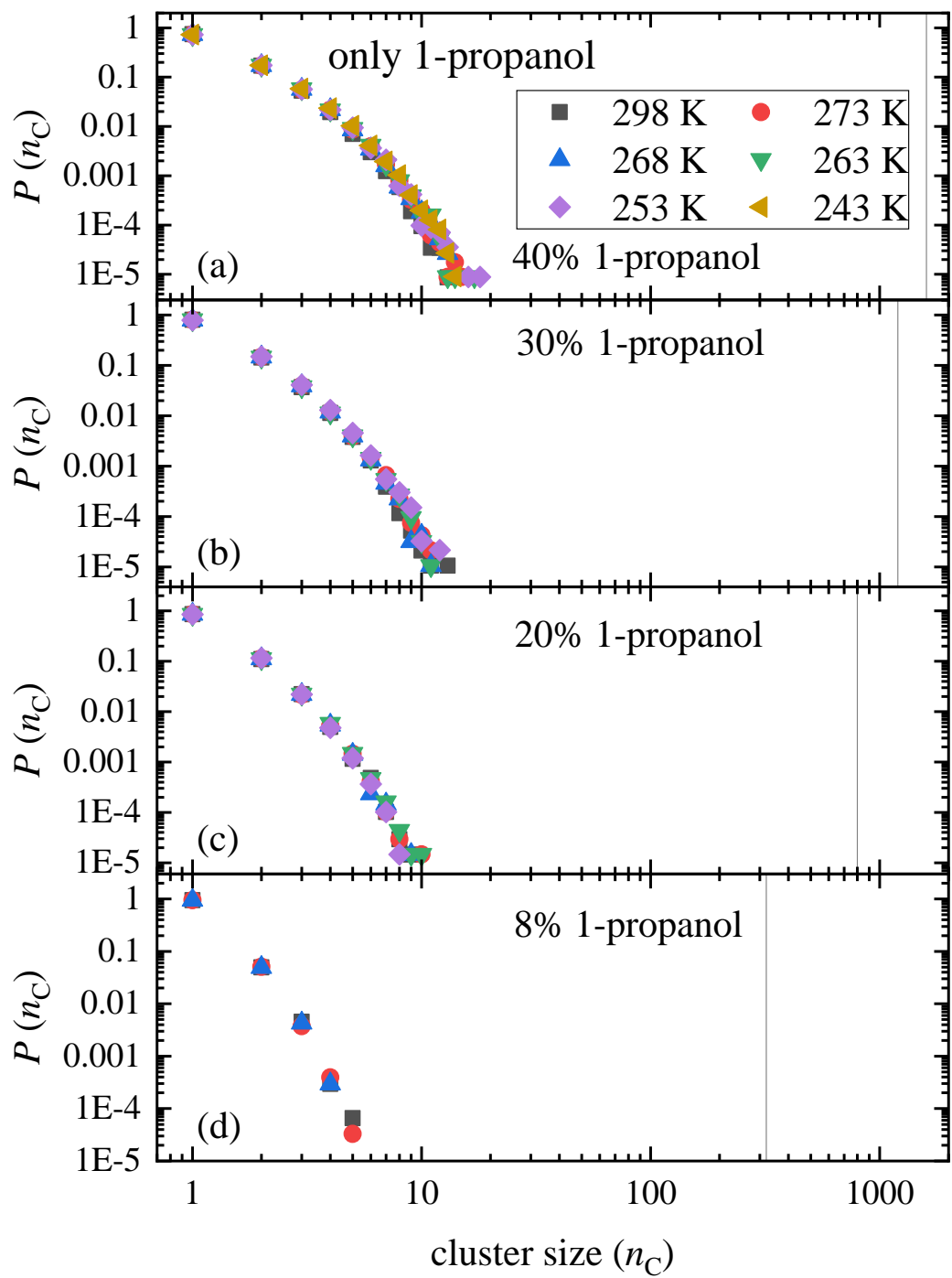

**Figure S55** Cluster size distributions within the 1-propanol subsystem in water-rich 1-propanol – water mixtures at different temperatures (H-bonds between 1-propanol molecules are considered only.) The vertical lines show the numbers of 1-propanol molecules in the simulation boxes.

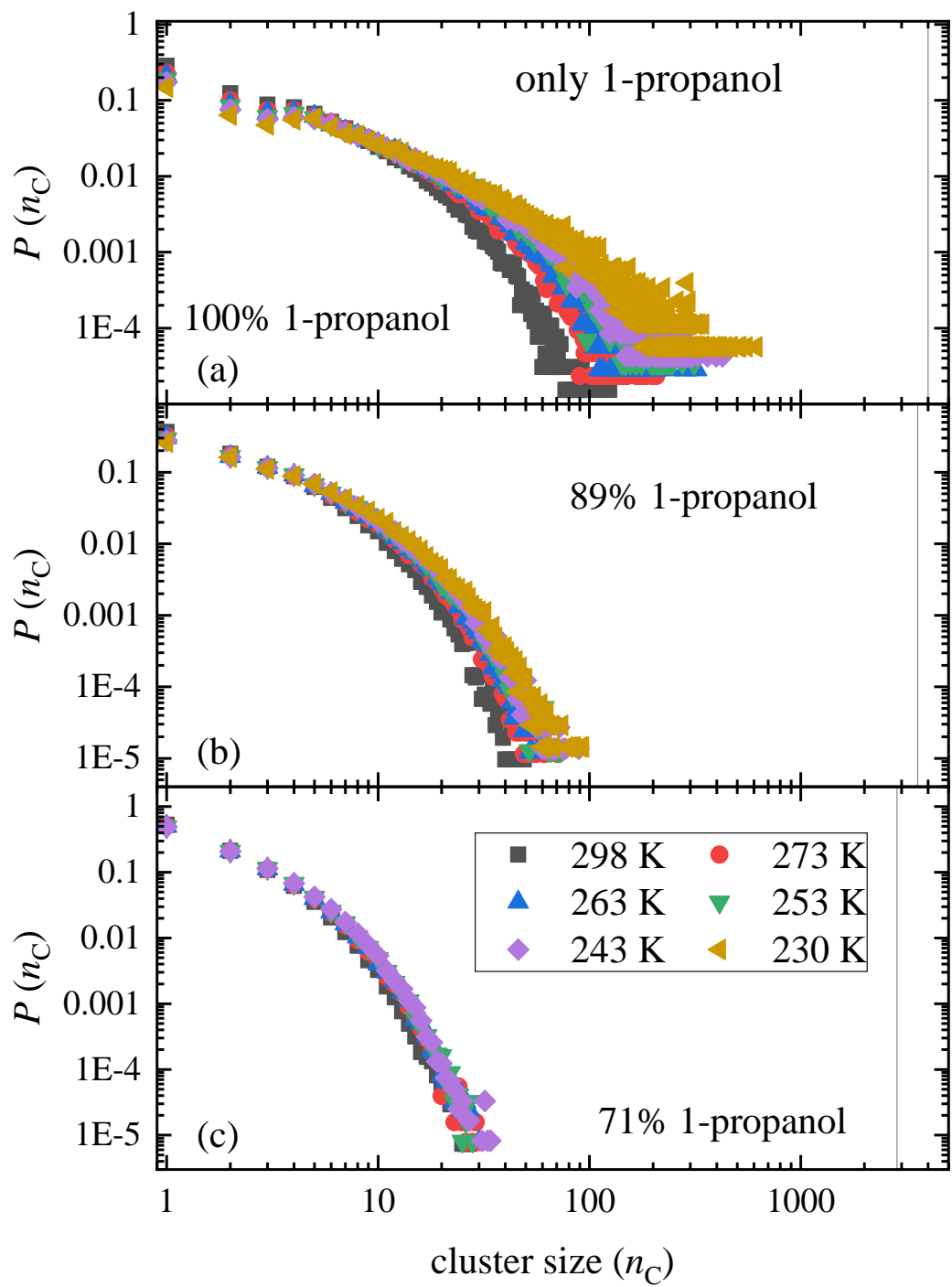

**Figure S56** Cluster size distributions within the 1-propanol subsystem in 1-propanol-rich 1-propanol – water mixtures at different temperatures (H-bonds between 1-propanol molecules are considered only.) The vertical lines show the numbers of 1-propanol molecules in the simulation boxes.

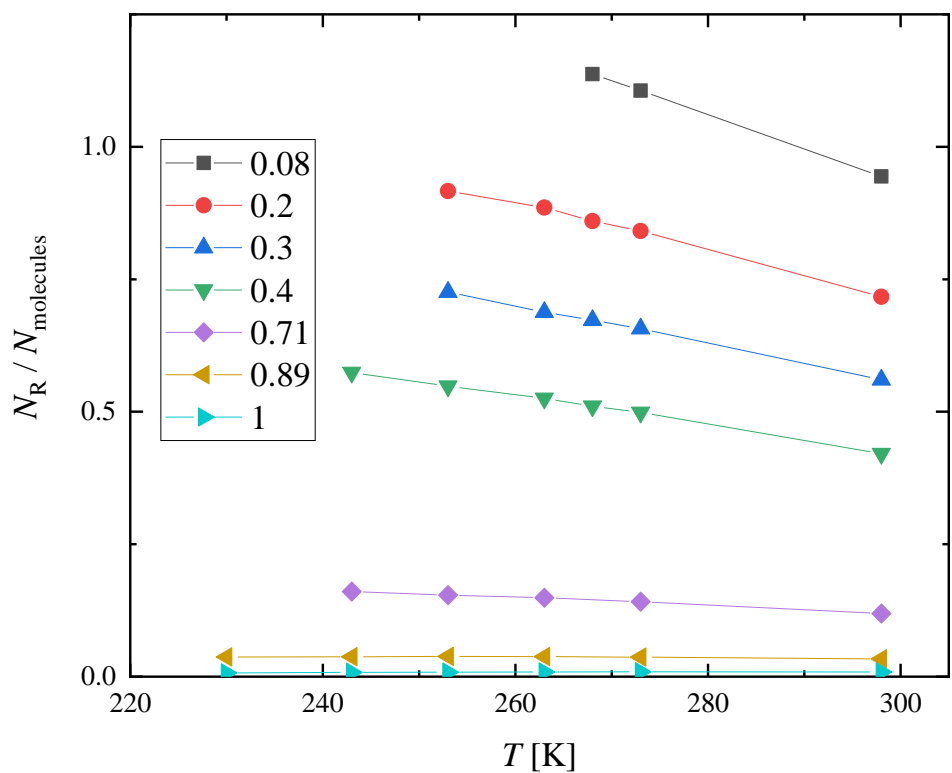

**Figure S57** Average number of primitive rings per molecules in 1-propanol – water mixtures as a function of temperature at different concentrations, obtained by MD simulations using the TIP4P/2005 water model. Rings are calculated up to 8 membered ones.

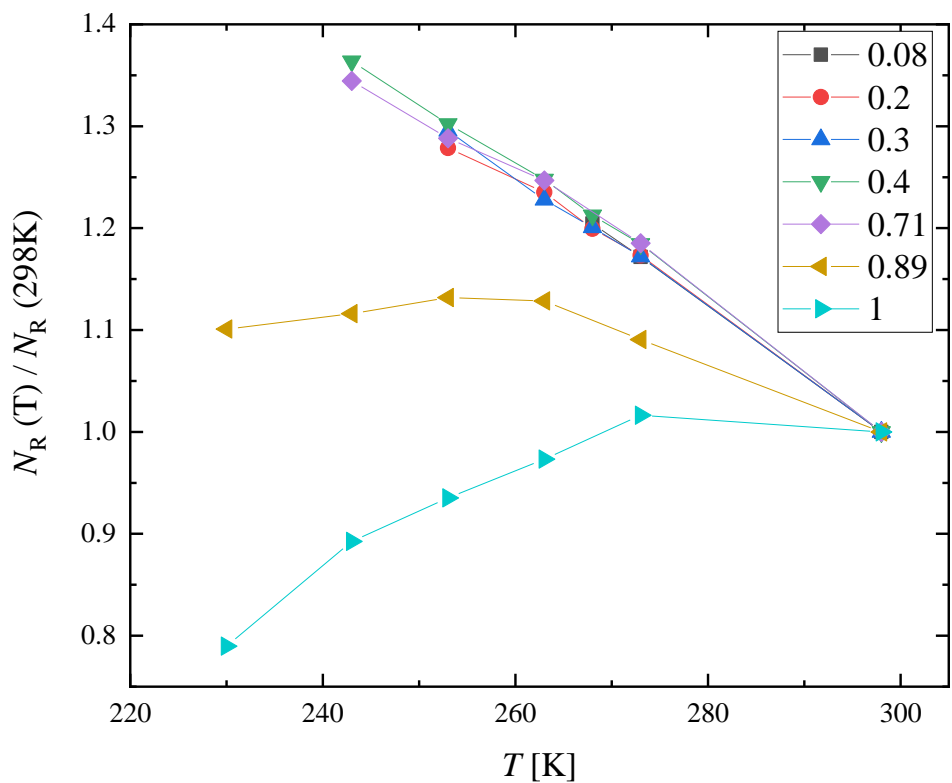

**Figure S58** Temperature dependence of the average number of rings normalized by the 298 K value. (Note: black squares are covered by red circles and blue triangles.)

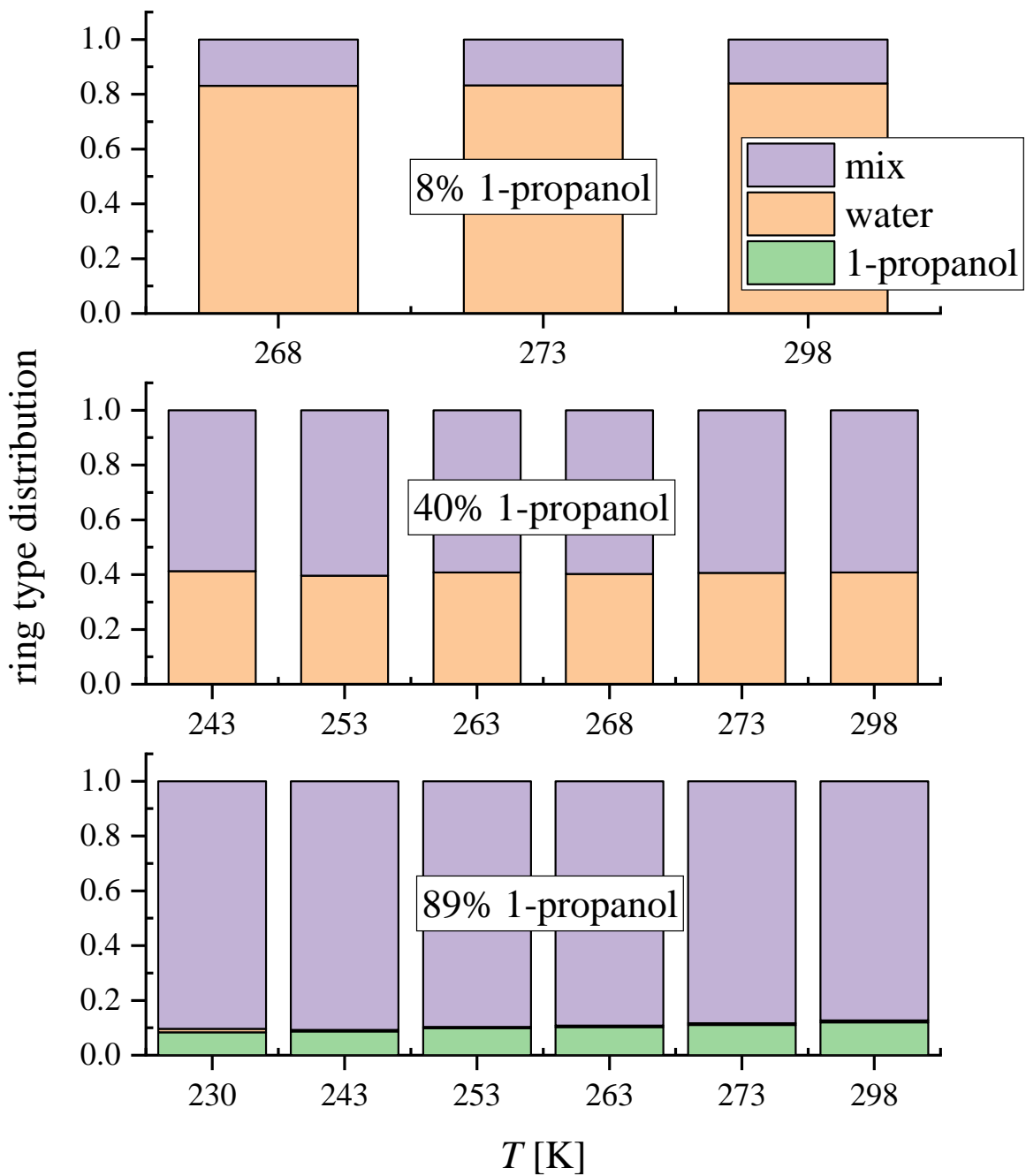

**Figure S59** Temperature dependence of the distribution of different ring types at three selected 1-propanol concentrations.

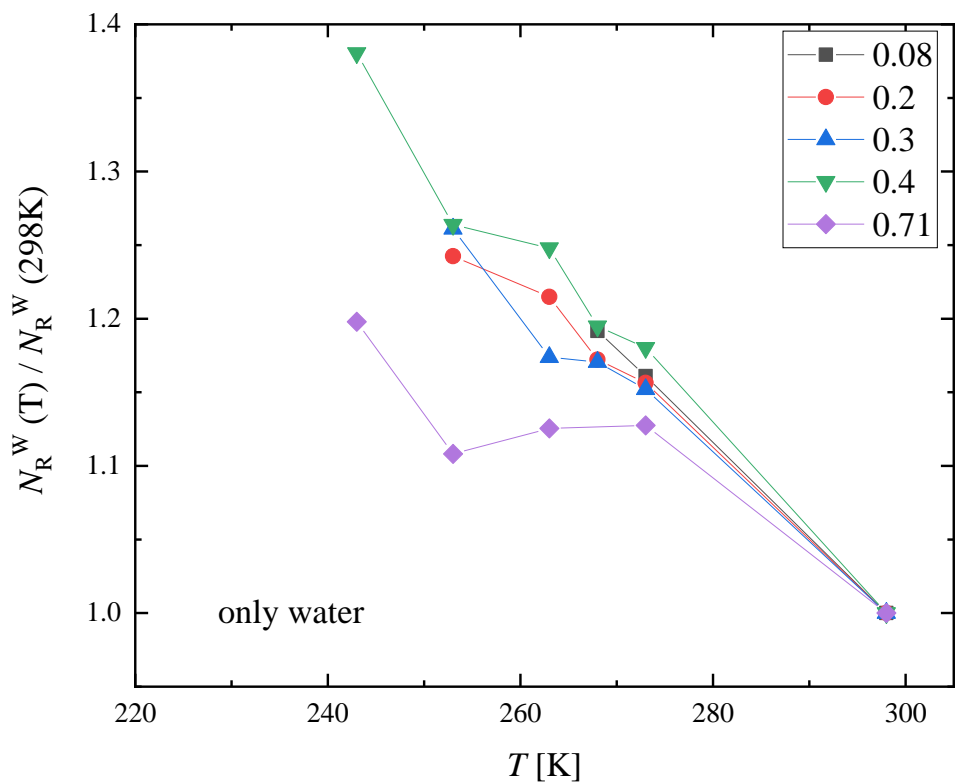

**Figure S60** Temperature dependence of the average number of water rings, normalized by the value at 298 K, at different 1-propanol concentrations. (In the $x_P = 0.89$ mixture the number of water rings is very low and the ratio has high uncertainty, not shown here.)

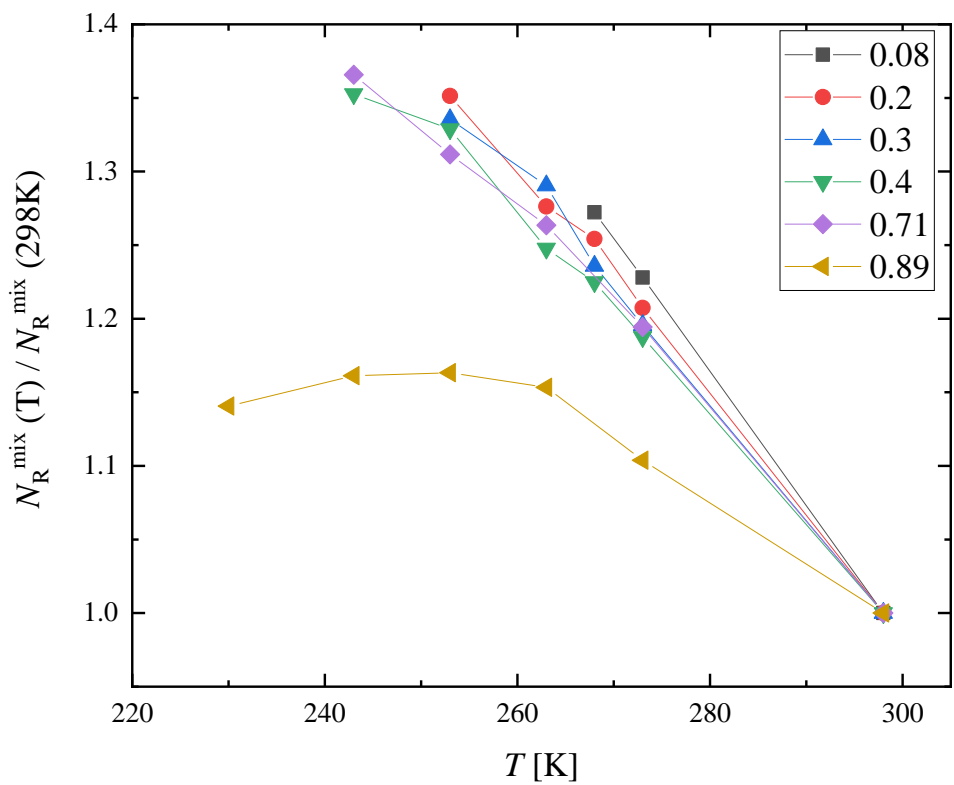

**Figure S61** Temperature dependence of the average number of rings containing both 1-propanol and water molecules, normalized by the 298 K value.

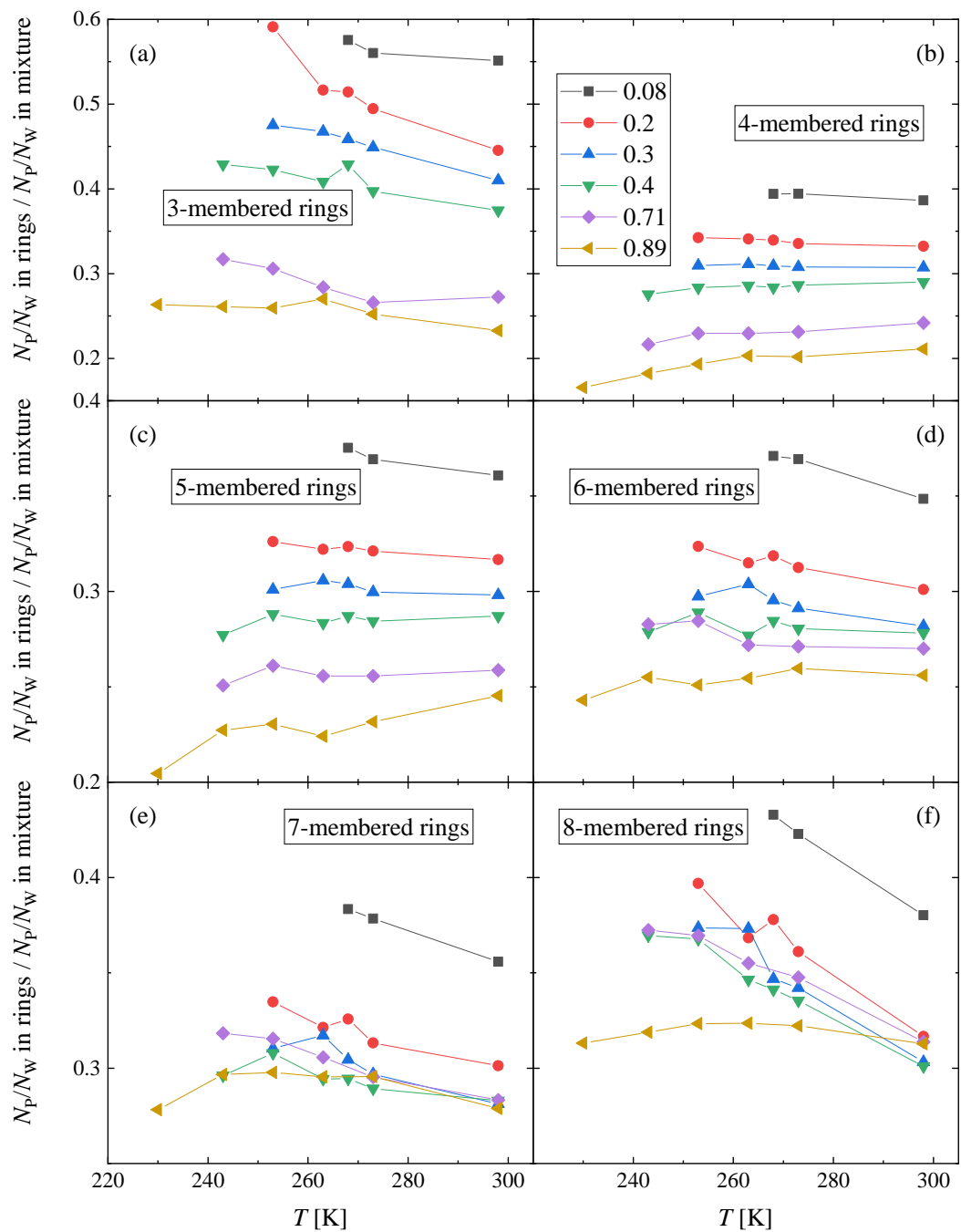

**Figure S62** Temperature and concentration dependence of the participation of 1-propanol and water molecules in different rings: (a) 3-membered, (b) 4-membered, (c) 5-membered, (d) 6-membered, (e) 7-membered and (f) 8-membered rings. The ratio of 1-propanol and water molecules in the rings is compared to the ratio of 1-propanol and water molecules in the mixture.

**H-bond related figures obtained by simulations using the SPC/E water model**

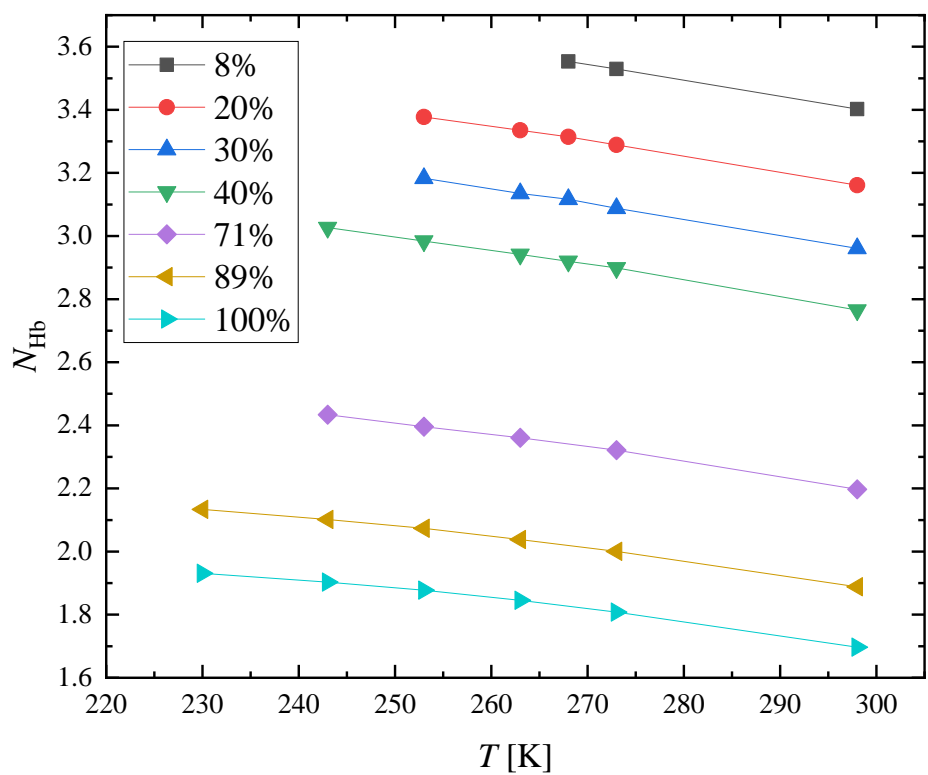

**Figure S63** Average number of hydrogen bonds per molecule in 1-propanol – water mixtures as a function of temperature at different concentrations, obtained by MD simulations using the SPC/E water model.

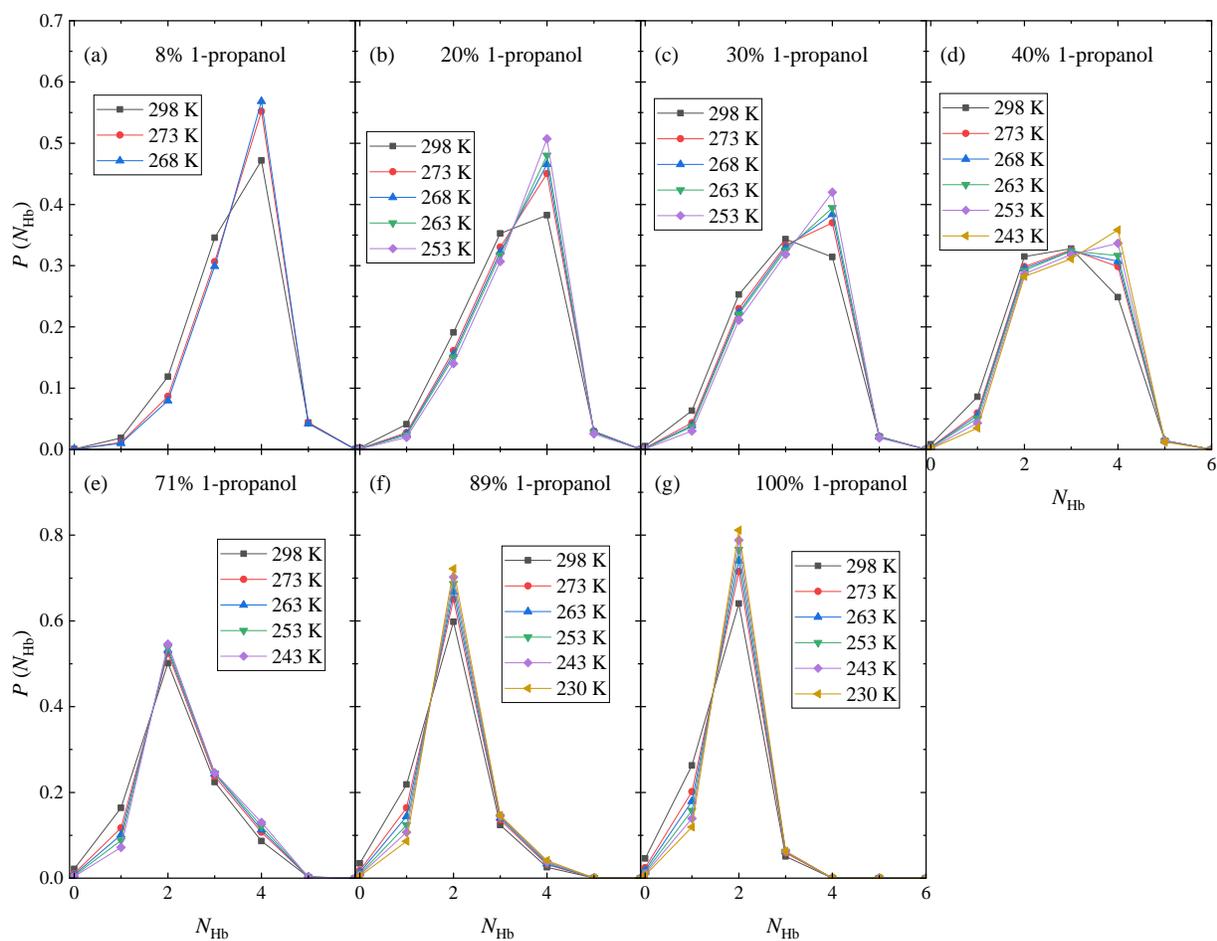

**Figure S64** Distribution of the number of H-bonds at different temperatures and 1-propanol concentrations in 1-propanol – water mixtures, as obtained by MD simulations using the SPC/E water model. (a) 8 %, (b) 20%, (c) 30%, (d) 40%, (e) 71%, (f) 89% and (g) 100% 1-propanol content.

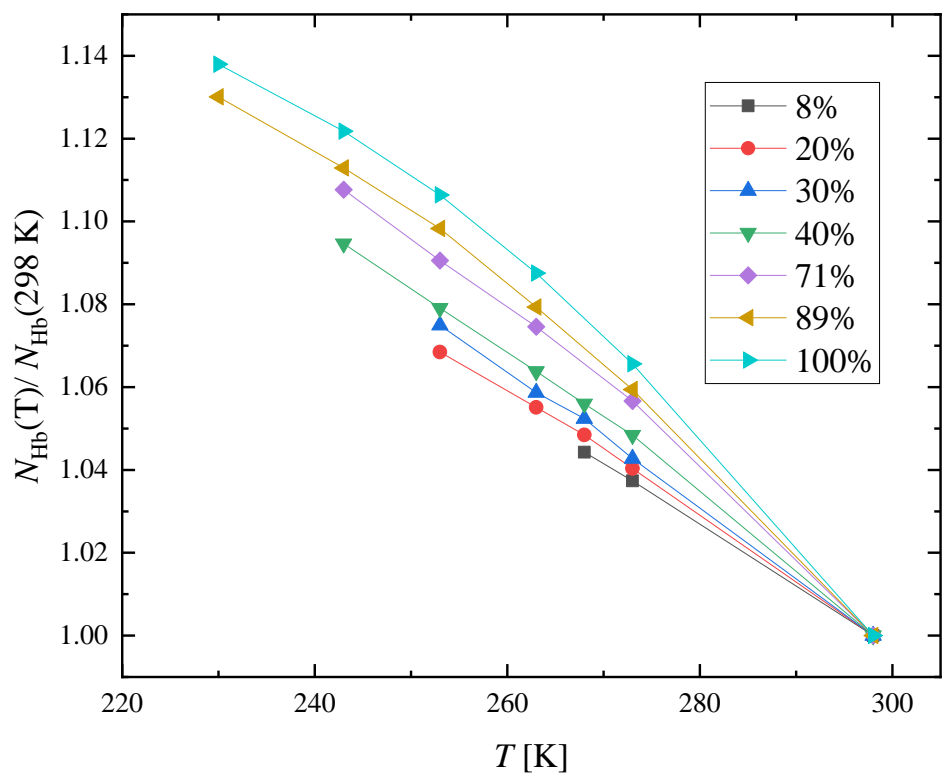

**Figure S65** Temperature dependence of the average number of hydrogen bonds per molecule, normalized to the 298 K value, at different 1-propanol concentrations.

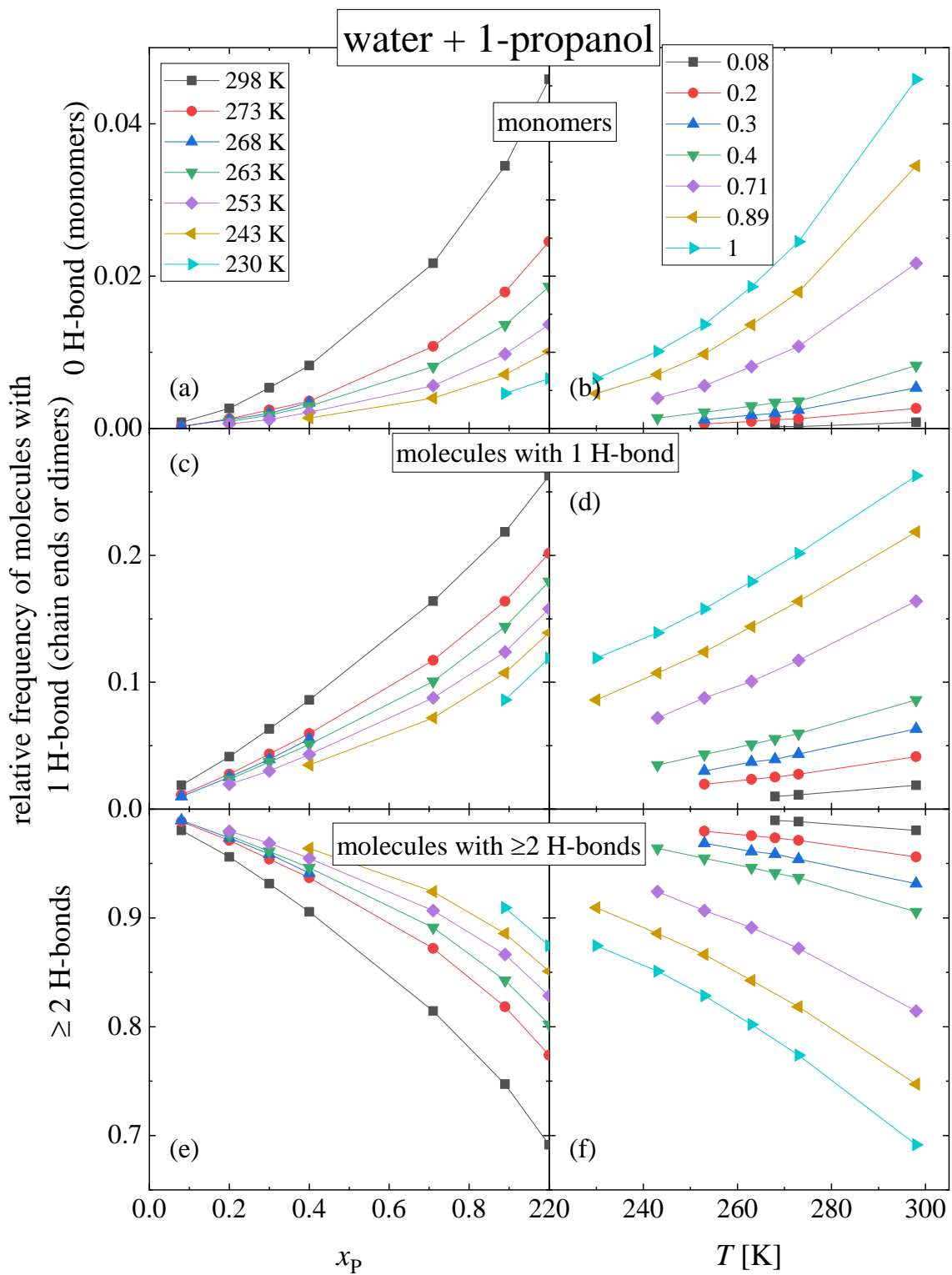

**Figure S66** (a,c,e) Concentration and (b,d,f) temperature dependence of the relative frequencies of molecules with (a,b) 0, (c,d) 1 or (e,f) ≥ 2 H-bonds. All types of molecules and all types of bonds (bonds between like and unlike molecules, as well) are taken into account.

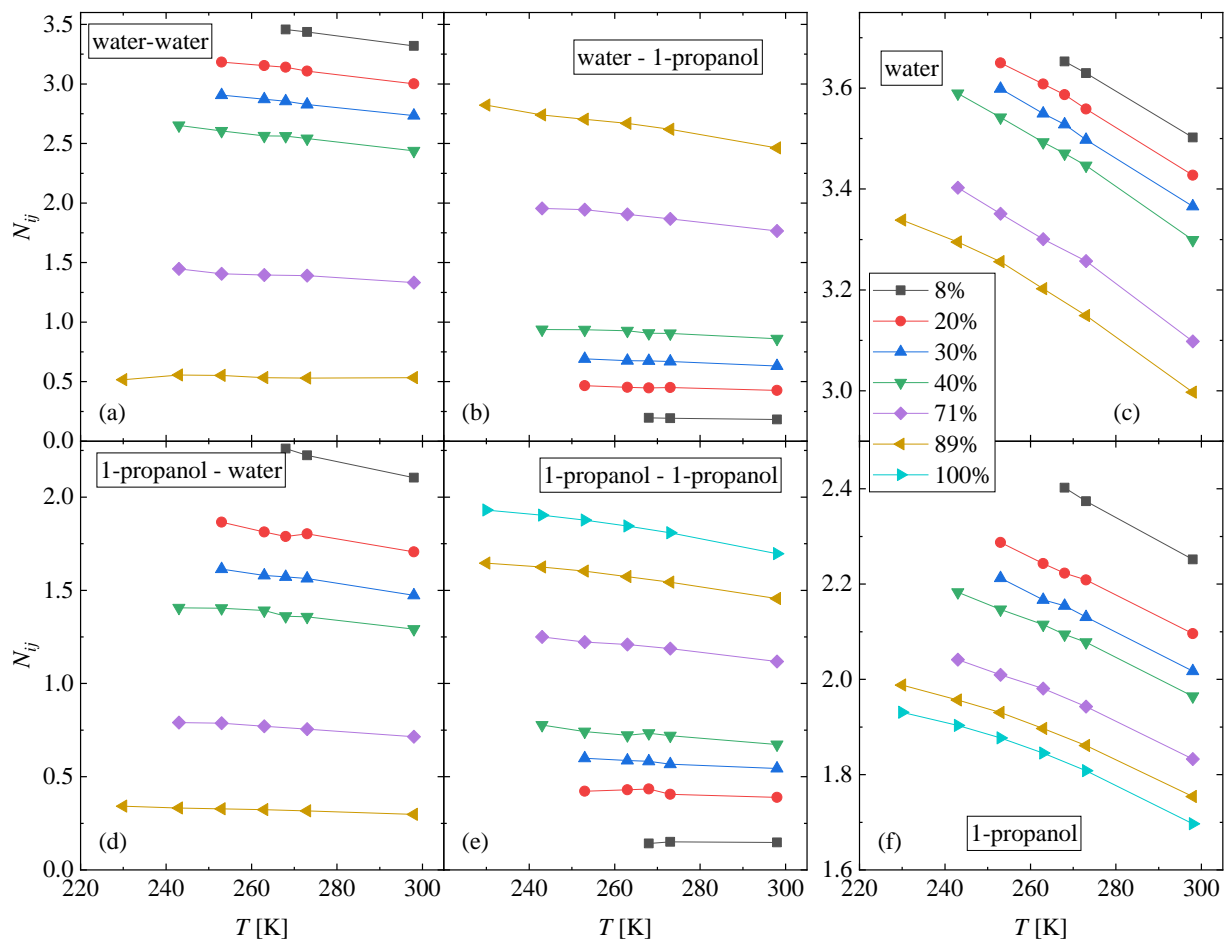

**Figure S67** Temperature dependence of the number of hydrogen bonds at different concentrations, as obtained by MD simulations using the SPC/E water model: (a) average number of H-bonded water molecules around water, (b) average number of H-bonded 1-propanol molecules around water, (c) average number of H-bonded (water and 1-propanol) molecules around water, (d) average number of H-bonded water molecules around 1-propanol, (e) average number of H-bonded 1-propanol molecules around 1-propanol, (f) average number of H-bonded (water and 1-propanol) molecules around 1-propanol.

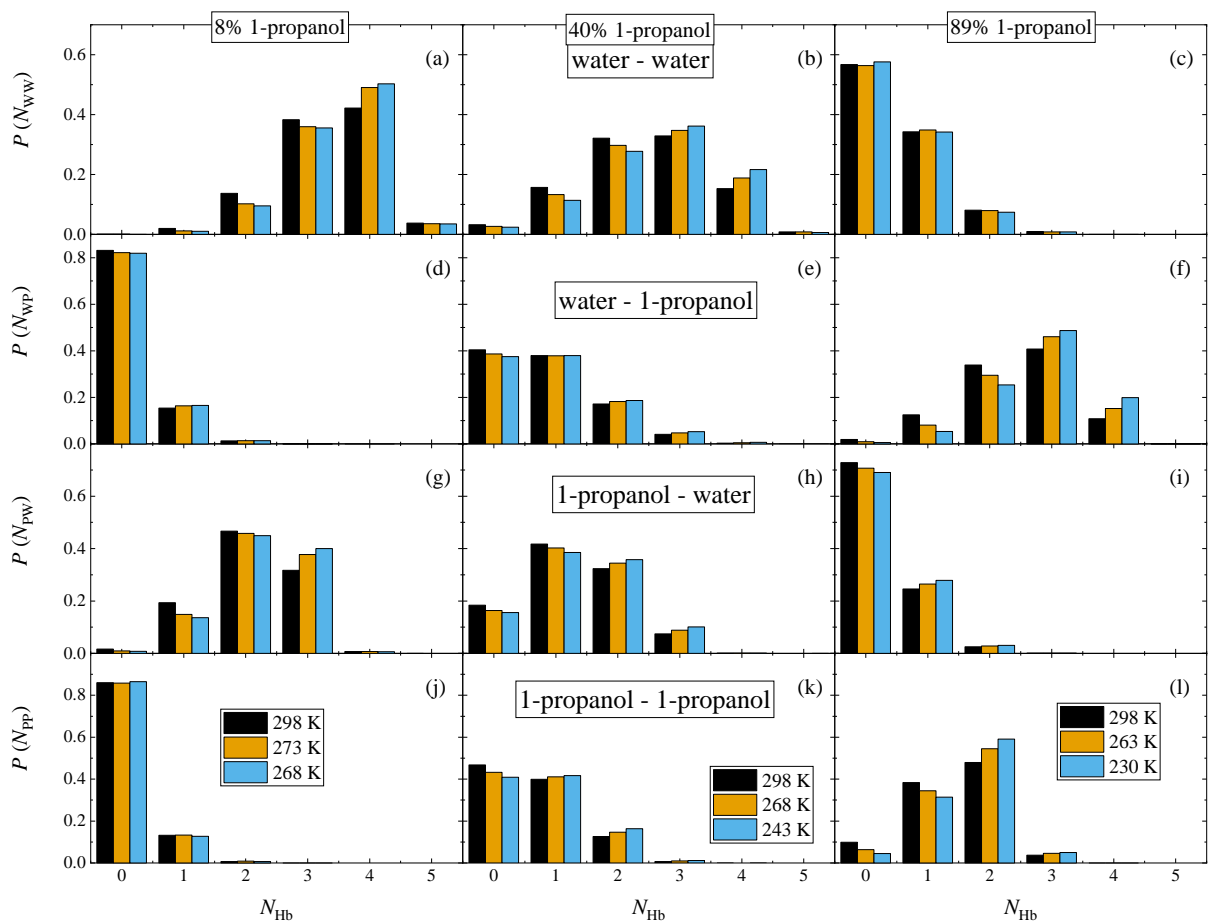

**Figure S68** Distribution of the number of hydrogen bonds at selected temperatures and 1-propanol concentrations (at (a,d,g,i) 8 mol%, (b,e,h,k) 40 mol% and (c,f,l) 89 mol% 1-propanol contents), as obtained by MD simulations using the SPC/E water model: (a,b,c) H-bonded water molecules around water, (d,e,f) H-bonded 1-propanol molecules around water, (g,h,i) H-bonded water molecules around 1-propanol, (j,k,l) H-bonded 1-propanol molecules around 1-propanol.

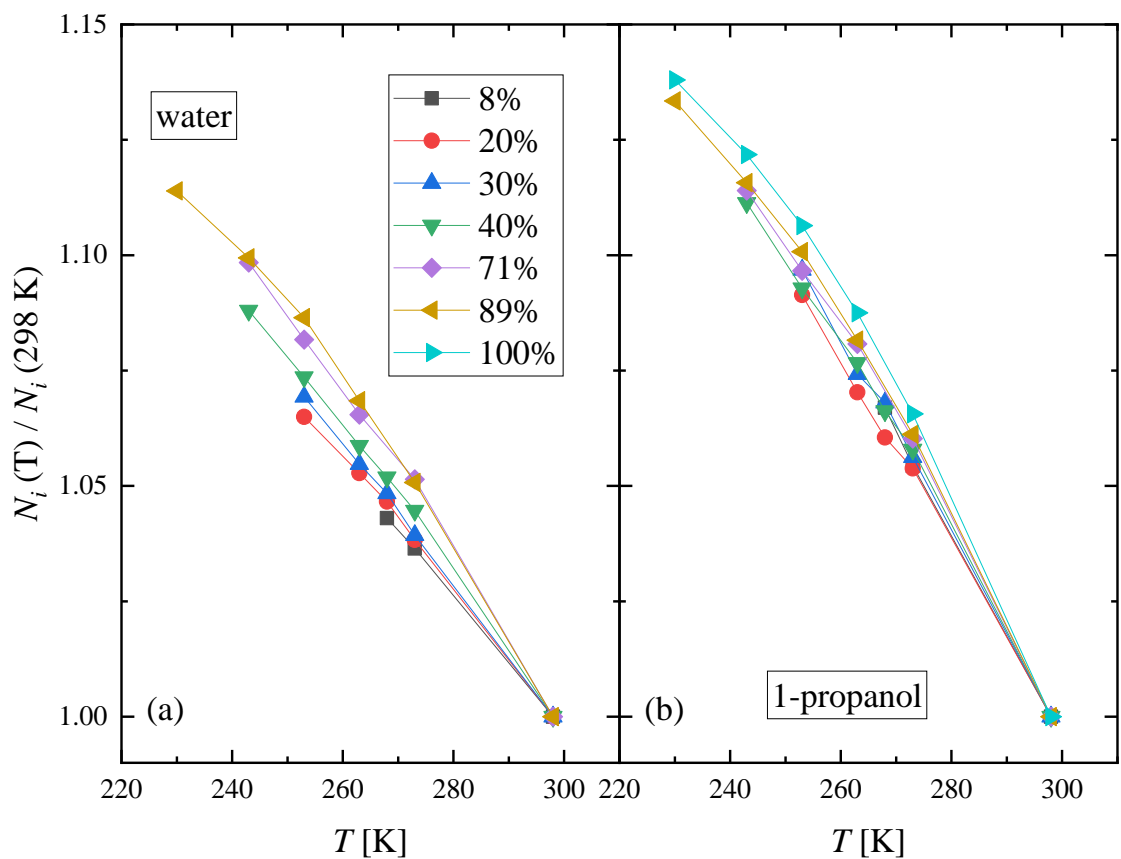

**Figure S69** Temperature dependence of the average number of H-bonded molecules (water and 1-propanol) around (a) water and (b) 1-propanol, normalized to the 298 K value, at different 1-propanol concentrations.

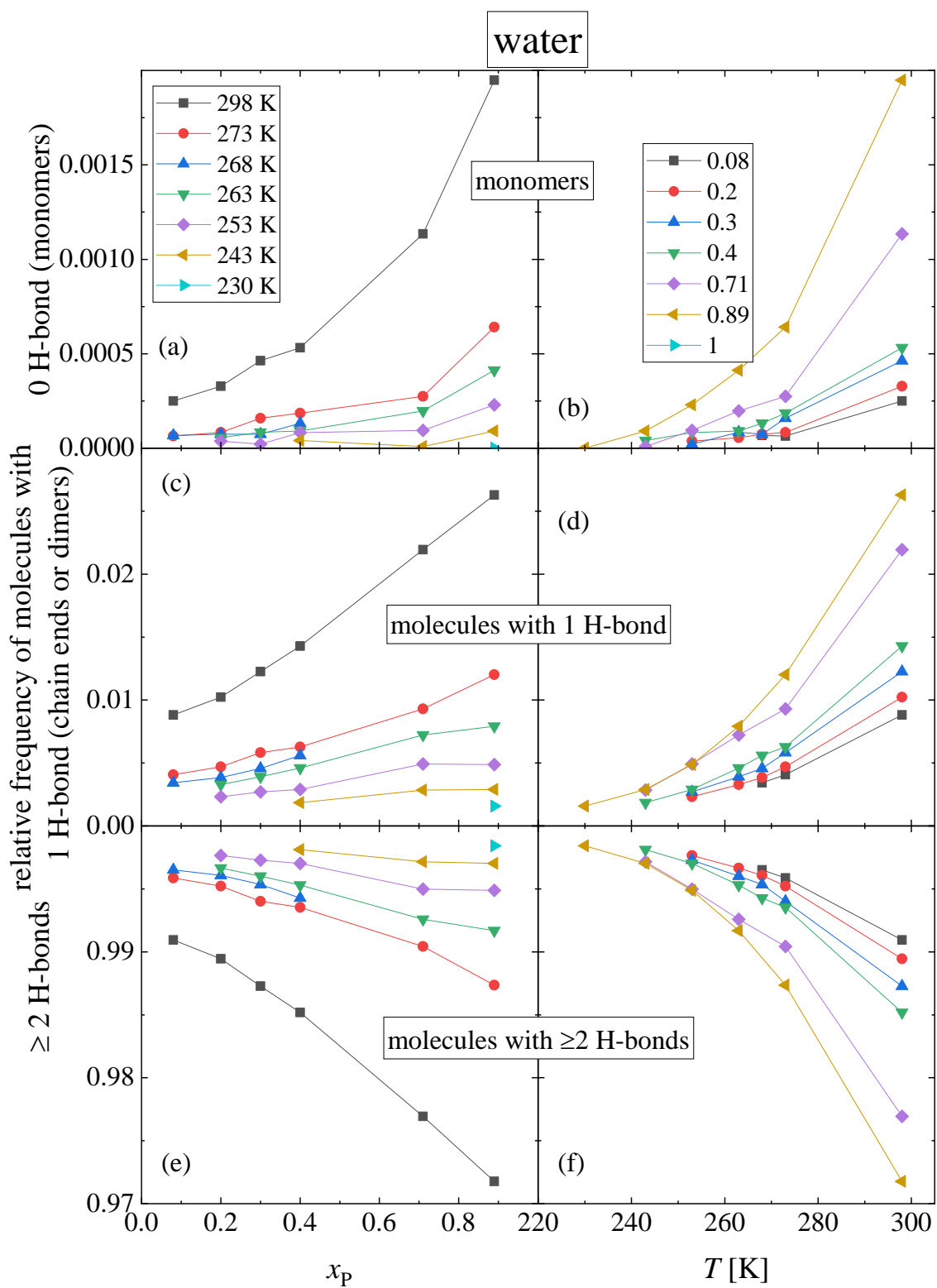

**Figure S70** (a,c,e) Concentration and (b,d,f) temperature dependence of the relative frequencies of water molecules with (a,b) 0, (c,d) 1 or (e,f) ≥ 2 H-bonds. All type of bonds (water – water and water – 1-propanol) are taken into account.

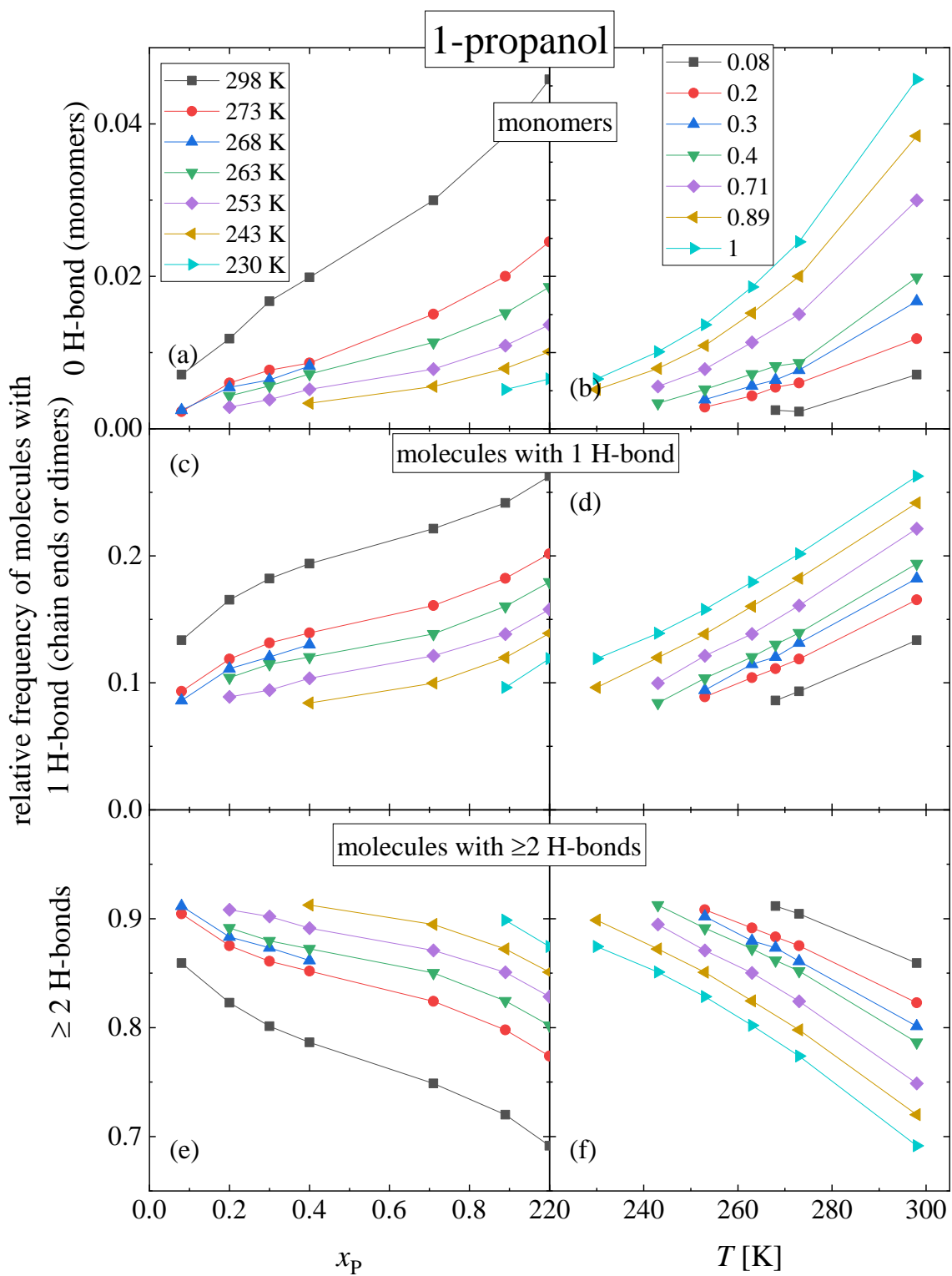

**Figure S71** (a,c,e) Concentration and (b,d,f) temperature dependences of the relative frequencies of 1-propanol molecules with (a,b) 0, (c,d) 1 or (e,f) ≥ 2 H-bonds. All type of bonds (1-propanol – 1-propanol and 1-propanol – water ) are taken into account.

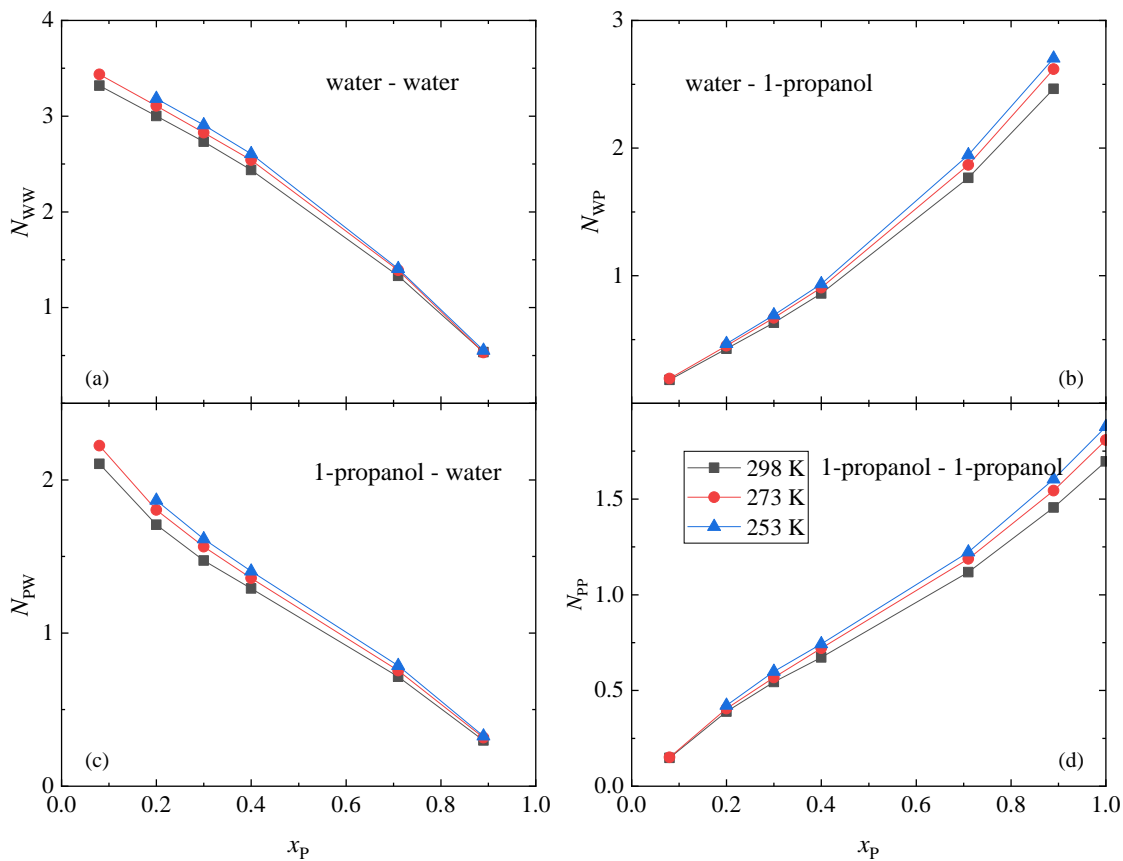

**Figure S72** Concentration dependence of the number of hydrogen bonds at different temperatures, as obtained by MD simulations using the SPC/E water model: (a) average number of H-bonded water molecules around water, (b) average number of H-bonded 1-propanol molecules around water, (c) average number of H-bonded water molecules around 1-propanol, (d) average number of H-bonded m1-propanol molecules around 1-propanol.

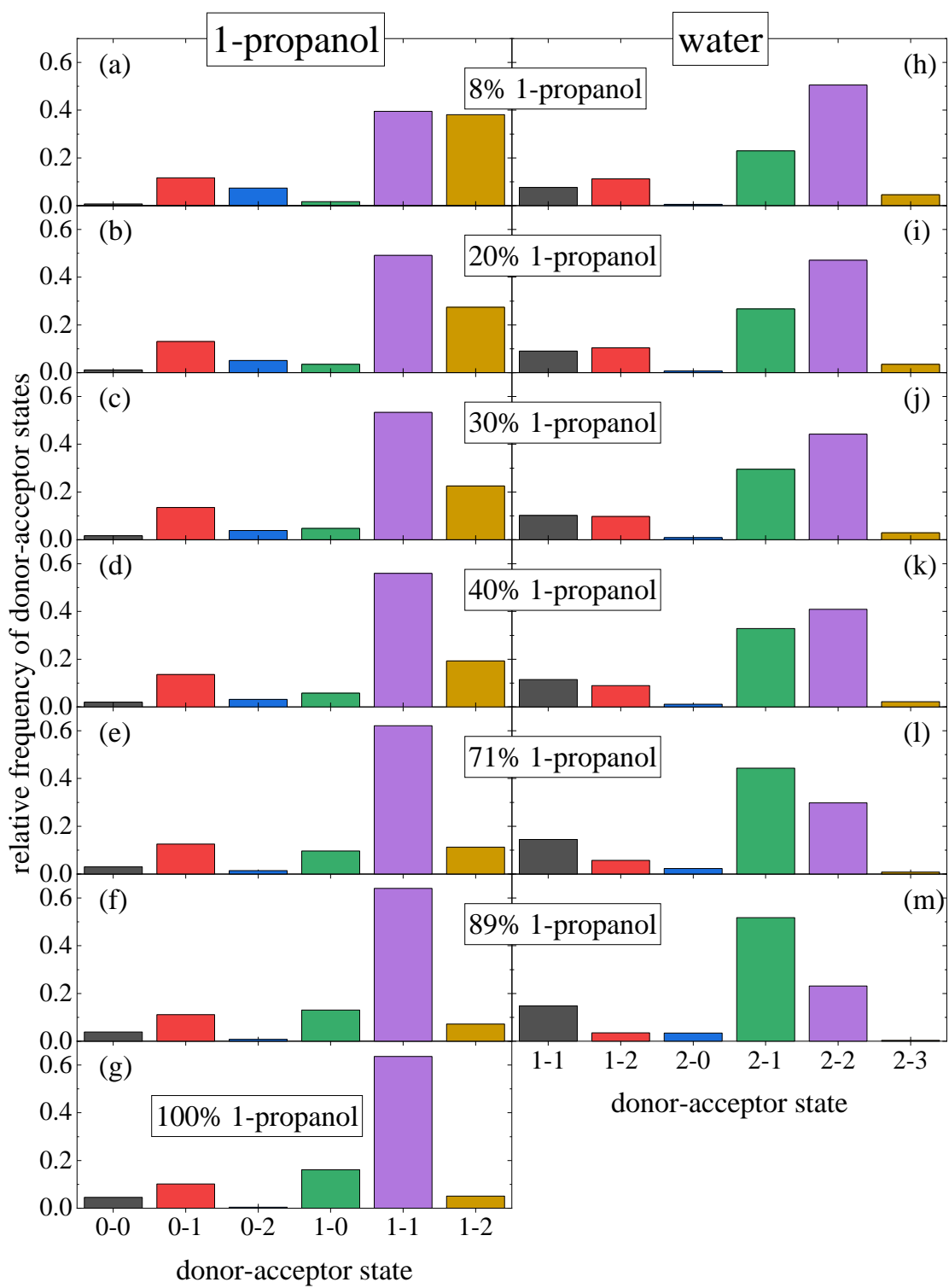

**Figure S73** Relative frequencies of different donor-acceptor states of (a-g) 1-propanol and (h-m) water molecules in 1-propanol – water mixtures at 298 K, as obtained by MD simulations using the SPC/E water model.

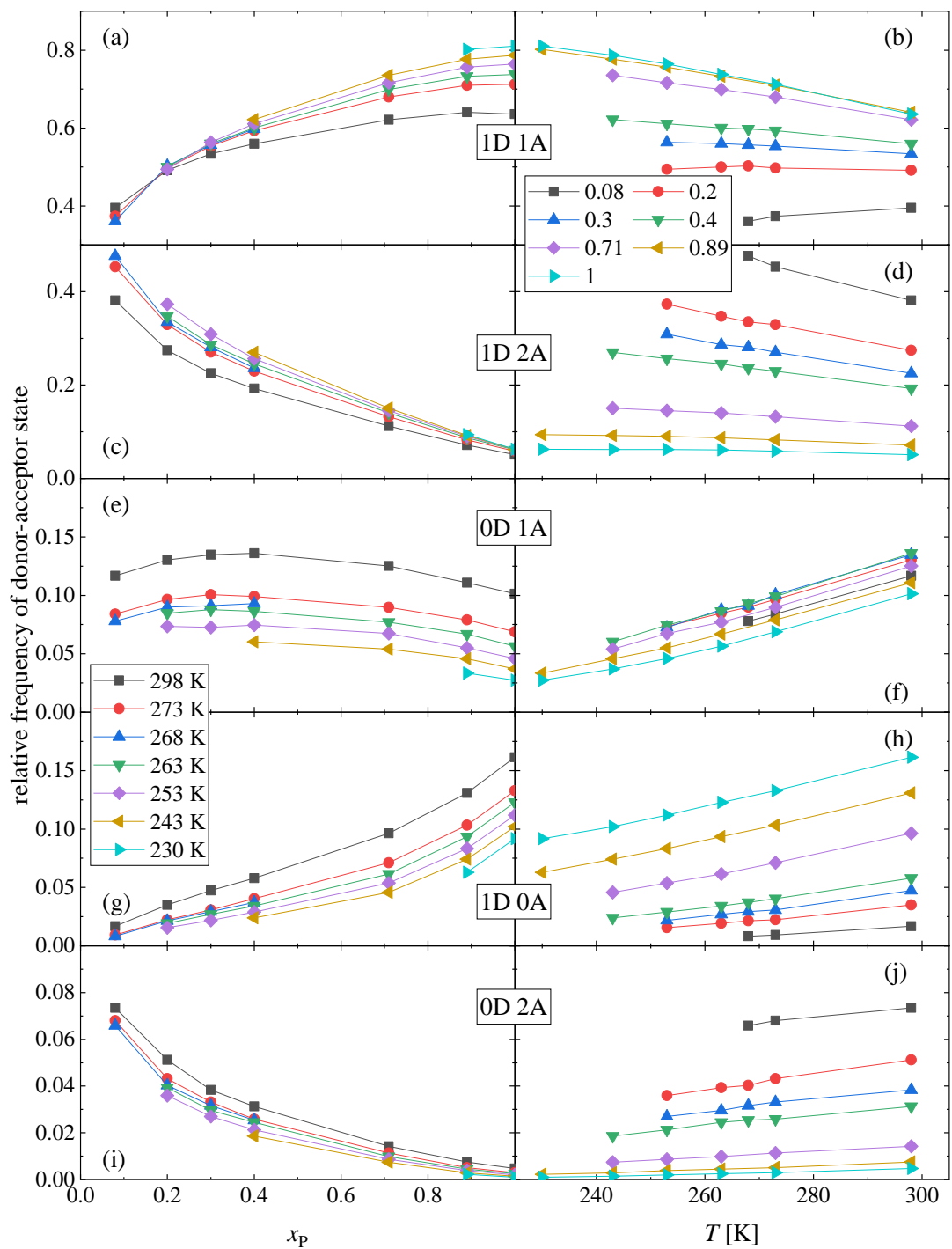

**Figure S74** (left side) Concentration dependence and (right side) temperature dependence of the most frequent donor-acceptor states of 1-propanol molecules: (a,b) 1 donor 1 acceptor state, (c,d) 1 donor 2 acceptor state, (e,f) 0 donor 1 acceptor state, (g,h) 1 donor 0 acceptor state, and (i,j) 0 donor 2 acceptor state.

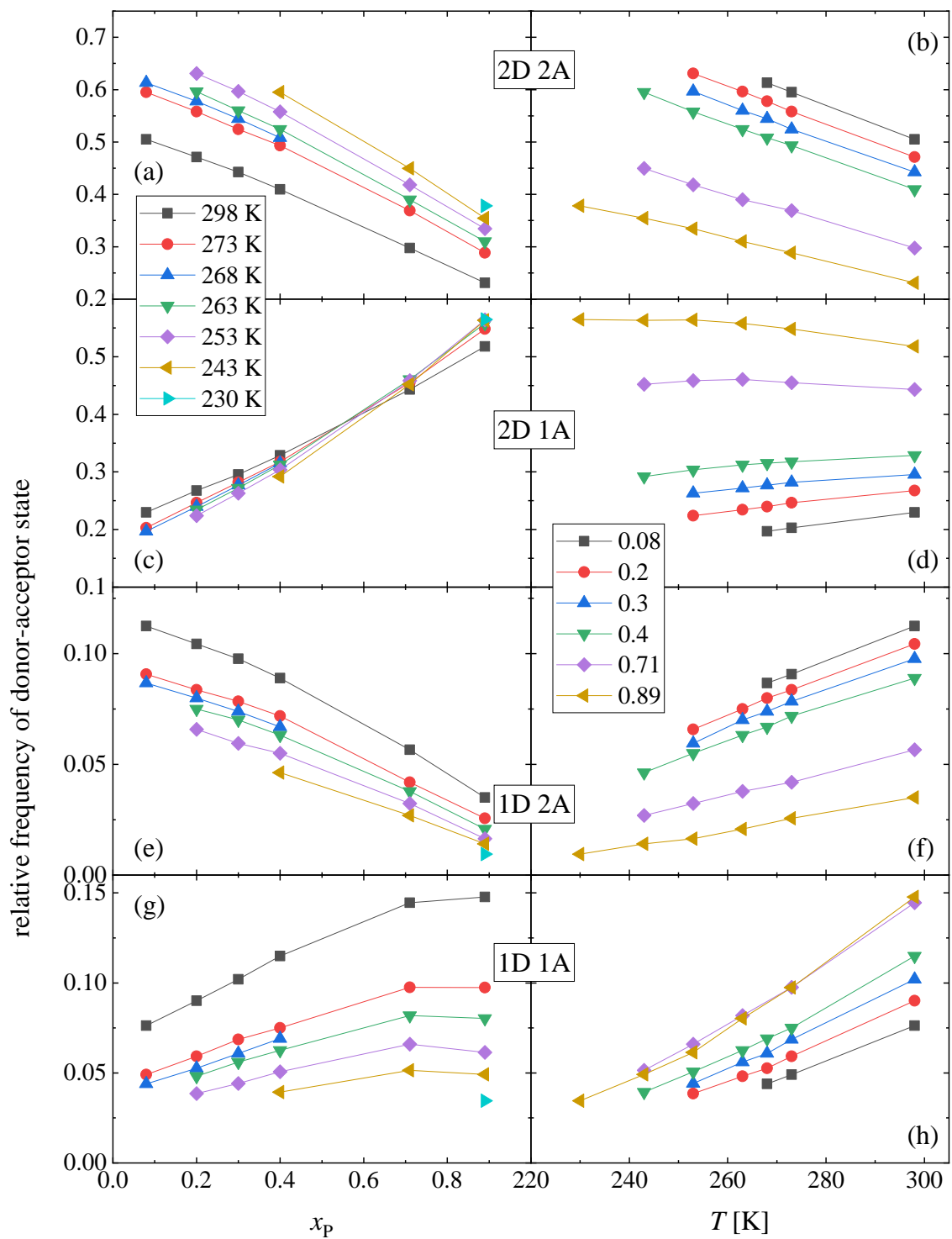

**Figure S75** (left side) Concentration dependence and (right side) temperature dependence of the most frequent donor-acceptor states of 1-propanol molecules: (a,b) 2 donor 2 acceptor state, (c,d) 2 donor 1 acceptor state, (e,f) 1 donor 2 acceptor state, and (g,h) 1 donor 1 acceptor state.

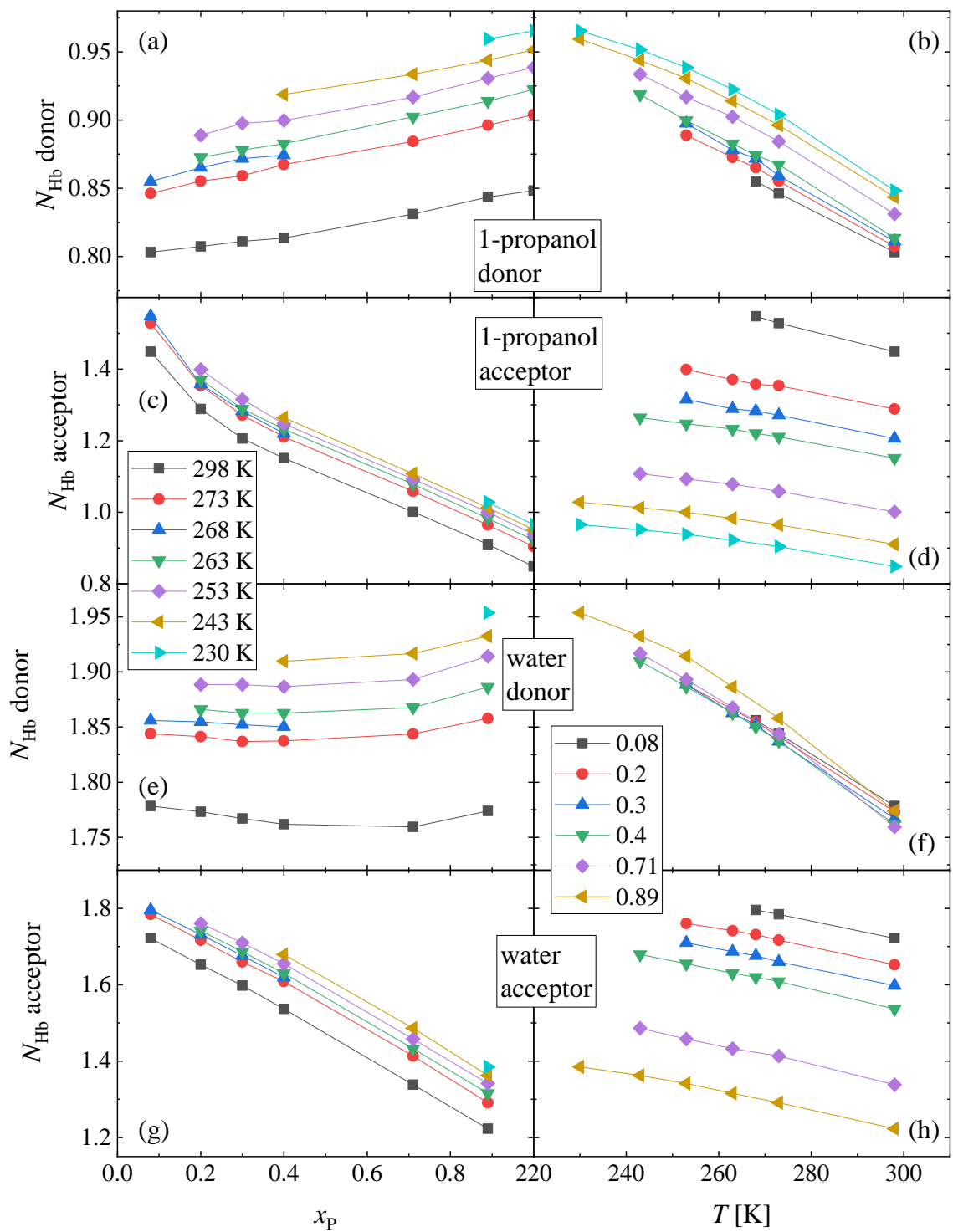

**Figure S76** (left side) Concentration and (right side) temperature dependence of the average number of H-bonds of (a-d) 1-propanol and (e-h) water molecules as (a,b,e,f,) donors and as (c,d,g,h) acceptors.

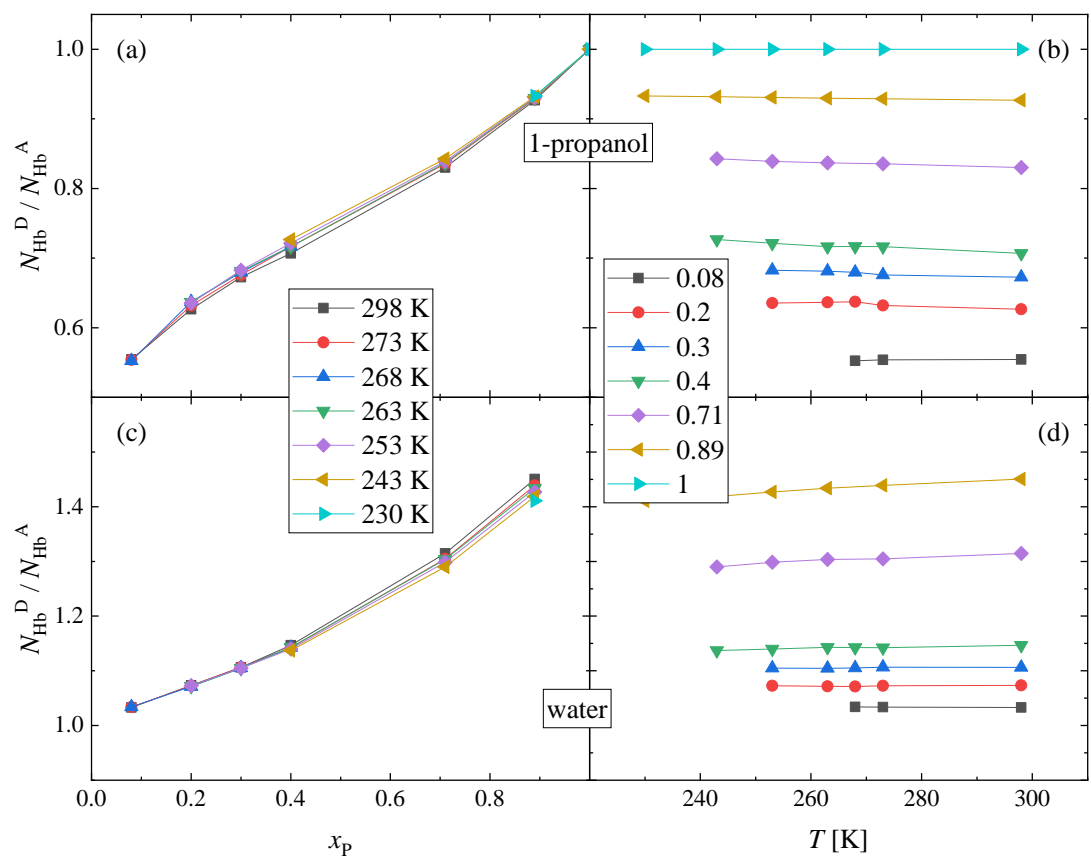

**Figure S77** (left side) Concentration and (right side) temperature dependence of the donor/acceptor ratios of (a,b) 1-propanol and (c,d) water molecules.

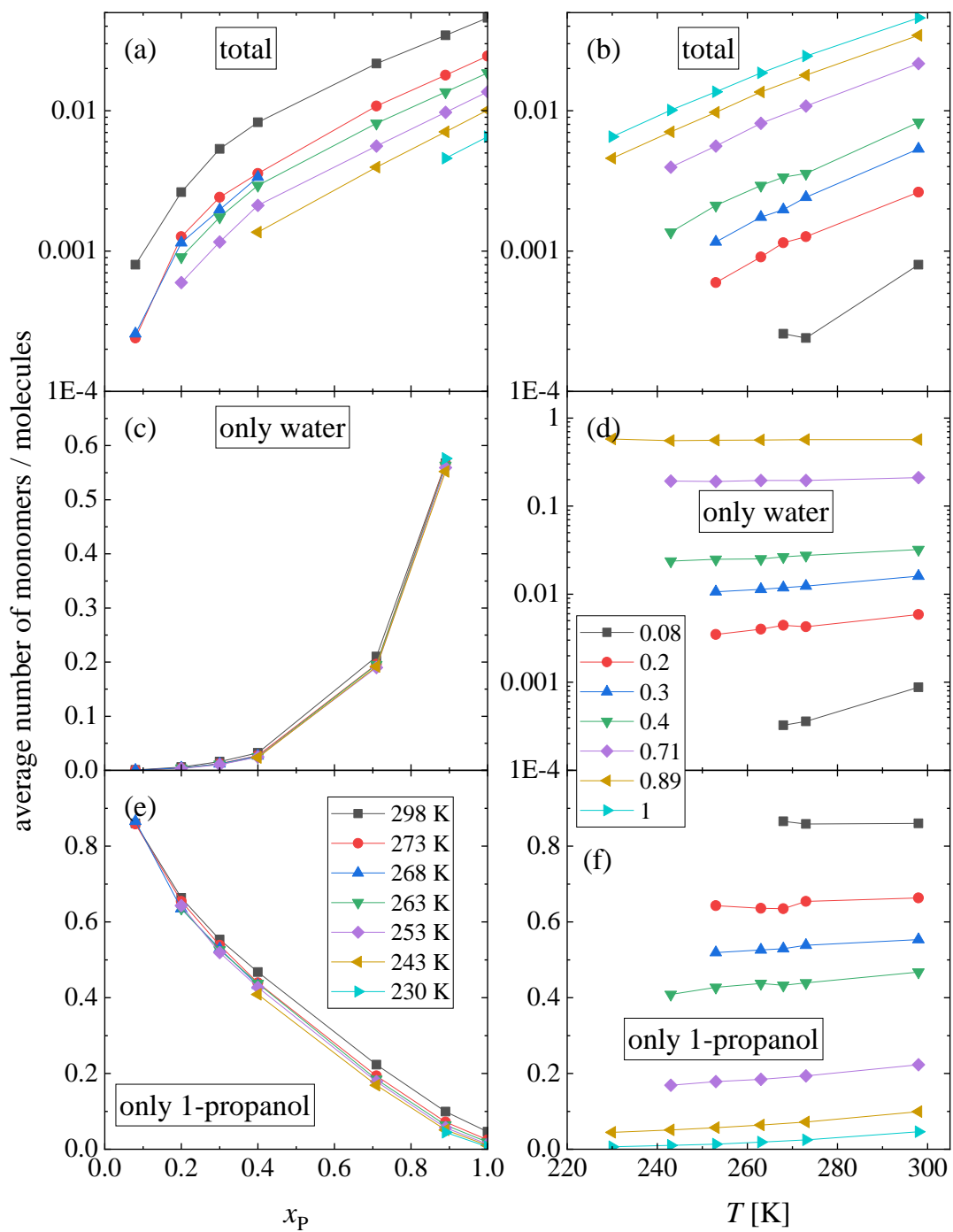

**Figure S78** (left side, a,c,e) Concentration and (right side, b,d,f) temperature dependence of the average number of monomer molecules considering (a,b) all molecules, (c,d) water subsystem, (e,f) 1-propanol subsystem. The number of monomers are normalized with the number of the corresponding molecules in the simulation box. (Note the $y$-scale of (c, e, f) is linear, while the others are logarithmic!)

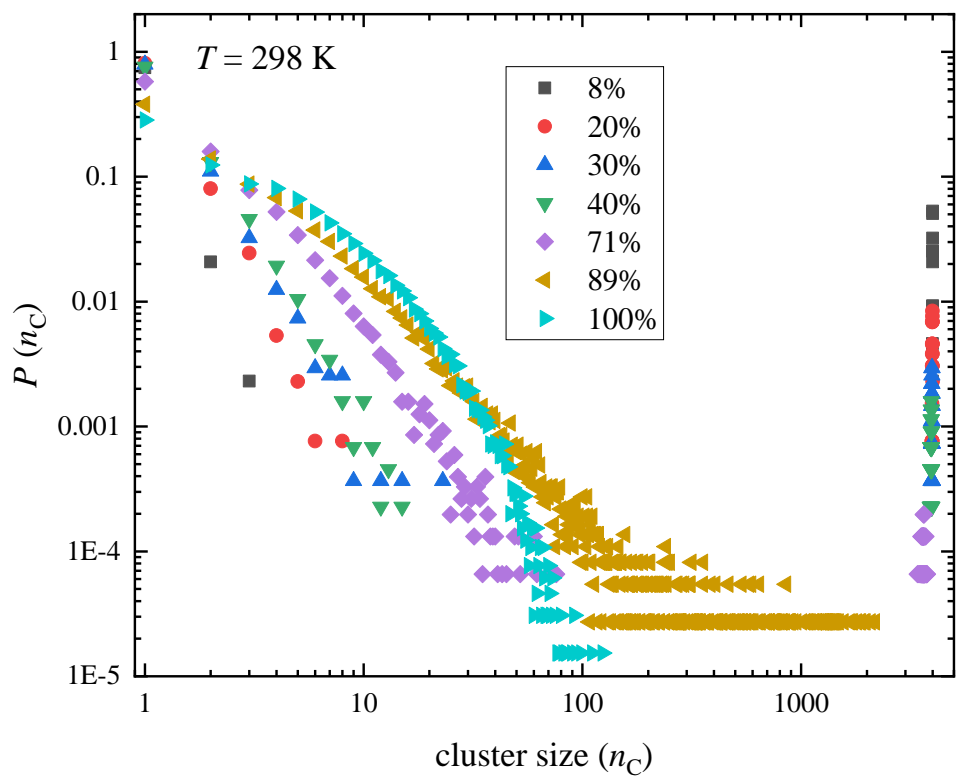

**Figure S79** Cluster size distributions of 1-propanol – water mixtures at $T = 298$ K, considering H-bonds between any types of molecules.

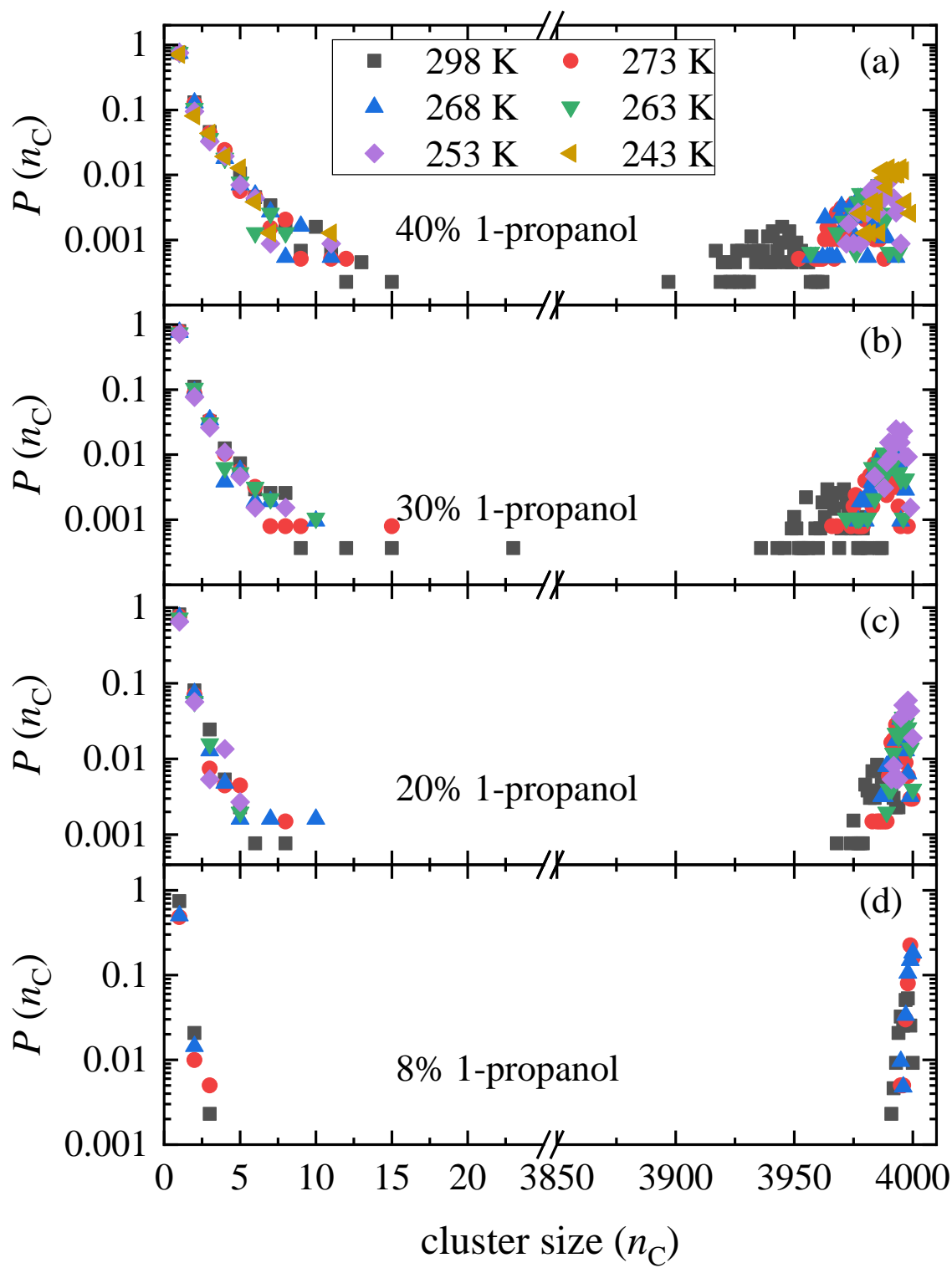

**Figure S80** Cluster size distributions of water-rich 1-propanol – water mixtures at different temperatures, considering H-bonds between any types of molecules.

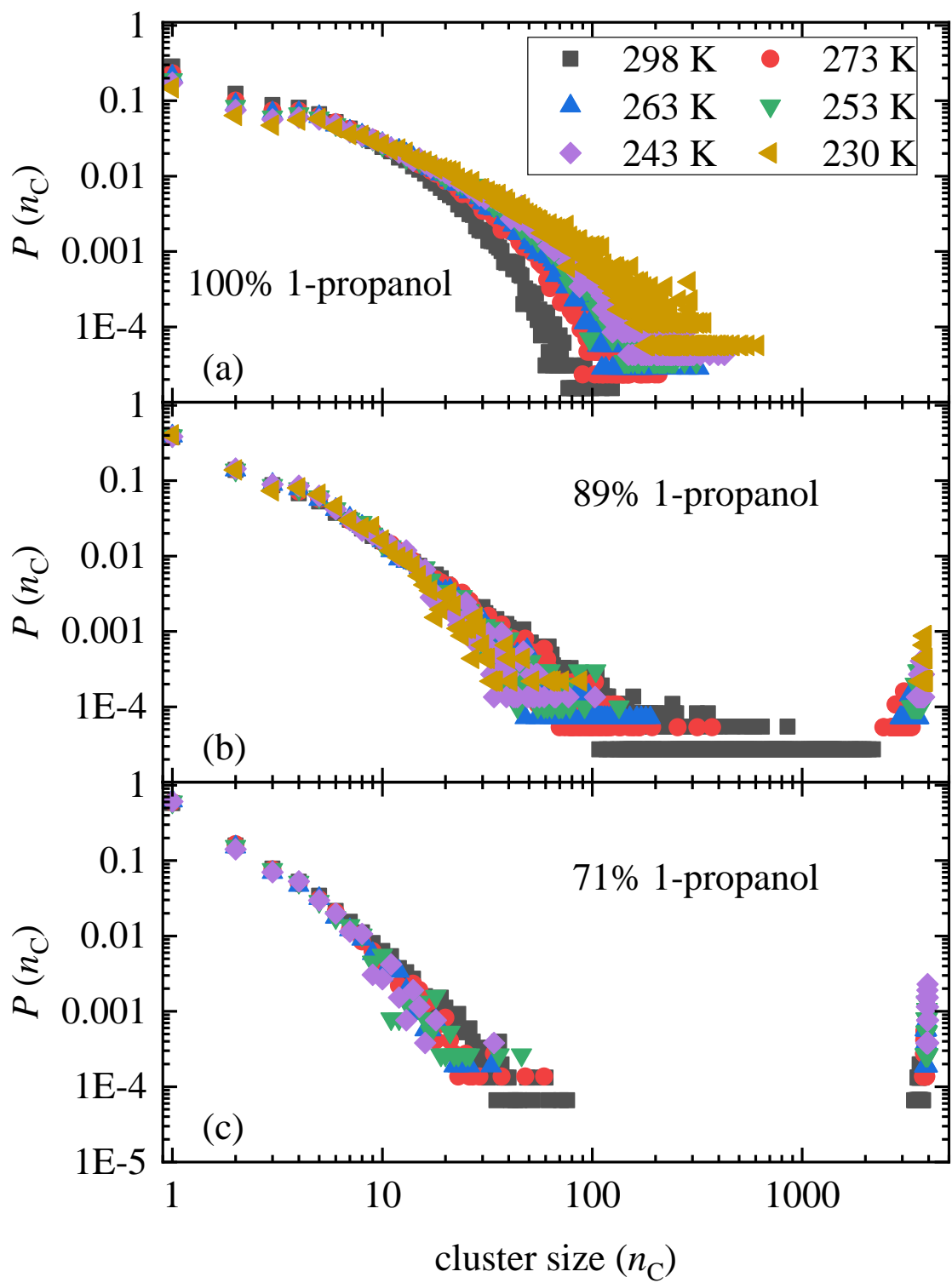

**Figure S81** Cluster size distributions of alcohol-rich 1-propanol – water mixtures at different temperatures, considering H-bonds between any types of molecules.

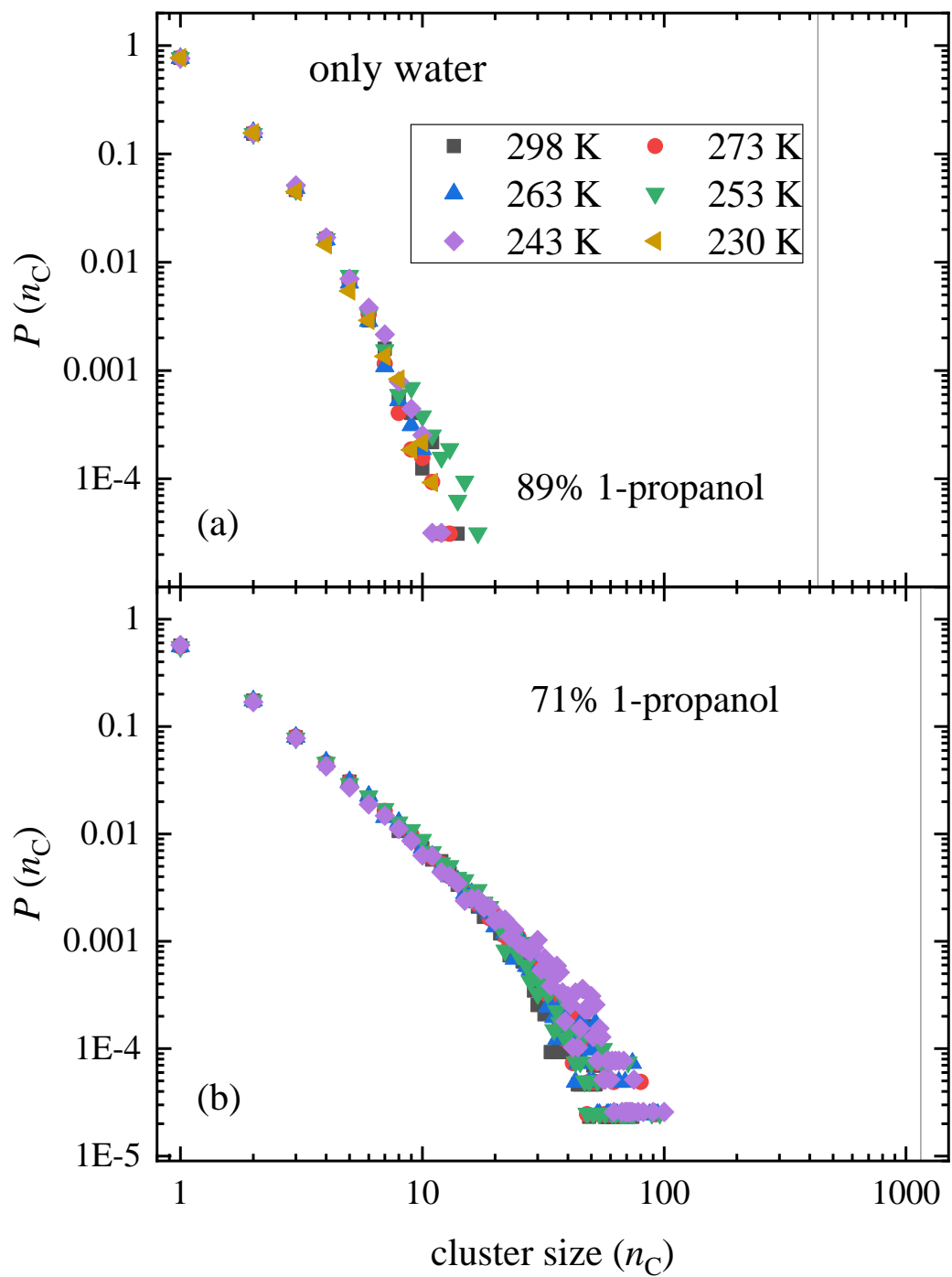

**Figure S82** Cluster size distributions within the water subsystem in 1-propanol – rich 1-propanol – water mixtures at different temperatures (H-bonds only between water molecules are considered.) The vertical lines show the numbers of water molecules in the simulation boxes.

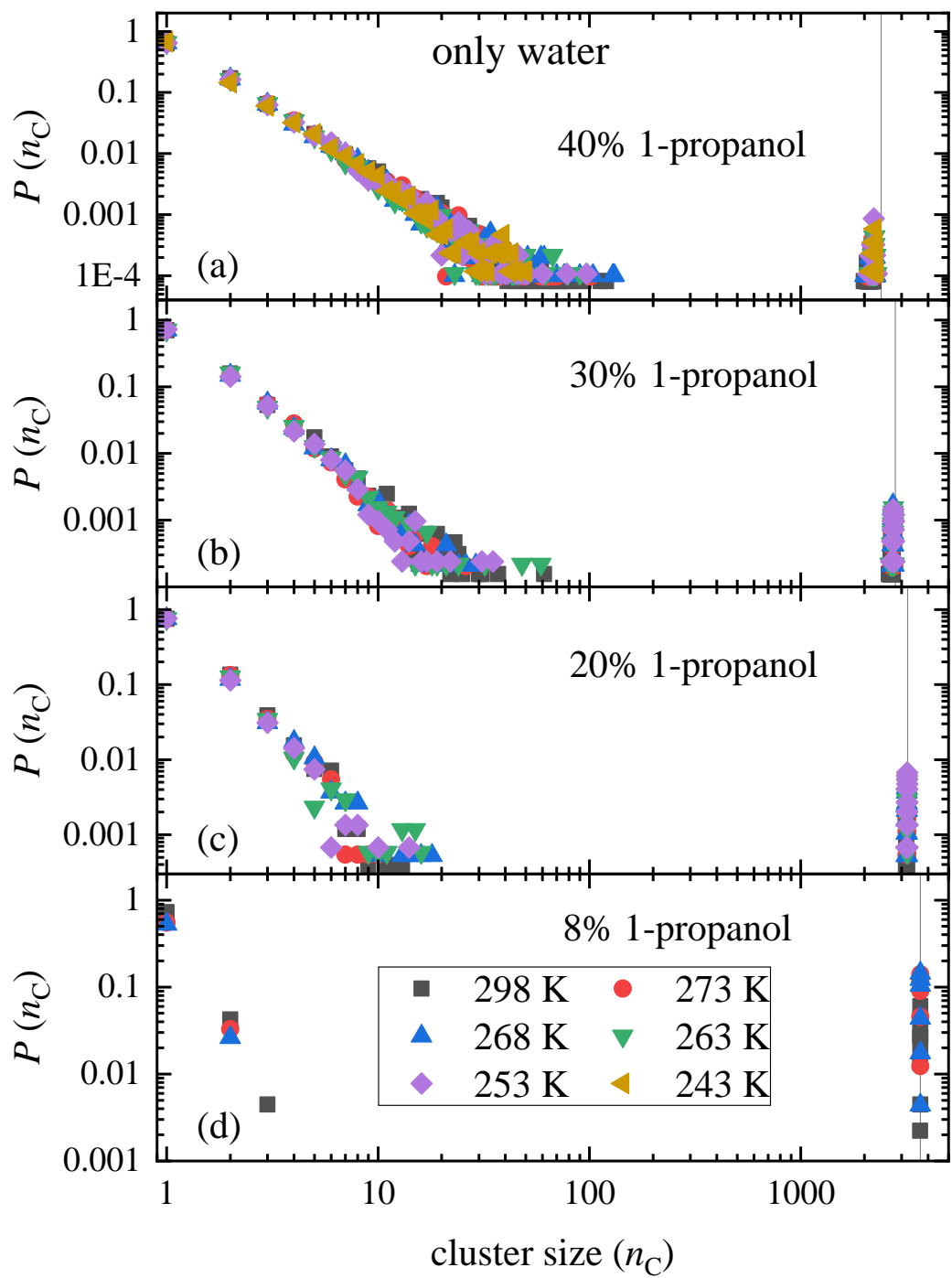

**Figure S83** Cluster size distributions within the water subsystem in water-rich 1-propanol – water mixtures at different temperatures (H-bonds only between water molecules are considered.) The vertical lines show the numbers of water molecules in the simulation boxes.

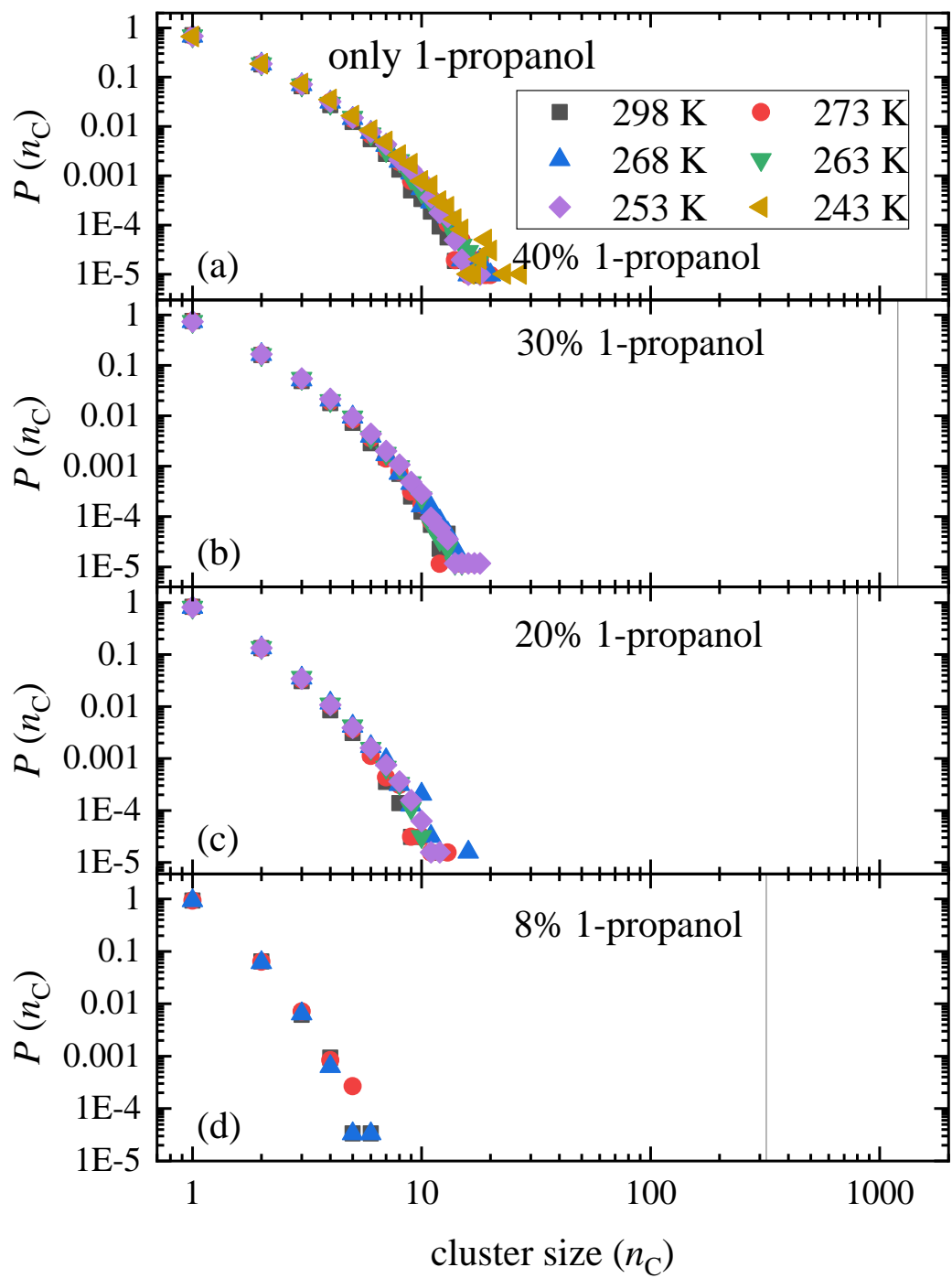

**Figure S84** Cluster size distributions within the 1-propanol subsystem in water-rich 1-propanol – water mixtures at different temperatures (H-bonds only between 1-propanol molecules are considered.) The vertical lines show the numbers of 1-propanol molecules in the simulation boxes.

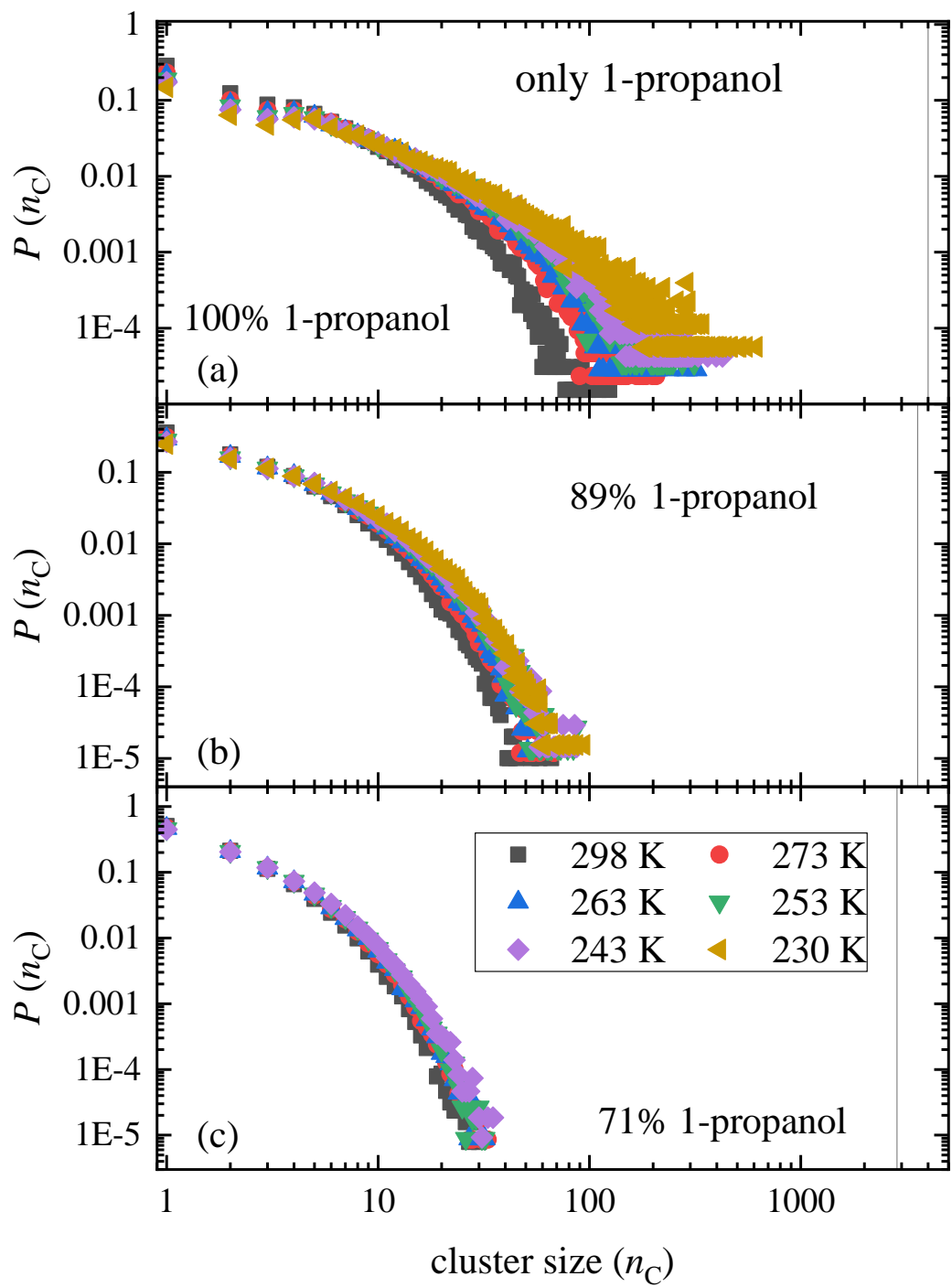

**Figure S85** Cluster size distributions within the 1-propanol subsystem in 1-propanol - rich 1-propanol – water mixtures at different temperatures (H-bonds only between 1-propanol molecules are considered.) The vertical lines show the numbers of 1-propanol molecules in the simulation boxes.

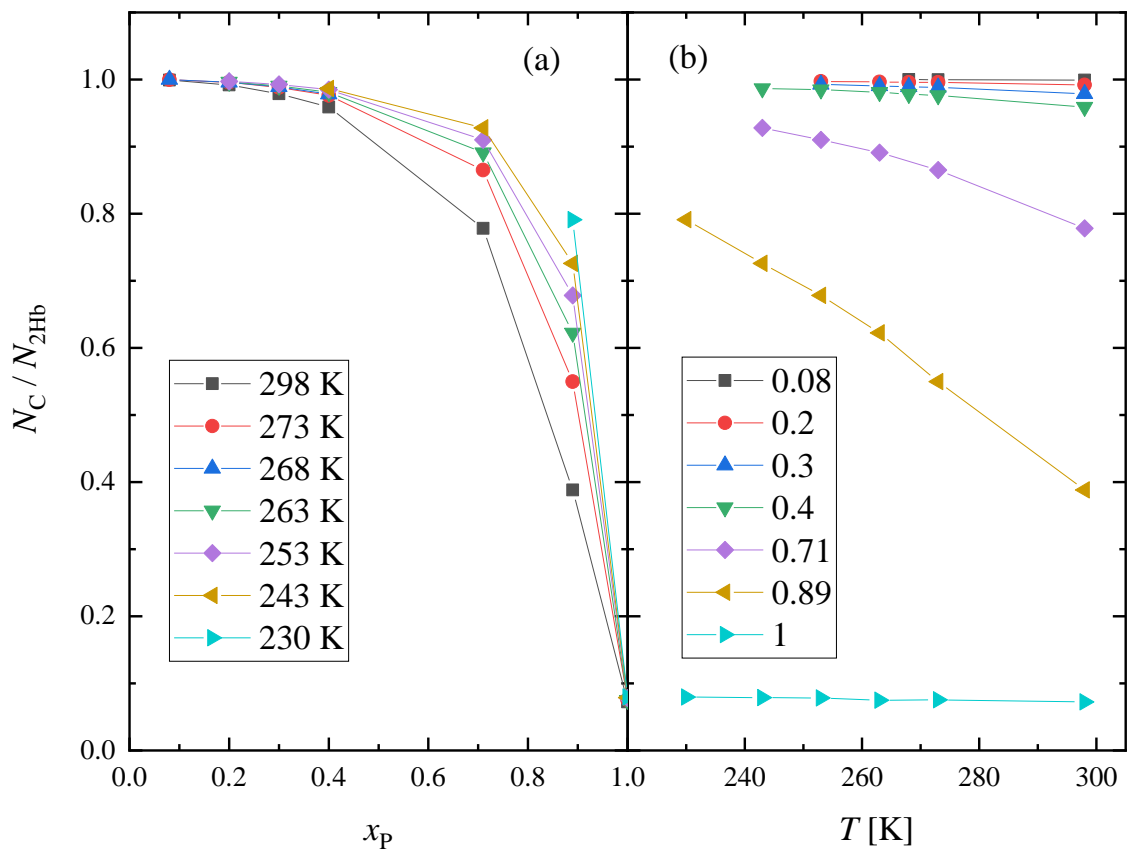

**Figure S86** (a) Concentration and (b) temperature dependence of the ratio of molecules participating in cycles ($N_C$) and molecules with at least 2 H-bonds ($N_{2Hb}$). (All molecules and all bond types are taken into account.)

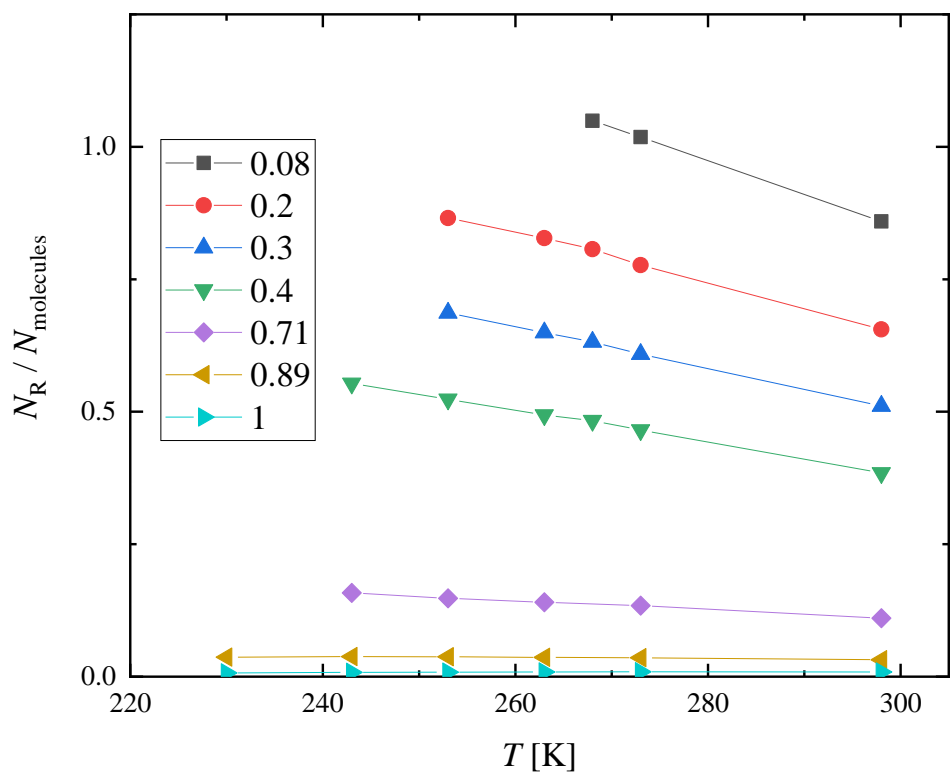

**Figure S87** Average number of primitive rings per molecule in 1-propanol – water mixtures as a function of temperature at different concentrations, as obtained by MD simulations using the SPC/E water model. Rings are calculated up to 8 membered ones.

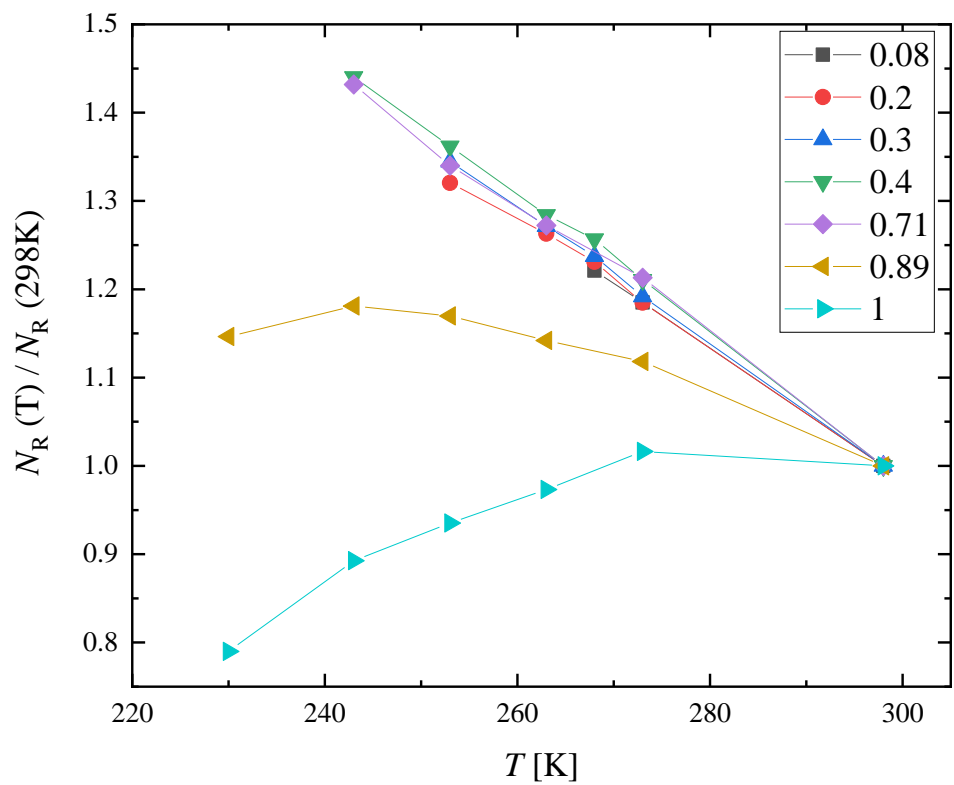

**Figure S88** Temperature dependence of the average number of rings normalized by the 298 K value.

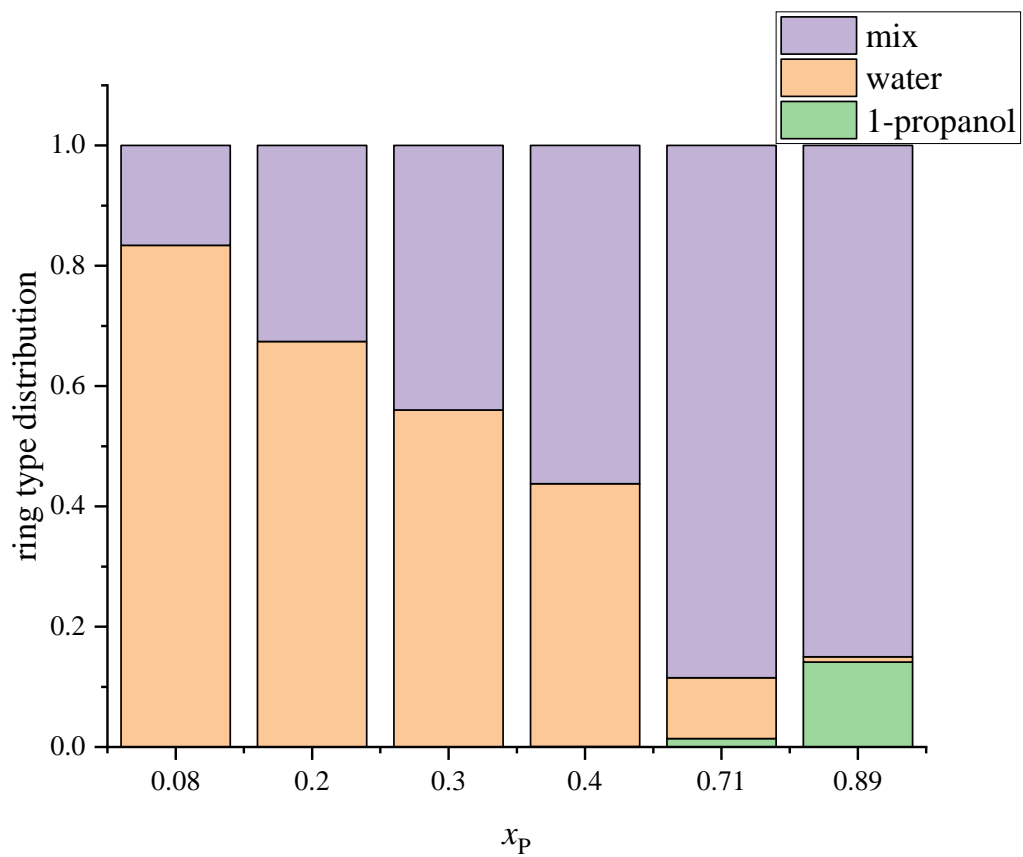

**Figure S89** Concentration dependence of the distribution of different ring types (rings contains only water molecules, only 1-propanol molecules, or both water and 1-propanol molecules) at 298 K.

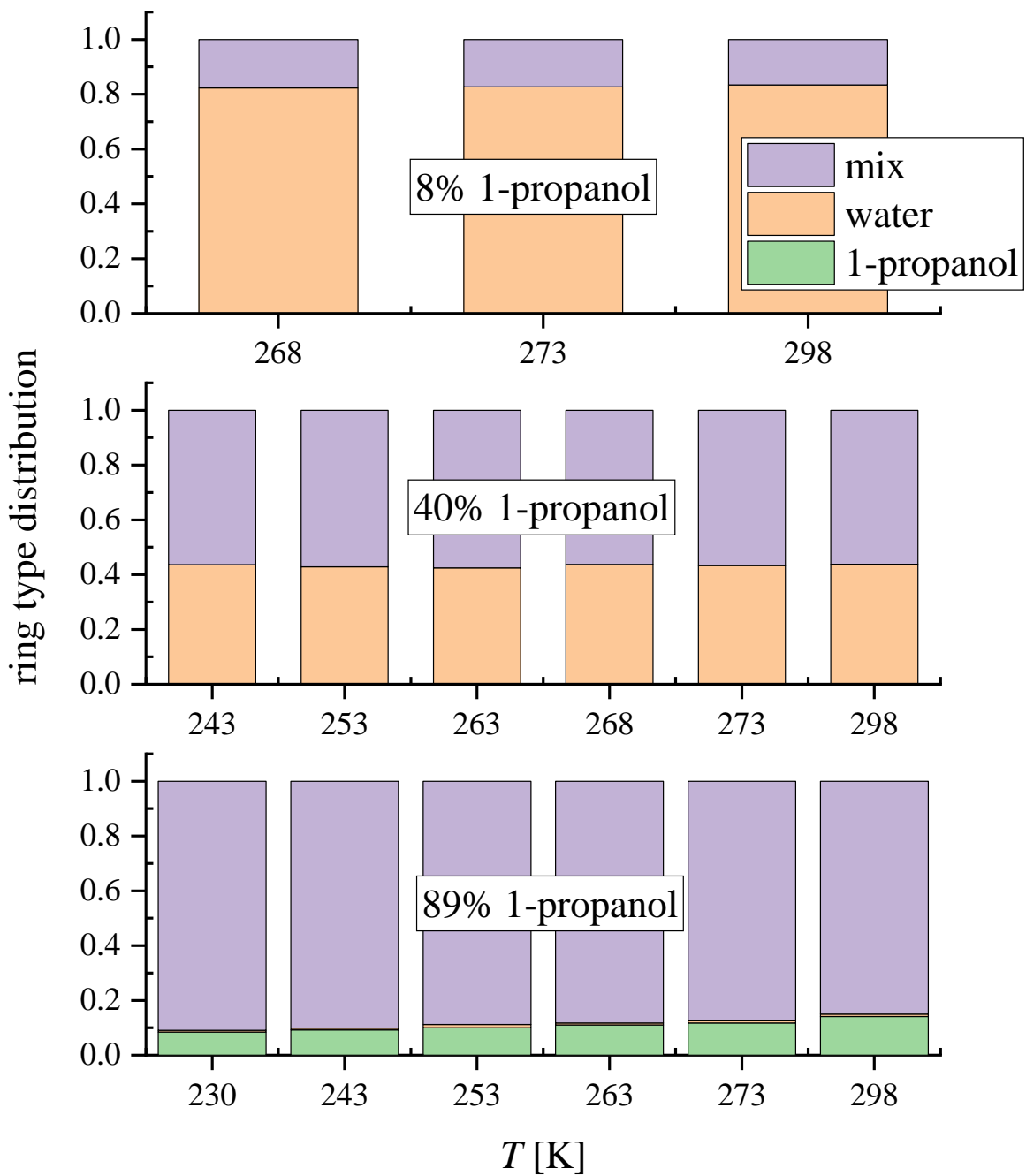

**Figure S90** Temperature dependence of the distribution of different ring types at three selected 1-propanol concentrations.

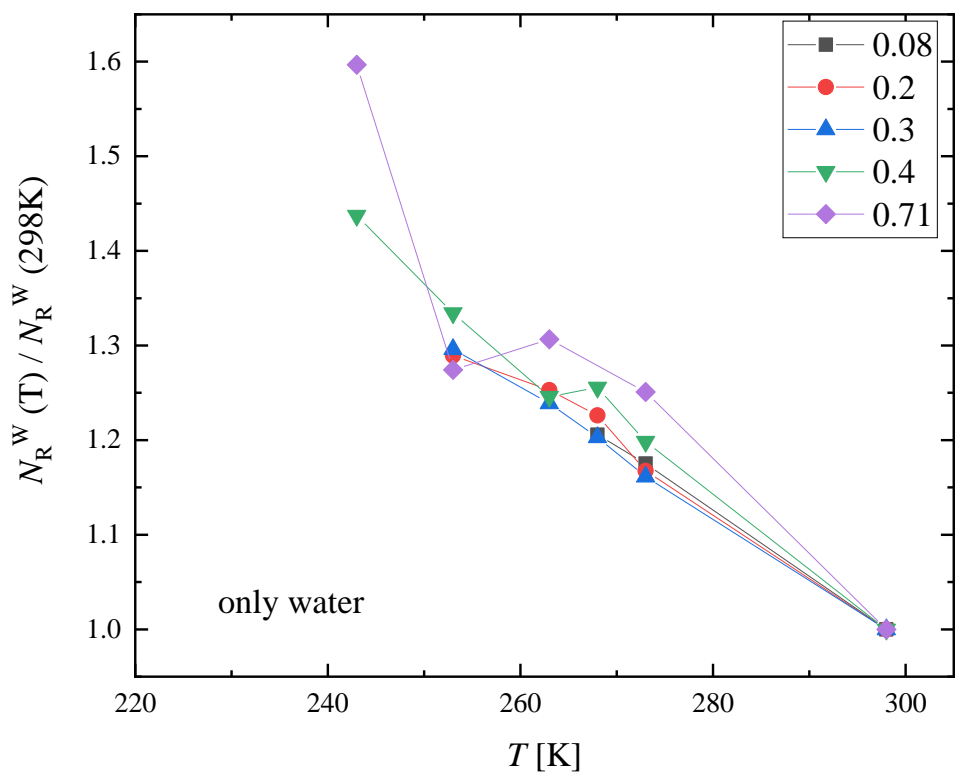

**Figure S91** Temperature dependence of the average number of water rings, normalized by the value at 298 K, at different 1-propanol concentrations. (In the $x_P = 0.89$ mixture the number of water rings is very low and the ratio has high uncertainty, so it is not shown here.)

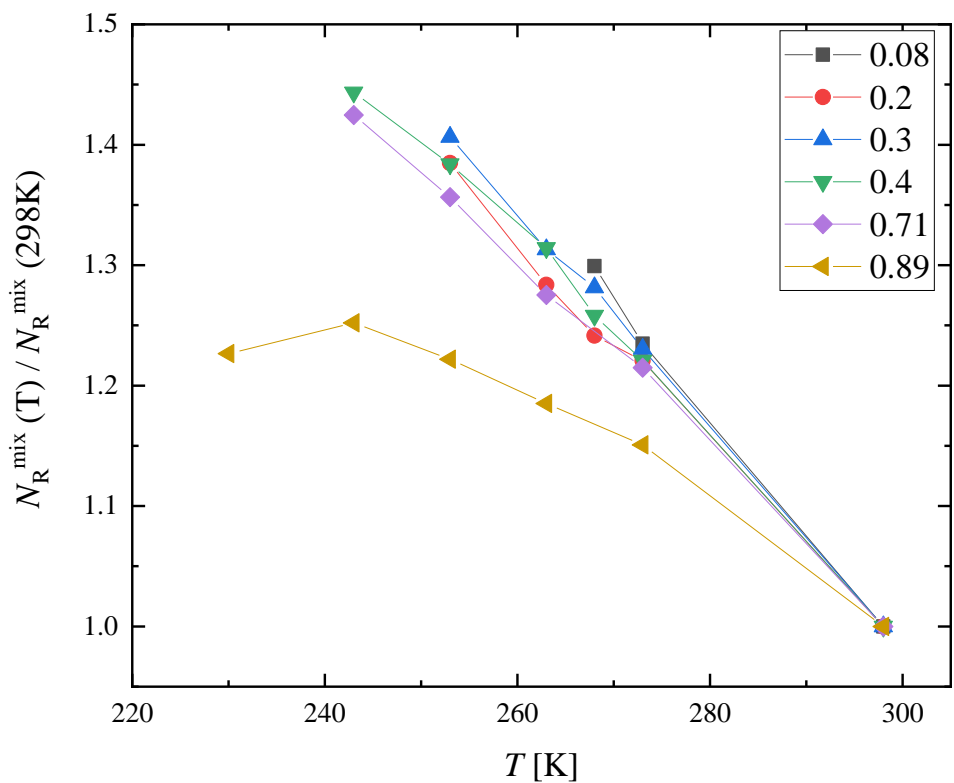

**Figure S92** Temperature dependence of the average number of rings containing both 1-propanol and water molecules, normalized by the 298 K value.

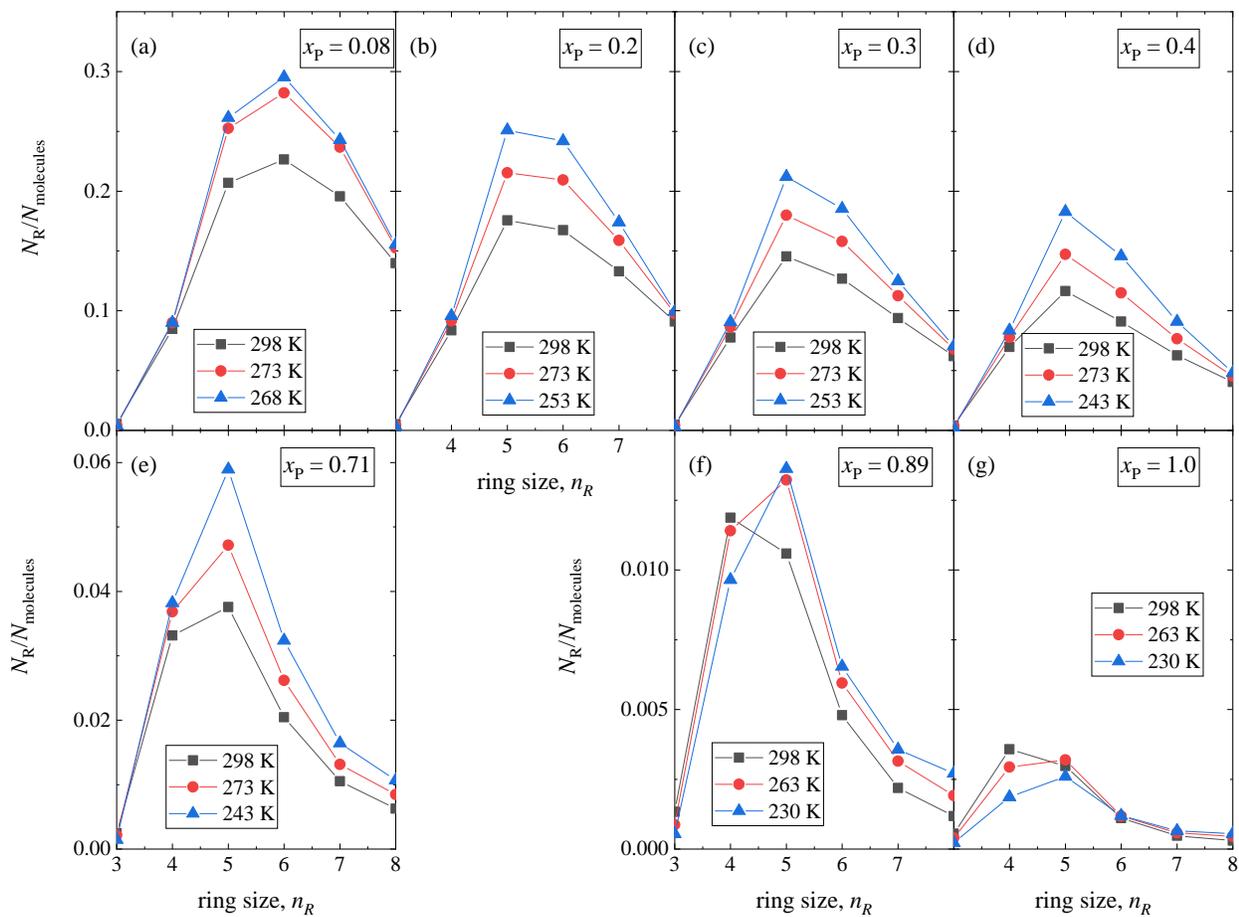

**Figure S93** Ring size distributions normalized by the number of molecules, as a function of temperature at different 1-propanol concentrations.

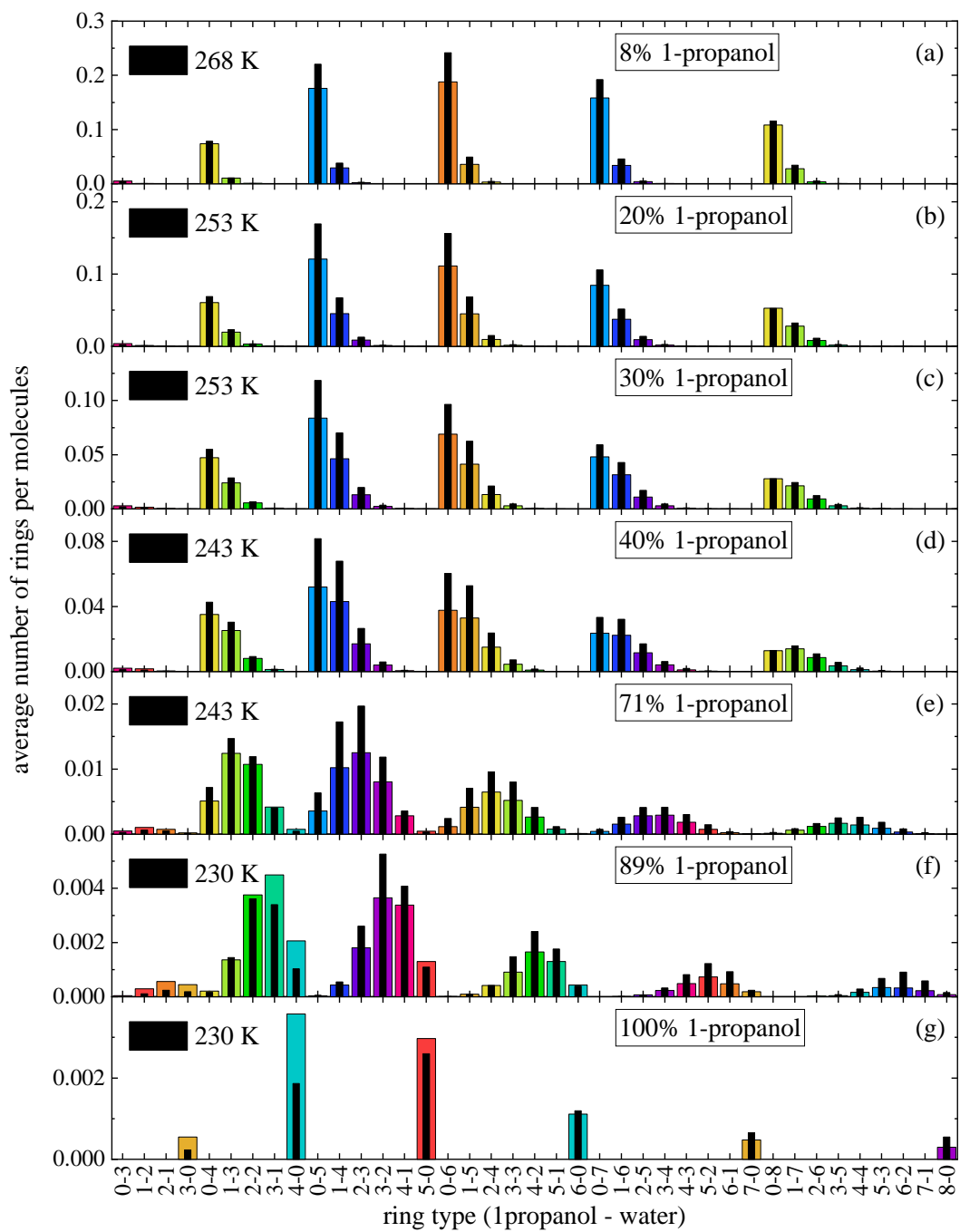

**Figure S94** Temperature and concentration dependence of the ring type distributions in 1-propanol – water mixtures. The number of rings is normalized by the number of molecules in the simulation box. Colored bars show values at 298 K, while black bars are the values at the lowest investigated temperature: (a) 268 K, (b,c) 253 K, (d,e) 243 K, (f, g) 230 K.

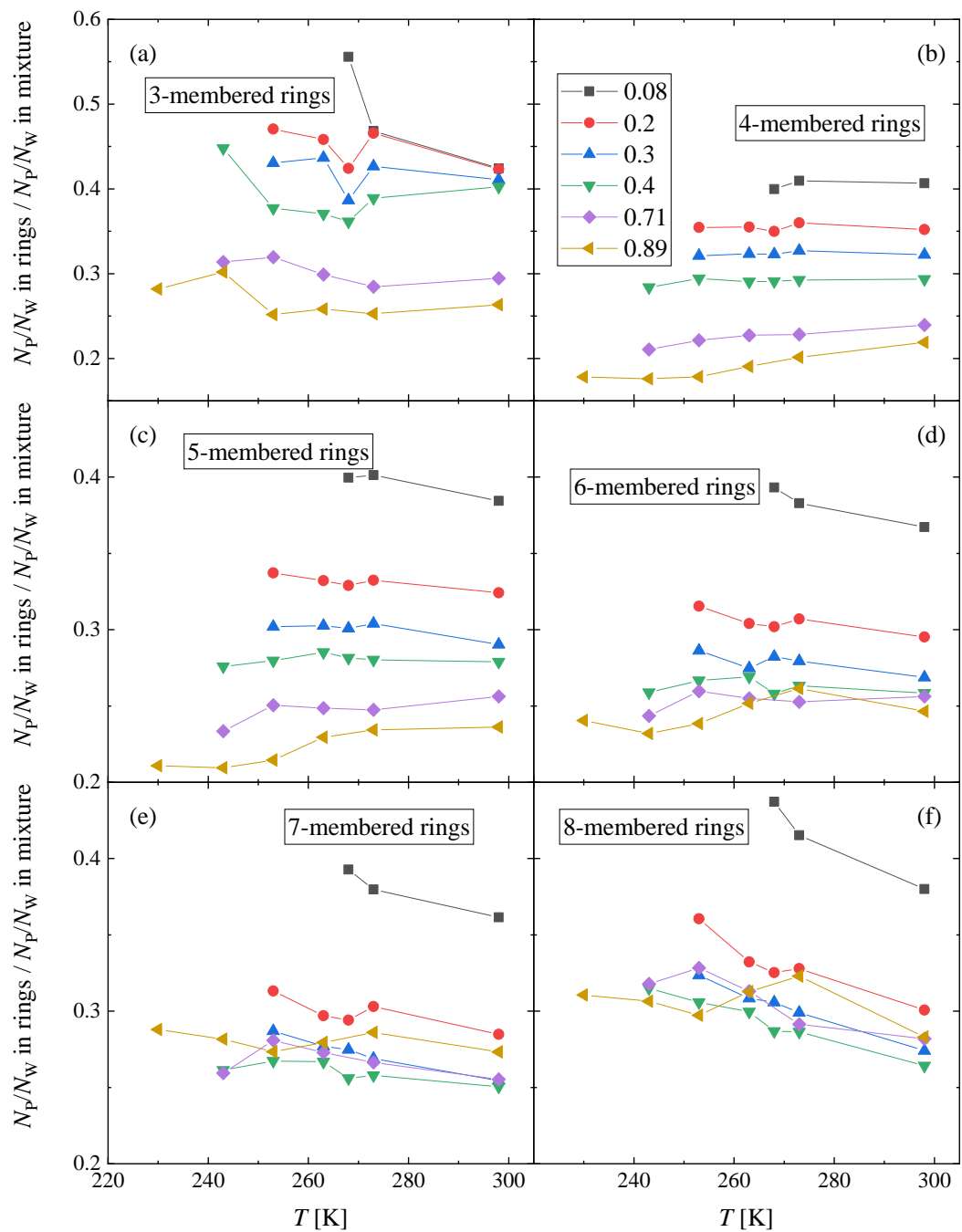

**Figure S95** Temperature and concentration dependence of the participation of 1-propanol and water molecules in different rings: (a) 3-membered, (b) 4-membered, (c) 5-membered, (d) 6-membered, (e) 7-membered and (f) 8-membered rings. The ratio of 1-propanol and water molecules in the rings is compared to the ratio of 1-propanol and water molecules in the mixture.